\pdfoutput=1
\RequirePackage{ifpdf}
\ifpdf 
\documentclass[pdftex]{sigma}
\else
\documentclass{sigma}
\fi

\usepackage{mathrsfs}

\usepackage{array}
\usepackage[mathscr]{euscript}

\usepackage{slashed}
\usepackage{youngtab}

\numberwithin{equation}{section}
\numberwithin{table}{section}
\numberwithin{figure}{section}

\newtheorem{Theorem}{Theorem}[section]
\newtheorem*{Theorem*}{Theorem}

 { \theoremstyle{definition}

\newtheorem{Example}[Theorem]{Example}
\newtheorem{Remark}[Theorem]{Remark} }

\newcommand{\half} {\frac 12}
\newcommand{\pa} {\partial}

\newcommand {\la} {\left \langle}
\newcommand {\ra} {\right \rangle}
\newcommand {\lb} {\left (}
\newcommand {\rb} {\right )}

\newcommand{\stimes} {{|}\kern -.05in\times}

\newcommand {\y} {\mathscr Y}

\newcommand {\ac} {\mathfrak{a}}
\newcommand {\fb} {\mathfrak{b}}
\newcommand {\fc} {\mathfrak{c}}
\newcommand {\fC} {\mathfrak{C}}
\newcommand {\qe} {\mathfrak{q}}
\newcommand {\tqe}{\tilde{\mathfrak{q}}}
\newcommand {\fe} {\mathfrak{f}}
\newcommand {\fu} {\mathfrak{u}}
\newcommand {\ma} {\mathfrak{m}}
\newcommand {\zb} {{\bar z}}
\newcommand {\xb} {{\bar x}}

\newcommand {\CalA} {\mathcal A}
\newcommand {\CalB} {\mathcal B}
\newcommand {\CalC} {\mathcal C}
\newcommand {\obs} {\mathscr C}

\newcommand {\CalE} {\mathcal E}
\newcommand {\CalF} {\mathcal F}
\newcommand {\CalG} {\mathcal G}
\newcommand {\CalH} {\mathcal H}
\newcommand {\CalI} {\mathcal I}
\newcommand {\CalJ} {\mathcal J}
\newcommand {\CalK} {\mathcal K}
\newcommand {\CalL} {\mathcal L}
\newcommand {\CalM} {\mathcal M}
\newcommand {\CalN} {\mathcal N}
\newcommand {\CalO} {\mathcal O}
\newcommand {\CalP} {\mathcal P}

\newcommand {\CalR} {\mathcal R}
\newcommand {\CalS} {\mathcal S}
\newcommand {\CalT} {\mathcal T}
\newcommand {\CalU} {\mathcal U}
\newcommand {\CalV} {\mathcal V}
\newcommand {\CalX} {\mathcal X}
\newcommand {\CalW} {\mathcal W}

\newcommand {\ec} {\mathscr E}

\newcommand {\mv} {{\mathfrak M}}
\newcommand {\mve} {{\mathfrak M}^{\mathrm{ext}}}
\newcommand {\pv} {{\mathfrak P}}
\newcommand {\pve} {{\mathfrak P}^{\mathrm{ext}}}

\newcommand {\fib} {{\mathscr A}_{u}}
\newcommand {\crf} {{\mathscr X}}

\newcommand{\crfDvec}{{}_{2}\tilde {\crf}^{\hat D}}
\newcommand{\crfDspin}{{}_{1}\tilde {\crf}^{\hat D}}

\newcommand{\Cox}{\mathrm{Cox}}
\newcommand{\McKay}{\mathrm{McKay}}

\newcommand {\pp} {{\mathscr F}}
\newcommand {\rl} {{\rm Q}}

\renewcommand {\sc} {\mathcal Q}

\newcommand {\Gli} {\widehat{{\rm GL}}_{\infty}}
\newcommand {\gli} {\widehat{\mathfrak{gl}}_{\infty}}

\newcommand {\BB} {\mathbb B}
\newcommand {\BD} {\mathbb D}
\newcommand {\BC} {\mathbb C}
\newcommand {\BF} {\mathbb F}
\newcommand {\BI} {\mathbb I}
\newcommand {\BO} {\mathbb O}
\newcommand {\BH} {\mathbb H}

\newcommand {\BL} {\mathbb L}
\newcommand {\BR} {\mathbb R}
\newcommand {\BP} {\mathbb P}
\newcommand {\BQ} {\mathbb Q}

\newcommand {\bG} {\mathbf{G}}

\newcommand {\bR} {\mathbf{R}}
\newcommand {\bq} {\mathscr P}

\newcommand {\bT} {\mathbf{T}}
\newcommand {\ba} {\mathbf{a}}
\newcommand {\ex} {\mathbf{e}}

\newcommand {\bg} {\mathbf{g}}
\newcommand {\bbr}{ \mathbf{r}}
\newcommand {\bm}{ \mathbf{m}}
\newcommand {\bv}{ \mathbf{v}}
\newcommand {\bw}{ \mathbf{w}}
\newcommand {\bu}{ \mathbf{u}}
\newcommand {\bx}{ \mathbf{x}}
\newcommand {\bX}{ \mathbf{X}}
\newcommand {\bz}{ \mathbf{z}}
\newcommand {\xr}{ \mathrm{x}}
\newcommand {\BS} {\mathbb S}
\newcommand {\BT} {\mathbb T}
\newcommand {\BW} {\mathbb W}
\newcommand {\BZ} {\mathbb Z}

\newcommand {\CP} {\mathbb C \mathbb P}
\newcommand {\WP} {\mathbb W \mathbb P}

\newcommand {\al} {\alpha}

\newcommand {\alc}{\check{\alpha}}
\newcommand {\be} {\beta}

\newcommand {\de} {\delta}
\newcommand {\Ga} {\Gamma}
\newcommand {\ve} {\varepsilon}
\newcommand {\ep} {\epsilon}

\newcommand {\lam} {\lambda}
\newcommand {\cla} {\check{\lambda}}
\newcommand {\cmu} {\check{\mu}}
\newcommand {\si} {\sigma}

\newcommand {\ze} {\zeta}

\newcommand {\om} {\omega}

\newcommand{\g}{\mathfrak{g}}
\newcommand{\h}{\mathfrak{h}}

\newcommand{\fl}{\mathfrak{l}}

\DeclareMathOperator{\Hom}{Hom}

\DeclareMathOperator \tr {tr}
\DeclareMathOperator{\Tr} {Tr}

\DeclareMathOperator{\rk} {rk}

\DeclareMathOperator{\diag}{diag}

\newcommand{\ch}{\mathrm{ch}}

\renewcommand{\hat}{\widehat}

\newcommand{\Gg}{G_{\text{g}}}

\newcommand{\GM}{G_{M}}

\newcommand{\iw}{{^{i}{\CalW}}}
\newcommand{\Gq}{{\mathbf{G}}_{{\rm q}}}
\newcommand{\Gad}{{\mathbf{ G}}^{{\rm ad}}}
\newcommand{\Tad}{{\mathbf{ T}}^{{\rm ad}}}
\newcommand{\Gqad}{{\mathbf{ G}}^{{\rm ad}}_{{\rm q}}}
\newcommand{\Tqad}{{\mathbf{ T}}^{{\rm ad}}_{{\rm q}}}
\newcommand{\Bg}{B({\g})}
\newcommand{\Bq}{B({\gq})}
\newcommand{\Bgad}{B^{{\rm ad}}({\g})}
\newcommand{\Bqad}{B^{{\rm ad}}({\gq})}

\newcommand{\Tq}{{\bf T}_{{\rm q}}}

\newcommand{\gq}{\mathfrak{g_{{\rm q}}}}

\newcommand{\Ver}{\mathrm{Vert}_{\gamma}}

\newcommand{\Edg}{\mathrm{Edge}_{\gamma}}

\newcommand{\cc} {\rm C}
\newcommand{\ct}{\check{t}}

\newcommand{\bt}{\mathscr T}

\newcommand{\Bun}{\mathrm{Bun}}
\newcommand{\Bunss}{\mathrm{Bun}^{ss}}

\newcommand{\Bnq}{{\Bun}_{\bG}({\ec})}

\newcommand{\ii}{\mathrm{i}}

\newcommand{\Cx}{\mathbf{C}_{\la x \ra}}
\newcommand{\Cpx}{{\BC\BP}^{1}_{\la x \ra}}
\newcommand{\Ct}{\mathbf{C}_{\la t \ra}}

\begin{document}
\allowdisplaybreaks

\renewcommand{\thefootnote}{}

\newcommand{\arXivNumber}{1211.2240}

\renewcommand{\PaperNumber}{047}

\FirstPageHeading

\ShortArticleName{Seiberg--Witten Geometry of Four-Dimensional $\mathcal N=2$ Quiver Gauge Theories}

\ArticleName{Seiberg--Witten Geometry of Four-Dimensional\\ $\boldsymbol{\mathcal N=2}$ Quiver Gauge Theories\footnote{This paper is a~contribution to the Special Issue on Differential Geometry Inspired by Mathematical Physics in honor of Jean-Pierre Bourguignon for his 75th birthday. The~full collection is available at \href{https://www.emis.de/journals/SIGMA/Bourguignon.html}{https://www.emis.de/journals/SIGMA/Bourguignon.html}}}

\Author{Nikita NEKRASOV~$^{\rm a}$ and Vasily PESTUN~$^{\rm b}$}

\AuthorNameForHeading{N.~Nekrasov and V.~Pestun}

\Address{$^{\rm a)}$~Simons Center for Geometry and Physics, Stony Brook University,\\
\hphantom{$^{\rm a)}$}~Stony Brook, NY 11794-3636, USA}
\EmailD{\href{mailto:nnekrasov@scgp.stonybrook.edu}{nnekrasov@scgp.stonybrook.edu}}
\URLaddressD{\url{https://scgp.stonybrook.edu/people/faculty/bios/nikita-nekrasov}}

\Address{$^{\rm b)}$~Institut des Hautes Etudes Scientifiques, 91440 Bures-sur-Yvette, France}
\EmailD{\href{mailto:vasily.pestun@gmail.com}{vasily.pestun@gmail.com}}

\ArticleDates{Received December 19, 2022, in final form June 20, 2023; Published online July 16, 2023}

\Abstract{Seiberg--Witten geometry of mass deformed ${\CalN}=2$ superconformal ADE quiver gauge theories in four dimensions is determined. We solve the limit shape equations derived from the gauge theory and identify the space ${\mv}$ of vacua of the theory with the moduli space of the genus zero holomorphic (quasi)maps to the moduli space $\Bnq$ of holomorphic $G^{\BC}$-bundles on a (possibly degenerate) elliptic curve $\ec$ defined in terms of the microscopic gauge couplings, for the corresponding simple ADE Lie group $G$. The integrable systems~$\pv$ underlying the special geometry of~${\mv}$ are identified. The moduli spaces of framed $G$-instantons on ${\BR}^{2} \times {\BT}^{2}$, of $G$-monopoles with singularities on ${\BR}^{2} \times {\BS}^{1}$, the Hitchin systems on curves with punctures, as well as various spin chains play an important r\^ole in our story. We also comment on the higher-dimensional theories.}

\Keywords{low-energy theory; instantons; monopoles; integrability}

\Classification{81T12; 81T13; 81T70}

\medskip

\noindent
{\bf A note added ten years later.} This is a minimally edited version of the manuscript~\cite{NP2012}, which
was never previously submitted to a peer-reviewed journal for no reason in particular.
We can observe that~\cite{NP2012} laid foundation to several important developments in supersymmetric gauge theories and two-dimensional conformal/integrable theories, in BPS/CFT correspondence, including connections to quantum groups and integrability~\cite{Costello:2018txb, Nekrasov:2013xda}, generalizations~\cite{Kimura:2017hez,Kimura:2016dys, Kimura:2015rgi,Nieri:2019mdl} of $W$-algebras, generalizations~\cite{Nekrasov:2015wsu, Nekrasov:2016qym} of $q$-characters~\cite{Frenkel:1998}, gauge origami~\cite{Nekrasov:2016ydq}, studies of surface defects~\cite{Jeong:2021rll,Jeong:2017mfh,Nekrasov:2017rqy,Nekrasov:2017gzb,Nekrasov:2021tik}, connections to (and generalizations of)
geometric Langlands program~\cite{Kapustin:2006pk} and representation theory, refinements of Donaldson theory~\cite{Manschot:2019pog}, and many many more. The connection of geometry and physics explored here was always appreciated by our former Director Jean-Pierre Bourguignon. We are happy to dedicate our work to his anniversary.

\newpage

\setcounter{tocdepth}{1}

{\small \tableofcontents}

\renewcommand{\thefootnote}{\arabic{footnote}}
\setcounter{footnote}{0}

\section{Introduction}

In this work a class of quiver ${\CalN}=2$ supersymmetric theories in four dimensions is analyzed. The first problem of this sort was solved in~\cite{Seiberg:1994rs,Seiberg:1994aj}
for the ${\rm SU}(2)$ gauge theory in four dimensions with eight supercharges.

We study mass perturbed ${\CalN}=2$ superconformal theories,
and compute the exact metric
\[
{\rm d}s_{\mv}^{2} = G_{{\CalI}{\bar \CalJ}} {\rm d}u^{\CalI}{\rm d}{\bar u}^{\bar\CalJ}
\]
on the moduli space
${\mv}$ of vacua of the low-energy effective theory. We also compute the vacuum expectation values
\[
\langle {\CalO}_{i,n} \rangle_{u}
\]
of all gauge invariant ${\CalN}=2$ chiral operators.

Our theories have the gauge group $\Gg$ which is a product of a finite number of special unitary groups.
The technique we use is the saddle point approach to the
calculation of the supersymmetric partition function of the theory in $\Omega$-background~\cite{Nekrasov:2002qd}. The partition function is given by the sum over special instanton configurations. In the limit, where the $\Omega$-deformation is removed so that the theory approaches the original flat space theory, the sum over the special instantons is dominated by the contribution of one particular special instanton configuration, of a very large instanton charge (with the expected small effective density of instanton charge). This configuration, the so-called limit shape, is found in this work using a novel approach, built on the analytic techniques of~\cite{Nekrasov:2003rj}. Namely,
we interpret the limit shape equations as the conditions defining the analytic continuations of the generating functions
\begin{equation*}
{\y}_{i}(x) = \exp \langle \tr \log ( x - {\Phi}_{i} ) \rangle_{u},
\end{equation*} where $i$ labels the simple factors in the gauge group $\Gg$, and ${\Phi}_{i}$ is the corresponding complex adjoint Higgs scalar field.
We get the system of (algebraic) equations determining these functions by fixing the set of basic invariants of the monodromy of the analytic continuation.

Recall that a complex Lie group $\Gq$ is naturally associated with the quiver gauge theory. This group is different from the original gauge group $\Gg$ of the theory. Roughly speaking the Dynkin diagram of $\Gq$ is the universal cover of the quiver of the gauge theory.
The group $\Gq$ may be infinite-dimensional. In fact,
for the ${\CalN}=2$ superconformal theories the corresponding Lie algebra $\gq$
 is the finite-dimensional simple Lie algebra
of the ADE type~$\g$, or its affine version~$\hat\g$, or the algebra~$\gli$.

Our main construct is the $x$-dependent
element $g(x)$ of the maximal torus $\Tqad$ of $\Gqad$,
which can also be viewed as the multi-valued $\Tq$-valued function. The group element is locally analytic in $x$,
\begin{equation}
g(x) = \prod_{i \in \widetilde{\Ver}} {\bq}_{i}(x + {\mu}_{i})^{-{\lam}^{\vee}_{i}}
{\y}_{i}(x+{\mu}_{i})^{{\al}_{i}^{\vee}},
\label{eq:bgofx}
\end{equation}
where $i$ runs over the set $\widetilde{\Ver}$ of vertices of the universal cover of the
quiver graph $\gamma$, the polynomials ${\bq}_{i}(x)$ and the complex parameters ${\mu}_{i}$ are determined by the gauge couplings and the masses of matter hypermultiplets, and ${\al}_{i}^{\vee}$ and ${\lam}_{i}^{\vee}$ are the simple coroots and the fundamental
coweights of $\gq$.
It is also convenient to introduce another group element
\begin{equation}
g_{\infty}(x) = \prod_{i \in \widetilde{\Ver}} {\bq}_{i}(x + {\mu}_{i})^{-{\lam}^{\vee}_{i}}.
\label{eq:bgiofx}
\end{equation}

{\em The notation $z_i^{{\alpha}^{\vee}_{i}}$, $w_i^{{\lam}^{\vee}_{i}}$ used in \eqref{eq:bgofx}, \eqref{eq:bgiofx} and below should be understood with the help of the exponential map $\operatorname{Lie}({\Tq}) \to {\Tq}$, see Appendix~{\rm \ref{se:Lie}}. It is well-defined for the coroots $\alpha^{\vee}$, while for the coweights $\lam^{\vee}$ it is well-defined as valued in the conformal extension discussed below}.

Our main claim is that the conjugacy class $[ g(x) ] \in {\Tqad}/W({\gq})$
is holomorphic in $x$, so that the basic adjoint invariants of $\Gq$, evaluated on $[g(x)]$ (up to some twist discussed further) are polynomials of $x$, leading to a system of
equations relating ${\y}$ and $x$:
\begin{equation}
{\crf}_{i} ({\y}(x)) = T_{i}(x) = T_{i,0}x^{{\bv}_{i}} +
T_{i,1} x^{{\bv}_{i}-1} + \dots + T_{i,{\bv}_{i}},
\label{eq:camcurv}
\end{equation}
which define what we call the \emph{cameral curve}
\begin{equation*}
{\CalC}_{u} \subset {\Cx} \times \big( {\BC}^{\times} \big)^{\Ver}.
\end{equation*}
The invariants ${\crf}_{i}$ are normalized characters of $g(x)$ in the fundamental
representations~${\bR}_i$ of~$\Gq$, of the highest weight $\lam_i$:
\[
{\crf}_{i} ({\y}(x)) = g_{\infty}(x)^{-\lam_{i}} \Tr_{R_{i}} g(x).
\]
Moreover, from the work of Steinberg~\cite{Steinberg:1965} (see also~\cite{Morgan:2002math}) we know that for the finite-dimen\-sion\-al~$\Gq$ one can conjugate~$g(x)$ in $\Gq$ to obtain a smooth $\Gq$-valued
function ${\bf g}(x)$ of $x$. Further inspection shows that ${\bf g}(x)$ is a quasi-classical limit of an element of the Yangian algebra $Y({\gq})$, built on the Lie algebra $\gq$ of
$\Gq$. Hopefully the analogous statements hold for all $\Gq$'s we encounter.

In this way one recovers all known results about the Seiberg--Witten geometries of the
${\CalN}=2$ theories in
four dimensions (we do not review all of them in this work) as well as finds new results. {\em We do not claim to reproduce all conjectured Seiberg--Witten geometries, as, e.g., theories with non-classical gauge groups are outside the realm of our methods.} In particular, we find the families of curves describing the geometry of the moduli space of vacua for the theories which were previously believed not to have such description. We also find that the special geometry of the quiver theories with unitary groups is captured in general by a polylogarithmic system of differentials on these curves.

\subsubsection*{Higher dimensions}
The gauge theories we discuss can be also lifted to five-dimensional
theories compactified on a~circle~$\BS^{1}_{\la \beta \ra}$ of
circumference
 $\beta$, or even to the six-dimensional theories, compactified on a two-torus, of the area $\beta^2$. In the limit $\beta \to 0$
one recovers the original four-dimensional theory.
In the five-dimensional case the polynomials $T_i(x)$ in
equation~(\ref{eq:camcurv}) are replaced by Laurent polynomials in~${\rm e}^{\ii
 \beta x}$, while in the six-dimensional case the functions $T_i(x)$ become elliptic.

\subsubsection*{Defreezing}
One of the initial questions which led us to the subject of this work was the following. Consider the ${\rm SU}(2)$ theory with $N_f = 4$ hypermultiplets in the fundamental
representation, with the coupling $\qe$.
By now there is an overwhelming evidence~\cite{Alday:2009aq} of connection this theory has to Liouville conformal blocks on a sphere with four punctures. The momenta of Liouville vertex operators
at the punctures are related to the masses of the hypermultiplets, the locations of the vertex operators are, e.g., $0$, $1$, $\qe$, $\infty$, and the momentum at the intermediate channel
is the Coulomb parameter $a$.

Let us view this theory as a $U(2)$ theory, and let us single out the maximal torus
$U(1)^4$ of the ${\rm Spin}(8)$ flavor symmetry group. Let us gauge these $U(1)$ groups. This gauging is possible in the noncommutative geometry setup. One acquires four additional
coupling constants. What will happen to the Liouville theory?

Upon some reflection one concludes that the resulting theory is a particular case of the
${\hat D}_{4}$ theory, with ${\bv}_{0} = {\bv}_{1} = {\bv}_{3} = {\bv}_{4} = 1$, and ${\bv}_{2} = 2$. We then decided to solve the general quiver superconformal theory which led us to discover many other interesting things.

\begin{figure}[t]
 \centering
 \includegraphics[width=5.5cm]{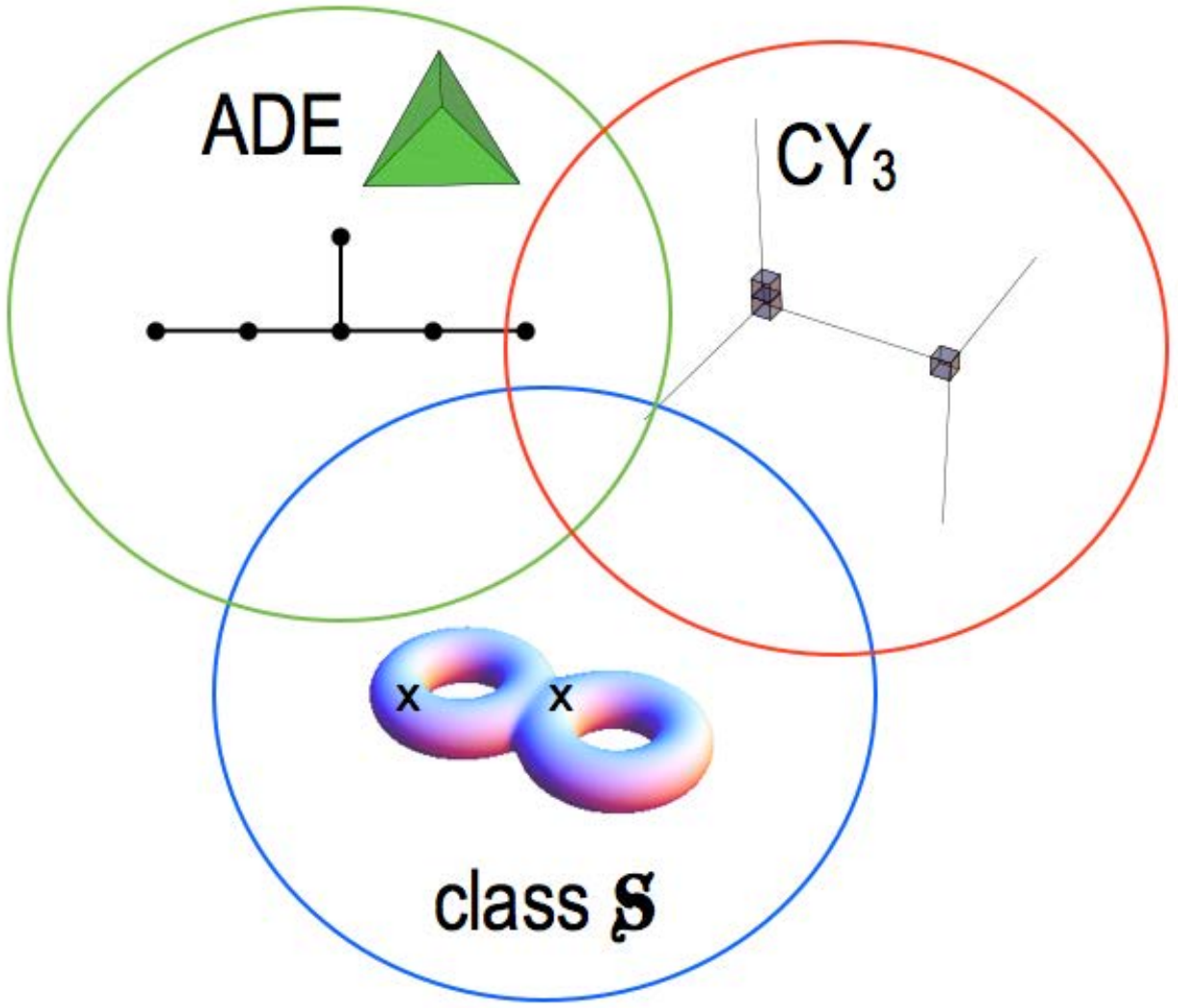}
 \caption{The three major ways to construct $\CalN=2$ theories.}
 \label{fig:world1}
\end{figure}

\subsubsection*{Classification}
Another motivation comes from the question whether Hitchin's system exhaust the list of all reasonable Seiberg--Witten integrable systems. From the early discovery~\cite{Donagi:1995cf} that the ${\CalN}=2^*$ theory with ${\Gg} = {\rm SU}(N)$ is governed by the ${\rm SU}(N)$ Hitchin system on a one-punctured torus (which is nothing but the elliptic Calogero--Moser system, as shown previously in~\cite{Gorsky:1994dj}), proposals in~\cite{Martinec:1996by}, and subsequent developments culminating in the introduction of the ``S-class'' theo\-ries~\cite{Alday:2009aq,Gaiotto:2009we,Gaiotto:2009hg,Witten:1997sc} there was a lot of activity with experimental evidence suggesting that ${\CalN}=2$ theories can be described by some version of Hitchin's system. The underlying construction in these approaches is the
compactification of the six-dimensional superconformal $(0,2)$-theory on some Riemann surface embedded as a supersymmetric cycle in some ambient geometry, and it is believed
that the global features of the embedding should play virtually no r\^ole in the effective gauge theory dynamics.

Another way of engineering ${\CalN}=2$ theories, using string theory, is the so-called geometric engineering~\cite{Katz:1996fh, Lerche:1996xu}, which is the study of the gravity-decoupled limit of the IIA compactification on a Calabi--Yau threefold, with the Calabi--Yau becoming effectively non-compact. A large class of models comes from toric Calabi--Yau's. One then employs the local mirror symmetry to generate curves with differentials, whose periods capture the special geometry of the ${\CalN}=2$ theory.\looseness=-1

In our work, we presented another characterization of the integrable systems underlying the special geometry of the ${\CalN}=2$ theories with the superconformal ultraviolet limit.
Namely, we identify these systems with the moduli spaces of some
gauge/Higgs configurations, such as monopoles or instantons, with the
gauge group $\Gq$ corresponding to the quiver diagram encoding, among
other things, the matter sector of the theory. Unlike all previous
approaches, see Figure~\ref{fig:world1}, which involved some reference to the non-perturbative dualities, or even embedding of the gauge theory to string theory and M-theory, we derive these statements within the quantum field theory, by analyzing the instanton contributions to the low-energy effective action. \looseness=-1

In some cases (e.g., in a simple fashion for the $A_r$ type theories, in a more subtle way
for the~$D_r$ type theories) our phase spaces can be identified with the
phase spaces of Hitchin systems on the low genus curves with punctures,
using some version of Nahm--Fourier--Mukai transform, but in general we
don't have such a duality.
Provided a complete description of the Seiberg--Witten curves and
algebraic integrable systems for the $\CalN=2$ ADE quiver theories it
would be interesting to further investigate this ADE quiver class,
along the lines of~\cite{Gaiotto:2010be,Gaiotto:2009hg} or
\cite{Alim:2011ae} for the ``S-class''.

In the other cases, e.g., the class II $E_r$ type theories we can use the relation between
the moduli of del Pezzo surfaces and the moduli of $E$-bundles on elliptic curve to assign to our version of the Seiberg--Witten curve a one-parametric family of del Pezzo surfaces, which can be viewed as an example of the mirror noncompact threefold of~\cite{Katz:1997eq}.

\subsubsection*{Outlook}
\begin{table}\centering \renewcommand{\arraystretch}{1.15}
 \begin{tabular}{c l c c |c|c|c}
 & & & & classical & quantum & double quantum \\
& & & & $\begin{array}{c}
 \ep_1 = 0\\
 \ep_2 = 0
 \end{array}$ & $\begin{array}{c}
 \ep_1 \neq 0\\
 \ep_2 = 0
 \end{array}$
 & $\begin{array}{c}
 \ep_1 \neq 0\\
 \ep_2 \neq 0
 \end{array}$ \\
\hline
4d& $\Cx = \BC$ & XXX & rational & $\Gq(\BC)$ & $Y(\gq)$ & \dots \\
\hline
5d& $\Cx = \BC^{\times}$ & XXZ & trigonometric & $\Gq
(\BC^\times)$ & $U_q(\hat \gq)$ & \dots \\
\hline
6d& $\Cx = E_T$ & XYZ & elliptic & $\Gq(E) $ & $E_{T,\eta}(\gq)$ & \dots
\end{tabular}
\caption{The (rational/trigonometric/elliptic) by
(classical/quantum/double quantum).}\label{tab:table33}
\end{table}
In the companion paper~\cite{Nekrasov:2013xda} the
 connection between the class of ADE quiver gauge theories
 and \emph{quantum} integrable ADE spin chains was studied in some detail.
In particular we explain there that the five-dimensional version of the ADE quiver gauge
theory on the twisted bun\-dle~$\BR^{4} {\tilde\times} \BS^{1}_{\la \ep_1, \ep_2 ; \beta \ra}$
\cite{Nekrasov:2002qd} with the equivariant parameters set to $\ep_1 = \ep$, $\ep_2 = 0$ as in
\cite{Nekrasov:2009ui} is associated with the XXZ spin chain $\gq$. The theory is solved by the quantum
version of the master equation \eqref{eq:camcurv}: the group
$\Gq$ is replaced by the \emph{quantum affine algebra} $U_{q}( \hat \gq)$
with the quantum parameter $q = {\rm e}^{\ii \beta \ep}$, while the
characters $\crf_i$ are promoted to the $q$-characters of
Frenkel--Reshetikhin~\cite{Frenkel:1998}. (If $\Gq$ is itself affine Kac--Moody
group $\Gq = \hat \bG$ then $U_q(\hat \gq)$ is naturally \emph{quantum
 toroidal algebra}). In the
four-dimensional limit the XXZ $\gq$ spin chain turns into the~XXX~$\gq$
spin chain, the quantum affine algebra $U_q(\hat \gq)$
degenerates into Yangian $Y(\gq)$, and the gauge theory on the twisted bundle becomes the four-dimensional theory subject to a~two-dimensional $\Omega$-background. Finally, the six-dimensional theory
 compactified on a~torus~$E_T$ corresponds to the XYZ $\gq$ spin chain, with
the quantum affine group $U_{q}(\hat \gq)$
elevated to the quantum elliptic group $E_{T, {\eta}} (\gq)$
\cite{Drinfeld_1986,Felder_1995,Felder_1996}, with ${\eta} =
{\be}{\ep}/2\pi$. \looseness=1

It is clear that there is an even larger picture in which the algebraic integrable
systems we encountered in this work are quantized, or
$\ep_2$-deformed,
cf.~Table~\ref{tab:table33}, with the rational/trigonomet\-ric/elliptic
trichotomy in the vertical direction
established in~\cite{Belavin:1982,Faddeev:1982,Sklyanin:1979} and connected with the gauge theories in
\cite{Nekrasov:1996cz}.
It would
be exciting to explore the connection with H.~Nakajima's work~\cite{Nakajima:2001q}
on quiver varieties and quantum affine algebras as well as the
connection with elliptic cohomology
\cite{Ginzburg_1995,Grojnowski_2007,Lurie_2009} of moduli spaces. Notice that
the quantum or double quantum exploration of the ADE quiver world is in a sense
orthogonal to the approach of~\cite{Alday:2009aq} dealing with the
``S-class'' world in Figure~\ref{fig:world1}. Classically, on the
overlap, the relation between the corresponding algebraic
integrable systems comes from the Nahm--Fourier--Mukai/Corrigan--Goddard/ADHM reciprocity
 relating the moduli space of $\bG$-bundles and Hitchin systems. The (doubly) quantum version of this Nahm transform, if it exists, seems to cover the ``quantum'' geometric Langlands duality, separation of variables for quantum systems~\cite{Enriquez:1996xc,Feigin:1994in,Frenkel:1995zp,Sklyanin:1987ih,Sklyanin:1995bm}. The new ingredient~\cite{Nekrasov:2009rc}
in this relatively classic field of research are the supersymmetric gauge theories in four dimensions.
Table~\ref{tab:table33} has been basically filled in recent years.

 \subsection{Organization of the material}

 Section~\ref{se:quiver} introduces the quiver supersymmetric gauge theories which we shall study.

 Section~\ref{se:adequiver} presents the classification of the gauge
 theories which are ${\CalN}=2$ superconformal in the ultraviolet. We
 distinguish three classes of such theories, I, II, and II*. The I and
 II classes have an ADE classification so that for class I $\Gq = \bG$
 and for class II $\Gq = \hat \bG$ where $\bG$ is ADE group, the II*
 theories correspond to $\Gli$ group.

 Section~\ref{se:low-energy} reviews the special K\"ahler geometry of the vectormultiplet moduli spaces $\mv$ of vacua of ${\CalN}=2$ theories. We also recall the relation of $\mv$ to the algebraic integrable systems and the hyperK\"ahler manifolds. We give some examples to be used later.

 Section~\ref{se:limit-shape} introduces our main tool: the limit shape equations, which summarize the microscopic gauge theory calculation leading to the effective low-energy action, i.e., the prepotential~${\pp}$.

 Section~\ref{se:limit-shape-sol} presents the solution of the limit shape equation. We reformulate the equations as the Riemann--Hilbert problem for the set of functions ${\y} (x)$ and solve it by equating the invariants ${\crf}({\y}(x))$ of the monodromy group, the \emph{iWeyl group} which we attach to every ${\CalN}=2$ gauge theory, to some polynomials $T(x)$. In this manner we find an (algebraic) curve ${\CalC}$ and a~system of differentials, whose periods give the special coordinates~$\ac$ and the derivatives~${\pa}{\pp}/{\pa}{\ac}$ of the prepotential $\pp$.

 Section~\ref{se:Seiberg--Witten} analyzes the solution in some detail.
 We interpret the data for the solution of the class I theories as describing a holomorphic map with prescribed singularities of ${\BC\BP}^{1}$ to the space of conjugacy classes ${\bT}/W({\bg})$ in a complex Lie group $\bG$, which can be also viewed as the moduli space of holomorphic $\bG$-bundles on a degenerate elliptic curve. For the class II theories the analogous data parametrizes (quasi)maps to the moduli space ${\Bun}_{\bG}({\ec})$ of holomorphic $\bG$-bundles on elliptic curve.
 In some cases we relate the curve $\CalC$ to the more familiar Seiberg--Witten curves. For the theories corresponding to the $A$ series we manage to relate our curves $\CalC$ to the spectral curves of rational and elliptic Gaudin models (the Hitchin system on the genus zero and one curves with punctures), and also reproduce the results of~\cite{Shadchin:2005cc,Witten:1997sc}. For the class~II~$D$ type theories we reproduce the results of~\cite{Kapustin:1998fa}.
 For the class II $E$ type theories we find yet another interpretation
 of our solution, in terms of families of del Pezzo surfaces. In this
 way we get a~field theory understanding of some of the local mirror
 symmetry predictions~\cite{Katz:1997eq} and brane construction~\cite{Kapustin:1998xn,Kapustin:1998pb}.

 Section~\ref{se:integrable} discusses the moduli spaces $\pv$ of vacua of the gauge theory compactified on a~circle~${\BS}^{1}$. We don't present the full analysis of the hyperK\"ahler metric
 on ${\pv}$ in this work. Instead, we focus on the geometry of ${\pv}$ in the complex structure inherited from four dimensions (this complex structure is sometimes called the complex structure $\bf I$), in which it presents itself as an (algebraic) integrable system. Our solution of the four-dimensional theory comes in a form which leads to a natural guess for the phase space $\pv$ of the integrable systems corresponding to our theories. For the class I theories it is the moduli space of $G$-monopoles on ${\BR}^{2} \times {\BS}^{1}$ with singularities, for the class II theories it is the moduli space of $G$-instantons on~${\BR}^{2} \times {\BT}^{2}$, and for the class II* ${\hat A}_r$ theories it is the moduli space of noncommutative $U(r+1)$ instantons on~${\BR}^{2} \times {\BT}^{2}$. Of course these spaces have a natural hyperk\"ahler structure which depends in the expected fashion on all the parameters of the theory and its compactification. Although our motivation comes from the field theory analysis in the previous chapters, our results confirm the conjectures of~\cite{Chalmers:1996xh,Cherkis:2000cj,Cherkis:1997aa,Cherkis:2001gm,Hanany:1996ie,Kapustin:1998fa,Kapustin:1998xn,Kapustin:1998pb} which are motivated by the string theory analysis, and in particular by the brane constructions.

 Section~\ref{se:higher} discusses the modifications of our solutions in the five and six-dimensional cases.

In Appendix~\ref{se:mckay}, we review the affine ADE graphs, the
McKay correspondence and the M-theory/D-brane picture for the present
work; in Appendix~\ref{se:partitions}, we put our conventions on the
partitions and representations by free fermions;
in Appendix~\ref{se:Lie},
 we review some standard material on Lie groups and Lie algebras which
 we use in solving our theories. We recall the notions of the (co)root
 and the (co)weight lattices, Weyl groups, and the integrable highest
 weight representations; in Appendix~\ref{se:elliptic}, we collect
 our conventions for elliptic functions; in Appendix~\ref{se:appendixE}, we give some technical details on spectral curves of
 affine E-series.

 \newpage

 \subsection{Notations\label{se:sumpro}}
{\bf Quivers}, Section~\ref{se:quiver}
\begin{alignat*}{3}
&{\Ver} \quad && \text{set of vertices} &\\
&{\Edg} \quad && \text{set of edges} &\\
&s(e) \in \Ver \quad && \text{the source of the edge $e \in \Edg$} &\\
&t(e) \in \Ver \quad && \text{the target of the edge $e \in \Edg$} &\\
&\qe_{i}={\rm e}^{2 \pi \ii \tau_i} \quad && \text{gauge coupling constants}&\\
&{\tqe} \quad && ({\qe}_{i})_{i\in \Ver}& \\
&\bv_i \quad && \text{number of colors for $i$-th node gauge group ${\rm SU}(\bv_i)$}&\\
&\bw_i \quad && \text{number of flavors for $i$-th node fundamental
 matter}&\\
&\ac_{i,\ba} \quad && \text{eigenvalues of the complex scalars} &\\
&\ac_{i,\ba} \leftrightarrow \ac^{\CalI} \quad && \text{the special coordinates on Coulomb moduli
 space}&\\
&C_{ij} \quad && \text{Cartan matrix associated to the quiver by its Dynkin graph}&\\
&r = \rk(C) \quad && \text{$|\Ver|$ if $\gamma$ is finite ADE
or $|\Ver|-1$ if $\gamma$ is affine ADE}&
 \end{alignat*}
{\bf Lie groups}
\begin{alignat*}{3}
 &\ii \quad &&\sqrt{-1} &\\
& \Gg \quad && \text{Gauge group of the four-dimensional theory }& \\
& \Gq \quad && \text{Kac--Moody group associated with quiver Dynkin diagram}& \\
& {\GM}\quad && \text{the flavor group}& \\
& {\bG} = G^{\BC} \quad && \text{finite-dimensional complex Lie group}&\\
& {\hat\bG} \quad && \text{affine Kac--Moody group for $\bG$}&\\
& {G} \quad && \text{maximal compact subgroup of $\bG$} &\\
& {\bT} \quad && \text{maximal torus of $\bG$}&\\
& T \quad && \text{maximal torus of $G$} &\\
& Z \quad && \text{the center of both $G$ and $\bG$} &\\
& {\Gad} = {\bG}/Z \quad && \text{adjoint form of the complex Lie group $\bG$} & \\
& {\Tad} = {\bT}/Z \quad && \text{the maximal torus of $\Gad$}& \\
& {\Gqad} = {\Gq}/Z \quad && \text{adjoint form of the complex Lie group $\Gq$}&\\
& {\Tqad} = {\Tq}/Z \quad && \text{its maximal torus}&\\
\end{alignat*}
{\bf Lie algebras}
 \begin{alignat*}{3}
& \gq \quad && \text{Kac--Moody Lie algebra associated with quiver Dynkin diagram} &\\
 &{\g} \quad && {\rm Lie}({\bG})& \\
& {\h} \quad && {\rm Lie}({\bT})&
\end{alignat*}
{\bf Representation theory}, Appendix~\ref{se:Lie}
 \begin{alignat*}{3}
&a_i \quad && \text{Kac--Dynkin marks}& \\
& {\rl} , \ {\Lambda}^{\vee}\quad && \text{root lattice, coweight lattice}&\\
& {\Lambda}, \ {\rl}^{\vee} \quad && \text{weight lattice, coroot lattice}& \\
& R_{i} \quad && \text{$i'$th fundamental representation of $\bG$}& \\
& {\hat R}_{i} \quad && \text{$i'$th fundamental representation of $\hat\bG$} &\\
&{\CalR}_{i} \quad && \text{$i'$th fundamental representation of $\Gli$}&
\end{alignat*}
{\bf Spaces}
 \begin{alignat*}{3}
&{\Bg} \quad && \text{${\bT}/ W({\g})$ the space of conjugacy classes in $\bG$ } &\\
&{\Bq} \quad && \text{${\Tq}/ W({\gq})$ the space of conjugacy classes in $\Gq$} &\\
&{\Bgad} \quad && \text{${\bT}/( Z \times W({\g}))$ the space of conjugacy classes in $\Gad$} &\\
&{\Bqad} \quad && \text{${\Tq}/( Z \times W({\gq}) )$ the space of conjugacy classes in $\Gqad$} &\\
& {\Cx} \quad && \text{complex plane $\BC$ in 4d, cylinder $\BC^{\times}$ in 5d, torus $E$ in 6d} &\\
&{\Cpx} \quad && = {\Cx} \cup \{ \infty \} &\\
& {\ec}(\qe) \quad && \text{elliptic curve $\BC^{\times}/\qe^{\BZ}$} &\\
&\qe \quad && \text{$\prod_{i \in \Ver} \qe_i^{a_i}$ for class II theories}&\\
&\Bun_{\bG} (\ec) \quad && \text{coarse moduli space of semistable holomorphic $\bG$-bundles on $\ec$} &\\
&\mv \quad && \text{the Coulomb moduli space of the 4d $\times {\rm SU}(\bv_i)$ gauge theory }&\\
&\mve \quad && \text{the Coulomb moduli space of the 4d $\times U(\bv_i)$ gauge theory}&\\
&\pv \to \mv \quad && \text{the algebraic integrable system $\dim_{\BC}\pv =2 \dim_{\BC} \mv$} &\\
&\pve \to \mve \quad && \text{the complex integrable system $\dim_{\BC}\pve = 2 \dim_{\BC} \mve$}&
\end{alignat*}
{\bf Seiberg--Witten curves}
 \begin{alignat*}{3}
& \CalC_u \quad && \text{cameral curve: Section~\ref{se:cameral}}&\\
& C_u \quad && \text{spectral curve: Section~\ref{se:spectral}}&\\
& \obs \quad && \text{obscure curve: Section~\ref{sec:obscura}}&\\
& x \quad && \text{flat coordinate on $\Cx$}&\\
& \y_i(x) \quad && \text{\emph{amplitudes} (the solution of the theory): Section~\ref{se:amplitude}} &\\
& T_i(x) \quad && \text{\emph{gauge polynomials} of degree $\bv_i$}&\\
& \bq_i(x) \quad && \text{\emph{matter polynomials} of degree $\bw_i$}&\\
& g(x) \quad && \text{$\Tqad$ valued analytic function on $\Cx$}&\\
&\mathbf{g}(x) \quad && \text{$\Gq$ valued analytic function on $\Cx$}&\\
& \crf_j \quad && \text{$\Gq$ character (or Weyl invariant) for $i$-th fundamental weight of $\Gq$}&
 \end{alignat*}
{\bf Partitions}
 \begin{alignat*}{3}
& {\lam}\quad && \text{partition ${\lam}_{1} \geq {\lam}_{2} \geq \cdots \geq {\lam}_{{\ell}({\lam})} > 0 $, ${\lam}_{i} \in {\BZ}_{\geq 0}$} &\\
& {\ell}({\lam})\quad && \text{the length of the partition $\lam$}&\\
& | {\lam} | \quad && \text{the size of the partition $|\lam | = \sum_{i=1}^{{\ell}({\lam})} {\lam}_{i}$ }&
\end{alignat*}

Let $x \in R^{\BZ}$ be a sequence $(x_{i})_{i\in \BZ}$, with $x_{i}$ in some ring $R$.

\noindent
{\bf Consecutive products}
 \begin{gather*}
 x^{[i]}_{j} \equiv \frac{\prod_{k=-\infty}^{k=j+i -1} x_{k}}{\prod_{k=-\infty}^{j-1} x_{k}} , \\
 x^{[i]} = x^{[i]}_{1}, \\
 x^{[i+1]} = x^{[i]} x_{i+1},
\end{gather*}
for example
\begin{gather*}
 x^{[0]} = 1, \\
 x^{[2]}_{4} = x_{4}x_{5}, \\
 x^{[-1]} = x_{0}^{-1}, \\
 x_{5}^{[-4]} = x_{1}^{-1}x_{2}^{-1}x_{3}^{-1}x_{4}^{-1}.
 \end{gather*}
{\bf Consecutive sums}
\begin{gather*}
 x_{(i)} = \sum_{j=-\infty}^{i} x_{j} - \sum_{j=-\infty}^{0} x_{j}, \\
 x_{(i+1)} = x_{(i)} + x_{i+1},
\end{gather*}
for example
\begin{gather*}
 x_{(0)} = 0, \\
 x_{(3)} = x_{1}+x_{2}+x_{3}, \\
 x_{(-2)} = - x_{0} - x_{-1}, \\
{\lam}_{({\ell}({\lam}))} = |{\lam}|.
\end{gather*}

\section{Supersymmetric quiver theories}\label{se:quiver}

Consider any
${\CalN}=2$ supersymmetric field theory in four dimensions
whose gauge group is a~product of special unitary groups,
while the matter hypermultiplets are in the
fundamental, bi-fundamental, and adjoint representations. The field content, the parameters of the Lagrangian, and the choice of the vacuum are conveniently encoded in the quiver data, which is
\begin{enumerate}\itemsep=0pt
\item
An oriented graph $\gamma$ with the set $\Ver$ of vertices and the set
$\Edg \subset \Ver \times \Ver$ of oriented edges. Let $s,t\colon \Ver
\times \Ver \to \Ver$ by the projections onto the first and the second
factors. They define the two maps $s, t\colon \Edg \to \Ver$ which assign to
an oriented edge its source and the target, respectively.
In what follows we shall use the notation
\[
|\Ver| = \# \Ver
\]
for the number of vertices in the quiver.

\item
An assignment of polynomials to the vertices:
\begin{equation*}
{\bt}, {\bq}\colon \ \Ver \to {\BC}[x], \qquad i \mapsto {\bt}_{i}(x), {\bq}_{i}(x)
\end{equation*}
and
${\bv}, {\bw}\colon \Ver \to {\BZ}_{+}$, where
\[
{\bv}_{i} = \deg {\bt}_{i}, \qquad {\bw}_{i} = \deg {\bq}_{i}, \qquad i \in \Ver.
\]
The polynomials ${\bt}_{i}(x)$ are monic, the highest-order term coefficients ${\qe}_{i}$ of the polynomials~${\bq}_{i}(x)$ are required to obey: $| {\qe}_{i} | < 1$.

\item
A $1$-cocycle ${\bf m} \in {\CalC}^{1}({\gamma}, {\BC})$, in other words
an assignment
\begin{equation*}
e \in \Edg \mapsto m_{e} \in {\BC}.
\end{equation*}

\end{enumerate}
We now proceed with the explanation of the r\^oles of the polynomials ${\bt}$, ${\bq}$, as well as that of the cocycle $\bf m$.

\subsection{Gauge group, matter fields, couplings, parameters}

\subsection{The gauge group} We denote the gauge group by $\Gg$. It is the product
\begin{equation}
{\Gg} = \prod_{i \in \Ver} {\rm SU}({\bv}_{i}).
\label{eq:gagg}
\end{equation}
The vector multiplet therefore splits into a collection of vector multiplets
for the ${\rm SU}({\bv}_{i})$ gauge factors
\[
{\Phi}_{\text{g}} = ( {\Phi}_{i})_{i \in \Ver}.
\]
We have a gauge coupling $e_{i}$ and the theta angle $\vartheta_{i}$ for each $i \in \Ver$. As usual, we combine them into the complexified gauge couplings,
\begin{equation*}
{\tau}_{i} = \frac{1}{2\pi \ii} \log ({\qe}_{i}) = \frac{{\vartheta}_{i}}{2\pi} + \frac{4\pi \ii}{e_{i}^{2}}.
\end{equation*}
The bosonic part of the action for gauge fields is given by
\begin{equation*}
{\CalL}_{\text{YM}} = \sum_{i \in {\Ver}} \left( \frac{1}{e_{i}^{2}} \int \tr_{{\bv}_{i}} F_{A_{i}} \wedge \star F_{A_{i}} + \frac{ {\ii} {\vartheta}_{i}}{8\pi^{2}}
\int \tr_{{\bv}_{i}} F_{A_{i}} \wedge F_{A_{i}} \right),
\end{equation*}
where $ \tr _{v}$ denotes the trace of a $v \times v$ matrix.
The exponentiated coupling
\[
{\qe}_{i} = {\rm e}^{2\pi \ii \tau_i}
\]
enters the path integral measure. The perturbative effects do not depend on $\vartheta_{i}$, while the non-perturbative effects, which are the contributions of the gauge fields with non-trivial instanton charge, depend on ${\qe}_{i}$, ${\bar \qe}_{i}$. In other words, the partition function is expected to be invariant under the shifts
\begin{equation*}
{\tau}_{i} \longrightarrow {\tau}_i + 1.
\end{equation*}
For $i \in \Ver$ let ${\Phi}_{i}$
denote the corresponding complex scalar in the adjoint representation.
The bosonic potential of the vector multiplet field $\Phi_i$ contains a universal term
\[
 \tr _{{\bv}_{i}} \big[ {\Phi}_{i}, {\Phi}_{i}^{\dagger} \big]^{2}
\]
plus some possible non-negative terms coming from interactions with matter fields. If the matter fields are massive then
this term alone forces ${\Phi}_{i}$ to commute with its conjugate at low energies. Therefore, at low energy the field $\Phi_i$
can be diagonalized:
\begin{equation}
{\Phi}_{i} \longrightarrow \operatorname{diag}(\mathfrak{a}_{i, {\al}})_{{\al} = 1}^{{\bv}_{i}}.
\label{eq:aeig}
\end{equation}
The gauge invariant order parameters are the vacuum expectation values of the coefficients
of the characteristic polynomial ${\bt}_{i}$ of ${\Phi}_{i}$:
\begin{equation}
{\bt}_{i}(x) = \left\langle \det\nolimits_{{\bv}_{i}}( x - {\Phi}_{i} ) \right\rangle_{u},
\label{eq:vev}
\end{equation}
where we assume the normalization $\langle 1 \rangle_{u} = 1$, and
$\det\nolimits_{v}$ denotes the determinant of a $v \times v$-matrix.
Therefore the polynomials ${\bt}_{i}(x)$ in (\ref{eq:vev}) are monic.

Thus, a collection of polynomials ${\bt}_{i}(x)$, $i \!\in\! {\Ver}$ fixes the choice of the vacuum ${u = ( u_{i , {\ba}} ) \!\in\! {\mv}}$:%
\begin{gather}
{\bt}_{i}(x) = x^{{\bv}_{i}} + \sum_{{\ba}=2}^{{\bv}_{i}} u_{i , {\ba}} x^{{\bv}_{i}-{\ba}}.
\label{eq:ppol}
\end{gather}
Because of the non-perturbative (instanton) effects the relation between $u_{i,{\ba}}$ and $\mathfrak{a}_{i,{\ba}}$ is not polynomial, and for the same reason
${\bt}_{i}(x) \neq {\y}_{i}(x)$.

\subsection{The hypermultiplets in the bi-fundamental, or adjoint
 representations} The bifundamental or adjoint
hypermultiplet $H_{e}$, $e \in {\Edg}$ transforms
in the following ${\Gg}$ representation:
\begin{gather*}
\big({\overline {\bv}_{s(e)}}, {\bv}_{t(e)} \big) \oplus \big({\overline {\bv}_{t(e)}}, {\bv}_{s(e)} \big) \qquad {\rm for} \quad s(e) \neq t(e),\\
{\rm Adj}(su({\bv}_{i})) \qquad {\rm for} \quad s(e) = t(e) = i.
\end{gather*}
The masses of the bi-fundamental hypermultiplets are conveniently represented by the $1$-cocycle: ${\bf m} \in {\CalC}^{1}({\gamma}, {\BC})$,
$e \mapsto {\bf m}_{e}$.
Let $[ {\bf m} ] \in H^{1}({\gamma}, {\BC})$ be the corresponding cohomology class. If we denote by $m^{*}$
a particular representative of $[ {\bf m} ]$ in ${\CalC}^{1}({\gamma}, {\BC})$, then
\begin{equation}
{\bf m} = m^{*} + {\delta}{\mu}
, \qquad {\mu} \in {\CalC}^{0}({\gamma}, {\BC})
\label{eq:bmpm}
\end{equation}
or, in components,
\[
{\bf m}_{e} = m^{*}_{e} + {\mu}_{t(e)} - {\mu}_{s(e)}.
\]

\subsection{The hypermultiplets in the fundamental representations} These are assigned to the vertices $i \in \Ver$. We have ${\bw}_{i}$ such multiplets.
Write
\begin{equation*}
{\bq}_{i}(x) = {\qe}_{i} \prod_{{\fe}=1}^{{\bw}_{i}} ( x - m_{i,{\fe}} ).
\end{equation*}
Then
$m_{i,{\fe}}$ are the masses of the fundamental hypermultiplets, charged
under ${\rm SU}({\bv}_{i})$. A ${\bw}_i$-tuplet of
fundamental hypermultiplets can be thought as a bifundamental
$(\mathbf{v}_i, \bar{\mathbf{w}}_{i})$ for~${\rm SU}({\bv}_i)$ and an auxiliary frozen $U({\bw}_i)$, so that $m_{i,{\fe}}$ can be
interpreted as the values of the frozen scalar field in the vector
multiplet
of $U({\bw}_i)$.

\section[The ADE classifications of superconformal ${\CalN}=2$ theories]{The ADE classifications of superconformal $\boldsymbol{{\CalN}=2}$ theories} \label{se:adequiver}

In quantum gauge theory the coupling constants $e_{i}$ are subject to the renormalization which leads to their dependence on the energy scale at which one measures
the interaction between the charged particles. The consistent theories
have the gauge couplings which tend to zero as the scale approaches ultraviolet, or approach some fixed values. These theories are called asymptotically free and asymptotically conformal, respectively. Moreover, starting with the asymptotically conformal theory, one can perturb it by the mass terms. Then, by tuning the masses and the bare couplings, one arrives at the asymptotically free theory.
All asymptotically free quiver theories arise in this way. Therefore it
suffices to solve the asymptotically conformal theories.

From the perspective of geometrical engineering the ADE quiver theories
were studied in~\cite{Katz:1997eq}, and three-dimensional ADE quiver theories were studied in
\cite{Gaiotto:2008ak}.

\subsection{Beta functions and Cartan matrix}

The running of the gauge coupling ${\tau}_{i}$ is described by the Gell-Mann--Low equations which are one-loop exact for the ${\CalN}=2$ supersymmetric theories, the result of~\cite{Novikov:1983uc}. The actual contributions of the matter and gauge multiplets to the gauge couplings are
\begin{equation}
{\be}_{i} = {2\pi \ii} \frac{{\rm d} {\tau}_{i} }{{\rm d} \log {\Lambda}} = {\bw}_{i} - 2{\bv}_{i} + \sum_{e\colon t(e) = i} {\bv}_{s(e)} + \sum_{e\colon s(e) = i} {\bv}_{t(e)},
\label{eq:betf}
\end{equation}
where ${\Lambda}$ is the energy scale. The theory is asymptotically
conformal if ${\be}_{i}=0$ vanishes for all~$i \in \Ver$. The theory is asymptotically free if ${\be}_{i} < 0$.

Let us define the incidence of the pair of vertices $I_{ij}$ to be the number of edges $e$ connecting the vertices $i$ and $j$:
\begin{gather*}
I_{ij} = \# \{ e \mid s(e) = i, \, t(e) =j, \ {\rm or}\ s(e) = j,\, t(e) = i \}
\end{gather*}
with the understanding that if the vertex $i \in \Ver$ is connected to itself by a loop, then the corresponding edge contributes $2$ to the incidence matrix element $I_{ii}$.
Define, for all quivers,
the Cartan matrix of size $|\Ver| \times |\Ver|$
\begin{equation}
C_{ij} = 2 {\de}_{ij} - I_{ij}.
\label{eq:cartan}
\end{equation}
Then
\[
{\be}_{i} \propto ({\bw}- C{\bv})_{i},
\]
where
\[
(C{\bv})_{i} = \sum_{j\in \Ver} C_{ij} {\bv}_{j}.
\]
Let us solve the ${\be}_{i} = 0$ conditions (cf.~\cite{Howe:1983wj,Katz:1997eq,Lawrence:1998ja}.
It is convenient to separate the solutions into three cases, which we shall call
the theories of class I, the
theories of class II
and the theories of class II*. By $r$ we shall
denote the rank of the Cartan matrix $C$
\begin{equation*}
 r = \rk (C).
\end{equation*}
 The main difference between the class I and class II, II* theories is that the Cartan matrix
 of class I theories has the maximal rank
 \begin{equation*}
 r(C_{\text{$\gamma$ of class I}}) = |\Ver|
 \end{equation*}
 while for the
 theories of class II and class II* the Cartan matrix has one-dimensional kernel,
 \begin{equation*}
 r(C_{\text{$\gamma$ of class II}}) = |\Ver| -1.
 \end{equation*}

\subsection{Class I theories}

The solutions to the equations ${\beta}_{i}=0$ with ${\bw} \neq 0$ are the theories of class I.
It is well-known that the graph $\gamma$ is in this case
a Dynkin diagram of a finite-dimensional simple simply-laced Lie algebra $\gq$, of the ADE type, with $\Ver$ labeling the simple roots of $\gq$:
\[
i \in \Ver \mapsto {\al}_{i}.
\]
To solve the equation ${\be}_{i}=0$ is equivalent to finding two vectors
\begin{equation*}
{\bv} = \sum_{i \in \Ver} {\bv}_{i} {\al}_{i}^{\vee}, \qquad
{\bw} = \sum_{i \in \Ver} {\bw}_{i} {\al}_{i}^{\vee}
\end{equation*}
with non-negative components ${\bv}_{i}, {\bw}_{i} \in {\BZ}_{\geq 0}$,
such that (cf.~(\ref{eq:cartan})
\begin{equation}
{\bw} = C^{\gq} {\bv},
\label{eq:betz}
\end{equation}
where $C^{\gq}$ is the Cartan matrix of the corresponding
finite-dimensional Lie algebra $\gq$ of the~ADE type.
Equivalently
\[
{\bv} = \sum_{i \in \Ver} {\bw}_{i} {\lam}_{i}^{\vee},
\]
where ${\lam}_{i}^{\vee}$ are the fundamental coweights of $\gq$.

For class I theories we set $\Gq = \bG$ where $\bG$ is finite
dimensional complex ADE group.

\begin{Remark}
In the case of $\gq$ of the $A_{r}$ type
the dimensions
${\bv}_{i}$ must be a convex function of~$i$. In particular,
they grow with $i$, for $i = 1, \dots, i_{*}$, and then decrease:
\begin{equation}
{\bv}_{1} \leq {\bv}_{2} \leq \cdots \leq {\bv}_{i_{*}-1} \leq
{\bv}_{i_{*}} \geq {\bv}_{i_{*}+1} \geq \cdots \geq {\bv}_{r}.
\label{eq:gorka}
\end{equation}
\end{Remark}

\begin{Remark}
The graphs of the $D_r$ and $E_r$ Dynkin type have a single tri-valent vertex, let us call it $i_{*}$. One can easily show using the ${\be}_{i}=0$ equations that
${\bv}_{i_{*}}$ is the maximal value of ${\bv}_{i}$ on~$i \in \Ver$, and that ${\bv}_{i}$ decrease
along each leg emanating from
 the tri-valent vertex $i_{*}$.
\end{Remark}

\subsection{Class II theories}

The class II theories have ${\bw} \equiv 0$, and $[ {\bf m} ] =0$. It is well-known that
the graphs $\gamma$, such that the corresponding Cartan matrix $C$ has a
zero eigenvector with positive integer entries are in one-to-one
correspondence with
the simply laced affine Dynkin diagrams (see Appendix~\ref{se:mckay} for
our conventions on ADE graphs and McKay correspondence):
\begin{enumerate}\itemsep=0pt
\item[1)]
${\hat A}_{r}$, $r \geq 2$,

\item[2)]
${\hat D}_{r}$, $r \geq 5$,

\item[3)]
${\hat E}_{r}$, $r = 7,8,9$.
\end{enumerate}
These Dynkin diagrams correspond to the affine Lie algebras $\hat \g$
associated to finite-dimensional Lie algebras $\g$ of rank $r$.
We set $\Gq = \hat \bG$ and $\gq = \hat \g$.
We discuss the relevant aspects of the theory of affine Kac--Moody algebras in the following subsections.

{}Note that the ${\hat A}_{0}$ case (its quiver has one vertex and one edge connecting it to itself), given our constraint $[ {\bf m} ] =0$ for the class II theory,
corresponds to the ${\CalN}=4$ superconformal theory. It is well known that the classical moduli space of vacua gets no quantum corrections in this theory.

{}The dimensions ${\bv}$ are uniquely specified, up to a single multiple:
\begin{equation*}
{\bv}_{i} = N a_{i},
\end{equation*}
where $a_{i}$ are the so-called Dynkin labels. We shall recall several interpretations
of these numbers below.

\subsection{Class II* theories}

The class II* theories have ${\bw}=0$ and $[ {\bf m} ] \neq 0$.
The first condition reduces our choice of $\gamma$ to the affine Dynkin diagrams (including the ${\hat A}_{0}$ case of the quiver with one vertex and one loop connecting this vertex with itself). The second condition
implies that $\gamma$ is the Dynkin diagram of the ${\hat A}_{r}$ type for some $r \geq 1$. Indeed, only in this the affine Dynkin diagram has $H^{1}({\gamma}, {\BC}) = {\BC}$, the diagram being a regular $(r+1)$-gon. The dimensions ${\bv}_{i}$ are all
equal to $N$, a non-negative integer.

In particular, the class II* $r=0$, ${\hat A}_{0}$-theory with $[ {\bf m} ] \neq 0$ is the celebrated ${\CalN}=2^{*}$ theory, the ${\rm SU}(N)$ theory with massive adjoint hypermultiplet.

We shall see that the Kac--Moody Lie algebra which corresponds to the theories of class II* is the
$\widehat{{\g\fl} ({\infty})}$ algebra, which contains
$\widehat{\mathfrak{u}(r)}$ as a subalgebra of $r$-periodic
matrices.

\section{Low-energy effective theory}\label{se:low-energy}

We now can proceed with the main subject of our study.
Our goal is to determine the two-derivative/four-fermions terms in the low-energy effective action of our theory.

The low-energy effective theory of the ${\CalN}=2$ supersymmetric quiver theory with generic masses $( m_{e}), (m_{i, {\fe}})$ is the abelian ${\CalN}=2$ theory of ${\bbr}$ vector multiplets,
\begin{equation*}
{\bbr} = \sum_{i \in {\Ver}} ({\bv}_{i} - 1).
\end{equation*}
For generic masses the theory has the manifold $\mv$ of vacua, which is a complex variety of complex dimension $\bbr$:
\[
\dim _{\BC} {\mv} = {\bbr}.
\]
The effective theory is a sigma model on $\mv$,
interacting with $\bbr$ abelian gauge fields $A_{\mu}^{\CalI}$, ${\CalI}= 1, \dots , {\bbr}$, and
some fermionic fields. Our goal is to determine the metric on $\mv$, the effective gauge couplings $\operatorname{Im}{\tau}_{\CalI\CalJ}$ and the effective theta-angles $\operatorname{Re}{\tau}_{\CalI\CalJ}$ of these gauge fields.

\subsection{Special K\"ahler geometry}

One can interpret the eigenvalues (\ref{eq:aeig}) obeying
\begin{equation*}
\sum_{{\ba} =1}^{{\bv}_{i}} {\ac}_{i, {\ba}} = 0
\end{equation*}
as the \emph{special coordinates} on the moduli space $\mv$ of vacua. As is well known, $\mv$ is a K\"ahler manifold, with a peculiar metric, and a rigid system of local coordinate systems. The corresponding geometry is called the \emph{rigid special geometry}, and it is a limit~\cite{Seiberg:1994rs} of the \emph{special geometry} of~${\CalN}=2$ supergravity, studied in
\cite{Cremmer:1984hj}.

Let us label the effective abelian vector multiplets
${\CalA}^{\CalI}$ by ${\CalI} = (i, {\ba})$, $i \in \Ver$, ${\ba} = 2, \dots , {\bv}_i$. In components
\[
{\CalA}^{\CalI} = {\ac}^{\CalI} + {\vartheta} {\psi}^{\CalI} + {\vartheta}{\vartheta} \big(F_{A}^{\CalI}\big)^{-} + \cdots,
\]
where
\begin{equation*}
\big(F_{A}^{\CalI}\big)^{\pm} = \frac 12 \big( F_{A}^{\CalI} \pm \star F_{A}^{\CalI} \big).
\end{equation*}
The scalar components ${\ac}^{\CalI} \leftrightarrow {\ac}_{i, {\ba}}$, ${\ba}= 2, \dots, {\bv}_{i}$, more precisely their vacuum expectation values are the local \emph{special coordinates}.
Globally they are subject to monodromy transformations, unlike the global coordinates $( u_{i,{\ba}} )$ in equation~(\ref{eq:ppol}), which are defined via the expectation values of the gauge invariant local operators of the microscopic theory.

The monodromy transformations act by symplectic transformations mixing the special coordinates ${\ac}^{\CalI}$ and their duals ${\ac}_{\CalI}^{D}$ together with the masses $( m_{e} )$, $e \in \Edg$ and $(m_{i, {\fe}})$, $i \in \Ver$, ${\fe} = 1, \dots , {\bw}_{i}$.
The dual coordinates are the derivatives of the prepotential ${\pp}$,
\begin{equation}
{\ac}^{D}_{\CalI} = \frac{{\partial}{\pp}}{{\partial}{\ac}^{\CalI}}.
\label{eq:spcf}
\end{equation}
The prepotential is a multi-valued analytic function of ${\ac}^{\CalI}$, it is the superspace action which determines the low-energy effective action in the approximation we are working:
\begin{equation}
{\CalL}^{\rm eff} = \frac \ii{4\pi} \int {\tau}_{\CalI\CalJ} \big(F_{A}^{\CalI}\big)^{-} \wedge \big(F_{A}^{\CalJ}\big)^{-} + {\bar\tau}_{\CalI\CalJ} \big(F_{A}^{\CalI}\big)^{+} \wedge \big(F_{A}^{\CalJ}\big)^{+} - \ii \operatorname{Im}{\tau}_{\CalI\CalJ}\, {\rm d}{\ac}^{\CalI} \wedge \star {\rm d}{\bar {\ac}}^{\CalJ},
\label{eq:boseff}
\end{equation}
where
\begin{equation}
{\tau}_{\CalI\CalJ} = \frac{{\pa}^2{\pp}}{{\pa}{\ac}^{\CalI}{\pa}{\ac}^{\CalJ}}.
\label{eq:tauij}
\end{equation}
The invariant formulation of equation~(\ref{eq:spcf}) is that the two-form
\[
\sum_{\CalI} {\rm d}{\ac}^\CalI \wedge {\rm d}{\ac}^D_\CalI
\]
identically vanishes on $\mv$. The proper formulation of this condition uses the additional structure which we review below.

\subsection{Extended moduli space}

In our solution of the theories of class I and class II
 it would be sometimes convenient to trade the bifundamental masses formally
with the $U(1)$ factors as explained in \eqref{eq:bmpm} if one
considers~$\times_{i \in \Ver} U(\bv_i)$ gauge group instead of
$\times_{i \in \Ver} {\rm SU}(\bv_i)$. The $|\Ver|-1$ bifundamental
masses\footnote{Recall that in the class II $\hat A_{r}$ theory there are
 $r$ mass parameters that can be traded for the $U(1)$ scalars
 and $1$~additional ``twist'' mass parameter $m^{*}$ promoting the class II
 to class II*.} and one overall $U(1)$ factor add $|\Ver|$ parameters
to $\mv$. We set
\begin{equation}
\label{eq:mve}
 \mve = \mv \times \BC^{\Ver}
\end{equation}
with
\begin{equation*}
 \dim_{\BC} \mve = \dim_{\BC} \mv + |\Ver| = \sum_{i \in \Ver} \bv_i.
\end{equation*}
For the class II theories
\begin{equation*}
 \dim_{\BC} \mve = \sum_{i \in \Ver} \bv_i = N h,
\end{equation*}
where $h= \sum_{i \in \Ver} a_i$ is the Coxeter number of $\bG$. Recall that we only encounter simply-laced Lie algebras for which $h = h^{\vee}$.

We should emphasize that only the true moduli space $\mv$ of vacua has the special geometry with \eqref{eq:tauij} defining a positive (outside the loci of singularities which signal the appearance of massless BPS particles) metric in the appropriate duality frame.
On the extended moduli space~$\mve$ the prepotential $\pp$ still
defines some kind of metric, but it cannot be positive everywhere
throughout the variety of masses. This is because the dependence on
masses is purely perturbative. Once we gauge the flavor symmetry (an
example of such gauging, promoting the~$A_1$ class I theory to the $\hat
D_4$ class II theory, will be discussed in Section~\ref{se:formDhattoA}), we correct the metric by the instanton contributions.

\subsection{Finite size effects}

Subjecting the gauge theory to some boundary conditions reveals more structure.

For example, we can compactify the four-dimensional ${\CalN}=2$ gauge theory on a~circle~${\BS}^{1}$ of radius~$R$. The resulting theory looks like a three-dimensional sigma model with the target space~${\pv}$
which is a hyperk\"ahler manifold of real dimension $4{\bbr}$. The hyperk\"ahler metric on
$\pv$ contains a~lot of interesting information about the particle content of the original four-dimensional theory.

The hyperk\"ahler structure on $\pv$
is a triplet of integrable complex structures, ${\bf I}$, ${\bf J}$, ${\bf K}$, such that
every linear combination $(a {\bf I} + b {\bf J} + c {\bf K})$ for $a^2 + b^2 + c^2 = 1$ is also an integrable complex structure, and a triplet of the corresponding
symplectic forms ${\om}_{\bf I}$, ${\om}_{\bf J}$, ${\om}_{\bf K}$ which are the K\"ahler forms for the metric $g$ on $\pv$ in the corresponding complex structures.

 Among the two-sphere of complex structures, one complex structure, which is usually called~$\bf I$, plays a special r\^ole. This complex structure and the corresponding $(2,0)$ symplectic form ${{\Omega}_{\bf I} = {\om}_{\bf J} + \ii {\om}_{\bf K}}$ are visible in the limit $R \to \infty$, where $\pv$ as a metric space collapses to $\mv$.
For very large but finite $R$ the manifold
$\pv$ looks like a fibration over $\mv$ whose fibers ${\fib}$, $u \in {\mv}$
are the abelian varieties (complex tori, which we describe in more detail momentarily) of diameter which scales like $R^{-1}$.

These fibers ${\fib}$ parametrize the ${\BS}^{1}$ holonomy of the abelian gauge fields $A^{\CalI}$ and their duals~$A^{D}_{\CalI}$. The reduction of the action (\ref{eq:boseff}) on ${\BS}^{1}$ gives
\begin{gather*}
 {\CalL}^{\rm eff3d} = \int \ii \operatorname{Re}{\tau}_{\CalI\CalJ} \, {\rm d}{\al}^{\CalI} \wedge B^{\CalJ}\nonumber \\
\hphantom{{\CalL}^{\rm eff3d} =}{} + \frac 12 \operatorname{Im}{\tau}_{\CalI\CalJ}\big( R {\rm d}{\ac}^{\CalI} \wedge \ast {\rm d}{\bar {\ac}}^{\CalJ} + R^{-1} {\rm d}{\al}^{\CalI} \wedge \ast {\rm d}{\al}^{\CalJ} + R B^{\CalI} \wedge \ast B^{\CalJ} \big),
\end{gather*}
where we denote by $\ast$ the three-dimensional Hodge star, and by $B^{\CalI}$ the curvature of the three-dimensional gauge field $B^{\CalI} = dA^{\CalI}_{\rm 3d}$ which is obtained by decomposing $A^{\CalI}_{\rm 4d} = {\al}^{\CalI} d{\theta} + A^{\CalI}_{\rm 3d}$. The scalar ${\al}^{\CalI}$ is actually circle-valued, since the gauge transformations
${\rm e}^{2\pi \ii n_{\CalI} {\theta}}$ shift it by $2\pi \ii n^{\CalI}$, $n^{\CalI} \in {\BZ}$. Next we dualize the three-dimensional abelian gauge field, by promoting $B^{\CalI}$
to the independent $2$-form, and coupling it to the dual scalar ${\be}_{\CalI}$, which is also circle-valued, in order to ensure the flux quantization of the original gauge curvature $B_\CalI$:
\begin{gather}
{\CalL}^{\rm eff3dd} = \int \ii \big( {\rm d}{\be}_{\CalJ} + \operatorname{Re}{\tau}_{\CalI\CalJ} \, {\rm d}{\al}^{\CalI}\big) \wedge B^{\CalJ} \nonumber\\
\hphantom{{\CalL}^{\rm eff3dd} =}{} + \frac 12 \operatorname{Im}{\tau}_{\CalI\CalJ} \big( R \, {\rm d}{\ac}^{\CalI} \wedge \ast {\rm d}{\bar {\ac}}^{\CalJ} + R^{-1} {\rm d}{\al}^{\CalI} \wedge \ast {\rm d}{\al}^{\CalJ} + R B^{\CalI} \wedge \ast B^{\CalJ} \big) \nonumber\\
\hphantom{{\CalL}^{\rm eff3dd}}{} \longrightarrow \frac{R}{2} \int \operatorname{Im}{\tau}_{\CalI\CalJ} \, {\rm d}{\ac}^{\CalI} \wedge \ast {\rm d}{\bar {\ac}}^{\CalJ} +
\frac{1}{2R} \int \big(\operatorname{Im}{\tau}^{-1}\big)^{\CalI\CalJ} {\rm d}z_{\CalI} \wedge \ast {\rm d}{\bar z}_{\CalJ}.
\label{eq:boseff3dd}
\end{gather}
In the last line we have integrated out the unconstrained Gaussian field $B_\CalI$. We also
 introduced the holomorphic coordinates
 \begin{equation*}
 z_{\CalI} = {\be}_{\CalI} + {\tau}_{\CalI\CalJ} {\al}^{\CalJ}, \qquad I = 1, \dots , {\bbr}
 \end{equation*}
 on the fibers ${\fib}$ of the fibration ${\pv} \to {\mv}$. Both
 ${\ac}_\CalI$ and $z_\CalI$ are the ${\bf I}$-holomorphic coordinates
 on~$\pv$. By construction, the coordinates $z_\CalI$ are subject to the periodic identifications:
 \begin{equation}
 z_{\CalI} \to z_{\CalI} + 2\pi \ii \big( n_{\CalI} + {\tau}_{\CalI\CalJ}m^{\CalJ} \big) , \qquad n_{\CalI}, m^{\CalI} \in {\BZ},
 \label{eq:ab}
 \end{equation}
 which confirm our assertion that the fibers ${\fib}$ of the map
 $\pv \to \mv$ are abelian varieties (recall that the metric
 $\operatorname{Im}{\tau} d{\ac} \otimes d{\bar {\ac}}$ is positive definite, the unitarity requirement).
The coordinates~${\ac}_{\CalI}, z_{\CalI}$ are the Darboux coordinates for the $(2,0)$ form ${\Omega}_{\bf I}$:
\begin{equation}
 {\Omega}_{\bf I} = \sum_{{\CalI}=1}^{{\bbr}} {\rm d}{\ac}^{\CalI} \wedge {\rm d}z_{\CalI} = \sum_{{\CalI}=1}^{{\bbr}} {\rm d}{\ac}^{D}_{\CalI} \wedge {\rm d}z_{D}^{\CalI}
 \label{eq:dadz}
 \end{equation}
 as well as the electric-magnetic duals ${\ac}^{D}_{\CalI}$ and
 $z_{D}^{\CalI} = \big({\tau}^{-1}\big)^{\CalI\CalJ} {\be}_{\CalJ} + {\al}^{\CalI}$.
 The fibers ${\fib}$ are Lagrangian with respect to $\Omega_{\bf I}$.

The metric on $\pv$, which enters the kinetic term in the
equation~(\ref{eq:boseff3dd}) is actually not the correct
hyperk\"ahler metric on $\pv$ for finite $R$. It receives corrections which are exponentially small with $R$,
\begin{equation}
\sim {\rm e}^{- M({\ac})R},
\label{eq:expbps}
\end{equation}
where $M({\ac})$ is the mass of a BPS particle in the
Hilbert space of the theory in four dimensions built over the vacuum $u \in \mv$. As is well-known, the masses
of some BPS particles vanish along some loci in $\mv$, where the corrections (\ref{eq:expbps}) become significant. One can show, however, that~$\Omega_{\bf I}$ does not get corrected by the finite size effects of these BPS particles.

One can also compactify the theory on a two-dimensional Riemann surface $\Sigma$ (with a partial twist along $\Sigma$, to preserve some supersymmetry). For $\Sigma$ other then two-torus this leads to the two-dimensional theory with ${\CalN}=2$ supersymmetry.
One has various sectors labeled by the electric and magnetic fluxes ${\bf e} = (e_{\CalI})$, ${\bf m} = \big(m^{\CalI}\big)$ through $\Sigma$. In the sector where $({\bf e},{\bf m}) \neq (0,0)$ one gets an effective superpotential~\cite{Losev:1997tp}:
\begin{equation*}
W_{({\bf e},{\bf m})} = \sum_{\CalI = 1}^{\bbr} e_{\CalI} \mathfrak {a}^{\CalI} + m^{\CalI} {\mathfrak{a}}^{D}_{\CalI},
\end{equation*}
which in four-dimensional theory is the central charge of the ${\CalN}=2$ superalgebra.
It is also equal to one of the \emph{action variables} of the Seiberg--Witten integrable system~\cite{Donagi:1997sr,Donagi:1995cf,Gorsky:1995zq}.

If one compactifies on a two-torus, then the resulting two-dimensional theory is the ${\CalN}=4$ supersymmetric sigma model whose target space $\pv$ is the hyperk\"ahler manifold.

It turns out to be quite useful to interpret the ${\CalN}=2$ theory on a four-dimensional manifold~$X$ which can be viewed as a two-torus fibration over some base $B$, as an effective sigma model with $B$ as a world sheet. In case where the fibration has singularities of real codimension one (for example, if $X$ is a product of a disk and a cylinder), then $B$ has a boundary, and the smoothness of the four-dimensional field configurations translates to particular boundary conditions in the two-dimensional sigma model~\cite{Nekrasov:2010ka}. An interesting class of such boundary conditions come from the so-called canonical coisotropic branes~\cite{Kapustin:2001ij, Kapustin:2006pk,Nekrasov:2010ka}. The algebra of the open string vertex operators corresponding to such a brane turns out to be the deformation quantization~\cite{MR2062626} of the algebra of holomorphic (in the appropriate complex structure) functions on $\pv$. Remarkably, when $\pv$ is an algebraic integrable system in one of the complex structures, one can apply
the fiberwise T-duality along the Liouville fibers, leading to the mirror perspective
on the quantization procedure. First of all, in the case of the Hitchin system the mirror manifold turns out to be the Hitchin system for the Langlands dual group. In the general case the mirror~${\pv}^{\vee}$ of the original hyperk\"ahler manifold $\pv$ is also expected to be an integrable system. The mirror of the canonical coisotropic brane is believed to be a holomorphic (in appropriate complex structure) Lagrangian brane. In the case of Hitchin system this brane is argued~\cite{Kapustin:2006pk} to be the so-called brane ${\CalB}_{\CalO}$ of opers, with evidence supported by exact computations in~\cite{Jeong:2023qdr,Jeong:2021rll,Jeong:2018qpc,Jeong:2020uxz,Lee:2020hfu, Nekrasov:2020qcq, Nekrasov:2021tik}.

\subsection{The appearance of an integrable system}\label{subsec:appear}

The complex symplectic manifold $({\pv}, {\Omega}_{\bf I})$, its projection ${\pi}\colon \pv \to \mv$ with Lagrangian fibers ${\fib} = {\pi}^{-1}(u)$, $u \in \mv$, which are principally polarized abelian varieties (the principal polarization comes from the restriction of ${\om}_{\bf I}$ onto the fibers) define what is known as the \emph{algebraic integrable system}~\cite{Donagi:1997sr,Donagi:1995am,Donagi:1995cf}. It is one of the possible complexifications of the familiar notion of the completely integrable system in the classical mechanics.

The other possibility, namely a complex symplectic manifold with the Lagrangian fibration whose fibers are the complex tori $( {\BC}^{\times})^{\bbr}$, is also realized
in the context of gauge theories. However, the base of such a system typically parametrizes
the space of mass parameters of the gauge theory.

The fibers ${\fib}$ are the Liouville tori, while $({\ac}_{\CalI}, z_{\CalI})$ are the
action-angle variables. The novelty of the complex case is the doubling of the possible choices of the action-angle variables with fixed Liouville fibration. Indeed, the fibers ${\fib}$ are the $2{\bbr}$-real-dimensional tori, therefore
in producing the action variables as in the Arnol'd--Liouville theorem one has a choice of $\bbr$ out of $2\bbr$ cycles in $H_1 ({\fib}, {\BZ})$. The lattice
$H_1 ({\fib}, {\BZ})$ has a symplectic form $\varpi$, which comes from the polarization, i.e., a properly normalized class of the restriction ${\om}_{\bf I} \vert_{{\fib}}$. It turns out that any Lagrangian sublattice $L$ in $H_1 ({\fib},{\BZ})$ defines a system of local coordinates $({\ac}_I)$ on the base $\mv$ near the point $u \in \mv$, as well as the conjugate angle-like coordinates $(z_{\CalI})$ on the fiber ${\fib}$ itself. Let ${\CalA}_{\CalI}$ be the integral basis of this sublattice $L \subset H_1 ({\fib},{\BZ})$. Then
\begin{equation}
{\rm d}{\ac}^{\CalI} = \oint_{{\CalA}_{\CalI}} {\Omega}_{\bf I}.
\label{eq:aper}
\end{equation}
One can also define
\begin{equation}
{\rm d}{\ac}^{D}_{\CalI} = \oint_{{\CalB}^{\CalI}} {\Omega}_{\bf I},
\label{eq:bper}
\end{equation}
where ${\CalB}^{\CalI}$ is the basis in the dual sublattice $L^{\vee} \subset H_1 ({\fib},{\BZ})$, such that
\begin{equation}
{\varpi}( {\CalA}_{\CalI}, {\CalA}_{\CalJ} ) = {\varpi}\big( {\CalB}^{\CalI}, {\CalB}^{\CalJ} \big) = 0, \qquad {\varpi}\big( {\CalA}_{\CalI}, {\CalB}^{\CalJ} \big) = {\de}_{\CalI}^{\CalJ}.
\label{eq:dualba}
\end{equation}
One then shows that
\begin{equation}
\sum_{{\CalI}=1}^{\bbr} {\rm d}{\ac}^{\CalI} \wedge {\rm d}{\ac}_{\CalI}^{D} \equiv 0
\label{eq:symlag}
\end{equation}
on $\mv$, which, in turn, implies (\ref{eq:spcf}).
The coordinates $z_{\CalI}$ along ${\fib}$ are defined using (\ref{eq:dadz}) with the normalization (\ref{eq:ab}) that half of the periods of $z_{\CalI}$ are in $2\pi \ii {\BZ}$.

The integrable systems which one encounters in the classical mechanics are rarely given in the form of the action-angle variables. Usually one has the phase space
$\pv$, the symplectic form~${\Omega}_{\bf I}$, perhaps some Darboux coordinates
\[
{\Omega}_{\bf I} = \sum_{{\CalI}=1}^{\bbr} {\rm d}p_{\CalI} \wedge {\rm d}q^{\CalI}
\]
 and the collection
of Poisson-commuting functionally independent
Hamiltonians $U_{1}(p,q), \dots, \allowbreak U_{\bbr}(p,q)$. One then looks for the action-angle coordinates, i.e., the Darboux coordinates $({\ac}, z)$, such that the Hamiltonians $U_{\CalI}(p,q) = u_{\CalI}({\ac})$ depend only on ${\ac}$, the action variables. The Hamiltonian evolution then linearizes on the fibers ${\fib}$, which are the level sets of the Hamiltonians. The motion is a constant velocity motion in the $z$ coordinates:
\[
z_{\CalI}(t) = z_{\CalI}(0) + \sum_{\CalJ = 1}^{\bf r} t_{\CalJ} \frac{{\partial}u_{\CalJ}}{{\partial}{\ac}_{\CalI}}.
\]
It is interesting to study the level sets ${\fib}$ of the Hamiltonians, the Liouville tori. The algebraic integrable systems are such, that the fibers can be compactified to become the
polarized abelian varieties. Where do the polarized abelian varieties come from?

\subsection{Integrable systems from classical gauge theories}

One source of the polarized abelian varieties are the Jacobians of the algebraic curves. The Liouville tori of algebraic integrable systems can be often found inside the Jacobians of the algebraic curves, constructed while solving some classical gauge field equations.

\subsubsection{Hitchin system}

There is an interesting class of algebraic integrable systems for which the
Liouville tori are precisely these Jacobians. Take the $U(N)$ Hitchin system
on a genus $g$ Riemann surface. The phase space $\pv$ is
the cotangent bundle (up to a birational transformation) to the moduli space~$M_{N,c}$
of holomorphic rank $N$ vector bundles $E$ over $\Sigma$ with fixed first Chern class $c = c_1 (E)$. It is convenient to take $(c,N) =1$ to avoid complications coming from the reducible connections.

In the complex structure $\bf I$ the holomorphic coordinates on $\pv$
are $\big({\bar A}, {\Phi}\big)$, where ${\bar\partial} + {\bar A}$ is the $(0,1)$-connection on the smooth vector bundle $E$ which endows it with the complex
structure, and $\Phi \in \operatorname{End}(E) \otimes {\Omega}^{1,0}({\Sigma})$ is
the holomorphic Higgs field
\begin{equation}
{\bar\partial}{\Phi} + \big[ {\bar A}, {\Phi} \big] = 0.
\label{eq:hitcheq}
\end{equation}
The symplectic form on $\pv$ comes from the $(2,0)$ symplectic form on the space of all smooth pairs $({\bar A}, {\Phi})$
\begin{equation*}
{\Omega}_{\bf I} = \int_{\Sigma} \tr {\de} {\Phi} \wedge {\de} {\bar A}
\end{equation*}
by the symplectic reduction with respect to the action of the gauge group:
\[
g\colon \ \big({\bar A}, {\Phi}\big) \longrightarrow \big( g^{-1} {\bar A} g + g^{-1}{\bar\partial}g , g^{-1} {\Phi}g\big).
\]
The set of Poisson-commuting Hamiltonians is given by
\begin{equation}
U_{i,{\ba}} = \int_{\Sigma} {\nu}_{i,{\ba}} \tr {\Phi}^{i}, \qquad i = 1, \dots , N,
 \label{eq:hitcham}
\end{equation}
where ${\nu}_{i,{\ba}} \in H^{0,1} \big( {\Sigma}, K_{\Sigma}^{\otimes (1-i)} \big)$, ${\ba} = 1, \dots , (2i-1) (g-1) + {\delta}_{i,1}$ form a basis in the space of holomorphic $(1-i,1)$-differentials. Fixing the values $u_{i, {\ba}}$ of all the Hamiltonians $U_{i, {\ba}}$ gives us a~point $u \in {\mv}$ in the vector space
\begin{equation*}
{\mv} = \bigoplus_{i} H^{0,1} \big( {\Sigma}, K_{\Sigma}^{\otimes (1-i)} \big).
\end{equation*}
One defines the \emph{spectral curve} $C_{u} \subset T^{*}{\Sigma}$
as the zero locus of the characteristic polynomial of~$\Phi$:%
\begin{equation}
\operatorname{Det}( {\Phi} - {\lam} ) = 0.
\label{eq:detphi}
\end{equation}
It is a holomorphic curve thanks to (\ref{eq:hitcheq}), which is invariant under the Hamiltonian flows generated by the Hamiltonians (\ref{eq:hitcham}). The curve $C$ is an $N$-sheeted cover of $\Sigma$
\begin{equation*}
{\pi}\colon \ C_{u} \to \Sigma.
\end{equation*}
Its genus can be computed using the Riemann--Hurwitz formula
\[
2 - 2g_{C_{u}} = N ( 2 - 2g_{\Sigma}) - {\delta},
\]
where ${\delta} = 2N(N-1) (g_{\Sigma} - 1)$ is the number of branch points. The latter is the number of zeroes of the discriminant of the polynomial (\ref{eq:detphi}), which is a holomorphic $N(N-1)$-differential on~$\Sigma$. Thus
\[
g_{C_{u}} = N^2 (g_{\Sigma} -1 ) + 1.
\]
The Jacobian of $C$ is thus an abelian variety of dimension
\begin{equation}
g_{C_{u}} = g_{\Sigma} + \sum_{j=2}^{N} (2j-1) (g_{\Sigma}-1),
\label{eq:gc}
\end{equation}
which is equal to the dimension of the base $\mv$ of the Hitchin fibration. The fibers $\fib$ of the Hitchin fibration are thus the Jacobians of the corresponding spectral curves.

One generalization is to study the ${\rm SL}(N)$ Hitchin system. In this case the corresponding rank~$N$ vector bundles have the trivial determinant, and the corresponding Higgs field is traceless. The base of the Hitchin fibration now has the dimension
$\big(N^2-1\big)(g_{\Sigma}-1)$, the equation (\ref{eq:detphi}) has vanishing $\propto {\lam}^{N-1}$ term, and the fibers $\fib$ are not the full Jacobians of the spectral curve~$C_{u}$, which still has the genus $g_{C_{u}}$ (\ref{eq:gc}) but the kernel $J_0$ of the map ${\pi}_{*}\colon \operatorname{Jac}(C_{u}) \to \operatorname{Jac}({\Sigma})$, which sends the degree zero line bundle $L$ on $C_{u}$ to the line bundle ${\mathscr L} = \operatorname{Det}{\pi}_{*}L$ on $\Sigma$, whose fiber ${\mathscr L}_{z}$ over the point $z \in \Sigma$ is the tensor product of the fibers $L_y$ of $L$ over all preimages of $z$:
\begin{equation*}
{\mathscr L}_{z} = \bigotimes_{y \in {\pi}^{-1}(z)} L_{y}.
\end{equation*}
The Hitchin system can be defined~\cite{Hitchin:1987mz} for any algebraic Lie group $G$, with the maximal torus~$T$. Let $\g = \operatorname{Lie}(G)$, $\h = \operatorname{Lie}(T)$. The Hitchin space is the moduli space of stable pairs $({\CalP}, {\Phi})$, where~${\CalP}$ is a holomorphic $G$-bundle over $\Sigma$, and $\Phi$ is a holomorphic $(1,0)$-form on $\Sigma$, valued in the bundle of Lie algebras $\g$, associated with $\CalP$ via the adjoint representation:
\[
{\Phi} \in H^{0}({\Sigma}, K_{\Sigma}\otimes {\rm ad}({\CalP})).
\]
The
Hitchin fibration is 
defined by fixing
the gauge-invariant polynomials $P_{j}({\Phi}) \in H^0 \big( {\Sigma}, K_{\Sigma}^{\otimes d_{j}} \big)$ of the $\g$-valued Higgs field $\Phi$:
\[
u = ( P_{j}({\Phi}) )_{j=1}^{r}
\in \bigoplus_{j=1}^{r} {\BC}^{(2d_{j}-1)(g_{\Sigma}-1)} = {\mv},
\]
where $d_j$'s are the
 degrees of basic Ad-invariant polynomials on $\g$.

 The fibers of the Hitchin fibration are now trickier to define. First of all, there is no preferred notion of the spectral curve. For some gauge groups one can use the minuscule representation, but this is not always available.

One option
is to consider the so-called \emph{cameral curve} ${\CalC}_{u}$, which is a $W({\g})$-cover of the base curve $\Sigma$.
The points of the cameral curve ${\CalC}_{u}$ are, over generic $z$, the pairs
$( {\varphi}, z)$, where
$z \in \Sigma$ and $\varphi \in \h$ is the element of the fixed Cartan subalgebra $\h \subset \g$ which is conjugate to the Higgs field~$\Phi (z)$. This definition makes sense for the points $z \in \Sigma$ for which $\Phi (z)$ is semi-simple, i.e., belongs to the $\operatorname{ad}(G)$-orbit
of an element in $\h$. If this is not the case (e.g., $\Phi(z)$ is conjugate to a Jordan block in the ${\rm GL}(N)$ case), one can find an appropriate representative in~$\h$ by modifying the equivalence relation (e.g., two matrices are equivalent if their characteristic polynomials coincide).
To stress the fact that ${\CalC}_{u}$ depends on $u$ which is the set of
 holomorphic ${\rm d}_j$-differentials $P_{j}({\Phi})$ .

Over ${\CalC}_{u}$ so defined one has $r$ line bundles, ${\CalL}_i$, $i=1, \dots, r$, which correspond to the fundamental weights $\lam_{i} \in \h^*$. The line bundle ${\CalL}_i$ is a subbundle in the holomorphic vector bundle~${\bf R}_{i} = R_{i} \times_{G} {\CalP}$, associated with $\CalP$ via the $i$-th fundamental representation $R_i$ of $G$. The fiber of ${\CalL}_i \subset {\bf R}_{i}$ over
$({\varphi},z)$ is the eigenspace corresponding to the eigenvalue $ {\lam}_{i}( {\varphi})$.

{}To any weight vector ${\lam} \in \Lambda$ a line bundle ${\CalL}_{\lam}$ over $\CalC_{u}$ can be associated:
\begin{equation*}
{\lam} = \sum_{i=1}^{r} n_{i} {\lam}_{i} \mapsto {\CalL}_{\lam} = \bigotimes_{i=1}^{r} {\CalL}_{i}^{\otimes n_{i}}.
\end{equation*}
In a more physical language, the Hitchin moduli space
is the quotient of the space of pairs~$\big({\bar A}, {\Phi}\big)$,
where ${\bar A}$ is a $(0,1)$-connection on smooth principal $G$-bundle ${\CalP}$ over $\Sigma$, and $\Phi$ is a $(1,0)$ $\g$-valued form, which are compatible,
i.e., solve the equation~(\ref{eq:hitcheq}),
and are considered up to the $G$-gauge transformations:
\[
g\colon \ \big({\bar A}, {\Phi}\big) \mapsto \big( g^{-1}{\bar\pa}g + {\rm Ad}_{g} {\bar A} , {\rm Ad}_{g} {\Phi}\big).
\]
By fixing the partial gauge ${\Phi} = {\varphi} \in \h$ for fixed $\h \subset \g$, one reduces the gauge invariance from~$G$ to
$N(T)$. The equation~(\ref{eq:hitcheq}) imply that in this gauge $\bar A$ is a $T$-connection $\bar A = \bar a$, with the~$T$ subgroup of $N(T)$ acting by the $T$-gauge transformations $\bar a \mapsto \bar a + {\bar\pa} {\chi}$, ${\rm e}^{\chi} \in T$. On $\Sigma$ the $T$-valued gauge field and the $\h$-valued Higgs field $\varphi$ are not well-defined, since there are the $W({\g}) = N(T)/T$ remaining gauge transformations.
On $\CalC$, however, both $\varphi$ and $\bar a$ are well-defined. In fact, ${\bar a}$ defines on $\CalC_{u}$ a holomorphic principal $T$-bundle $\CalT$, so that
$L_i = {\CalT}^{{\lam}_{i}}$. The $T$-bundle $\CalT$ is $W({\g})$-equivariant. This is the translation of the fact that the Weyl group $W({\g})$ acts simultaneously on $\varphi$ and $\bar a$. The isomorphism (properly understood at the ramification points)
\[
\hbox{\vbox{\hbox{Holomorphic}
\hbox{principal $G$-bundles $\CalP$ on $\Sigma$,}
\hbox{holomorphic Higgs fields}
\hbox{$\Phi \in H^{0}({\Sigma}, K_{\Sigma}\otimes {\rm ad}({\CalP}))$}}\vbox{\hbox{}\hbox{
\qquad $\Leftrightarrow$\qquad}\hbox{}}\vbox{\hbox{$W({\g})$-covers $\CalC$ of $\Sigma$,}\hbox{$\CalC \subset T^*{\Sigma}\otimes \h$,}\hbox{holomorphic $W({\g})$-equivariant}
\hbox{principal $T$-bundles on $\CalC$}}}
\]
allows to represent the Hitchin moduli space as a fibration over the vector space $\mv$, whose points are the $W({\g})$-invariant curves $\CalC_{u}$ sitting in the tensor product $T^*\Sigma \otimes \h$ (this is almost a~tautology: a $W({\g})$-invariant curve in
$T^*\Sigma \otimes \h$ is a curve in $T^*\Sigma \otimes \h /W({\g})$, i.e., a~holomorphic section of the vector bundle $T^*{\Sigma} \otimes {\BC}[{\h}]^{W({\g})}$).

The fiber $\fib$ of the Liouville fibration (which is called Hitchin's fibration
in this case) is a~generalized Prym variety, which is, roughly speaking,
\begin{equation}
{\fib} \approx {\Hom}_{W({\g})} \left( {\Lambda}, {\rm Pic}({\CalC}_{u}) \right) =
{\Bun}_{T}({\CalC}_{u})^{W({\g})}.
\label{eq:prymcc}
\end{equation}
The papers~\cite{Donagi:1995alg,Donagi:1998vx,Donagi:2000dr}
correct the equation~(\ref{eq:prymcc}) in a couple of subtle points as well as provide the additional theory.

\subsubsection{Instanton moduli spaces as integrable systems}

Hitchin's equations (\ref{eq:hitcheq}), for flat $\Sigma$, are the dimensional reduction of the instanton (or anti-self-duality) equations from four dimensions. It turns out that one can get an integrable system directly from the moduli spaces of four-dimensional instantons, or
three-dimensional monopoles (examples of integrable systems on moduli spaces of instantons were found in~\cite{NikThesis:1996}).

We only briefly sketch the constructions here.

Let $S$ be an elliptic K3 manifold, i.e., an algebraic surface, with the holomorphic ${\om}_{S}^{2,0}$ form, and with the projection ${\pi}\colon S \to {\BC\BP}^1$ whose fibers ${\pi}^{-1}(z)$, $z \in {\BC\BP}^{1}$ are the elliptic curves ${\ec}_{z}$ (generically nonsingular). One can endow $S$ with the hyperk\"ahler metric. Consider the moduli space ${\pv} = {\CalM}_{N} (G)$
of charge $N$ $G$-instantons on $S$, i.e., the solutions to the system of partial differential equations
\begin{gather*}
 F_{A} \wedge {\omega}_{\bf I} = F_{A} \wedge {\omega}_{\bf J} = F_{A} \wedge {\omega}_{\bf K} = 0 , \\
 F^{0,2}_{A} = 0
\end{gather*}
(the last equation is a linear combination of the $\bf J$ and $\bf K$ equations from the first line)
of fixed instanton charge $N \geq 0$:
\begin{equation*}
-\frac{1}{8\pi^2} \int_{S} \tr F_{A} \wedge F_{A} = N.
\end{equation*}
Here $G$ is some compact simply-connected simple Lie group, which has a simply-laced Lie algebra~$\g$. The moduli space ${\pv} = {\CalM}_{N} (G)$ is also hyperk\"ahler, in particular it is holomorphic symplectic, with the $(2,0)$-form given by
\begin{equation*}
{\Omega}_{\bf I}^{N, G} = \int_{S} {\om}_{S}^{2,0} \wedge \tr {\de}{\bar A} \wedge {\de} {\bar A}.
\end{equation*}
The integrable system structure is obtained by studying the restriction
of the instanton gauge field on the elliptic fibers, where generically
they define a point in the coarse moduli space $\Bun_{\bf G}({\ec}_{z})$ of semi-stable principal holomorphic $\bf
G$-bundles on the fiber, see Appendix~\ref{se:conjugacy}. Thanks to E.~Loojienga's theorem, this moduli
space is a weighted projective space, which can be identified for
different non-singular fibers. One gets thus a section of the locally trivial bundle~of
\[
{\Pi}\colon \ {\mathscr P} = \bigcup_{z \in {\BC\BP}^{1}} \Bun_{\bf G}({\ec}_{z}) \longrightarrow {\BC\BP}^{1}.
\]
One has to be careful at the singular fibers. The base $\mv$ of the integrable systems
is the properly compactified moduli space of the holomorphic sections
${\si}\colon{\BC\BP}^{1} \longrightarrow
 {\mathscr P}$ of appropriate degree with some ramification conditions at the discriminant locus of the original elliptic fibration $\pi$.

 In this work we shall not encounter these difficulties.

 In fact, as we shall explain in more detail in Section~\ref{se:integrable}, the moduli spaces of vacua of the quiver gauge
 theories we study lead to the integrable systems which arise from the
 the moduli spaces of $G$-monopoles on ${\BR}^{2} \times
 {\BS}^{1}$ for class I theories with $\Gq = \bG$,
or from the moduli spaces of $G$-instantons on ${\BR}^{2} \times {\BT}^{2}$ for
 class II theories with $\Gq = \hat \bG$.
 Here $G$ is a compact Lie group, whose complexification is the complex
 simple Lie group $\bG$.

 The moduli space $\pv$ of $G$-instantons, viewed
 in the complex structure where ${\BR}^{2} \times {\BT}^{2} = {\BC}^{1} \times {\ec}$,
 is birational to the moduli space of semi-stable holomorphic $\bG$-bundles on ${\Cpx} \times {\ec}$, with fixed trivialization at ${\infty} \times {\ec}$. The moduli space ${\pv}$ projects down to the moduli space $\mv$ of quasimaps from $\Cpx$ to
 the moduli space of semi-stable holomorphic bundles $\Bun_{\bG}(\ec)$ on a fixed elliptic
 curve $\ec$. The moduli space of monopoles maps to the moduli space
 of quasimaps with prescribed singularities on ${\Cpx}$ to $\Bg =
 {\bG}/{\rm Ad}({\bG}) = {\bT}/W({\g})$.

\subsection{Extended moduli space as a complex integrable system}

The extended moduli space $\mve$ is a base of a complex, but not algebraic, integrable system ${\pve} \to {\mve}$. The Liouville tori of this integrable system
are acted on by an algebraic torus~$({\BC}^{\times})^{\Ver}$, so that the quotients are the compact abelian varieties, the Liouville tori fibered over~$\mv$.
The symplectic quotient of $\pve$ with respect to $({\BC}^{\times})^{\Ver}$ at some level
of the moment map, which is linearly determined by the values of the bi-fundamental masses, gives $\pv$. Recall that Duistermaat--Heckmann theorem~\cite{Duistermaat:1982vw} then implies that the cohomology class $[{\Omega}_{\bf I}]$ of the $(2,0)$-symplectic form ${\Omega}_{\bf I}$ on $\mv$ is linear with masses.

{\em The physics behind the reason $\pve \to \mve$ is not an algebraic integrable system is that the kinetic term for the instanton/monopole zero modes in the
$\big({\BR}^{2} \times T^{2}\big)/\big({\BR}^{2} \times S^{1}\big)$ geometry diverges. These modes are non-dynamical in the effective three dynamical theory obtained by compactifying our $\CalN=2$ theory from four to three dimensions. The electric-magnetic duality of dynamical vector multiplets in four dimensions leading to the algebraic integrable system on the moduli space of vacua of the corresponding three-dimensional theory is therefore broken}.

\section{The limit shape equations}\label{se:limit-shape}

In this section we return to the microscopic analysis of our gauge theory. Recall that the ${\CalN}=2$ supersymmetry algebra is generated by four supercharges ${\sc}_{{\al}i}$, ${\al}=1,2$, $i=1,2$ of the left and by four supercharges ${\bar\sc}_{\dot\al}^{i}$, ${\dot\al}=1,2$ of the right chirality. The prepotential ${\CalF}({\CalA})$ of the theory is a~function of the superfield ${\CalA}$ which is annihilated by ${\sc}_{{\al}i}$'s. We shall now focus on the observables which are in the cohomology of one of the ${\sc}_{{\al}i}$ supercharges, which we shall call simply $\sc$.

\subsection{The amplitude functions}\label{se:amplitude}

The basic such observable is the scalar ${\Phi}_{i}$ in the vector multiplet. More precisely, any gauge invariant functional, in particular the local operator $P({\Phi}_{i}({\bf x}))$, where $P$ is some invariant polynomial on the Lie algebra of ${\rm SU}(\BB_{i})$,
and $\bf x$ is a point in space-time, is annihilated by $\sc$. Moreover, the observables
$P({\Phi}_{i}({\bf x}))$ and $P({\Phi}_{i}({\bf x}'))$ for two different points ${\bf x}$ and
${\bf x}'$ are in the same $\sc$-cohomology class. Therefore, one may talk about the
vacuum expectation value of~$P({\Phi}_{i})$ without specifying the point $\bf x$.

Consider the observables ${\CalO}_{n, i} = \tr _{{\bv}_{i}} {\Phi}_{i}^{n}$.
Form the generating function
\begin{equation}
{\y}_{i} (x) = x^{{\bv}_{i}} \exp \left( - \sum_{n=1}^{\infty}
\frac {\langle {\CalO}_{n,i}\rangle_{u}}{n} x^{-n} \right),
\label{eq:asyofx}
\end{equation}
 which turns out to be well-behaved for sufficiently large $x$. We shall denote the $x$-plane
 where~${\y}_{i}(x)$ are defined, by $\Cx$. Actually, the
 analytic continuation in the $x$ variable gives us
the set~${\bf \y}(x)$ of multi-valued analytic functions on
 $\Cx$,
\[
{\bf \y}(x) = ( {\y}_{i}(x) )_{i \in \Ver}.
\]
This set of multi-valued functions
captures the vacuum expectation values of all the local gauge invariant observables commuting with the supercharge ${\sc}$.

\begin{Remark}
The general relation between the amplitude functions ${\y}_{i}(x)$ and the polynomials~${\bt}_{i}(x)$ generating the first non-trivial Casimirs of the gauge group $\Gg$ is
\[
{\bt}_{i}(x) =( {\y}_{i}(x) )_{+},
\]
where $(\dots )_{+}$ denotes the polynomial part.

In what follows we shall use another set of $(T_{i}(x))_{i \in \Ver}$ polynomials, $\deg T_{i}(x) = {\bv}_{i}$, which are not monic. The coefficients of $T_{i}(x)$ are related to the coefficients of ${\bt}_{i}(x)$ by a ``mirror map'' change of variables, which will become clear in the course of our exposition.
\end{Remark}

\begin{Remark}
In the way we defined these functions, the information about the vacuum expectation values
of these observables is contained in the expansion of ${\y}_{i}(x)$ near $x = \infty$ on the physical sheet of these functions. It would be interesting to see whether the expansion
at $x = \infty$ of the branches of ${\y}_{i}(x)$ contains information about the vevs of the chiral observables of the theories, related to the one we started with via some version of $S$-duality.
\end{Remark}

The functions ${\y}_{i}(x)$ are the integral transforms of the densities
\[
{\rho}_{i}({\xr}), \qquad {\xr} \in {\BR}, \quad i \in \Ver,
\]
which describe the combinatorics of the set of fixed points
of the symmetry group action on the instanton moduli space used in the localization approach to the calculation
of the supersymmetric partition function of the gauge theory.
For the introduction to the subject see~\cite{Losev:2003py, Nekrasov:2002qd, Nekrasov:2002dd, Nekrasov:2003zj} and for the novel applications and refinements~\cite{Alday:2009aq, Pestun:2007rz}.

We now write down the equations obeyed by the amplitude functions, the so-called \emph{limit shape equations}, generalizing the limit shape equations studied in~\cite{Nekrasov:2003rj, Nekrasov:2004vw, Shadchin:2005mx, Shadchin:2005cc, Shadchin:2005hp}.
We shall solve the limit shape equations using
the analytic properties of the amplitude functions. One finds that the analytic continuation of these functions is governed by the monodromy group, which we shall call the iWeyl group (the \emph{instanton Weyl group}).

The iWeyl group is the Weyl group $W(\gq)$. For the class I theories
$W(\gq)$ is the finite Weyl group $W({\g})$ of the corresponding ADE
simple Lie algebra $\g$, for the class II theories the iWeyl group turns
out to be the affine Weyl group $W(\hat \g )$ of
the corresponding affine Lie algebra $\hat\g = \gq $. The Weyl group of $\Gli$ shows
up in the class II* theories.

We solve the limit shape equations by constructing the iWeyl invariants ${\crf}_{j} ({\y}(x))$
of ${\y}_{i}(x + {\mu}_{i})$, for the appropriate shifts ${\mu}_{i}$,
and showing that these invariants are polynomials in $x$,
\begin{equation*}
{\crf}_{j}({\y}(x)) = T_{j}(x), \qquad j \in \Ver.
\end{equation*}
For the class II theories the invariants ${\crf}_j$ are convergent power series in ${\qe}_j$. Moreover, in each order
in expansion in ${\qe} = \prod_i {\qe}_{i}^{a_i}$ they are finite Laurent polynomials in
${\y}_j$'s. For the class II* theories
the invariants ${\crf}_j$ are convergent power series in ${\qe}_j$,
and finite Laurent polynomials in ${\y}_{j}(x + {\mu}_{j} + l m^* )$,
for a finite collection of integers $l \in {\BZ}$, again in every order in $\qe$ expansion.
For the class I theories
the functions ${\crf}_j$ are polynomials in ${\bq}_{j}(x)$ and Laurent polynomials in~${\y}_{j}(x)$.

\subsection{The densities and the amplitude functions}

The amplitude functions ${\y}_{i}(x)$ are the multi-valued
analytic functions, which we defined, for large $x$, via
equation~(\ref{eq:asyofx}):
\begin{equation}
{\y}_{i}(x) = \exp \langle \tr _{{\bv}_{i}}\log ( x - {\Phi}_{i} ) \rangle_{u}.
\label{eq:yiofx}
\end{equation}
One shows, using the fixed point techniques that
\begin{equation}
{\y}_{i}(x) = \exp \int_{\BR} {\rm d}{\xr} {\rho}_{i}({\xr}) \log (x - {\xr}),
\label{eq:rhoy}
\end{equation}
where the \emph{density} function ${\rho}_{i}(x)$ has
compact support which consists of ${\bv}_i$ intervals
\[
\operatorname{supp} {\rho}_{i} = \bigcup_{{\ba}=1}^{{\bv}_{i}}
I_{i, {\ba}}.
\]
The intervals $I_{i, {\ba}}$ should be thought of as the ``fattened'' versions of the
eigenvalues ${\ac}_{i, {\ba}}$. More precisely,
\begin{equation}
1 = \int_{I_{i, {\ba}}} {\rho}_{i}({\xr}) \, {\rm d}{\xr}, \qquad
{\ac}_{i, {\ba}} = \int_{I_{i, {\ba}}} {\xr} {\rho}_{i}({\xr}) \, {\rm d}{\xr}.
\label{eq:aia}
\end{equation}
The functions ${\y}_{i}(x)$ have, therefore, the cuts at the intervals $I_{i, {\ba}}$, with the limit values
${\y}^{\pm}_{i}$ of the ${\y}_{i}$ function at the top
and the bottom banks of the interval $I_{i, {\ba}}$
being related via
\[
{\y}_{i}^{+}(x)/{\y}_{i}^{-}(x) = \exp \bigg(2\pi \ii
\int^{x}_{-\infty} {\rho}_{i}({\xr})\, {\rm d}{\xr}\bigg).
\]
One then analytically continues ${\y}_{i}(x)$ across the cuts, which leads to the set of the multi-valued analytic
functions. We shall describe this analytic continuation in detail in the coming section.

\subsection{The special coordinates}

\begin{figure}
 \centering
 \includegraphics[width=10cm]{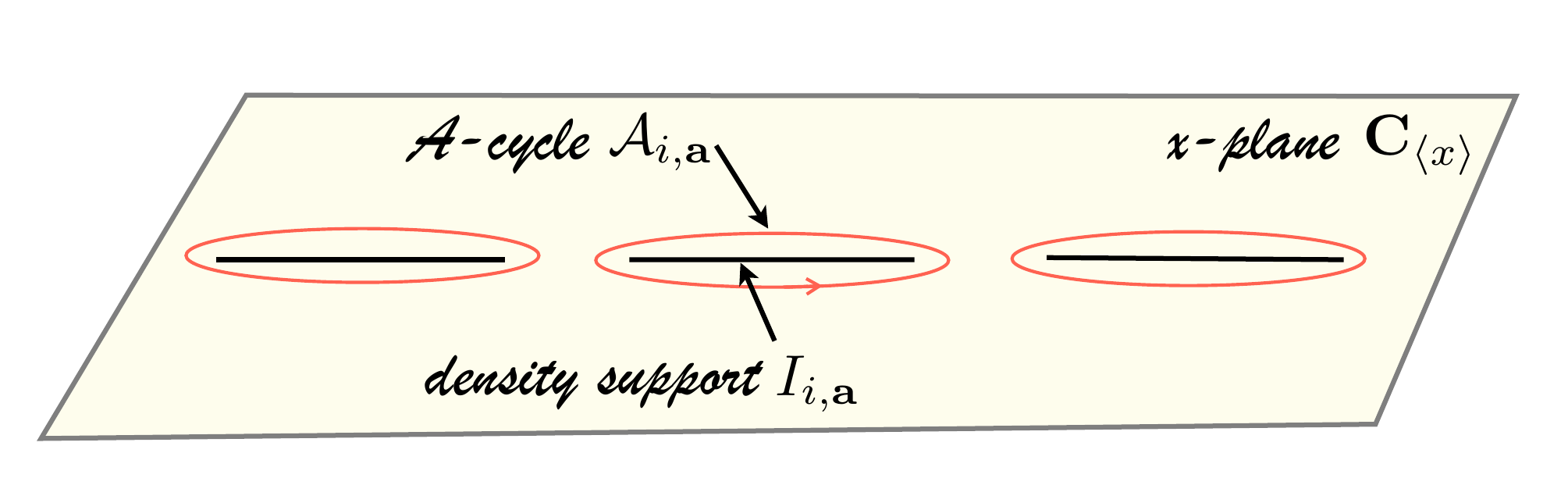}
 \caption{Support intervals $I_{i,\ba}$ of the densities $\rho_{i}$
 and the cycles $A_{i\ba}$.
}
 \label{fig:acycle}
\end{figure}

From the equations~\eqref{eq:yiofx} and \eqref{eq:aia} one derives
\begin{equation}
{\ac}_{i, {\ba}} = \frac{1}{2\pi \ii} \oint_{A_{i{\ba}}} x \, {\rm d} \log {\y}_{i}(x),
\label{eq:spco}
\end{equation}
where $A_{i{\ba}}$ is a small loop surrounding the cut
$I_{i,{\ba}}$, see Figure~\ref{fig:acycle}.

\section{The limit shape prepotential}

The prepotential of the low-energy theory is expressed in terms of the densities as follows:
\begin{align*}
{\CalF}({\ac}; m ; {\tau}) = & - \iint_{{\BR}^2} {\rm d}{\xr}' {\rm d}{\xr}''
\sum_{i \in \Ver} {\rho}_{i}({\xr}') {\rho}_{i}({\xr}'') {\CalK}({\xr}' - {\xr}'') \\
& + \iint_{{\BR}^2} {\rm d}{\xr}' {\rm d}{\xr}'' \sum_{e \in \Edg}
{\rho}_{t(e)}({\xr}') {\rho}_{s(e)}({\xr}'') {\CalK}({\xr}' - {\xr}'' + m_{e}) \\
& + \sum_{i \in \Ver} \int_{{\BR}} {\rm d}{\xr} \, {\rho}_{i}({\xr}) \Bigg(
 ( \log {\qe}_{i}) \frac{{\xr}^{2}}{2} + \sum_{{\fe} =1}^{{\bw}_{i}}
 {\CalK}({\xr} - m_{i, {\fe}}) \Bigg) \\
& + \sum_{i,\ba} \mathfrak{b}_{i,\ba}
\Bigg( 1- \int_{I_{i, \ba}} \rho_i({\xr}) \, {\rm d}{\xr} \Bigg)
 + \sum_{i, \ba} \ac_{i,\ba}^{D} \Bigg( \ac_{i,\ba} - \int_{I_{i,\ba}} {\xr}
\rho_i({\xr}) \, {\rm d}{\xr} \Bigg),
\end{align*}
where the constraints (\ref{eq:aia}) are incorporated by the last two
lines via Lagrangian multipliers~$\mathfrak{b}_{i,\ba}$,~$\ac_{i,\ba}^{D}$,
\begin{equation*}
{\CalK}(x) = \frac{x^{2}}{2} \left( \log\left (\frac{x}{\Lambda_{\rm UV}} \right) - \frac 32 \right)
\end{equation*}
and $\Lambda_{\rm UV}$ is the UV cutoff scale. In fact, the $\Lambda_{\rm UV}$-dependence drops out for the theories solving the ${\be}_{i}=0$ equations. However, in the intermediate formulae we keep the explicit $\Lambda_{\rm UV}$-dependence.

\subsection{The limit shape equations}
Originally the limit shape equations were derived as the variational equations on $\CalF$ \cite{Nekrasov:2003rj, Nekrasov:2004vw}. These are linear integral equations on the densities ${\rho}_{i}(x)$: for any $x \in I_{i,{\ba}}$ the following
should hold
\begin{gather}
 -2 \int_{\BR} {\rho}_{i}({\xr}) {\CalK}(x-{\xr}) \, {\rm d}{\xr} + \sum_{e\colon t(e) = i}
\int_{\BR} {\rho}_{s(e)}({\xr}) {\CalK}(x-{\xr}+ m_{e} ) \, {\rm d}{\xr} \label{eq:limshape}\\
 \qquad {}+ \sum_{e\colon s(e) = i} \int_{\BR} {\rho}_{t(e)}({\xr})
{\CalK}(x-{\xr}- m_{e} ) \, {\rm d}{\xr} + \sum_{{\fe}=1}^{{\bw}_{i}}
{\CalK} (x - m_{i, \fe}) + \frac{x^2}{2} \log ({\qe}_{i}) = x {\ac}_{i, {\ba}}^{D} + {\mathfrak {b}}_{i, {\ba}},\nonumber
\end{gather}
where ${\ac}_{i, {\ba}}^{D}$, ${\mathfrak {b}}_{i, {\ba}}$ are some constants, the Lagrange multipliers for the conditions (\ref{eq:aia}) which are determined from the solution.
Actually, ${\ac}_{i, {\ba}}^{D}$ \emph{is} the dual special coordinate, cf.~(\ref{eq:spcf}),
\begin{equation}
\frac{{\partial}{\CalF}}{{\partial}{\ac}_{i, {\ba}}} = {\ac}_{i, {\ba}}^{D}.
\label{eq:duala}
\end{equation}
We find it useful to rewrite the second derivative with respect to $x$ of the linear integral equations~(\ref{eq:limshape}) on $\rho_{i}({\xr})$ as the non-linear polynomial difference equations on the amplitudes ${\y}_{i}(x)$:%
\begin{equation}
{\y}^{+}_{i}(x) {\y}^{-}_{i}(x) = {\bq}_{i}(x) \prod_{e\colon t(e) = i} {\y}_{s(e)} ( x + m_{e}) \prod_{e\colon s(e) = i}
{\y}_{t(e)}(x - m_{e})
\label{eq:rhy}
\end{equation}
for $x \in I_{i, {\ba}}$, ${\ba} = 1, \dots , {\bv}_{i}$, where we used the notation:
\begin{equation*}
{\y}^{\pm}_{i}(x) = {\y}_{i}(x \pm \ii 0), \qquad x \in I_{i, {\ba}}
\qquad \text{and}\qquad
 {\bq}_{i}(x) = \prod_{{\fe}=1}^{{\bw}_{i}} (x - m_{i,{\fe}}).
 \end{equation*}

\subsection{The mass cocycles}
In what follows we shall redefine the amplitude functions
and the ${\bq}$-polynomials
 \begin{equation}
 {\y}_{i}(x)\to {\y}_{i}(x + {\mu}_{i}), \qquad {\bq}_{i}(x) \to {\bq}_{i}(x+{\mu}_{i}),
 \label{eq:yimu}
 \end{equation}
so as to simplify the shifts of the arguments by the masses $m_{e}$ of the bi-fundamental hypermultiplets:
\[
m_{e} \longrightarrow m_{e} + {\mu}_{t(e)} - {\mu}_{s(e)}.
\]
The equations (\ref{eq:rhy}) are the main equations which determine the low-energy effective action as well as the expectation values of all gauge invariant chiral observables. One can view the equations~(\ref{eq:rhy}) as a Riemann--Hilbert problem. They are also similar, but not identical, to the so-called Y-systems
and discrete Hirota equations.

 For the class I and class II theories the shift (\ref{eq:yimu}) maps the
 equations (\ref{eq:rhy}) to
\begin{equation}
{\y}^{+}_{i}(x) {\y}^{-}_{i}(x) = {\bq}_{i}(x) \prod_{j \in \Ver, j \neq i} {\y}_{j}(x)^{I_{ij}}, \label{eq:rhyi}
\end{equation}
where for the class II theories ${\bq}_{i}(x) = {\qe}_{i}$. For the
class II* theory $\hat A_{r}^{*}$, with the clockwise, say, orientation of the quiver, we can make all masses $m_e$ to be equal, $m_{e}=\frac 1{r+1} {\ma}$, by using the shift~(\ref{eq:yimu}).
More precisely, in writing (\ref{eq:rhyiis})
we chose the representative $m^*$ such that if all the edges are oriented so that $t(e) = (s(e) + 1)\, {\rm mod} \, (r+1)$, then
 $m^{*}_{e} = \frac 1{r+1} {\ma}$. Then,
 \begin{equation}
{\y}^{+}_{i}(x) {\y}^{-}_{i}(x) = {\qe}_{i} {\y}_{i-1}\bigg(x-\frac{\ma}{r+1} \bigg) {\y}_{i+1} \bigg( x+\frac{\ma}{r+1} \bigg),
\label{eq:rhyiis}
\end{equation}
where ${\y}_{i+r+1}(x) = {\y}_{i}(x)$.

\subsection{Analytic continuation}

We use equation~(\ref{eq:yiofx}) to analytically continue the functions
${\y}_{i}(x)$ through the cuts $I_{i, {\ba}}$: for the class I and II theories, after the redefinition (\ref{eq:yimu}) which eliminates $m_{e}$,
\begin{equation}
r_{i}\colon \ {\y}_{i}(x) \longrightarrow {\y}_{i}(x) {\bq}_{i}(x) \prod_{j \in \Ver} {\y}_{j}(x)^{- C_{ij}}.
\label{eq:analytic}
\end{equation}
 For the class II* theories after the redefinition (\ref{eq:yimu}), we have the following analogue of (\ref{eq:analytic}):
\begin{equation}
r_i\colon \ {\y}_{i}(x) \longrightarrow {\qe}_{i} \frac{{\y}_{i-1}\big(x-\frac{\ma}{r+1} \big) {\y}_{i+1} \big( x+\frac{\ma}{r+1} \big) }{{\y}_{i}(x)}.
\label{eq:classis}
\end{equation}

 \section{The iWeyl group}

The transformations (\ref{eq:analytic}) and (\ref{eq:classis}) generate a group, which we shall call the \emph{instanton Weyl group}, or \emph{iWeyl group}, $\iw$, for short. This group can be defined for a much larger class of ${\CalN}=2$ theories, not necessarily of the superconformal quiver type we study in this work.

It is clear that the transformations $r_i$ are reflections
$r_i \circ r_i = \operatorname{Id}$, so the iWeyl group is the group, generated by reflections.

Now, by comparing equation~(\ref{eq:analytic}) and equations~(\ref{eq:riz}),
(\ref{eq:twwac}), we see that for the class~I theories the iWeyl group is
the finite Weyl group $W({\g})$. Similarly, for the class II theories
the iWeyl group coincides with the affine Weyl group $W(\hat \g)$. For the class II* theory the iWeyl group is the Weyl group $W\big({\gli}\big)$ of the group $\Gli$.

The groups $\bG$, $\hat \bG$, $\Gli$ and their Weyl groups are discussed in Appendix~\ref{se:Lie}.

\section{Moduli of vacua and mass parameters}\label{subsec:moduli}

After all the redefinitions (\ref{eq:yimu}) the original mass parameters $m_{e}$, the moduli $\big({\ac}^{\CalI}\big)_{{\CalI}=1}^{\bf r}$ of the vacua $u \in \mv$, and the derivatives of the prepotential $\big({\partial}{\CalF}/{\partial}{\ac}^{\CalI}\big)$ are recovered from the study of periods of certain differentials on the curve ${\CalC}_{u}$ defined as follows.

\subsection{The first glimpses of the cameral curve}

The functions ${\y}_{i}(x)$, $i \in \Ver$, after the maximal analytic continuation through the cuts $I_{i, {\ba}}$ form
a local $\iw$-system. It is easy to see that, as long as $| {\qe}_j | \ll 1$ for all $j \in \Ver$, there is exactly one branch of ${\y}_{i}(x)$ as $x \to \infty$ which behaves as
\begin{equation*}
{\y}_{i}^{\rm phys} (x) \sim x^{{\bv}_{i}} + \cdots.
\end{equation*}
The other branches behave as
\begin{equation*}
{\y}_{i}^{\rm unphys}(x) \sim \bigg( \prod_{j\in \Ver} {\qe}_{j}^{n_{ji}}\bigg) x^{{\bv}_{i}} + \cdots
\end{equation*}
with $n_{ji} \geq 0$, and
\[
 \sum_{j} n_{ji} > 0.
\]
Now, the branches meet at the cuts
\begin{equation*}
\bigcup_{i \in \Ver ; \, {\ba} =1, \dots , {\bv}_{i}} I_{i, {\ba}}
\end{equation*}
for the class I and II theories, and at the cuts
\begin{equation*}
\bigcup_{i \in \Ver; \,
{\ba}=1, \dots , N} I_{i, {\ba}} + \frac{\ma}{r+1} {\BZ}
\end{equation*}
for the class II* theories.

The collection of these branches defines a curve ${\CalC}_{u}$ which we shall describe explicitly in the next section. The curve ${\CalC}_{u}$ is a $\iw$-cover of the $x$-plane $\Cx$, with the branch points at the ends of the cuts $I_{i,{\ba}}$. Because of the $\iw$-action on ${\CalC}_{u}$ and the relation to the Weyl groups which permute Weyl cameras, the curve ${\CalC}_{u}$ will be called the \emph{cameral curve}, following~\cite{Donagi:1995alg}.

\subsection{The special coordinates and the mass parameters}
\begin{figure}[t]
 \centering
 \includegraphics[width=10cm]{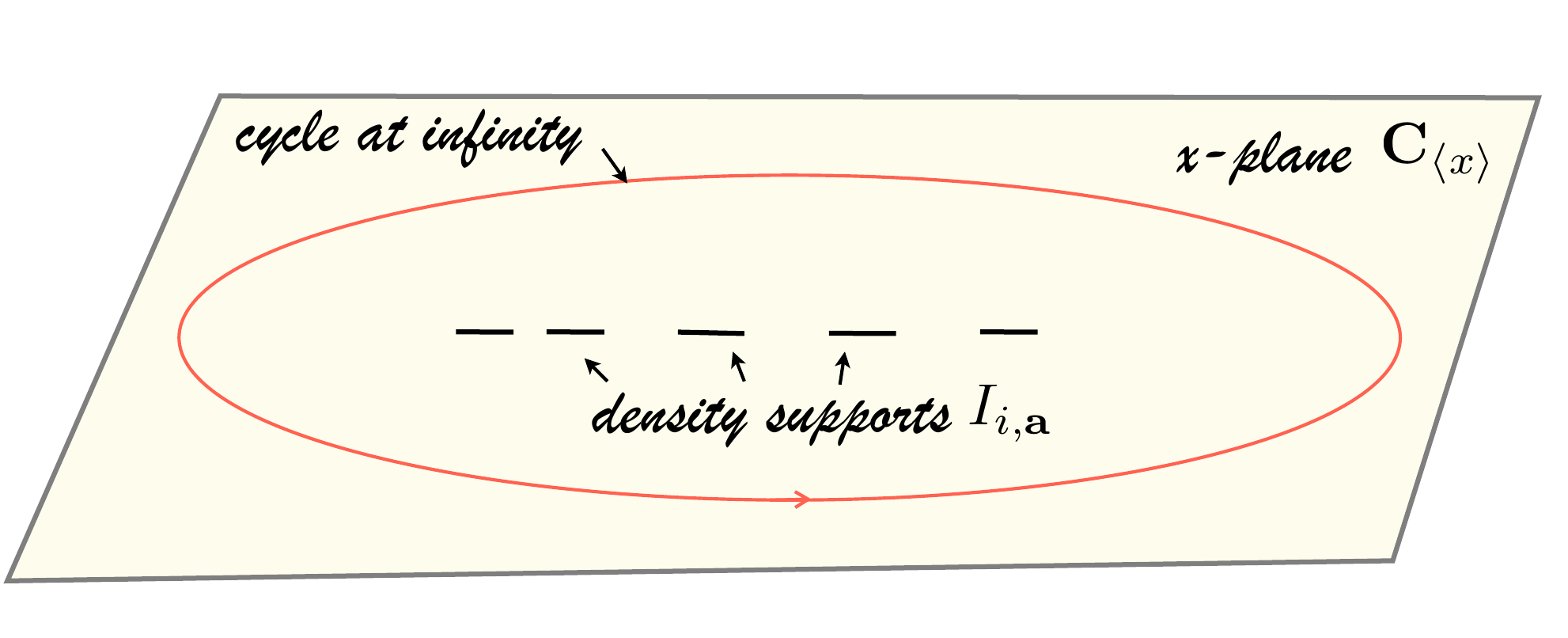}
 \caption{The cycle at $x = \infty$ surrounding
the branch cuts of $\y_i^{\text{phys}}(x)$.}
 \label{fig:infcycle}
\end{figure}

Take the physical branch and expand it at $x = \infty$. Then
the next-to-leading term gives ${\mu}_{i}$, the mass shift which determines (or partially determines, in the II* case) the bi-fundamental masses:
\begin{equation}
{\y}^{\rm phys}_{i}(x) \sim x^{{\bv}_{i}} + {\bv}_{i}{\mu}_{i} x^{{\bv}_{i}-1} + \cdots.
\label{eq:massh}
\end{equation}
The equation~(\ref{eq:spco}) is modified by the ${\mu}_{i}$-shift:
\begin{equation}
{\ac}_{i,{\ba}} + {\mu}_{i} = \frac{1}{2\pi \ii} \oint_{A_{i{\ba}}} x \, {\rm d} \log {\y}_{i}(x),
\label{eq:acco}
\end{equation}
where $A_{i{\ba}}$ is a loop on the physical sheet of ${\CalC}_{u}$ which surrounds
the cut $I_{i,{\ba}}$. Note that the only singularities of ${\y}_{i}(x)$ on the physical
sheet are at $x = \infty$ and at the cuts. Therefore (see Figure~\ref{fig:infcycle}),
\begin{equation*}
\sum_{{\ba}=1}^{{\bv}_{i}} \frac{1}{2\pi \ii} \oint_{A_{i{\ba}}} x \, {\rm d} \log {\y}_{i}(x)
 = \frac{1}{2\pi \ii} \oint_{\infty}
 x \, {\rm d} \log {\y}_{i}(x) = {\bv}_{i}{\mu}_{i},
\end{equation*}
which is consistent with equations~\eqref{eq:massh}, \eqref{eq:acco} thanks to \eqref{eq:gagg}, i.e.,
\begin{equation}
\sum_{{\ba}=1}^{{\bv}_{i}} {\ac}_{i,{\ba}} = 0.
\label{eq:suv}
\end{equation}

\begin{Remark}
The residues
\begin{equation*}
U_{ji} = \frac{1}{2\pi \ii} \oint_{\infty}
 x^{j} \, {\rm d} \log {\y}_{i}(x)
\end{equation*}
determine the ``mirror map'', the change of variables $( T_{i}(x) )_{i \in \Ver}
\longrightarrow( {\bt}_{i}(x))_{i \in \Ver}$ we talked about earlier.
\end{Remark}
\subsection{The dual special coordinates}

Now let us discuss the dual coordinates ${\ac}_{i, {\ba}}^{D}$.
First of all, the equation~(\ref{eq:duala}) does not quite make sense in view of (\ref{eq:suv}). Suppose we relax (\ref{eq:suv}) by absorbing ${\mu}_{i}$ into the definition of ${\ac}_{i,{\ba}}$. We can then analyze the limit shape problem in the usual fashion.
We should keep in mind, however, that only the ${\rm SU}({\bv}_{i})$ part of the gauge group is dynamical. The trace part of the dual special coordinates ${\ac}_{i,{\ba}}^{D}$,
\[
{\mu}^{D}_{i} = \sum_{{\ba}=1}^{{\bv}_{i}} {\ac}_{i,{\ba}}^{D}
\]
is ambiguous. The traceless part, i.e.,
\[
\sum_{{\ba}=1}^{{\bv}_{i}} \lam_{\ba} {\ac}_{i,{\ba}}^{D}
\]
for any weight vector $({\lam}_{\ba})$,
\[
\sum_{\ba} {\lam}_{\ba} = 0,
\]
should be well-defined. Let us now see how this works in detail.

 By
differentiating (\ref{eq:limshape}) with respect to $x$, we find
\begin{gather}\label{eq:ad-long-formula}
-2 \int_{\BR} {\rho}_{i}({\tilde x}) {\CalK}'(x-{\tilde x}) \, {\rm d}{\tilde x} + \sum_{e\colon t(e) = i} \int_{\BR} {\rho}_{s(e)}({\tilde x}) {\CalK}'(x-{\tilde x}+ m_{e} ) \, {\rm d}{\tilde x} \\
+ \sum_{e\colon s(e) = i} \int_{\BR} {\rho}_{t(e)}({\tilde x})
{\CalK}'(x-{\tilde x}- m_{e} )\, {\rm d}{\tilde x} + \sum_{{\fe}=1}^{{\bw}_{i}}
{\CalK}' (x - m_{i, \fe}) + x \log {\qe}_{i} = {\ac}_{i,
 {\ba}}^{D}, \qquad x \in I_{i,{\ba}}, \nonumber
\end{gather}
where
\begin{equation*}
 {\CalK}'(x) = x \log \bigg( \frac x{\tilde \Lambda_{\rm UV}} \bigg) \equiv \int_{\tilde \Lambda_{\rm UV} }^x \CalK''({\tilde x})\, {\rm d}{\tilde x}
\end{equation*}
with
\begin{equation*}
 \CalK''(x) = \log \bigg( \frac{ x}{ \Lambda_{\rm UV}} \bigg)
\end{equation*}
and $\tilde \Lambda_{\rm UV} = \exp(1) \Lambda_{\rm UV}$. Using the definition (\ref{eq:rhoy}) of
$\y_i(x)$ functions, we find
\begin{gather}
 {\ac}_{i, {\ba}}^{D} = \tilde \Lambda_{\rm UV} \log \qe_{i} - \int_{\tilde \Lambda_{\rm UV}}^x \, {\rm d}{\tilde x}
\log \y_i^{\mathrm{phys}}({\tilde x}) + \int_{\tilde \Lambda_{\rm UV}}^{x} {\rm d}{\tilde x} \Biggl( - \log \y_{i}^{\mathrm{phys}}({\tilde x}) \nonumber\\
\hphantom{{\ac}_{i, {\ba}}^{D} =}{} + \sum_{e\colon t(e)=i} \log
 \y^{\mathrm{phys}}_{s(e)}({\tilde x}+m_{e}) + \sum_{e\colon s(e) = i} \log \y^{\mathrm{phys}}_{t(e)}({\tilde x} - m_{e})
+ \log {\bq}_{i} ({\tilde x}) \Biggr).
\label{eq:adual}
\end{gather}
The integration contour in the above formula runs over a physical sheet
from a marked point $p_{*} \in {\CalC}_{u}$ which sits over the point $\tilde \Lambda_{\rm UV} \in \Cx$ to a point $x \in I_{i, \ba}$ which we view as sitting on $\CalC_{u}$. The choice of $x$ is irrelevant\footnote{Physically the meaning of
 the integrand in the effective electrostatic problem is the force acting on
elementary charge, and the integral is the chemical potential for the
charge, or the energy required to move an elementary charge from the
density support to infinity. The force vanishes on the support of the
charge in the stationary charge distribution.} as long as $x \in I_{i, \ba}$ precisely due to the critical point equations~\eqref{eq:rhy}.
The above expression for $ {\ac}_{i, {\ba}}^{D}$ can be converted into much
nicer form by noting that the second integral in (\ref{eq:adual}) is
in fact
\begin{equation}
 \int_{\tilde \Lambda_{\rm UV}}^{x} \log \big({} ^{r_{i}}\y_i^{\mathrm{phys}}({\tilde x}) \big) {\rm d}{\tilde x},
 \label{eq:intref}
\end{equation}
where $r_i$ is the $i$-th reflection (\ref{eq:analytic}). In other words, (\ref{eq:intref}) is the integral of the analytic continuation of
the function $\y_i({\tilde x})$ onto the mirror sheet of ${\CalC}_{u}$ obtained from the
physical sheet by the $\iw$-reflection $r_i$, i.e., by continuing across any of the cuts $I_{i, \ba}$, ${\ba} = 1, \dots , {\bv}_{i}$, supporting the density
$\rho_i(x)$. Thus the integral of the expression in the brackets on the last two lines in (\ref{eq:adual}) is equal to the integral
\begin{equation*}
 \int_{r_i(p_{*})}^{x} \log {\y}_{i}({\tilde x}) \, {\rm d}{\tilde x}.
\end{equation*}
Thus we conclude
\begin{equation}
 {\ac}_{i, {\ba}}^{D} = \tilde \Lambda_{\rm UV} \log \qe_{i} -
\int_{B_{i \ba}} \log \y_{i} ({\tilde x})\, {\rm d}{\tilde x},
\label{eq:acba}
\end{equation}
where the contour $B_{i\ba}$ starts at the point $p_{*}$ which sits over $x = \tilde \Lambda_{\rm UV}$ on the physical sheet, runs
through the cut $I_{i, \ba}$ to the mirror sheet $r_i(\mathrm{phys})$ and
terminates at the point $r_i(p_{*})$, which sits over $x = \tilde \Lambda_{\rm UV}$ on the mirror
sheet.

It is tempting to send $\Lambda_{\rm UV}$ to
infinity. However, there is a subtlety which we already discussed in the beginning of this section. The integral (\ref{eq:acba}) diverges for ${\Lambda}_{\rm UV} \to \infty$. The linear divergence is canceled by the constant term ${\tilde\Lambda}_{\rm UV} \log {\qe}_{i}$ due to
\begin{equation*}
 \log \big( {} ^{r_{i}}\y_{i}^{\mathrm{phys}}({\tilde x}) \big) - \log \big( \y_{i}^{\mathrm{phys}}({\tilde x}) \big)
 = \log \qe_i, \qquad {\tilde x} \to \infty .
\end{equation*}
However the subleading logarithmic divergence does not, in general, cancel. The simplest way to calculate it is to compute the logarithmic derivative ${\Lambda}_{\rm UV} {\rm d} {\ac}_{i,{\ba}}^{D}/{\rm d} {\Lambda}_{\rm UV}$ and then send $\Lambda_{\rm UV} \to \infty$:%
\begin{equation}
{\Lambda}_{\rm UV} \frac{{\rm d} {\ac}_{i,{\ba}}^{D}}{{\rm d} {\Lambda}_{\rm UV}} = - \sum_{{\mathfrak f}=1}^{{\bw}_{i}} m_{i, {\mathfrak {f}}} + \sum_{e\colon t(e) =i} {\bv}_{s(e)} m_{e} - \sum_{e\colon s(e)=i} {\bv}_{t(e)} m_{e}.
\label{eq:adma}
\end{equation}
Luckily the right-hand side of (\ref{eq:adma}) does not depend on ${\ba}$.

We can use the formal expression
\begin{equation*}
 {\ac}_{i, {\ba}}^{D} = \int_{B_{i \ba}} x \, {\rm d} \log \y_i,
 \end{equation*}
where the contour
\begin{equation*}
 B_{i \ba}\colon \ \infty_{\mathrm{phys}}
\to \stackrel{I_{i,\ba}} \qquad \to r_{i}(\infty_{\mathrm{phys}})
\end{equation*}
starts at the point $x = \infty_{\mathrm{phys}}$ on the physical sheet,
then runs through the cut $I_{i,\ba}$ to the mirror sheet
$r_i(\mathrm{phys})$ and finishes at the point
$r_i(\infty_{\mathrm{phys}})$ on this mirror sheet.

The canonical contour $B_{i \ba}$ computing $\ac_{i \ba}^{D}$ is an open
contour, and, as we said above, the integral of $x \, {\rm d} \log {\y}_i$ is divergent. However the variation of Coulomb parameters in ${\rm SU}({\bv}_i)$ concerns only the differences
\begin{equation}
\label{eq:adual-diff}
 \ac_{i ,\ba'}^{D} - \ac_{i, \ba''}^{D} = \int_{B_{i; \ba'}^{\ba''}}
 x\, {\rm d} \log \y_i
\end{equation}
computed by the closed contour $B_{i; \ba'}^{\ba''} = B_{i \ba'} - B_{i \ba''}$ running on physical sheet through the cut~$I_{i\ba'}$ to the mirror sheet $r_i(\mathrm{phys})$ and then through the
cut $I_{i \ba''}$ back to the physical sheet. The divergence (\ref{eq:adma}) cancels in the
integration over $B_{i{\ba}'} - B_{i{\ba}''}$.

\begin{figure}
 \centering
 \includegraphics[width=11cm]{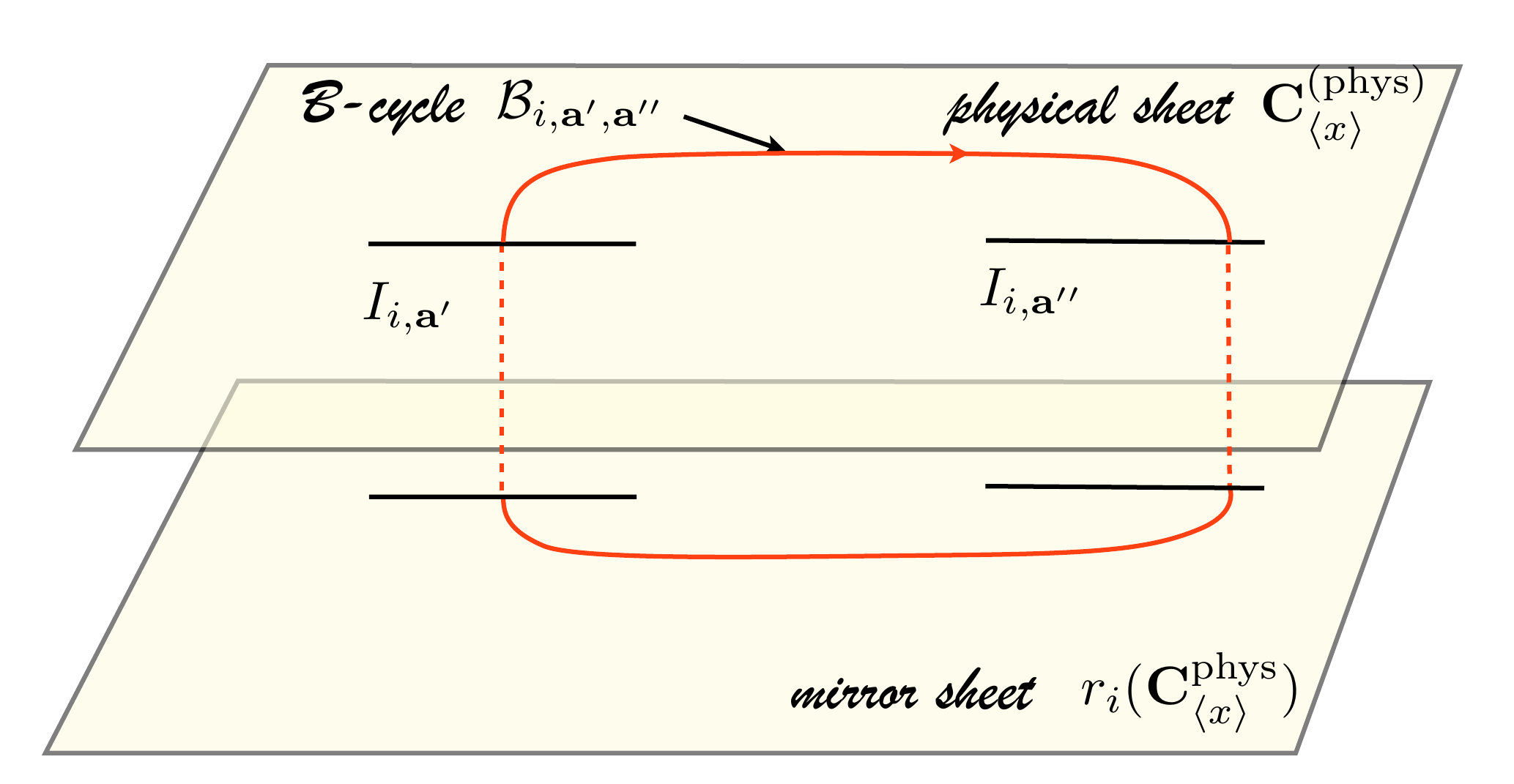}
 \caption{The cycles $B_{i; \ba'}^{\ba''}$.
}
 \label{fig:bcycle}
\end{figure}

In fact, the difference (\ref{eq:adual-diff}) is represented
as the integral over the closed contour $B_{i; \ba'}^{\ba''}$ without any divergent quantities, as follows immediately
from the (\ref{eq:ad-long-formula}), by connecting two points ${x' \in I_{i,
 \ba'}}$ and $x'' \in I_{i, \ba''}$ on the physical
sheet and replacing the integrand as in the second integral of
(\ref{eq:adual}) over the physical sheet by an integral of $-\log \y(x')
dx'$
over the return segment from $x'' \in I_{i, \ba''}$ to $x' \in I_{i, \ba'}$
on the mirror sheet $r_i(\mathrm{phys})$, see Figure~\ref{fig:bcycle}

In the weakly coupled regime we have the following BPS particles in the gauge theory:
for each gauge group factor $i$
the $W$-bosons associated with the breaking ${\rm SU}({\bv}_{i}) \to U(1)^{{\bv}_{i}-1}$, which correspond to the roots of the ${\rm SU}({\bv}_{i})$, and magnetic monopoles, which correspond
to the fundamental weights. Accordingly, it seems natural to define the following cycles on the cameral curve: the $\CalA$- and $\CalB$-cycles, more precisely
 ${\CalA}_{\CalI}$, ${\CalB}^{\CalI}$, labelled by
${\CalI} = (i, {\ba})$, with $i \in \Ver$, and ${\ba} = 1, \dots , {\bv}_{i}-1$:
\begin{equation*}
{\CalA}_{\CalI} = A_{i{\ba}} - A_{i({\ba}+1)}, \qquad
{\CalB}^{\CalI} = \sum_{{\ba}'=1}^{\ba} B_{i{\ba}'} -\frac{\ba}{{\bv}_{i}} \sum_{{\ba}''=1}^{{\bv}_{i}} B_{i{\ba}''}.
 \end{equation*}
The cycles ${\CalA}_{\CalI}$, ${\CalB}^{\CalI}$ determine the special coordinates
\begin{equation}
{\ac}^{\CalI} = \oint_{{\CalA}_{\CalI}} x\, {\rm d} \log {\y}_{i}, \qquad
{\ac}_{\CalI}^{D} = \oint_{{\CalB}^{\CalI}} x\, {\rm d} \log {\y}_{i}.
\label{eq:aadc}
\end{equation}
In the weak coupling regime the pairing \eqref{eq:aadc} between the
cycles and the differentials is non-zero only for ${\CalI} = (i, {\ba})$, for
some ${\ba} = 1, \dots , {\bv}_{i}-1$. The main property of these cycles
is the vanishing of the following two-form on the space $\mv$ of $u$-parameters:
\begin{equation*}
 0 = \sum_{\CalI} {\rm d}{\ac}^{\CalI} \wedge {\rm d}{\ac}_{\CalI}^{D},
\end{equation*}
which follows simply from the equation~\eqref{eq:duala}.

\section{Solution of the limit shape equations}\label{se:limit-shape-sol}

In this section, we solve the equations~\eqref{eq:rhy},~\eqref{eq:rhyi},~\eqref{eq:rhyiis}, and get an explicit formula for the curve~${\CalC}_{u}$.

\subsection{From invariants to the curve}

Our strategy is to define a set of basic invariants ${\crf}_{i}({\y}(x))$ of the $\iw$ group. We shall find the basic invariants which are power series in ${\qe}_{j}$'s, and which are
normalized in such a way that, for the class I and II theories:
\begin{gather}
 {\crf}_{i}({\y}(x)) = {\y}_{i}(x) + \sum_{ {\nu}, \, | {\nu} | > 0}
\prod_{j\in \Ver} {\bq}_{j}^{{\nu}_{j}}(x) {\Psi}_{i\nu}({\y}_{1}(x), \dots , {\y}_{r}(x)), \nonumber \\
 {\nu} = ( {\nu}_{1}, \dots , {\nu}_{r} ) ,
 \qquad | {\nu} | = \sum_{j\in \Ver} {\nu}_{j},
\label{eq:invpsi}
\end{gather}
where ${\Psi}_{\nu}({\y}(x)) \in {\BC} \big[ {\y}_{j}(x), {\y}_{j}(x)^{-1} \big]$ are quasi-homogeneous Laurent polynomials
\begin{equation*}
\sum_{j \in \Ver} \bigg( {\bv}_{j} {\y}_{j} \frac{\pa}{{\pa}{\y}_{j}} + {\bw}_{j} {\nu}_{j} \bigg) {\Psi}_{i\nu} =
 {\bv}_{i}
 {\Psi}_{i\nu}.
\end{equation*}
For the class II* theories there is one modification
\begin{gather}
{\crf}_{i}({\y}(x)) = {\y}_{i}(x) + \sum_{ {\nu}, \, | {\nu} | > 0}
{\tqe}^{\nu} {\tilde\Psi}_{i\nu}({\y}(x)), \nonumber \\
{\nu} = ( {\nu}_{0}, \dots , {\nu}_{r} ) ,
 \qquad | {\nu} | = \sum_{j=0}^{r} {\nu}_{j},
\label{eq:invpsii}
\end{gather}
where
\[
{\tqe}^{\nu} = \prod_{j=0}^{r} {\qe}_{j}^{{\nu}_{j}}
\]
and
${\tilde\Psi}_{i\nu}({\y}(x)) \in {\BC} \Big[ {\y}_{j} (x + \frac{n}{r+1} {\ma}), {\y}_{j}^{-1} ( x + \frac{n}{r+1} {\ma} ) \Big]$
with $- |{\nu} | \leq n \leq | {\nu} |$. The functions $\tilde\Psi_{i\nu}$ are
quasi-homogeneous:
\begin{equation*}
\sum_{j \in \Ver} a_{j}{\y}_{j} \frac{\pa}{{\pa}{\y}_{j}} {\tilde\Psi}_{i\nu} = a_{i} {\tilde\Psi}_{i\nu}.
\end{equation*}

\subsection{Master equations}

Now, the $\iw$-invariance of ${\crf}_{j}({\y}(x))$ implies that
they are continuous across all the cuts, that is they are single-valued
analytic functions of $x$. Given their large $x$ asymptotics, they are polynomials in $x$:
\begin{equation}
\label{eq:master}
{\crf}_{j}({\y}(x)) = T_{j}(x),
\end{equation}
where
\begin{equation}
T_{j}(x) = T_{j,0}({\tqe}) x^{{\bv}_{j}} + T_{j,1}({\tqe}; m) x^{{\bv}_{j}-1} + \dots + T_{j, {\bv}_{j}}.
\label{eq:cxtj}
\end{equation}
The coefficients $T_{j,0}({\tqe})$ are determined by the gauge couplings ${\qe}_{k}$
\[
T_{j,0}({\tqe}) = 1 + \sum_{{\nu}, \, | {\nu} | > 0} {\tqe}^{\nu} {\Psi}_{j\nu} ( 1, \dots , 1 ).
\]
The coefficients $T_{j,1}({\tqe}; m)$ are determined by the
masses $m_{e}$ from~\eqref{eq:massh},~\eqref{eq:invpsi},~\eqref{eq:invpsii}.

The rest of the coefficients
\begin{equation*}
 \left( T_{j,{\ba}}({\tqe}; m, u) \right)^{j \in\Ver}_{{\ba} = 2, \dots , {\bv}_{j}}
\end{equation*}
is determined by $u = (u_{i, {\ba}})_{i \in \Ver}^{{\ba}=2, \dots , {\bv}_{i}}$.
The equations~\eqref{eq:master} and~\eqref{eq:cxtj} define an analytic curve
\[
{\CalC}_{u}^{\circ} \subset {\Cx} \times ({\BC}^{\times})^{\Ver},
\]
which can be compactified to the \emph{cameral curve} ${\CalC}_{u}$ which is the $\iw$-cover
of ${\Cpx} = {\Cx} \cup \{ \infty \}$:
\[
{\CalC}_{u} \longrightarrow {\Cpx}
\]
unramified (for generic $\tqe$) over $x = \infty$.
The details of the compactification of ${\Cx} \times ({\BC}^{\times})^{\Ver}$ and ${\CalC}_{u}^{\circ} \subset {\CalC}_{u}$ will be discussed elsewhere. In what follows we drop the superscript $\circ$ in the definition of $\CalC_{u}$.

\subsection{The periods}

The cameral curve ${\CalC}_{u}$ depends on $u$, forming a family ${\CalC}$ of curves
parametrized by the ``u-plane''~$\mv$. The family $\CalC$ depends on the microscopic
couplings ${\qe}_{i}$ and on the mass parameters. Let us keep the masses fixed.
When $|{\qe}_{i}| \ll 1$ for all $i \in \Ver$
we have a well-separated system of cycles~${\CalA}_{\CalI}$
and ${\CalB}^{\CalI}$, which we defined in Section~\ref{subsec:moduli}.
We transport this system of cycles throughout the moduli space of gauge couplings
via Gau{\ss}--Manin connection.

\subsection{Vector-valued Seiberg--Witten differential}

Let us introduce the following vector-valued $1$-differential which is schematically given by
\begin{equation*}
{\rm d}{\BS} = x \sum_{i \in \Ver} \big( {\rm d} \log {\y}_{i} {\al}^{\vee}_{i} - {\rm d} \log {\bq}_{i}(x) {\lam}^{\vee}_{i} \big),
\end{equation*}
where ${\al}_{i}^{\vee}$ and ${\lam}_{i}^{\vee}$ are the simple coroots and the fundamental coweights of $\gq$. We shall have more specific notations for each class.

The differential ${\rm d}{\BS}$ takes values in the vector space ${\BC}^{\Ver}$ which is acted upon by the $\iw$-group. The group $\iw$ also acts on the curve $\CalC_{u}$. It is clear
from our construction that ${\rm d}{\BS}$ is $\iw$-equivariant:
\begin{equation*}
w^{*}{\rm d}{\BS} = w\cdot {\rm d}{\BS}
\end{equation*}
for any $w \in \iw$.

\subsection{Degeneration and filtration}
In this section we consider the theories of class I and class II,
and the extended Coulomb moduli space $\mve$ (which includes the masses of
bifundamental hypermultiplets, recall equation~\eqref{eq:mve}).
Consider conformal quiver with assigned dimensions $(\gamma, \bv, \bw)$
at vertices and recall that they satisfy $ \bw = C \bv$, $\bw_{i} \geq 0$,
$\bv_{i} \geq 0$ where $C$ is the Cartan matrix of $\gamma$.
 We say that the theory $(\gamma,\bv,\bw)$ strictly \textit{contains} the
theory $(\gamma, \bv', \bw')$ if $0 \leq \bv' \leq \bv$, $0 \leq \bw' \leq \bw$ and $ C
\bv' = \bw'$, and $(\gamma, \bv',\bw') \neq (\gamma,
\bv,\bw)$. The extended Coulomb moduli space $\mve$ of theory $(\gamma,\bv,\bw)$
contains a~locus related to the Coulomb moduli ${\mve}'$ space of $(\gamma,\bv',\bw')$ as follows. Suppose that
given a~point $\ma_{f} \in \Cx$
the polynomials $T_i(x)$, $\bq_i(x)$, $i \in \Ver$ factorize as follows:
\begin{gather}
 T_i(x) = T'_i(x) \big(x-\ma_{f}\big)^{\bv_i - \bv'_i} , \qquad
 \bq_i(x) = \bq'_i (x) \big(x - \ma_{f}\big)^{\bw_i - \bw'_i} .
\label{eq:degeneration}
 \end{gather}
Then it is clear that the character equations \eqref{eq:master}
factorize as well, and the $\y_i(x)$ functions
solving the theory $(\gamma,\bv,\bw)$ are expressed in terms of
$\y_i'(x)$ functions solving the theory $(\gamma,\bv',\bw')$ as
\begin{equation*}
 \y_i(x) = \big(x - \ma_{f}\big)^{(\bv_i - \bv'_i)} \y_i'(x).
\end{equation*}
Suppose that the degeneration equation~\eqref{eq:degeneration} is minimal,
i.e., there is no intermediate and different
$(\gamma, \bv'', \bw'')$ such that $\bv' \leq \bv'' \leq \bv$ and $\bw'
\leq \bw'' \leq \bw$. Then we see that $\mve$ includes the loci ${\mve}'
\times \Cx$
\begin{equation*}
 {\mve}' \times \Cx \subset \mve ,
\end{equation*}
where $\Cx$ parameterizes the location of $\ma_f$.

In the monopole picture of Section~\ref{se:monopole} such degeneration
corresponds to the complete screening of the point-like
non-abelian monopoles by several Dirac monopoles.

Recall that dimensions $\bv$ for the theories of class II parametrized by a
single integer $N$ such that $\bv_i = N a_i$ where $a_i$ are Dynkin
marks. Therefore, the above inclusion is
\begin{equation*}
 \mve_{N-1} \times \Cx \subset \mve_{N}.
\end{equation*}
Geometrically, such inclusion for theories of class II is associated
with freckled (point) instantons described in more details after
equation~\eqref{eq:freckled}.

\subsection{The cameral curve as a modular object}\label{se:cameral}

In this section we give the modular interpretation of the curve $\CalC_{u}$.

\subsubsection{The class I theories}

Let $\bG= \Gq$ be the simple complex Lie group corresponding to the quiver of the class I theory.
Let $Z$ be its center, and let C$\bG$ be the conformal extension of $\bG$. Let ${\cla}_{i}^{\vee}$, ${\alc}_{i}^{\vee}$, $i=1, \dots , r$ be the fundamental coweights and the simple coroots in C$\h \subset$ C$\g$. Let
\begin{equation}
g(x) = \prod_{i=1}^{r} {\bq}_{i}(x)^{-{\cla}^{\vee}_{i}} {\y}_{i}(x)^{{\alc}_{i}^{\vee}} \in {\rm C}{\bT}
\subset {\rm C}{\bG}.
\label{eq:gkacm}
\end{equation}
In the notations of \eqref{eq:param},
\[
g(x) = {\bg}_{{\bq}(x), {\y}_{1}(x), \dots , {\y}_{r}(x)}.
\]
We also use
\[
g_{\infty}(x) = {\bg}_{{\bq}(x), 1, \dots , 1}.
\]
The importance of $g(x)$ is that it transforms by the reflection in the Weyl group $W(\g)$
when crossing the cuts $I_{i, {\ba}}$, cf.~(\ref{eq:twwac}),
\[
g_{+} (x) \longrightarrow g_{-}(x) = {} ^{r_{i}}g_{+}(x),
\]
which implies that for the class I theories the iWeyl group $\iw$ is the Weyl group $W(\g)$ of the corresponding simple Lie group $\bG$.
In order to construct the $\iw$-invariants one could take any C$\bG$-invariant function
on C$\bG$.
In fact, cf.~(\ref{eq:param}),
\begin{equation*}
{\crf}_{i}({\y}(x)) = g_{\infty}(x)^{-{\lam}_{i}} {\chi}_{i} ( g(x) ).
\end{equation*}
Using the formulae (\ref{eq:wmult}) and (\ref{eq:poswe}), we write
\begin{equation}
 {\crf}_{i}({\y}(x)) = {\y}_{i} \sum_{{\nu} = ({\nu}_{1}, \dots , {\nu}_r)} c_{\nu}^{i}
 \prod_{j=1}^{r} \Bigg( {\bq}_{j} \prod_{k=1}^{r} {\y}_{k}^{-C_{kj}^{\gq}} \Bigg)^{{\nu}_{j}},
\label{eq:cxiwts}
\end{equation}
where
\begin{equation*}
c_{\nu}^{i} =
{\chi}_{R_{i}, {\lam}_{i} - \sum_{j=1}^{r} {\nu}_{j} {\al}_{j}}, \qquad {\nu} \in {\BZ}_{+}^{r}.
\end{equation*}
In fact, the sum in (\ref{eq:cxiwts}) is finite, i.e., only for a finite number of vectors
$\nu$'s the multiplicity $c_{\nu}^{i}$ is non-zero.

We thus obtain the following geometric picture.
The solution of the class I theory is a $\Cx$-parametrized family $[g(x)]$ of
 conjugacy classes in C${\bG}$, which vary with $x$ polynomially, in the appropriate sense, and
such that the value of
the $D$-homomorphism on $[g(x)]$ is fixed for the theory:
\begin{equation*}
D( g(x)) = \Bigg( \prod_{j=1}^{r} {\bq}_{j}(x)^{-l_{j {\xi}}} \Bigg)_{{\xi}=1}^{z_{\g}}.
 \end{equation*}
Actually, as we explain in Section~\ref{se:conformalgroup}, the coweights ${\cla}_{i}^{\vee}$ are not uniquely specified. The group element $g(x)$ in (\ref{eq:gkacm}) defines a well-defined conjugacy class $[g(x)] \in {\Bqad}$ in ${\Gq}/Z$.
Its lift to~${\rm C}\bG $ can be twisted by any C-valued (meromorphic)
function of $x$. We shall use this freedom in our manipulations with spectral curves.

The cameral curve ${\CalC}_{u}$ can be viewed, geometrically, as the lift to $C\bT$ of the parametrized rational curve in
$C \bT/W({\g}) \approx ({\BC}^{\times})^{z_{\g}} \times {\BC}^{r}$:
\begin{equation*}
x \mapsto D(g(x)) \times ( T_{1}(x), \dots , T_{r}(x) ).
\end{equation*}

\subsubsection{The class II theories}\label{se:cameral2}

As we mentioned above, the quivers of the class II theories correspond
to the simply laced affine Kac--Moody algebras, i.e., $\gq = \hat\g$. Let $\hat\bG$
be the corresponding Kac--Moody group. Let
${\hat\lam}_{i}^{\vee}$, ${\hat\al}_{i}^{\vee}$ be the corresponding
affine coweights and coroots, $i = 0, 1, \dots , r$ (see Appendix~\ref{se:affineLie}). Define
\begin{equation}
g(x) = \prod_{i=0}^{r} {\qe}_{i}^{-{\hat\lam}_{i}^{\vee}} {\y}_{i}(x)^{{\hat\al}_{i}^{\vee}} \in {\hat \bT} \subset {\hat\bG}.
\label{eq:gkm}
\end{equation}
Again, strictly speaking $g(x)$ takes values in ${\hat\bG}/Z$ and so we
should consider the modification of~$\hat\bG$ corresponding to the
conformal extension $C\hat \bG$, but since the subtlety with the center $Z\subset \bG$ only involves the $x$-independent factor
\begin{equation*}
{\hat g}_{\infty} = \prod_{i=0}^{r} {\qe}_{i}^{-{\hat\lam}_{i}^{\vee}}
\end{equation*}
it will not affect the $x$-dependence of the invariants.
The limit shape equations, as in the class~I case, translate to the jump conditions
\[
g_{+}(x) \longrightarrow g_{-}(x) = {}^{r_{i}}g_{+}(x)
\]
for $x \in I_{i, {\ba}}$, with $r_i$ being the simple reflections generating the
affine Weyl group $W({\hat\g})$, which is the $\iw$ group for the class II theories.

The invariants of $W({\hat\g})$ are constructed using the characters $\hat\chi_{i}$ of the fundamental
representations ${\hat R}_{i}$ of $\hat\bG$:
\begin{equation}
{\crf}_{i}({\y}(x)) = \big( {\hat g}_{\infty}^{{\hat\lam}_{i}}\big)^{-1} {\hat\chi}_{i}(g(x)).
\label{eq:affcha}
\end{equation}
They can also be obtained by starting with ${\y}_{i}(x)$ and averaging
with respect to the $W({\hat\g})$-action. The $W({\hat\g})$-action
consists of the translations by the coroot lattice ${\rl}^{\vee}$ and
the $W({\g})$-transformations. The ${\rl}^{\vee}$-averaging produces
the lattice theta-functions of various characteristics, of the schematic
form (the details are given in Appendix~\ref{se:lattice-theta}):
\begin{equation*}
{\Theta}({\xi}, {\qe}) = \sum_{{\nu} \in {\rl}^{\vee}} {\qe}^{\frac 12 \langle {\nu}, {\nu} \rangle}
{\rm e}^{i \langle {\nu}, {\xi} \rangle},
\end{equation*}
where
\begin{equation}
{\qe} = \prod_{i=0}^{r} {\qe}_{i}^{a_{i}}.
\label{eq:quell}
\end{equation}
The affine analogue of the formula (\ref{eq:cxiwts}) is an infinite sum,
however, it is a power series in ${\qe}$. Using the fact that the weights $\hat\lam$ of the fundamental representation ${\hat R}_{i}$ differ from the highest weight $\hat\lam_i$ by a positive linear combination of simple roots, ${\hat \lam} = {\hat\lam}_{i} - {\hat\nu}$,
\[
{\hat\nu} = \sum_{j=0}^{r} {\nu}_{j} {\hat\al}_{j}, \qquad {\nu}_{j} \in {\BZ}_{+},
\]
we can write, with
\begin{gather}
{\tqe}^{\hat\nu} = \prod_{j=0}^{r} {\qe}_{j}^{\nu_{j}}, \nonumber\\
 {\crf}_{i}({\y}(x); {\tqe}) = {\y}_{i} \sum_{\hat\nu} {\hat c}_{\hat\nu}^{i} {\tqe}^{\hat\nu}
 \prod_{k,j=0}^{r} {\y}_{k}(x)^{-C_{kj}^{\hat\g}{\nu}_{j}},
\label{eq:affcxiwts}
\end{gather}
where we made the ${\tqe} = ({\qe}_{0}, \dots , {\qe}_{r})$ dependence explicit, and
\begin{equation*}
{\hat c}_{\hat\nu}^{i} =
{\chi}_{{\hat R}_{i}, {\hat\lam}_{i} - {\hat\nu}}.
\end{equation*}
Write ${\hat\nu} = n {\delta} + {\nu}$, where $n \in {\BZ}_{+}$, and ${\nu} \in {\rl}$ belongs to the root lattice of $\g$. Notice that the factor ${\tqe}^{\hat\nu}$
in (\ref{eq:affcxiwts}) depends on $n$ only via the ${\qe}^{n}$ factor. For fixed
$n$ the number of ${\nu} \in {\rl}$ such that ${\hat c}_{n {\delta} + {\nu}}^{i} \neq 0$ is finite.

The characters of $\hat\bG$ are well-studied~\cite{Kac:1984}. Physically they are the torus ${\ec} = {\BC}^{\times}/{\qe}^{\BZ}$ conformal blocks of the WZW theories with the group $G$, and levels $k = a_{i}$, $i=0,1, \dots , r$ (see~\cite{Dolan:2007eh} for recent developments). The argument of the characters can be viewed as the background $\bG$ $(0,1)$-gauge field $\bar{\bf A}$, which couples
to the holomorphic current ${\bf J} = g^{-1}{\partial}g$:
\begin{equation*}
Z_{k} \big( {\tau}; {\bar{\bf A}} \big) = \int Dg \exp k \left( S_{\rm WZW}(g) + \int_{\ec} \la {\bf J}, {\bar{\bf A}} \ra \right)
= \sum_{\hat \lam \text{ at level $k$}} c_{\hat\lam} \cdot {\hat\chi}_{\hat\lam} ({\hat t}; {\qe}).
\end{equation*}
The background gauge field has only $r$ moduli. In practice, one chooses the gauge ${\bar{\bf A}} = \frac{\pi}{\operatorname{Im}{\tau}} {\xi}$, where ${\xi} = {\rm const} \in {\h}$.

Technically, it is more convenient to build the characters using the free fermion theory, at least for the $A_r$, $D_r$ cases, and for the groups $E_6$, $E_7$, $E_8$ at level~$1$.
We review this approach in Appendix~\ref{sectionK}.

The master equations \eqref{eq:master} ${\crf}_{i}({\y}(x); {\tqe}) = T_{i}(x)$
describe a curve ${\CalC}_{u} \subset {\Cx} \times ({\BC}^{\times})^{r+1}$
 which is a $W({\hat\g})$-cover of the $x$-parametrized rational curve $\Sigma_{u}$ in ${\BC}^{r+1} = {\rm Spec} {\BC} [ {\hat\chi}_{0}, \dots , {\hat\chi}_{r} ]$, cf.~\eqref{eq:affcha}:
\begin{equation}
{\hat \chi}_{i} = \prod_{j} {\qe}_{j}^{-{\hat\lam}_{i}({\hat\lam}_{j}^{\vee})} T_{i}(x), \qquad i = 0, \dots , r.
\label{eq:hchi}
\end{equation}
Now, as we recall in Section~\ref{se:conjugacy}, the characters ${\hat\chi}_{i}$, $i = 0, \dots , r$ are the sections of the
line (orbi)bundle ${\CalO}(1)$ over the coarse moduli space $\Bnq$ of
holomorphic principal semi-stable $\bG$-bundles over the elliptic curve
$\ec$.
 Therefore, \eqref{eq:master}~and~\eqref{eq:hchi}
define for each $u$ a quasimap $U$ of the compactified $x$-plane
${\Cpx}$ to $\Bnq$, which is actually an honest map near $x = \infty$, whose image
approaches the fixed $\bG$-bundle ${\CalP}_{\tqe}$.
 This bundle can be described, e.g.,
by the transition function $g_{\infty}$, which is one of the $\bT$ lifts of
\begin{equation*}
{\tilde g}_{\infty} = \prod_{i=1}^{r} {\qe}_{i}^{-{\lam}^{\vee}_{i}} \in {\bT}/Z.
\end{equation*}
By definition, the local holomorphic sections of ${\CalP}_{\tqe}$ are
the ${\bG}$-valued functions ${\Psi}(z)$, defined in some domain in ${\BC}^{\times}$ such that
\[
{\Psi}({\qe}z) = g_{\infty} {\Psi}(z).
\]
 The complex dimension of the space of quasimaps $U$ with fixed $U({\infty})$ is the
 number of coefficients in the polynomials $(T_i(x))_{i \in \Ver}$ excluding the
 highest coefficients, that is (cf.\ equation~\eqref{eq:mve}),
\[
\dim_{\BC}
\mve = \sum_{i \in \Ver} \bv_i = N h.
\]
We say that $U$ is a quasimap, and not just a holomorphic map ${\Cpx} \to {\Bnq}$
for two reasons. Technically, a collection of ${\hat\chi}_{i}$ in \eqref{eq:hchi} defines a point in $\WP^{a_{0}, a_{1}, \dots , a_{r}}$ only if the polynomials~$T_{i}(x)$ don't have common weighted factors. If, however, for some
${\ma}_{f} \in {\Cx}$:
\begin{equation}
T_{i}(x) = {\tilde T}_{i}(x) ( x- {\ma}_{f})^{a_{i}}, \qquad \text{for all } i = 0, \dots , r,
\label{eq:freckled}
\end{equation}
then the map $\Cpx \to \Bnq$ is not well-defined at $x = {\ma}_f$. It is
trivial to extend the map there by removing the $(x-{\ma}_{f})^{a_{i}}$ factors.
This operation lowers $N \to N-1$. In a way, the point ${\ma}_f$ carries
a unit of the instanton charge. Such a configuration is called
a freckled instanton~\cite{Losev:1999tu}. Thus, the extended
moduli space equation~\eqref{eq:mve} of vacua ${\mve}_{N}$ of the
gauge theory with ${\Gg} = \times_{i} {\rm SU}(Na_{i})$,
contains the locus ${\mve}_{N-1} \times {\Cx}$. Allowing for several freckles
at the unordered points ${\ma}_{f1}, {\ma}_{f2}, \dots , {\ma}_{fi}$ we arrive at the hierarchy of embeddings of the moduli spaces of vacua of the gauge theories with different gauge groups $\Gg$:
\begin{gather*}
 {\mve}_{N} = \mathring{\mve}_{N} \cup \mathring{\mve}_{N-1} \times {\Cx} \cup \mathring{\mve}_{N-2} \times \operatorname{Sym}^{2}{\Cx} \cup \cdots
\\
\hphantom{{\mve}_{N} =}{} \cup \mathring{\mve}_{N-i} \times
\operatorname{Sym}^{i}{\Cx} \cup \dots \cup \operatorname{Sym}^{N}{\Cx},
\end{gather*}
where $\mathring{\mve}_{N}$ stands for the space of degree $N$ rational maps $U\colon {\Cpx} \to {\Bnq}$.

 This hierarchy of gauge theories is more familiar in the context of class I theories.
 Presently, the freckled instantons to $\Bnq$ correspond to the loci in $\mv$ where a
 Higgs branch of the gauge theory can open. Indeed, if \eqref{eq:freckled} holds, then
 we can solve the master equation \eqref{eq:master} by writing
 \begin{equation*}
 {\y}_{j}(x) = ( x - {\ma}_{f} )^{a_{j}} {\tilde\y}_{j}(x)
 \end{equation*}
 with ${\tilde\y}_{j}(x)$ solving the master equation \eqref{eq:master}
 of the
\[
 \times_{i \in \Ver} {\rm SU}( ( {N-1} ) a_{i} )
\]
 gauge theory. In the IIB string theory picture (see Appendix~\ref{sec:dbranes}) the full collection of fractional branes
 in the amount of $a_i$ for the $i$-th type recombine, and detach themselves from the fixed locus, moving away at the position ${\ma}_{f}$ on the transverse ${\BR}^{2} = {\Cx}$.

Now let us take $u \in \mathring{\mve}_{N}$. The corresponding map $U$ defines a rational curve $\Sigma_{u}$ in $\Bnq$ of degree $N$.
\begin{Remark}
Actually, there is another compactification of $\mathring{\mve}_{N}$, via genus zero Kontsevich stable maps of bi-degree $(1,N)$
to ${\BC\BP}^{1} \times \Bnq$ (see~\cite{Givental:1997}, where the space of quasimaps is called the toric map spaces). It would be interesting to study
its gauge theoretic meaning.
\end{Remark}

\begin{Remark}
The highest-order coefficients $T_{i,0}({\tqe})$ of the polynomials $T_{i}(x)$ depend only
on the gauge coupling constants, and determine the limit $U(x)$, $x \to \infty$
\begin{equation*}
U({\infty}) = [ {\CalP}_{\tqe} ] \in {\Bnq}.
\end{equation*}
The next-to-leading terms $T_{i,1}({\tqe}, m)$ depend only on the gauge couplings and the bi-fundamental masses. These define the first jet ${\bt}_{[ {\CalP}_{\tqe} ]}{\Sigma}_{u}$ of the rational curve $\Sigma_{u}$
at $x = \infty$.
\end{Remark}
{}Summarizing, \emph{the moduli space ${\mv}_{N}$ of vacua of the class II theory with the gauge group
\[
{\Gg} = \times_{i \in \Ver} {\rm SU}(Na_{i})
\]
is the moduli space of degree $N$ finely framed at infinity quasimaps
\begin{equation*}
U\colon \ {\Cpx} \to {\Bnq} \approx {\WP}^{a_{0}, a_{1}, \dots , a_{r}},
\end{equation*}
where the fine framing is the condition that $U$ is the honest map near $x = \infty$, and the first jet $($the value and the tangent vector$)$ at $x = \infty$ are fixed:
\begin{equation*}
\big(U({\infty}) , U^{\prime}({\infty}) \big) \leftrightarrow ( {\tqe}, m ).
\end{equation*}
We also have the identification of the extended moduli space ${\mve}$ with the space
of framed quasimaps}

\subsubsection{The class II* theories}

The theories with the affine quiver of the ${\hat A}_{r}$ type
can be solved uniformly in both class II and class II* cases.
This is related to the fact that the current algebra ${\widehat{u(r+1)}}$, the affine Kac--Moody algebra based on
$U(r+1)$ is a subalgebra of $\gli$, consisting of the $(r+1)$-periodic infinite matrices.

Let $\gamma$ be the affine Dynkin graph of the ${\hat A}_{r}$ algebra. We have,
$\Ver = \Edg = \{ 0, 1 , \dots , r \}$.
Choose such an orientation of the graph $\gamma$ that for any $e \in \Edg$: $s(e) = e$, $t(e) = (e+1)$ mod~${r+1}$. Let $m_{e}$, $e = 0, \dots, r$
be the corresponding bi-fundamental multiplet masses, and
\begin{equation*}
{\mathfrak {m}} = \sum_{e=0}^{r} m_{e}.
\end{equation*}
We are in the class II* theory iff $\mathfrak{m} \neq 0$.

It is convenient to extend the definition of $m_e$ to the universal cover of $\gamma$.
Thus, we define
\begin{gather}
{\mathfrak {m}}_{i} = m_{i\operatorname{mod} (r+1)}, \qquad
 Y_{i} (x) = {\y}_{i \operatorname{mod} (r+1)} ( x - {\mathfrak{m}}_{(i)}),\qquad i \in \BZ.
\label{eq:extamp}
\end{gather}
The extended amplitudes $Y_{i}(x)$ obey
\begin{equation}
Y_{i + r+1}(x) = Y_{i} (x - {\mathfrak{m}}).
\label{eq:shfper}
\end{equation}
Define
\begin{equation}
t_{j}(x) = {\ct}_{j} \frac{Y_{j}(x)}{Y_{j-1}(x)},
\label{eq:tfromy}
\end{equation}
where
\begin{gather}
{\ct}_{j+1} = {\qe}_{j \operatorname{mod} (r+1)} {\ct}_{j}, \qquad
\prod_{j=0}^{r} {\ct}_{j} = 1 , \qquad
{\ct}_{j+r+1} = {\qe} {\ct}_{j}.
\label{eq:ctistar}
\end{gather}
Then
\[
t_{j + r+1}(x) = {\qe} t_{j}(x - {\mathfrak{ m}}),
\]
where for the $\hat A_r$-series,
\[
{\qe} = \prod_{j=0}^{r} {\qe}_{j}.
\]
Now, consider the following element of $\Gli$:
\begin{equation}
g(x) = {\y}_{0}(x)^{K} \times \prod_{i \in \BZ} t_{i}(x)^{E_{i,i}}
\label{eq:ginf}
\end{equation}
with $t_{i}(x)$ from \eqref{eq:tfromy}, and $E_{i,j}$ denoting the matrix with all entries zero except $1$ at the $i$-th row and $j$-th column.
A closer inspection shows \eqref{eq:ginf} is the direct generalization of \eqref{eq:gkm}
with the $(r+1)$-periodic matrix $g_{\infty}$, and $({\y}_{i}(x))_{i \in \Ver}$ replaced by the infinite array $(Y_{i}(x))_{i \in \BZ}$.
Indeed, the simple coroots of $\Gli$ are the diagonal matrices, shifted in the central direction
\begin{equation*}
{\al}_{i}^{\vee} = K {\delta}_{i,0} + E_{i,i}-E_{i+1,i+1}, \qquad i \in \BZ
\end{equation*}
so that
the analogue of \eqref{eq:cenele} holds
\[
K = \sum_{i \in \BZ} {\al}_{i}^{\vee}
\]
if we drop the telescopic sum $\sum_{i \in \BZ} E_{i,i}-E_{i+1,i+1} \sim 0$.

We do not need to deal with all the coweights of $\Gli$, only with the $(r+1)$-periodic ones, defined via:
\[
\prod_{j=0}^{r}{\qe}_{j}^{-{\tilde\lam}_{j}^{\vee}} = \prod_{b \in \BZ} \prod_{j=1}^{r+1} \big( {\qe}^{b}{\ct}_{j} \big)^{E_{i + b (r+1), i+ b(r+1)}}.
\]
These coweights are the coweights of the ${\hat A}_{r}$ Kac--Moody algebra, embedded into $\mathfrak{gl}_{\infty}$ as the subalgebra of $(r+1)$-periodic matrices
\[
\sum_{i,j \in \BZ} a_{i,j} E_{i,j}, \qquad a_{i+r+1, j+r+1} = a_{i,j}.
\]
We shall describe the solution of this theory in detail in the next section.

\subsection{Spectral curves}\label{se:spectral}

The cameral curve captures all the information about the limit shape, the special coordinates, the vevs of the chiral operators, and the prepotential. Its definition is universal.

However, the cameral curve is not very convenient to work with.
In many cases one can extract the same information from a ``smaller'' curve, the so-called \emph{spectral curve}.
In fact, there are several notions of the spectral curve in the literature.

Suppose ${\lam} \in \operatorname{Hom} \big(({\BC}^{\times})^{\Ver}, {\BC}^{\times}\big)$
is a dominant weight, i.e., ${\lam}({\al}_{i}^{\vee}) \geq 0$ for all $i
\in \Ver$. Let~$R_{\lam}$ be the irreducible highest weight module of
$\Gq$ with the highest weight $\lam$, and ${\pi}_{\lam}\colon {\Gq} \longrightarrow {\rm End}(R_{\lam})$ the corresponding homomorphism.
Then the spectral curve $C^{R_\lambda}_{u}$ in $\Cx \times \Ct$ is
\begin{equation}
\label{eq:spectral-curve}
 \det\nolimits_{R_{\lambda}} \big(1 - t^{-1}{\zeta}(x)^{-1} {\pi}_{\lam} (g(x)) \big) = 0,
\end{equation}
where
\begin{enumerate}\itemsep=0pt
\item for the class I theories we introduce the factor
\[
\zeta (x) = g_{\infty}(x)^{\lam} \times \text{a rational function of~$x$},
\]
having to do with the lift of the conjugacy class $[g(x)]$ from ${\Gad}$ to ${\rm C}{\bG}$.
The rational function is chosen so as to minimize the degree of the curve $C^{R_{\lam}}_{u}$, as we
explain in the examples below.

\item
for the class II, II* theories the factor $\zeta(x)$ is a constant.
\end{enumerate}

Generally, the curve $C^{R_\lambda}_{u}$ defined by \eqref{eq:spectral-curve} is not irreducible. The equation \eqref{eq:spectral-curve} factorizes into a product
 of components, one component for each Weyl orbit in the set of weights
 $\Lambda_{R_{\lambda}}$ for the module $R_{\lambda}$. Each
 Weyl orbit intersects dominant chamber at one point and therefore can
 be parametrized by dominant weights $\mu$. Therefore,
\begin{equation*}
 C^{R_\lambda}_{u} = \bigcup_{\mu \in \Lambda_{R_\lambda} \cap
 \Lambda^{+}} {\mathrm{mult}(\lambda:\mu)} \cdot (C_u^{\mu} ),
\end{equation*}
where $\mathrm{mult}(\lambda:\mu)$ denotes multiplicity of weight
$\mu$ in the module $R_{\lambda}$.
If $R_\lambda$ is minuscule module, then, by definition, the curve $C^{R_{\lambda}}_{u}$ is irreducible.

\begin{Example} Consider the $A_1$ theory
and take $\lambda = 3 \lambda_1$, i.e., the spin $\frac 32$
representation. If $T_1(x) = \tr_{R_{1}} g(x) = t(x) + t(x)^{-1}$ one finds that
\begin{gather*}
C^{R_{\lambda_1}}\colon \ 0 = 1 - T_{1}(x) t +t^2, \\
C^{R_{3\lambda_1}}\colon \ 0 = \big(1 - T_{1}(x) t +t^2 \big)\big(1 + 3 T_{1}(x) t -T_{1}(x)^{3} t + t^2\big).
 \end{gather*}
\end{Example}

Let $\iw_{\mu} \subset \iw$ be the stabilizer of $\mu$ in $\iw$, a subgroup of $\iw$.
Consider the map
\[
p_{\mu} \colon \ {\Cx} \times \left( {\BC}^{\times} \right)^{\Ver} \longrightarrow {\Cx} \times {\Ct}
\]
given by
\begin{gather*}
p_{\mu}\colon \ \big(x, ({\y}_{i})_{i \in \Ver} \big) \mapsto (x , t(x)) , \nonumber \\
t(x) = g(x)^{\mu}/g_{\infty}(x)^{\mu} =
\prod_{i\in \Ver} {\y}_{i}^{{\mu}({\al}_{i}^{\vee})}.
\end{gather*}
Under the map $p_{\mu}$ the curve ${\CalC}_{u}$ maps to $C_{u}^{\mu} = {\CalC}_{u}/{\iw}_{\mu}
\subset {\Cx} \times {\Ct}$, the irreducible $\mu$-component of the spectral curve.
This curve comes with the canonical differential, which is the restriction of the differential on ${\Cx} \times {\Ct}^{\times}$:
\begin{equation*}
{\rm d}S = x \frac{{\rm d}t}{t}.
\end{equation*}

Actually, in the case of the class II, II* theories the commonly used notion of the spectral curve differs from the one in \eqref{eq:spectral-curve}.

Although we suspect the study of spectral curves associated with the integrable highest weight representations of affine Kac--Moody algebras may be quite interesting, in this paper for the analysis of the class II and II* theories
we use the conventional notion of the spectral curve used for the study of families of $\bG$-bundles.

To define it, let us fix an irreducible representation $R$ of $\bG$,
${\pi}_{R}\colon {\bG} \to {\rm End}(R)$, and let us study the theory of a complex chiral fermion valued in $R$, more precisely, an $(1,0)$ $bc$ system in the representations $(R^{*}, R)$:
\begin{equation*}
{\mathscr L}_{bc} = \sum_{i=1}^{\dim R} \int b_{i} {\bar\pa} c^{i}
\end{equation*}
coupled to a background ${\bG} \times {\BC}^{\times}$ gauge field $\bar{\bf A} \oplus \bar A$, and compute its partition function on the torus $\ec$:
\begin{equation*}
Z ( {\bf t}, t , q ) = \Tr_{{\CalH}_{R}} \big( (-t)^J_{0} {\bf t}^{{\bf J}_{0}} q^{L_{0}} \big).
\end{equation*}

Mathematically, we consider the space
\begin{equation}
H_{R} = R \big[ z, z^{-1} \big] = H_{R}^{+} \oplus H_{R}^{-}
\label{eq:hr}
\end{equation}
of $R$-valued functions on the circle ${\BS}^{1}$. In \eqref{eq:hr},
we took Laurent polynomials in $z \in {\BC}^{\times}$, which correspond to Fourier polynomials on the circle. We may take some
completion of $H_R$ but we shall not do this in the definition of the spectral determinant below.
Consider an element ${\hat g} \in {\hat\bG}$ of the affine Kac--Moody group, i.e., the central extension of ${\widetilde{{\rm L}\bG}} = {\rm L}{\bG} \ltimes {\BC}^{\times}$, the loop group ${\rm L}{\bG}$ extended by the
$\BC^\times$ acting by loop rotations.
We have the canonical homomorphism-projection $f\colon {\hat\bG} \longrightarrow {\widetilde{{\rm L}\bG}}$ with the fiber ${\BC}^{\times}$, the center of the
central extension
\begin{equation}
f\colon \ {\hat g} \mapsto g(z) q^{z{\pa}_{z}}.
\label{eq:hgproj}
\end{equation}
The projection is topologically non-trivial.

Now, ${\widetilde{{\rm L}\bG}}$ acts in $H_R$ via rotation and evaluation,
and so does $\hat\bG$ thanks to~\eqref{eq:hgproj}:
for~${\Psi} \in H_{R}$:%
\begin{equation*}
\left( f({\hat g})\cdot {\Psi} \right) (z) = {\pi}_{R} (g(z)) \cdot {\Psi}(q z).
\end{equation*}
We would like to define the spectral determinant of $f({\hat g})$
in the representation $H_R$. The eigenvalues of $f({\hat g})$ are easy to compute
\begin{equation}
{\rm Eigen}(f ({\hat g})) = \big\{ {\bf t}^{\mu} q^{n} \mid {\mu} \in {\Lambda}_{R}, \, n \in \BZ \big\},
\label{eq:eighr}
\end{equation}
where we transformed $g(z)$ to a constant ${\bf t} \in \bT$ by means of a $z$-dependent $\bG$-gauge transformation:
\begin{equation*}
g(z) \mapsto h^{-1}(z) g(z) h ( q z) = {\bf t}.
\end{equation*}
The fibration $f\colon {\hat\bG} \to {\widetilde{{\rm L}{\bG}}}$, restricted onto ${\BC}^{\times}_{q} \times {\bT} \subset {\widetilde{{\rm L}{\bG}}}$ becomes trivial, $f^{-1} \left( {\BC}^{\times}_{q} \times {\bT} \right) \approx {\BC}^{\times}_{c} \times {\BC}^{\times}_{q} \times {\bT} $. Let us denote by $c$ the coordinate on the first factor.

The eigenvalues \eqref{eq:eighr}
concentrate both near $0$ and
$\infty$, so
we define
\begin{align}
\det\nolimits_{H_{R}} \big( 1 - t^{-1} {\hat g} \big) :={}& \det\nolimits_{H_{R}^{+}} \big( 1 - t^{-1} {\hat g} \big)
\det\nolimits_{H_{R}^{-}} \big( 1 - t {\hat g}^{-1} \big) \nonumber\\
={}& c^{{\kappa}_{R}} \prod_{{\mu} \in {\Lambda}_{R}} \prod_{n=0}^{\infty}
\big( 1 - q^{n} t^{-1} {\bf t}^{\mu} \big) \big( 1 - q^{n+1} t {\bf t}^{-{\mu}} \big).
\label{eq:dethr}
\end{align}
The expression \eqref{eq:dethr} is $W({\hat\g})$-invariant. The
shifts by $\rl$ act as follows, cf.~\eqref{eq:wshft}:
\begin{equation*}
( {\bf t}, c ) \mapsto \big( q^{\be} \cdot {\bf t} , {\bf t}^{\be} q^{\frac 12 \la \be, \be \ra} \cdot c \big),
\end{equation*}
where we view $\be \in \rl$ both as a vector in the root lattice and as a vector in the coroot lattice, and $\la \,,\, \ra$ is the Killing metric.
The level ${\kappa}_{R}$ in \eqref{eq:dethr} is defined as follows:
\begin{equation}
\sum_{{\mu} \in {\Lambda}_{R}} {\mu} \la \mu, \be \ra = {\kappa}_{R} {\be}
\label{eq:kapra}
\end{equation}
for any vector $\be \in \rl$.
Geometrically the spectral curve corresponding to $R$ is obtained as follows:
consider the universal principal $\bG$-bundle ${\CalU}$ over
$\Bnq \times \ec$, and associate the vector bundle~$\mathscr R$
with the fiber $R$:
\[
{\mathscr R} = {\CalU} \times_{\bG} R.
\]
Now restrict it onto the rational curve $\Sigma_{u} \subset \Bnq$. We get the $R$-bundle over
$\Sigma_{u} \times {\ec}$.

For generic point $x \in {\Cpx}$ over the corresponding point
$U(x) \in \Sigma_{u}$ we get the vector bundle~${\mathscr R}_{x}$ over $\ec$, which is semi-stable, and splits as a direct sum
of line bundles
\begin{equation*}
{\mathscr R}_{x} = \bigoplus_{{\mu} \in {\Lambda}_{R}} {\mathscr L}_{{\mu},x},
\end{equation*}
where the summands are the degree zero line bundles on $\ec$.
Under the identification $\operatorname{Pic}_{0} ({\ec})$ with $\ec$ the line bundle ${\mathscr L}_{{\mu},x}$
corresponds to the point ${\bf t}(x)^{\mu}\, {\rm mod}\, {\qe}^{\BZ}$
for some ${\bf t}(x) \in {\bT}/{\qe}^{{\rl}^{\vee}}$. The closure of the union
\begin{equation*}
\bigcup_{x \in {\Cpx}} \big\{ {\bf t}(x)^{\mu}\mid {\mu} \in {\Lambda}_{R} \big\} \subset {\Cpx} \times {\ec}
\end{equation*}
is the spectral curve $C^{R}_{u}\subset {\Cpx} \times {\ec}$.
It is given by the vanishing locus of the regularized determinant \eqref{eq:dethr}:
\begin{equation*}
c(x)^{{\kappa}_{R}} \prod_{{\mu} \in {\Lambda}_{R}} {\theta} \big( t^{-1} {\bf t}(x)^{\mu} ; q \big) = 0
\end{equation*}
the choice of the $x$-dependence of $c(x)$ seems immaterial at this point, as long
as $c(x) \in {\BC}^{\times}$.

\subsubsection*{Degree of the spectral curve}\label{se:Affine-degrees}
The $x$-degree of the spectral curve
for class II theories in representation $R$
is $ N \kappa_{R}$ where $\kappa_{R}$ is given by \eqref{eq:kapra}.
The $\kappa_{R}$ is
the proportionality constant for the second Casimir in representation $R$
$\tr_{R} (\cdot, \cdot) = \kappa_{R} (\cdot, \cdot)_2$ where
the $(\cdot,\cdot)_2$ is the canonical Killing form in which the long
roots have length square equal to $2$.
The standard computations leads to
\begin{equation*}
 \kappa_R = \frac{ \dim_R}{ \dim_{\g} } (\lambda_R, \lambda_R + 2 \rho)_{2},
\end{equation*}
where $\rho = \frac 1 2 \sum_{\alpha > 0} \alpha$ is the Weyl vector.
For fundamental representations $R_1$ we find for all cases
\begin{gather*}
\kappa_{R_1}(A_r) = 1, \qquad
\kappa_{R_1}(D_r) = 2, \qquad
\kappa_{R_1}(E_6) = 6, \qquad
\kappa_{R_1}(E_7) =12, \qquad
\kappa_{R_1}(E_8) =60.
\end{gather*}

\subsection{Obscured curve} \label{sec:obscura}

In the previous construction, in view of the identification ${\mathscr L}_{{\mu},x} \leftrightarrow {\bf t}(x)^{\mu}$ we can decompose, for each weight
\begin{gather*}
{\mu} = \sum_{i=1}^{r} {\mu}_{i} {\lam}_{i} \in {\Lambda}_{R}, \qquad
{\mathscr L}_{{\mu},x} = \bigotimes_{i=1}^{r} {\BL}_{i,x}^{\otimes {\mu}_{i}}
\end{gather*}
for some ``basic'' line bundles ${\BL}_{i,x}$ corresponding to the fundamental weights.
These basic line bundles are ordered, so they define a point
\[
\{ {\BL}_{1, x}, \dots , {\BL}_{r, x} \} \in \operatorname{Pic}_{0}({\ec})^{r} \approx {\ec}^{r} ,
\]
the Cartesian product of $r$ copies of the elliptic curve.
Taking the whole family and including the parametrization we obtain the \emph{obscured curve} ${\obs}_{u}$:
\begin{equation*}
{\obs}_{u} = \big\{ ( x ; {\BL}_{1, x}, \dots , {\BL}_{r, x} ) \mid x \in {\Cpx} \big\} \in
{\Cpx} \times {\ec}^{r}.
\end{equation*}
Let us present another simple construction of ${\obs}_{u}$.
Namely, let us use the fact~\cite{Donagi:1997dp,Friedman:1997yq,Friedman:2000ze,Friedman:1998si, Friedman:1997ih}, that
\begin{equation}
{\Bnq} = ({\ec} \otimes {\rl}) /W({\g}),
\label{eq:bnqec}
\end{equation}
where the tensor product is understood in the category of abelian groups. At the level of manifolds, \eqref{eq:bnqec} simply says that
\begin{equation*}
{\Bnq} = {\ec}^{r} / W({\g})
\end{equation*}
for some natural action of the Weyl group $W({\g})$ on the Cartesian product of
$r$ copies of $\ec$. Let us denote by ${\pi}_{W}$ the projection
\begin{equation*}
{\pi}_{W}\colon \ {\ec}^{r} \to {\Bnq} = {\ec}^{r}/W({\g}).
\end{equation*}
The rational curve $\Sigma_{u}$ in $\Bnq$ lifts to a curve in ${\ec}^{r}$, and the graph of the parametrized curve $\Sigma_{u} \in {\Cpx} \times {\Bnq}$ lifts
to the graph in
${\Cpx} \times {\ec}^{r}$ which is our friend \emph{obscured curve}~${\obs}_{u}$. It is the quotient
of the cameral curve by the lattice $\rl^{\vee}$:
\begin{equation*}
{\obs}_{u} = {\CalC}_{u}/{\rl}^{\vee}.
\end{equation*}
In Section~\ref{sec:double}, we shall present yet another construction of
${\BL}_{i,x}$'s, using gauge theory.

There is the so-called determinant line bundle $L$ over the
moduli space $\Bnq$,
whose sections are the fundamental characters $\hat\chi_{i}$, $i = 0, 1, \dots , r$.
In Loojienga's identification
${\Bnq} \approx {\WP}^{a_{0}, a_{1}, \dots , a_{r}}$
this line bundle is just the ${\CalO}(1)$
orbibundle over the weighted projective space.

We have then the line bundle ${\mathscr L}$ over ${\ec}^{r}$:
\begin{equation*}
{\mathscr L} = {\pi}_{W}^{*}L.
\end{equation*}
Let us call this line bundle \emph{the abelianized determinant line bundle}.

\section{The Seiberg--Witten curves in some detail}\label{se:Seiberg--Witten}

In this section, we shall discuss the geometry of curves describing the limit shape configurations and the special geometry of the gauge theories under consideration.
When possible we identify the cameral or the spectral curves with the analogous curves
of some algebraic integrable systems, namely the Hitchin systems on the
genus zero (i.e., Gaudin model) or genus one (i.e., spin elliptic
Calogero--Moser system) curves with punctures. These identifications are
less universal than the identification with the spectral curves of the
spin chains based on the Yangian algebra built on~$\g$,~$\hat \g$, or $\Gli$,
respectively. The latter identification is a subject of a separate venue of research which
touches upon various advances in geometric representation theory, study of the symplectic geometry of moduli spaces of instantons and monopoles, quantum cohomology of quiver varieties, to name just a few. We shall only mention the relation to spin chains in a~few examples, in this work.

Throughout this section we shall use the notation
\begin{equation*}
g_{\lam}(x) = {\zeta}(x)^{-1} {\pi}_{\lam} (g(x))
\end{equation*}
for the projectively modified operator in the representation $(R_{\lam}, {\pi}_{\lam})$ of
$\Gq$, corresponding to the group element $g(x) \in \Gq$.

\subsection[Class I theories of $A$ type]{Class I theories of $\boldsymbol{A}$ type}

This is the so-called linear quiver theory.
The set of vertices $\Ver = \{ 1, \dots, r \}$,
the set of edges $\Edg = \{ 1, \dots, r-1 \}$, the maps $s$, $t$ for a particular orientation are given by $s(e) = e$, $t(e) = e+1$. The bi-fundamental masses are a trivial cocycle:
\begin{equation*}
m_{e} = {\mu}_{e+1} - {\mu}_{e}.
\end{equation*}
 The corresponding conformal group ${\rm C}\bG = {\rm GL}(r+1, {\BC})$, the fundamental characters ${\chi}_{i}$
are the characters of the representations ${\Lambda}^{i}{\BC}^{r+1}$. We shall now describe
the spectral curve in the representation $R_{\lam_{1}} \approx {\BC}^{r+1}$.
The corresponding
group element $g_{\lam_{1}}(x)$ in~\eqref{eq:gkacm} is the diagonal matrix
\[
g_{\lam_{1}}(x) = \operatorname{diag}( t_{1}(x), \dots , t_{r+1}(x) )
\]
with
\begin{gather}
 t_{1}(x) = {\ze}(x){\y}_{1}(x), \qquad t_{r+1}(x) = {\ze}(x) {\bq}^{[r]} (x) {\y}_{r}(x)^{-1}, \nonumber\\
 t_{i}(x) = {\ze}(x) {\bq}^{[i-1]} (x) {\y}_{i}(x){\y}_{i-1}(x)^{-1}, \qquad i = 2, \dots , r,
\label{eq:glrel}
\end{gather}
with some normalization factor ${\ze}(x)$ which we choose shortly,
and the explicit formula for the invariants
${\CalX}_{i}({\y}(x))$ is (we omit the $x$-dependence in the right-hand side):
\begin{equation}
{\CalX}_{i}({\y}(x)) = \prod_{j=1}^{i-1} {\bq}_{j}^{j-i} e_{i} \big( {\y}_{1}, {\y}_{2}{\y}_{1}^{-1}{\bq}^{[1]}, \dots, {\y}_{i}{\y}_{i-1}^{-1} {\bq}^{[i-1]} , \dots , {\y}_{r}^{-1} {\bq}^{[r]} \big),
\label{eq:ainv}
\end{equation}
where $e_{i}$ are the elementary symmetric polynomials in $r+1$ variables.
Our master equations~\eqref{eq:master} equate the right-hand side of~\eqref{eq:ainv} with the
degree ${\bv}_{i}$ polynomial $T_{i}(x)$ in $x$, cf.~(\ref{eq:cxtj}).

It is convenient to organize the invariants (\ref{eq:ainv}) into a generating polynomial, which
is nothing but the characteristic polynomial of the group element $g(x)$ in some representation of~${\rm C}{\bG}$. The most economical is, of course, the defining fundamental representation ${\BC}^{r+1}$ with the highest weight ${\lam}_{1}$:
\begin{gather}
\det \left( t\cdot 1_{r+1} - g_{\lam_{1}}(x) \right) = t^{r+1} + \sum_{i=1}^{r} (-1)^{i}
t^{r+1-i} {\ze}(x)^{i} \prod_{j=1}^{i-1} {\bq}_{j}^{i-j}(x) {\CalX}_{i} ({\y}(x)) \nonumber \\
\hphantom{\det \left( t\cdot 1_{r+1} - g_{\lam_{1}}(x) \right) =}{} + (-{\ze}(x))^{r+1}
\prod_{j=1}^{r} {\bq}_{j}^{r+1-j} (x).
\label{eq:charar}
\end{gather}

The group $\iw$ is the symmetric group ${\CalS}_{r+1}$, which acts by permuting
the eigenvalues of~$g(x)$ in~\eqref{eq:glrel}. The cameral curve ${\CalC}_{u}$ is the $(r+1)!$-fold ramified cover of the compactified $x$-plane~$\Cpx$. The points in the fiber are the
\emph{ordered} sets of roots $(t_{1}(x), \dots , t_{r+1}(x))$ of the polynomial~\eqref{eq:charar}.

The curve ${\CalC}_{u}$ covers the \emph{spectral curve} $C_{u}$. The latter is defined as the zero locus of the characteristic polynomial~\eqref{eq:charar}.
The cover ${\CalC}_{u} \to C_{u}$ is $r! : 1$, it sends the ordered $(r+1)$-tuple of roots
$(t_{1}, \dots , t_{r+1})$ to the first root $t_{1}$. The cover $C_{u} \to \Cx$ is $(r+1) : 1$.

Explicitly, the curve $C_{u}$ is given by
\begin{equation}
0 = {\CalP}(t, x) = \sum_{i=0}^{r+1} (-1)^{i}
t^{r+1-i}{\ze}(x)^{i} {\prod_{j=1}^{i-1} {\bq}_{j}(x)^{i-j} } T_{i} (x).
\label{eq:curvu}
\end{equation}

\subsection{Relation to Gaudin model}\label{subsubsec:gaudin}
\begin{figure}
 \centering
 \includegraphics[width=5cm]{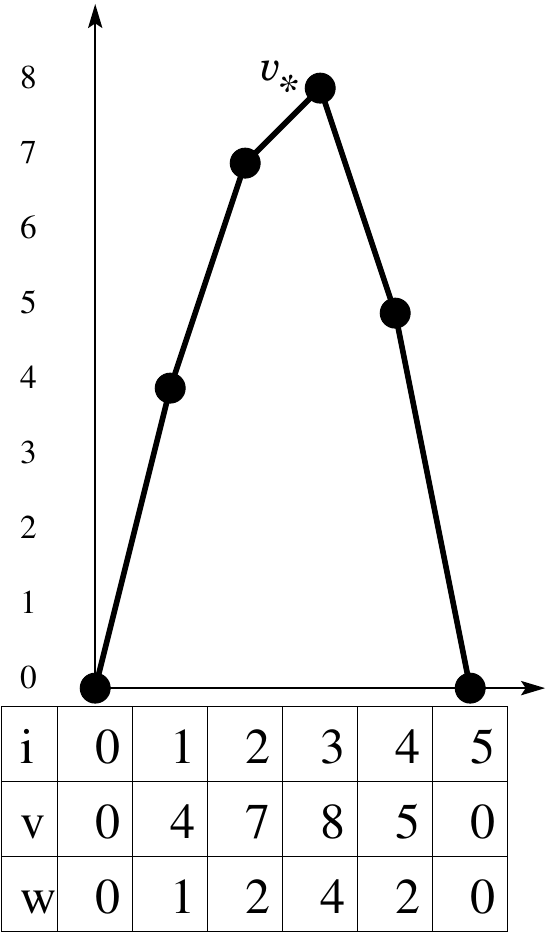}
 \caption{Degree profile example for $A_{4}$ theory and $(\bv_1, \bv_2,
 \bv_3, \bv_4) = (4,7,8,5)$. For convenience one can set boundary
 conditions $\bv_0 = \bv_{r+1} = \bw_{0} = \bw_{r+1} = 0$. }
\label{fig:gorka}
\end{figure}

It is easy to see, using the equations~\eqref{eq:betz},~\eqref{eq:gorka} and
Figure~\ref{fig:gorka} that
${\bw}_{i_{*}} = w_{+} + w_{-}$, where
\begin{equation*}
 w_{+} = {\bv}_{i_{*}}-{\bv}_{ i_{*}+1} \geq 0, \qquad
 w_{-} = {\bv}_{i_{*}}-{\bv}_{ i_{*}-1} \geq 0
\end{equation*}
and it useful to record
\begin{equation*}
 \begin{aligned}
 \bv_1 &= \bw_{1} + \dots +\bw_{i_{*}-1} + \bw_{-},\\
 \bv_{r}&=\bw_{r} + \dots + \bw_{i_{*}+1} + \bw_{+} ,\\
 \bv_{*}&= \sum_{i=1}^{i_{*}-1} i \bw_{i} + i_{*} \bw_{-},\\
 \bv_{*}&=\sum_{i=i_{*}+1}^{r} (r+1-i) \bw_{i} + (r+1-i_{*}) \bw_{+}.
 \end{aligned}
\end{equation*}
Accordingly, we can factorize the polynomial ${\bq}_{i_{*}}(x)$ as
\begin{equation*}
{\bq}_{i_{*}}(x) = {\qe}_{i_{*}} {\bq}^{+}(x) {\bq}^{-}(x),
\end{equation*}
where ${\bq}^{\pm}$ are monic polynomials of degrees
\[
\deg {\bq}^{\pm} = w_{\pm}.
\]
We can actually transform (\ref{eq:curvu}) into something nice, by
adjusting ${\ze}(x)$:
\begin{equation}
 {\ze}(x)^{-1} = {\bq}^{-}(x) {\bq}^{[i_{*}-1]}(x).
\label{eq:ttoze}
\end{equation}
Then $D(g(x))$ is given by
\begin{equation*}
\begin{aligned}
& D(g(x)) = \frac{P_{0}(x)}{P_{\infty}(x)}, \\
& P_{0}(x) = {\bq}^{+}(x)^{r+1-i_{*}} \prod_{j=i_{*}+1}^{r} {\bq}_{j}(x)^{r+1-j}, \\
& P_{\infty}(x) = {\bq}^{-}(x)^{i_{*}} \prod_{j=1}^{i_{*}-1} {\bq}_{j}(x)^{j}. \end{aligned}
\end{equation*}
Then ${\CalP}(t,x)$ can be written as
\[
{\CalP}(t,x) = \prod_{j=1}^{i_{*}-1} {\qe}_{j}^{j} \cdot \frac{P(t,x)}{P_{\infty}(x)},
\]
where $P(t,x)$ is a degree $N = {\bv}_{i_*}$ polynomial in $x$, and the degree $r+1$ polynomial in $t$, which is straightforward to calculate
\begin{gather}
(-1)^{i_{*}}
\prod_{j=1}^{i_{*}} {\qe}_{j}^{j} P(t,x) = (-{\qe}_{i_{*}})^{i_{*}} t^{r+1} P_{\infty}(x) + \sum_{i=1}^{i_{*}} t^{r+1-i} T_{i}(x) {\qe}_{i_{*}}^{i_{*}-i} ( -{\bq}_{*}^{-}(x) )^{i_{*}-i} \prod_{j=i}^{i_{*}-1} {\bq}_{j}^{j-i}(x) \nonumber \\
 \hphantom{(-1)^{i_{*}}\prod_{j=1}^{i_{*}} {\qe}_{j}^{j} P(t,x) =}{} + \sum_{i=i_{*}+1}^{r} t^{r+1-i} T_{i}(x) ( -{\bq}_{*}^{+}(x))^{i - i_{*}} \prod_{j=i_{*}+1}^{i-1}
{\bq}_{j}^{i-j}(x) \nonumber\\
\hphantom{(-1)^{i_{*}}\prod_{j=1}^{i_{*}} {\qe}_{j}^{j} P(t,x) =}{}+ (-1)^{r+1-i_{*}} P_{0}(x).
\label{eq:ptox}
\end{gather}
Now, recall that $T_{j,0}$ is fixed by the couplings ${\qe}$:
\begin{equation*}
T_{j,0}({\qe}) = \prod_{j=1}^{i-1} {\qe}_{j}^{j-i} e_{i}(1, {\qe}_{1},
{\qe}_{1}{\qe}_{2}, \dots , {\qe}_{1}{\qe}_{2}\dots {\qe}_{i}, \dots,
{\qe}_{1}\dots {\qe}_{r} )
\end{equation*}
and the coefficient $T_{j,1}$ is fixed by the masses $m_{i, {\mathfrak {f}}}$
and $m_{e}$.

Therefore, the coefficient of $x^{N}$ in $P(t,x)$ can be computed explicitly
\begin{equation*}
 \sum_{i=0}^{r+1} (-1)^{i} t^{r+1-i}
\prod_{j=1}^{i_{*}} {\qe}_{j}^{-i} e_{i}(1, {\qe}_{1},
{\qe}_{1}{\qe}_{2}, \dots , {\qe}_{1}{\qe}_{2}\dots {\qe}_{i}, \dots ,
{\qe}_{1}\dots {\qe}_{r} ) = \prod_{i=0}^{r}
\big( t - {\ct}_{i} \big),
\end{equation*}
where
\begin{equation*}
{\ct}_{i} = \frac{\prod_{j=1}^{i} {\qe}_{j}}{\prod_{j=1}^{i_{*}} {\qe}_{j}} , \qquad i = 0, \dots , r.
\end{equation*}
We thus rewrite the curve $C_{u}$ in the $(x,t)$-space, defined by the equation
\begin{gather}
 0 = {\CalR}_{A_{r}}(t, x) = \frac{P(t,x)}{\prod_{i=0}^{r}
\big( t - {\ct}_{i} \big)} = \prod_{l=1}^{N} ( x - x_{l}(t) )
 = x^{N} + \frac{1}{\prod_{i=0}^{r}
\big( t - {\ct}_{i} \big)} \sum_{j=1}^{N} p_{j}(t) x^{N-j},\!\!\!
\label{eq:spctrc}
\end{gather}
where
\begin{equation*}
N = {\bv}_{i_{*}}.
\end{equation*}
It is clear from the equation~(\ref{eq:spctrc}) that as $t \to {\ct}_{i}$ one of the roots $x_{l}(t)$ has a pole, while the other $N-1$ roots are finite.
Near $t = 0$ the polynomial $P(t,x)$ approaches:
\begin{equation*}
P(0,x) = ( -1)^{r+1 - i_{*}} P_{0}(x),
\end{equation*}
while near $t = \infty$
\begin{equation*}
P(t,x) t^{-r-1} \to
(-{\qe}_{i_{*}})^{i_{*}} P_{\infty}(x).
\end{equation*}
Let
\begin{equation*}
 {\rm d}S = x\frac{{\rm d}t}{t}.
 \end{equation*}
Then our discussion above implies that the differential $dS$ has the first-order poles on $C_{u}$:
at one of the $N$ preimages of the points ${\ct}_{i}$, $i = 0, 1, \dots ,r$, and
at all preimages of the points $t = 0$ and $t = \infty$. The residues of ${\rm d}S$ are linear combinations of the masses of the hypermultiplets, in agreement with the observations
in~\cite{Donagi:1995cf,Seiberg:1994aj}.

Remarkably, we can identify $C_{u}$ with the spectral curve of the meromorphic Higgs field ${\Phi}$:
\begin{equation}
{\Phi} = {\Phi}(t) {\rm d}t = \sum_{j=-1}^{r+1} {\Phi}_{j} \frac{{\rm d}t}{t - {\ct}_{j}},
\label{eq:phimj}
\end{equation}
where ${\ct}_{-1} = 0$, ${\ct}_{r+1} = {\infty}$, and ${\Phi}_j$
are $N \times N$ matrices, which have rank one for $j = 0, 1, \dots , r$, and have the maximal rank for $j = -1, r+1$. Moreover, the eigenvalues of ${\Phi}_j$ are all fixed in terms of the masses. The spectra of ${\Phi}_{j}$, $j = -1, \dots , r+1$ have specified multiplicity:
\begin{enumerate}\itemsep=0pt
\item
The matrix ${\Phi}_{-1}$ has $w_{+}$ eigenvalues of multiplicity $r+1-i_{*}$, and
${\bw}_{r+1-j}$ eigenvalues of multiplicity $j$, for $j = 1, \dots , r - i_{*}$; the eigenvalues are fixed by the masses.
\item
The matrices ${\Phi}_j$, $j = 0, 1, \dots, r$ has one non-vanishing eigenvalue each, and $N-1$ vanishing eigenvalues.
We can write
\[
( {\Phi}_{j} )_{a}^{b}= u_{a}^{j} v_{j}^{b}, \qquad a,b =1, \dots, N
\]
for some vectors $u^{j}, v_{j} \in {\BC}^{N}$, obeying
\begin{equation}
\sum_{a=1}^{N} u^{j}_{a} v_{j}^{a} = M_{j}
\label{eq:uvmom}
\end{equation}
and considered up to an obvious ${\BC}^{\times}$-action,
for some $M_j$ which is linear in the bi-fundamental and fundamental masses.
\item
The matrix
${\Phi}_{r+1}$ has $w_{-}$ eigenvalues of multiplicity $i_{*}$, and
${\bw}_{j}$ eigenvalues of multiplicity~$j$, for $j = 1, \dots , i_{*}-1$.
\end{enumerate}
Then
\begin{equation}
\bigg(\frac{{\rm d}t}{t}\bigg)^{N} {\CalR}_{A_{r}}(t,x) = \det \bigg( x\frac{{\rm d}t}{t} - {\Phi}\bigg).
\label{eq:specdet}
\end{equation}
We can make an ${\rm SL}(N)$ Higgs field out of $\Phi$ by shifting it by the scalar meromorphic
one-form~$\frac{1}{N} \Tr_{N} {\Phi}$, which is independent of the moduli $u$ of the curve $C_u$.

The moduli space of $(r+3)$-ples of matrices ${\Phi}_j$, obeying
\begin{equation}
\sum_{j=-1}^{r+1} {\Phi}_{j} = 0
\label{eq:momm}
\end{equation}
with fixed eigenvalues of the above mentioned multiplicity, considered
up to the simultaneous ${\rm SL}(N)$-similarity transformation,
is the phase space ${\pv}^{H}_{0,r+3}$ of the genus zero version of ${\rm SL}(N)$ Hitchin system, the classical Gaudin model on $r+3$ sites. The general Gaudin model has the residues ${\Phi}_j$ belonging to arbitrary conjugacy classes.

See~\cite{Kronheimer:1990a,Kronheimer:1990} for the geometry of complex coadjoint
orbits. The Hitchin system with singularities was studied in
\cite{Donagi:1995am, Gorsky:1994dj,Gukov:2006jk, Gukov:2008sn, Nekrasov:1995nq, Witten:2007td}. In~\cite{Gaiotto:2009we, Nanopoulos:2009xe, Nanopoulos:2009uw,Nanopoulos:2010zb, Nanopoulos:2010ga}
this Hitchin system with singularities was discussed from the point of view of brane constructions such as~\cite{Gaiotto:2009we,Witten:1997sc}.

\begin{Remark}
The curve $C_{u}$ is much more economical then ${\CalC}_{u}$. However,
the price we pay is the complexity of the relation between the special coordinates
${\ac}_{i{\ba}}$, ${\ac}_{i{\ba}}^{D}$ and the moduli $u$ of the curve $C_u$.
Roughly speaking, all special coordinates are linear combinations of the periods of the differential
\[
x \frac{{\rm d}t}{t}
\]
and the masses. The coordinates ${\ac}_{1{\ba}}$ come from the periods
\[
\oint x \, {\rm d} \log g_{1}(x) \sim \oint x \, {\rm d}t/t,
\]
the coordinates ${\ac}_{2{\ba}}$ come from the periods
\[
\oint x\, {\rm d} \log (g_{1}(x)g_{2}(x)) \sim \oint x \, {\rm d}t/t + \oint x \,{\rm d}t/t,
\]
the coordinates ${\ac}_{i{\ba}}$ come from the periods
\[
\oint x\, {\rm d} \log (g_{1}(x) \cdots g_{i}(x)) \sim \oint x \, {\rm d}t/t + \dots + \oint x \, {\rm d}t/t,
\]
etc.
\end{Remark}

\begin{Remark}
In the $A_{2}$ case our solution matches the one found in~\cite{Shadchin:2005cc}.
\end{Remark}

\begin{Remark}
We can connect the cameral curve ${\CalC}_{u}$ to the spectral curve $C_u$ via a tower of
ramified covers:
\begin{equation*}
{\CalC}_{u} \to C_{u}^{(r)} \to C_{u}^{(r-1)} \to \dots \to C_{u}^{(1)} = C_{u} \to {\Cpx},
\end{equation*}
which we can call the \emph{Gelfand--Zeitlin} tower of curves. The curve $C_{u}^{(i)}$ is the quotient of ${\CalC}_{u}$ by the subgroup $W(A_{r-i})$ of the Weyl group $W(A_{r})$, which acts on the amplitudes $({\y}_{i+1}, \dots , {\y}_{r})$ while preserving $({\y}_{1}, \dots , {\y}_{i})$.
\end{Remark}

\begin{Remark}
We should warn the reader that our cameral curves need not be the cameral curves of Hitchin systems~\cite{Donagi:1995alg}. We mapped the spectral curve of the family of conjugacy classes $[g(x)]$ corresponding to the fundamental representation $R_1$
to the spectral curve of the ${\rm GL}(N)$-Gaudin system, i.e., the genus zero Hitchin system, corresponding to the $N$-dimensional representation. One could then build the cameral curve for the ${\rm GL}(N)$-Gaudin system. This curve has all the reasons to differ from our cameral curve ${\CalC}_{u}$.

However, the identification of $\mv$ with the moduli spaces of curves describing the spectrum of the transfer matrix in the quasi classical limit of the $Y(A_{r})$ spin chain
is more natural, and carries over to the level of cameral curves.
\end{Remark}

\begin{Remark}
In view of~\cite{Gaiotto:2009we}, it is natural to identify the space of couplings ${\tqe} = ({\qe}_{1}, \dots , {\qe}_{r})$ with a coordinate patch in the moduli space $\overline{\CalM}_{0,r+3}$
of stable genus zero curves with $r+3$ punctures. In this fashion
the linear quiver theories (the class I type $A_r$ theories) can be analytically continued to other weakly coupled regions (weak coupling corresponds to the maximal degeneration of the stable curve). Most of these regions do not have a satisfactory Lagrangian description. Nevertheless, it would be interesting to try to generalize the limit shape equations even without knowing their microscopic origin. What would the iWeyl group
look like in this case?
\end{Remark}

\subsection{Quiver description} 

We have thus found that a particular subset of Gaudin--Hitchin models, with all but two residues of the minimal type, are the Seiberg--Witten integrable systems of the class I $A_r$ type theories. As a check, let us compute the dimension of the moduli space ${\pv}^{H}_{0,r+3}$ of solutions to the (traceless part of the) moment map equation~(\ref{eq:momm}) divided by the ${\rm SL}(N, {\BC})$-action is equal to
\begin{gather*}
 2(r+1) (N-1) - 2 ( N^2 - 1) + \bigg( N^2 - \sum_{j=1}^{i_{*}-1} j^{2} {\bw}_{j} - i_{*}^{2} w_{-} \bigg) \\
\qquad{}+\bigg( N^2 - \sum_{j=i_{*}+1}^{r} (r+1-j)^{2} {\bw}_{j} - (r+1-i_{*})^{2} w_{+} \bigg) = 2 \sum_{i=1}^{r} ({\bv}_{i}-1) = 2 \dim {\mv}.
 \end{gather*}

\begin{figure}
 \centering
 \includegraphics[width=12cm]{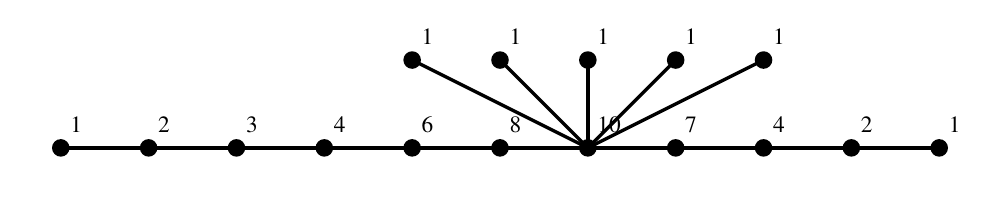}
\caption{The example quiver variety for $A_{4}$ quiver at $\bv=(7,10,8,5)$ and
 $\bw=(4,5,1,2)$ with $i_{*} =2$ and $\bw_{*} = \bw_{-} + \bw_{+}$ with
$ \bw_{-} = 3$ and $\bw_{+} = 2$. The labels at vertices denote the
dimensions in the pattern as explained. }
\label{fig:gaudin-pictures}
 \end{figure}

Actually, the moduli space ${\pv}^{H}_{0,r+3}$ can be described as a quiver variety. Its graph is an $(r+3)$-pointed star, with $r+1$ legs of length $1$, and two long legs, of the lengths
$l_{-1} = {\bv}_{r}-1$ and $l_{r+1} = {\bv}_{1}-1$, respectively. The dimensions of the vector spaces assigned to vertices are:
 the $(r+3)$-valent vertex (the star) has dimension $N$, the tails of the short legs all have dimension~$1$, the dimensions along the long legs start at~$1$ at the tails, then grow with the step~$1$ for the first~${\bw}_{1}$ (respectively,~${\bw}_{r}$) vertices, then grow with the step $2$ for the next ${\bw}_{2}$ (respectively,~${\bw}_{r-1}$) and so on. (See example in Figure~\ref{fig:gaudin-pictures}.)

The extended phase space $\pve$ for the class I $A_r$ type theories
is easy to describe. One just need to relax the ${\BC}^{\times}$ moment map constraints
 \eqref{eq:uvmom} as well as the analogous ${\BC}^{\times}$ constraints for the ${\Phi}_{-1}$, ${\Phi}_{r+1}$ residues. In the quiver description we make the quiver gauge group the
 product of the special unitary groups as opposed to the product of unitary groups.

\subsection{Reduction to the spin chain}

The simplest example of the class I theory of the $A$ type is, of course, the $A_1$ theory. This is the celebrated $N_{f}= 2N_c$ theory, with ${\bw}_{1} = N_{f}$, ${\bv}_{1} = N_{c} = N$, in our notation. Let ${\qe} = {\qe}_{1}$ and let $T(x) = T_{1,0}^{-1} T_{1}(x)$ denote the monic degree $N$ polynomial.

The reduced curve \eqref{eq:ptox} assumes a very simple form
\begin{equation}
\label{eq:redaone}
{\qe} {\bq}^{-}(x) t + {\bq}^{+}(x) t^{-1} = (1 + {\qe}) T(x).
\end{equation}
It is not difficult to recognize in this formula the quasiclassical limit of Baxter's $T$-$Q$ equation~\cite {Baxter:1985} for the XXX $\mathfrak{sl}_2$ spin chain.
In fact, it was observed already in~\cite{Gorsky:1996qp,Gorsky:1996hs} that the Seiberg--Witten curve of the ${\CalN}=2$ supersymmetric QCD can be interpreted using the integrable spin chain, albeit in a somewhat different fashion.
Note that a possible lift of $[g(x)]$ to $C {\bG} = {\rm GL}(2, {\BC})$ in this case is given by the
diagonal matrix
\begin{equation*}
g(x) = \left( \begin{matrix} {\qe} t {\bq}^{-}(x) & 0 \\
0 & t^{-1} {\bq}^{+}(x) \end{matrix} \right),
\end{equation*}
where $t$ solves \eqref{eq:redaone}.
However this choice of $g(x)$ is not continuous in $x$. As we cross the cuts $I_{1, {\ba}}$
the matrix $g(x)$ will have its diagonal entries exchanged.
We can conjugate $g(x) \to h^{-1} (x) g(x) h(x)$ into a form, e.g.,
\begin{equation}
\mathbf{g}(x) = \left( \begin{matrix} {\qe} T(x)&
1 \\
{\qe} \left( T^2 (x) - {\bq}^{+}(x) {\bq}^{-}(x) \right) & T(x) \end{matrix} \right),
\label{eq:lift1}
\end{equation}
whose entries are polynomials. This is a particular case of a general statement~\cite{Steinberg:1965}, lifting a~family of conjugacy classes in $\Gq$ to $\Gq$ itself (slightly adapted for the conformal extension~${\rm C}\bG $). The lift \eqref{eq:lift1} does not depend on the split ${\bq}(x)$ into the product of ${\bq}^{\pm}$ factors.

There is yet another lift of $[g(x)]$ to ${\rm C}\bG $, which does depend on the factorization, and makes closer contact with spin chains. We shall discuss it in the section devoted to the study of the phase spaces of the integrable systems corresponding to our gauge theories.

 \subsection{Duality}

In the mapping to the Gaudin--Hitchin system we employed a particular lift $g(x)$ of the conjugacy class $[g(x)]$ in ${\rm SL}(r+1,{\BZ})/{\BZ}_{r+1}$
to the conjugacy class in ${\rm GL}(r+1,{\BC})$ by a judicious choice of the normalization factor ${\ze}(x)$. More importantly, the spectral one-form describing the eigenvalues of the Higgs field, is equal to $x{\rm d}t/t$ where
$x$ is the argument of the amplitude function, and $t$ is the spectral variable describing the eigenvalues of $g(x)$. For the group ${\rm GL}(r+1,{\BC})$ the eigenvalues of~$g(x)$ in some representation take values
in ${\Ct} = {\BC}^{\times}$ which gets naturally compactified to~${\BC\BP}^{1}$
to allow the degenerations.

To summarize, the Lax operator of Gaudin--Hitchin system, the Higgs field ${\Phi}(t){\rm d}t$
lives on the curve ${\Ct}$ of the eigenvalues of the ``Lax operator'' $g(x)$ of the gauge theory.
Vice versa, the ``Lax operator'' $g(x)$ of the gauge theory lives on the curve ${\Cx}$
of the eigenvalues of the Higgs field of Hitchin system.

We shall encounter some versions of this ``eigenvalue -- spectral parameter'' duality in other sections of this work.

\subsection[Class I theories of $D$ type]{Class I theories of $\boldsymbol{D}$ type}

These are the ${\rm SU}({\bv}_{1}) \times \dots \times {\rm SU}({\bv}_{r})$ theories whose quiver contains a trivalent vertex which connects two one-vertex legs to a leg of the length $r-3$. The corresponding group~$\Gq$ is ${\rm Spin}(2r, {\BC})$, its conformal version ${\rm C}\bG $ is the extension of $\bG$ by ${\BC}^{\times}$ or ${\BC}^{\times} \times {\BC}^{\times}$, depending on the parity of~$r$.

Passing from the $A$ type theories to the $D$ type theories we encounter new phenomenon. In addition to the exterior powers $\wedge^i V$ of
the vector representation $V = {\BC}^{2r}$ of ${\rm Spin}(2r)$ the
fundamental representations of the group $\Gq$ come also from spin representations $S_{\pm}$. We should use the cameral curve ${\CalC}_{u}$ to get the special coordinates and the prepotential, however a lot of information is contained in the spectral curve $C_u^{R}$ in some fundamental representation $R$, which we shall take to be the vector $2r$-dimensional representation $V = R_{\lam_{1}} = {\BC}^{2r}$. In order to describe the spectral curve we need to know the characters of the group element $g(x)$ (\ref{eq:gkacm}) in the representations
$\wedge^{i}V$,
for $i = 1, \dots , 2r$.
When we deal with $V$ and its exterior powers only, we do not see the full conformal version of $\bG$, only its one-dimensional extension (which we shall denote simply by
C$\bG$) which consists of the matrices $g \in {\rm GL}(2r, {\BC})$, such that $g g^{t} = D(g) \cdot {\bf 1}_{2r}$, with $D(g) \in {\BC}^{\times}$ a scalar.

The spectral curve $C_{u} = C^{V}_{u}$ in the vector representation can be modified
by the transformation similar to (\ref{eq:ttoze}) to get the curve of
minimal degree in $x$. Let us label the vertices of the $D_r$ Dynkin diagram in such a way,
that the trivalent vertex is $r-2$, the tails are $r-1$, $r$, and the end vertex of the ``long leg''
has the label $1$, see Appendix~\ref{se:mckay}. Then the product of the matter polynomials ${\bq}_{r-1}$ and ${\bq}_{r}$
has degree
\[
\deg ( {\bq}_{r-1}{\bq}_{r} ) = 2({\bv}_{r-1} + {\bv}_{r} - {\bv}_{r-2}).
\]
Now we shall factorize ${\bq}_{r-1} {\bq}_{r}$ into a product of two
factors of equal degrees
\begin{equation}
{\bq}_{r-1} {\bq}_{r} = {\bq}^{+} {\bq}^{-} , \qquad \deg {\bq}^{+} = \deg {\bq}^{-} = {\bv}_{r-1} + {\bv}_{r} - {\bv}_{r-2}.
\label{eq:bqpm}
\end{equation}
There are many possible factorizations.
For example, if ${\bw}_{r-1} \leq {\bw}_{r}$, then we can take:
${\bq}_{r} (x) = {\bq}^{+} (x) S(x)$, ${\bq}^{-}(x) = S(x) {\bq}_{r-1}(x)$ for any degree
${\bv}_{r} + {\bv}_{r-1} - {\bv}_{r-2} \leq {\bw}_{r} = 2{\bv}_{r} - {\bv}_{r-2}$ subfactor~${\bq}^{+}(x)$ in ${\bq}_{r}(x)$. We shall normalize ${\bq}^{\pm}(x)$ so that the highest
coefficient in both polynomials equals
\[
\sqrt{{\qe}_{r-1}{\qe}_{r}}.
\]
That there exist different decompositions (\ref{eq:bqpm}) is a
generalization of $S$-duality of the ${\CalS}$-class ${\CalN}=2$ theories of the $A_r$ type studied in~\cite{Gaiotto:2009we}.
The spectral curve $C_{u}$ corresponding to the $2r$-dimensional vector representation of
$C{\rm Spin}(2r, {\BC})$
is mapped to the curve $P_{D_{r}}^{C}(t,x) = 0$ in the $(t,x)$-space,
where
\begin{equation}
 P_{D_r}^{C}(t,x) = t^{-r} P_{\infty}(x) \det\nolimits_{R_1}( t \cdot 1_{2r} - g(x))
 \label{eq:detvect}
\end{equation}
with some polynomial $P_{\infty}(x)$ to be determined below.
The group element $g(x)$ in the vector representation ${\BC}^{2r}$ of
$C\Gq$ is given by
\begin{equation*}
 g(x) = E^{-81} \diag(g_1(x), \dots, g_{2r}(x)) E
\end{equation*}
with $E$ being any matrix such that
\[
\big(EE^{t}\big)_{ij} = {\delta}_{i,2r+1-j}
\]
represents the symmetric bilinear form on ${\BC}^{2r}$
and
\begin{equation*}
 \begin{aligned}
 & g_1(x) = \zeta(x) \y_1(x), \\
 & g_i(x) = \zeta(x) \bq^{[i-1]}(x) \frac{\y_i(x)}{\y_{i-1}(x)}, \qquad i = 2, \dots, r-2, \\
 & g_{r-1}(x) = \zeta(x) \bq^{[r-2]}(x) \frac{\y_{r-1}(x)
 \y_{r}(x)}{\y_{r-2}(x)}, \\
 & g_{r}(x) = \zeta(x) \bq^{[r-2]}(x) \bq_{r-1}(x)\frac{\y_{r}(x)}{\y_{r-1}(x)},
 \\
 & g_{r+1}(x) = \zeta(x) \bq^{[r-2]}(x) \bq_{r}(x)
 \y_{r-1}(x)/\y_{r}(x), \\
 & g_{r+2}(x) = \zeta(x) \bq^{[r]}(x)
 \y_{r-2}(x)/(\y_{r-1}(x) \y_{r}(x)), \\
 & g_{2r+1-i}(x) = \zeta(x) \frac{\bq^{[r]}(x)\bq^{[r-2]}(x)}{{\bq}^{[i-1]}(x)} \frac{\y_{i-1}(x)}{\y_{i}(x)},
 \qquad i = 2,\dots, r-2, \\
& g_{2r} = \zeta(x) \bq^{[r]}(x)\bq^{[r-2]}(x) \frac{1}{
\y_{1}(x)}.
 \end{aligned}
\end{equation*}
The factor $\zeta(x)$ which likely gives the minimal degree curve is
\begin{equation*}
 \zeta(x)^{-1} = \bq^{+}(x) \bq^{[r-2]}(x).
\end{equation*}
Thus, the scalar $D(g(x))$ is equal to
\begin{equation*}
D(g(x)) = \frac{{\bq}^{-}(x)}{{\bq}^{+}(x)}
\end{equation*}
and the prefactor in (\ref{eq:detvect}) is
\begin{equation}
 P_{\infty}(x) = \bq^{+}(x)^{r} \prod_{j=1}^{r-2} \bq_{j}(x)^{j} .
 \label{eq:pinf}
\end{equation}
After some manipulations, we find
\begin{gather}
 (-1)^{r} P_{D_r}^{C}(t,x) = T_{r}^{2} {\bq}_{r-1} + T_{r-1}^{2} {\bq}_{r} - {\eta} T_{r-1} T_{r} + \sum_{l=1}^{\left[ \frac r2 \right]} T_{r-2l} \bigg( \prod_{j=r+1-2l}^{r-2} {\bq}_{j}^{j-r+2l} \bigg) {\xi}_{l}^{2} \nonumber \\
\hphantom{(-1)^{r} P_{D_r}^{C}(t,x) =}{} - \sum_{l=1}^{\left[ \frac{r-1}{2} \right]} T_{r-2l-1} \bigg( \prod_{j=r-2l}^{r-2} {\bq}_{j}^{j-r+2l+1} \bigg) {\xi}_{l} {\xi}_{l+1}, \nonumber \\
 {\xi}_{l} = ( {\bq}^{+} t )^{l} - \big({\bq}^{-} t^{-1} \big)^{l} , \qquad {\eta} = {\bq}^{+} t + {\bq}^{-} t^{-1} .
\label{eq:txdcase}
\end{gather}
This equation has degree $N = 2( {\bv}_{r} + {\bv}_{r-1}) - {\bv}_{r-2}$ in the $x$ variable.
Note
\begin{equation*}
{\bv}_{r-2} \leq N \leq 2 {\bv}_{r-2}.
\end{equation*}
As in the $A_r$ case, the curve $C_{u}$ has branches going off to infinity in the $x$-direction, over $2r$ points ${\ct}_{i}$, ${\ct}_{i}^{-1}$, $i = 1, \dots , r$ in the $t$-line ${\BC\BP}^{1}_{t}$ which correspond to the weights of $R_1$
\begin{equation*}
{\ct}_{i} = \frac{1}{\sqrt{{\qe}_{r-1}{\qe}_{r}}} \frac{{\qe}^{[i-1]}}{{\qe}^{[r-2]}}.
\end{equation*}
In addition, there are special points $t = 0, \infty$. Over these points the curve
$C_{u}$ has $N$ branches, where $x$ approaches one of the roots of the polynomial
$P_0(x)$
\begin{equation*}
P_{0}(x) = {\bq}^{-}(x)^{r} \prod_{j=1}^{r-2} {\bq}_{j}(x)^{j}.
\end{equation*}
and $P_{\infty}(x)$, cf.~(\ref{eq:pinf}), respectively.

The curve $C_{u}$ is invariant under the involution
\begin{equation}
t \mapsto \frac{{\bq}_{-}(x)}{{\bq}_{+}(x)} t^{-1}.
\label{eq:invcu}
\end{equation}
The fixed points of (\ref{eq:invcu}) are the points of intersection of
the curve $C_{u}$ and the curve
\begin{equation}
{\bq}^{+}(x) t - {\bq}^{-}(x) t^{-1} = 0.
\label{eq:zitwoc}
\end{equation}
The equations ${\CalR}_{D_{r}}(t,x) = 0$ (\ref{eq:txdcase}) and (\ref{eq:zitwoc}) imply
\begin{equation*}
 T_{r}^{2}(x) {\bq}_{r-1} (x) + T_{r-1}^{2}(x) {\bq}_{r}(x) = T_{r-1}(x)T_{r}(x) \big({\bq}^{+}(x) t + {\bq}^{-}(x) t^{-1}\big)
\end{equation*}
and
\begin{equation*}
T_{r}^{2}(x) {\bq}_{r-1} (x) = T_{r-1}^{2}(x) {\bq}_{r}(x).
\end{equation*}
Again, the curve $C_{u}$ is more economical then the full cameral curve ${\CalC}_{u}$. Again, the special coordinates ${\ac}_{i,{\ba}}$ and the duals ${\ac}_{i,{\ba}}^{D}$ are the linear combinations of the periods of the differential~$x\,{\rm d}t/t$ and the masses.

Let us map the curve $C_{u}$ to the curve $\Sigma_{u}$
in the space $S$ which is a ${\BZ}_{2}$-quotient of the (blowup of the) ${\Cx} \times {\BC\BP}^{1}_{t}$ space, parametrized by $(x,s)$, where
\[
s = \frac{{\bq}^{+}(x)}{{\bq}^{-}(x)} t^{2}.
\]
The curve $\Sigma_{u}$ is described by the equations $s + s^{-1} = 2c$ and
\begin{equation*}
P_{D_r}^{\Sigma}(x,c) \equiv {\bf A}(x,c)^{2} - 2{\bq}_{r}(x){\bq}_{r-1}(x) (c + 1) {\bf B}(x,c)^2 = 0,
\end{equation*}
where ${\bf A}$, ${\bf B}$ are the polynomials in $x$ and $c$
of bi-degrees $\big(N, \big[\frac{r}{2}\big]\big)$ and $\big({\bv}_{r-1}+{\bv}_{r}, \big[\frac{r-1}{2}\big]\big)$, respectively,
\begin{equation*}
\begin{aligned}
& {\bf A}(x,c) =
T_{r}^{2} {\bq}_{r-1} + T_{r-1}^{2} {\bq}_{r} + 2 \sum_{l=1}^{\left[ \frac r2 \right]} {\bf C}_{l}(c) T_{r-2l} {\bq}_{r-1}^{l}{\bq}_{r}^{l} \prod_{j=r+1-2l}^{r-2} {\bq}_{j}^{j-r+2l} , \\
& {\bf B}(x,c) = T_{r-1} T_{r} + 2\sum_{l=1}^{\left[ \frac{r-1}{2} \right]} {\bf D}_{l}(c) T_{r-2l-1} {\bq}_{r-1}^l{\bq}_{r}^{l} \prod_{j=r-2l}^{r-2} {\bq}_{j}^{j-r+2l+1} , \\
\end{aligned}
\end{equation*}
where the degree $l$ polynomials ${\bf C}_{l}(c)$, ${\bf D}_{l}(c)$ are defined as follows:
\begin{equation*}
\begin{aligned}
& {\bf C}_{l}(c) = \frac 12 \big(s^{l} + s^{-l}\big) - 1, \qquad s + s^{-1} = 2c, \\
& {\bf D}_{l}(c) = \frac{\big(s^{l} -1\big)\big(s^{l+1}-1\big)}{2 s^{l}(s+1)} = \sum_{j=0}^{l-1} (-1)^{j} {\bf C}_{l-j}(c).
\end{aligned}
\end{equation*}
Over the points $c=1$ and $c=-1$ the equation for $\Sigma_{u}$ becomes reducible
at $c=1$:
\begin{equation}
P_{D_r}^{\Sigma}(x,1) = \big({\bq}_{r-1}T_{r}^{2} - {\bq}_{r}T_{r-1}^{2}\big)^2
\label{eq:ceqo}
\end{equation}
and at $c= -1$:
\begin{equation}
P_{D_r}^{\Sigma}(x,-1) = {\bf A}(x, -1)^2.
\label{eq:ceqmo}
\end{equation}
It is easy to see that the curve $\Sigma_{u}$ has double points at
$(x,s)$ where either $s=1$ and $x$ being any of the $N$ roots of (\ref{eq:ceqo})
or $s = -1$ and $x$ is any of the $N$ roots of (\ref{eq:ceqmo}). The locations of these roots
are not fixed by the masses of the matter fields.

Let us normalize the equation of $\Sigma_u$ by dividing $P_{D_{r}}^{\Sigma}$
by the coefficient at $x^{2N}$:
\begin{equation}
{\CalR}_{D_{r}}(x,c) = \frac{ P_{D_{r}}^{\Sigma}(x,c)}{\prod_{i=1}^{r} \big( s - {\ct}_{i}^{2}\big)\big(1 - s^{-1} {\ct}_{i}^{-2}\big)}
\label{eq:crd}
\end{equation}
times a constant such that $\CalR_{D_{r}}(x,c)$ is monic in $x$ and a rational function of $c$.

We thus arrive at the following interpretation of the curve $\Sigma_{u}$.
It is the spectral curve
\begin{equation*}
{\CalR}_{D_{r}}\bigg(x, \frac{s^2 + 1}{2s} \bigg) \bigg( \frac{{\rm d}s}{s} \bigg)^{2N}
= \operatorname{Det}_{2N} \bigg( x\frac{{\rm d}s}{s} - {\Phi}(s) \bigg)
\end{equation*}
of the genus zero Higgs field
\begin{equation*}
{\Phi}(s) = \sum_{s_{j} \in J} {\Phi}_{j} \frac{{\rm d}s}{s - s_{j}},
\end{equation*}
where $J \subset {\BC\BP}^{1}_{s}$ is the set of $2r+2$ singularities
\[
J = \{ 0, \infty \} \cup \big\{ {\ct}_{i}^2, \, {\ct}_{i}^{-2} \mid i = 1, \dots , r \big\}.
\]
Let ${\si}\colon {\BC\BP}^{1}_{s} \to {\BC\BP}^{1}_{s}$ be the involution
${\si}(s) = s^{-1}$. The Higgs field must obey
\begin{equation}
{\si}^{*}{\Phi} = {\Omega} {\Phi}^{t} {\Omega}^{-1},
\label{eq:orbi}
\end{equation}
where $ {\Omega} $ is a constant anti-symmetric matrix (cf.~\cite{Kapustin:1998fa}), which defines the symplectic structure on
$V = {\BC}^{2N}$.
If we expand
\begin{gather*}
{\Phi}(s) = {\Phi}_0\frac{{\rm d}s}{s} + \sum_{i=1}^{r} {\Phi}_{i}^{+} \frac{{\rm d}s}{s - {\ct}_{i}^{2}} + \sum_{i=1}^{r} {\Phi}_{i}^{-} \frac{{\rm d}s}{s- {\ct}_{i}^{-2}} , \nonumber \\
{\Phi}_{\infty} = - {\Phi}_{0} - \sum_{i=1}^{r} \big({\Phi}_{i}^{+} + {\Phi}_{i}^{-}\big),
\end{gather*}
then \eqref{eq:orbi} implies
\begin{equation*}
\begin{aligned}
{\Phi}_{\infty} = {\Omega} {\Phi}_{0}^{t} {\Omega}^{-1}, \qquad
 {\Phi}_{i}^{+} = {\Omega} ({\Phi}_{i}^{-})^{t} {\Omega}^{-1}, \qquad i = 1, \dots , r. \end{aligned}
\end{equation*}
Also, the matrices ${\Phi}^{+}_{i}$, ${\Phi}^{-}_{i}$, $i = 1, \dots , r$, must have rank one, while the matrices ${\Phi}_{0,\infty. \pm 1}$ have rank $2N$.
We can interpret
\begin{gather*}
 {\mu} = {\Phi}_{0} + {\Phi}_{\infty} +
\sum_{i=1}^{r} \big({\Phi}_{i}^{+} + {\Phi}_{i}^{-}\big) , \qquad
 {\mu}^{t} = {\Omega}^{-1} {\mu} {\Omega}
\end{gather*}
as the moment map for the ${\rm Sp}(2N)$ group action on the product
of some orbits
\[
{\CalO}_{0} \times {\CalO}_{-1} \times {\CalO}_{1} \times_{i=1}^{r} {\CalO}_{i},
\]
which generates the action ${\Phi}_{j} \mapsto g^{-1} {\Phi}_{j} g$ of $g \in {\rm Sp}(2N)$, such that
\begin{equation*}
g {\Omega} g^{t} = {\Omega}.
\end{equation*}
It would be nice to develop further the theory of these orbifold Hitchin--Gaudin systems.
We shall encounter a genus one version of such theory in the class II $D_r$ section below.

The differential whose periods determine the special coordinates is equal to
\begin{equation*}
{\rm d}S = x \frac{{\rm d}s}{s}.
\end{equation*}

\subsection{Freezing example}\label{se:freezing}

Here we will illustrate how the $D_4$ theory with $v_1=v_3=v_4
= v$, $v_2 = 2 v$ and $w_1 = w_3 = w_4 = 0$, $w_2 = v$ reduces to $A_3$ with
$v_1 = v_3 = v$, $v_2 = 2 v$ and $w_2 = 2v$ when the node 4 freezes
under $\qe_4 \to 0$.
Keeping in mind unfreezing to the affine $\hat D_4$, let polynomial $Y_0$ of degree $v$ denote the fundamental matter polynomial
attached to the node ``2''.

The $D_4$ spectral curve for the node ``1'' from (\ref{eq:txdcase}) in
terms of variable $\eta$
\begin{equation}
\label{eq:eta}
 \eta = t + \frac{ \qe_1^2 \qe_2^2 \qe_3 \qe_4}{t},
\end{equation}
where $ \y_2 = Y_0 t$ is
\begin{gather}
 \CalR_{D_4}(\eta,x) =\eta ^4 Y_0^2 -\eta ^3 T_1 Y_0 +\eta ^2 \bigl(\qe_1 T_2-4 \qe_1^2 \qe_2^2 \qe_3
 \qe_4 Y_0^2\bigr)
 + \eta \bigl(-\qe_1^2 \qe_2 T_3 T_4+4 \qe_1^2 \qe_2^2 \qe_3 \qe_4 T_1 Y_0\bigr) \nonumber\\
\hphantom{\CalR_{D_4}(\eta,x) =}{}-4 \qe_1^3 \qe_2^2 \qe_3 \qe_4 T_2+\qe_1^3 \qe_2^2 \qe_4 T_3^2+\qe_1^3 \qe_2^2 \qe_3 T_4^2.
\label{eq:D4-curve}
\end{gather}
Notice that the curve is polynomial of degree $4$ in $\eta$ with
polynomial coefficients in $x$ of degree~$2 v$. In the limit $x \to
\infty$ we find the limiting values of $\eta$ are
\begin{equation*}
 1 + \qe_1^2 \qe_2^2 \qe_3 \qe_4, \qquad
 \qe_1 + \qe_1 \qe_2^2 \qe_3 \qe_4, \qquad
 \qe_1 \qe_2 + \qe_1 \qe_2 \qe_3 \qe_4, \qquad
 \qe_1 \qe_2 \qe_3 + \qe_1 \qe_3 \qe_4.
\end{equation*}
Notice
that the differential is
\begin{equation*}
 \lambda = x \frac{{\rm d}t}{t} = x \frac{ {\rm d}\eta} {\big(\eta^2 - 4 \qe_1^2 \qe_2^2 \qe_3
 \qe_4\big)^{\frac 1 2}}.
\end{equation*}
Also notice that at $\eta = \pm 2 \qe_1 \qe_2 \qe_3^{\frac 1 2}
\qe_{4}^{\frac 1 2}$ the curve factorizes
as
\begin{equation}
\label{eq:D4factorization}
 \CalR_{D_4}\big( \pm 2 \qe_1 \qe_2 \qe_3^{\frac 1 2}
\qe_{4}^{\frac 1 2} ,x\big) = \qe_1^{3} \qe_2^{2} ( \qe_3
 T_4(x) \mp \qe_{4} T_3(x) )^2
\end{equation}
as well as it factorizes at $\eta = \infty$
\begin{equation}
 \label{eq:D4factorization-inf}
 \CalR_{D_4}(\eta = \infty ,x) = Y_0(x)^2.
\end{equation}

We can interpret the multi-valued nature of $\lam$ on the $\eta$-plane as the
deformation of the punctured sphere underlying the $A_r$-type theories to the curve describing
the $D_r$-type theories, by opening punctures into cuts. Perhaps one can elevate this
observation to the corresponding deformation of the Liouville theory coupled to some conformal matter, along the lines of~\cite{Gerasimov:1988gy,Knizhnik:1987xp}.

We see that in the decoupling limit $\qe_4 = 0$ the above curve reduces to
\begin{equation}
\label{eq:A3-curve}
 \CalR_{A_3}(\eta,x) =\eta ^4 Y_0^2 -\eta ^3 T_1 Y_0 +\eta ^2\qe_1 T_2
 -\eta \qe_1^2 \qe_2 T_3 Y_4
+\qe_1^3 \qe_2^2 \qe_3 Y_4^2,
\end{equation}
where we just set that $\y_4$ freezes and converts
to a factor of degree $v$
contributing to the fundamental matter polynomial for the node ``2''; we denote
this factor by $Y_4 \equiv \y_4 = T_4$.
 The curve (\ref{eq:A3-curve}) is precisely the $A_{3}$ curve for the
 node ``1'' (\ref{eq:ptox}) in terms of the variable $\y_1 = Y_0
 \eta$. This curve corresponds to the ${\rm GL}(2)$ Hitchin system with punctures at
four punctures
 \begin{equation*}
 1, \quad \qe_1, \quad \qe_1 \qe_2, \quad \qe_1 \qe_2 \qe_3.
 \end{equation*}
Moreover, from the discussion after \eqref{eq:phimj} (we have $\bw_{0} =
0$, $\bw_{2} = 2$, $\bw_{3} = 0$ and $i_{*} = 2$ and $\bw_{+} = \bw_{-1}
=1$) it is clear the
eigenvalues of the Higgs field residues at $\eta = 0 $ and at $\eta =
\infty$ are doubly degenerate which effectively means that ${\rm SL}(2,\BC)$
part of the Higgs field does not have punctures at $\eta = 0$ and $\eta
=\infty$.
We can continue the freezing reduction and now we shall set $\qe_3 = 0$
declaring the function $\y_3$ as contributing to the fundamental matter at the
node ``2'', we denote $\y_3 = T_3 = Y_3$.
After factoring out $\eta$, the curve (\ref{eq:A3-curve}) reduces to the $A_2$ curve
\begin{equation}
\label{eq:A2-curve}
 \CalR_{A_2}(\eta,x) =\eta^3 Y_0^2 -\eta ^2 T_1 Y_0 +\eta \qe_1 T_2
 - \qe_1^2 \qe_2 Y_3 Y_4.
\end{equation}
The corresponding Gaudin system has punctures at $\eta = 0$ and $\eta =
\infty$ and at
\begin{equation*}
 1,\quad \qe_1, \quad \qe_1 \qe_2.
\end{equation*}
Finally, we can freeze the node ``1'' by sending $\qe_1$ to
zero and rescaling $\eta = \tilde \eta \qe_1$ so that the former
punctures $\qe_1$, $\qe_1 \qe_2$ on the $\tilde \eta$-plane in terms of
$\tilde \eta$
become
\begin{equation*}
 1, \quad \qe_2
\end{equation*}
while the puncture $\eta=1$ is send away to $\tilde \eta = \infty$. We set $Y_1
\equiv \y_1 = T_1$ and find that~\eqref{eq:A2-curve} reduces to the familiar $A_1$ curve with gauge
polynomial $T_2$ of degree $2v$ and four factors $(Y_0,Y_1,Y_3,Y_4)$ of
degree $v$ which make fundamental polynomial of degree $4 v$
\begin{equation*}
 \CalR_{A_2}(\eta,x) = -\tilde \eta^2 Y_1 Y_0 +\tilde \eta T_2
 - \qe_2 Y_3 Y_4.
\end{equation*}
The punctures of the corresponding Gaudin model in $\tilde \eta$ plane are at $( 0, \qe_2, 1,
\infty)$.

\subsection[Class I theories of $E$ type]{Class I theories of $\boldsymbol{E}$ type}
We are using Bourbaki conventions to label the nodes on the Dynkin graph
of $E_{r}$ series, see figures in Appendix~\ref{se:mckay}.
One can construct the analogues of the spectral curves $C_u$ or $\Sigma_{u}$ using the
minuscule representations in the $E_6$ and $E_7$ cases. For $E_8$ one
can construct the spectral curve using the adjoint representation $\bf
248$. However it seems more advantageous to use the degenerate version
of del Pezzo/$E$-bundle correspondence, which we review below in the
discussion of class~II theories of $E$ type. For the standard conformal
$E_r$ quivers, which are obtained by freezing of the node~``0'' in the affine $E_r$
quivers with ranks ${\bv}_i = N a_i$ where $a_i$ are Dynkin marks,
we find spectral curves of $(t,x)$-degree equal to $(27, 6N)$ for~$E_6$,
$(56, 12N)$ for $E_7$ and $(240, 60N)$ for~$E_8$. These degrees can be
understood from the degeneration of~$\hat E_r$ spectral curves computed
in Section~\ref{se:Affine-degrees}.

\subsection[The $E_6$ theory]{The $\boldsymbol{E_6}$ theory}
The spectral curve in the fundamental representation $R_{6} = \mathbf{27}$
associated with the node ``6'', in which the group element of the conformal
extension of $E_6$ is $g(x) = (\y_6(x), \dots )$ has the form
\begin{equation*}
 \mathcal{R}_{E_6}(t,x) = 0,
\end{equation*}
where the explicit expression is of the form\footnote{The explicit
 expression, which we do not list here, is available upon a request; it
 is computed by the straightforward expansion of the exterior powers
 $\bigwedge^{\bullet} R_{6}$ in the
 representation ring $\mathrm{Rep} (E_6)$ over the fundamental
representations $R_{1}, \dots, R_{6}$.}
\begin{gather}
\mathcal{R}_{E_6}(t,x) =\det\nolimits_{R_6} (t \cdot 1_{27} - g(x)) = t^{27}-t^{26} T_6+ t^{25} \bq_6 T_5-t^{24} \bq_5 \bq_6^2 T_4 \nonumber\\
 \hphantom{\mathcal{R}_{E_6}(t,x) =}{} + t^{23}\left(-\bq_2^2 \bq_3^2 \bq_4^4 \bq_5^4 \bq_6^4 T_1^2+\bq_1 \bq_2^2 \bq_3^2 \bq_4^4 \bq_5^4
 \bq_6^4 T_3+\bq_4 \bq_5^2 \bq_6^3 T_2 T_3 \right. \nonumber\\
 \hphantom{\mathcal{R}_{E_6}(t,x) =}{} \left. -\bq_2 \bq_3 \bq_4^2 \bq_5^2 \bq_6^3 T_1
 T_5+\bq_1^2 \bq_2^3 \bq_3^4 \bq_4^6 \bq_5^5 \bq_6^4 T_6\right) + \dots
 -\bq_1^{18} \bq_2^{27} \bq_3^{36} \bq_4^{54} \bq_5^{45} \bq_6^{36},\!\!\!\!\label{eq:E6-curve1}
\end{gather}
where we have omitted the explicit expressions for the terms from
 $t^{24}$ to $t^1$, and we omitted
 the dependence on $x$ in the notations for the
 polynomial coefficients so that $\CalP_i \equiv
 \CalP_i(x)$ and $T_i \equiv T_i(x)$. The curve~\eqref{eq:E6-curve1} has
 $x$-degree $27 v_6$, and, of course, is not the most economical. By
 rescaling $g(x) \to \zeta(x) g(x)$ with a suitably chosen $\zeta(x)$ of
 degree $-v_6$ made
 of some powers of the factors in fundamental polynomials we can reduce
 the degree of \eqref{eq:E6-curve1}.

 The most standard conformal $E_6$ quiver, which arises from the
degenerate limit $\qe_0 \to 0$ in the node ``0''
 of the affine $\hat E_6$ quiver, has matter polynomial $\bq_2 = \qe_2 Y_0$ of degree
$N$ only at the node~``2'' to which the affine node ``0'' was attached, while the degrees of the
 gauge polynomials are fixed by the Dynkin marks ${\bv}_i = N a_i$, that is
$(\bv_1,\dots, \bv_6) = (N,2N,2N,3N,2N,N)$. For such conformal $E_6$ quiver, the curve
\eqref{eq:E6-curve1} has canonical reduced form under
the choice ${\zeta^{-1}(x) =Y_0(x)}$ and the degree of the reduced curve is
$6N = 2 \bv_{*}$ where $\bv_{*} \equiv \bv_{4}= 3 N$ denotes the
rank in the trivalent node ``4''.
 The reduced curve of such special conformal $E_6$ quiver is $
 \mathcal{R}_{E_6}(t,x)$, with $\bq_i = \qe_i$, $i \neq 2$; $\bq_2 = \qe_2
 Y_0$ we find
 \begin{gather}
 \mathcal{R}_{E_6}(t,x) = t^{27} Y_0^6 -t^{26} Y_0^5 T_6+ t^{25} \qe_6 Y_0^4 T_5-t^{24}\qe_5 \qe_6^2 Y_0^3 T_4 + t^{23} \left(- \qe_2^2 \qe_3^2 \qe_4^4 \qe_5^4
 \qe_6^4 Y_0^4 T_1^2 \right. \nonumber\\
\hphantom{\mathcal{R}_{E_6}(t,x) =}{} +\qe_1 \qe_2^2 \qe_3^2 \qe_4^4 \qe_5^4 \qe_6^4 Y_0^4 T_3+ \qe_4 \qe_5^2 \qe_6^3 Y_0^2 T_2 T_3-\qe_2 \qe_3 \qe_4^2 \qe_5^2 \qe_6^3 Y_0^3 T_1 T_5 \nonumber\\
\hphantom{\mathcal{R}_{E_6}(t,x) =}{} \left.+ \qe_1^2 \qe_2^3 \qe_3^4 \qe_4^6 \qe_5^5 \qe_6^4 Y_0^5 T_6\right) + \dots -t^2 \qe_1^{15} \qe_2^{23} \qe_3^{30} \qe_4^{46} \qe_5^{39} \qe_6^{32}Y_0^4 T_3 \nonumber\\
\hphantom{\mathcal{R}_{E_6}(t,x) =}{} +t \qe_1^{16} \qe_2^{25} \qe_3^{33} \qe_4^{50} \qe_5^{42} \qe_6^{34}Y_0^5 T_1- \qe_1^{18} \qe_2^{27} \qe_3^{36} \qe_4^{54} \qe_5^{45} \qe_6^{36}Y_0^6,\label{eq:Ecurve-2}
 \end{gather}
where again we only indicated the middle terms but skipped the explicit
expressions. Indeed, one sees that the curve~\eqref{eq:Ecurve-2} of the $E_6$ quiver with the standard rank
 assignments $\bv_i = N a_i$ has degree $6N$.
At the limit $ x \to \infty$ the 27~roots of $\mathcal{R}_{E_6}(t,x)$ in
\eqref{eq:Ecurve-2} approach the set of points in the $t$-plane labeled
by the weights $\lambda$ in the $\mathbf{27}$ representation of $E_6$ and given
explicitly by $\prod_{i=1}^{6} \qe_i^{ (\lam_i, \lam -
 \lam_i)}$, or
\begin{equation}
\label{eq:E6-punctures}
 \left\{ \prod_{i=1}^{6} \qe_i^{n_i} \, \Bigg|\,
\sum_{i=1}^{6} n_i \alpha_i = \lam_6 - \lambda,\, \lambda \in
\mathrm{weights}(R_{6}) \right\},
\end{equation}
where $n_i$ are the coefficients of the expansion
 in the
basis of simple roots of the difference
between a given weight in $\mathbf{27}$ and the highest weight. One can
associate a Higgs field to the spectral curve~\eqref{eq:Ecurve-2} with
poles in the 27 punctures (\ref{eq:E6-punctures}) with certain
relations. In other words, the curve~\eqref{eq:Ecurve-2} realizes a
certain embedding of the standard conformal $E_6$ quiver theory
with gauge group ranks $\bv_i = (N,2N,2N,3N,2N,N)$ to some specialization
of the $A_{26}$ theory with ranks $(6N,6N,\dots, 6N)$,
and this embedding can be lifted to the Higgs field spectral
curve representation of (\ref{eq:Ecurve-2}).

For non-standard assignments of $\bw_i$ and $\bv_i$ for the conformal $E_6$
quiver we did not find a~simple choice of $\zeta(x)$ reducing the curve
\eqref{eq:E6-curve1} to the minimal degree. For small ranks $\bv_i$, $\bw_i$ we
can find the reduced curve using the brute search minimization problem on the
total degree of the reduced curve under $g(x) \to \zeta(x) g(x)$. We
have found different chambers in the space of parameters $\bw_i$, $\bv_i$ with
piece-wise linear dependence of the reduced degree of $\bw_i$ or $\bv_i$'s
but not a simple expression. For example, in several examples we find
\begin{center}
\begin{tabular}[center]{llc}
 $(w_i)$ & $(v_i)$ & reduced curve $x$-degree \\ \hline
 $ (0, 4, 0, 0, 0, 0)$ & $ (4, 8, 8, 12, 8, 4) $ & 24 \\
 $ (3, 0, 0, 0, 0, 3) $ & $ (6, 6, 9, 12, 9, 6) $ & 33 \\
 $ (6, 0, 0, 0, 0, 0) $ & $ (8, 6, 10, 12, 8, 4)$ & 40 \\
 $ (4, 0, 0, 0, 0, 1) $ & $ (6, 5, 8, 10, 7, 4) $ & 31 \\
 $ (6, 0, 0, 0, 0, 3) $ & $ (10, 9, 14, 18, 13, 8)$ & 53 \\ \hline
\end{tabular}
\end{center}

\noindent
where the first three lines list different conformal $E_6$ quivers
sharing the same $\bv_{*} = 12$, and one can see that the curve of the
minimal degree $2 \bv_{*}$ is obtained in the standard assignment $\bw_i =
0$, $i \neq 2$ associated to the degenerate limit of the affine $E_6$.

\subsection[$E_7$ theory]{$\boldsymbol{E_7}$ theory}
 We write the spectral curve in, for example, the $\mathbf{56}$
 representation of $E_7$ similar to the $E_6$ case. If $(\bv_0,\dots,\bv_7) = N a_i$
 where $a_i$ are Dynkin marks of $E_7$ quiver, again, similar to $E_6$
 quiver we find that the reduced curve of the standard conformal $E_7$
 quiver obtained from the degenerate limit of the affine theory has
 $x$-degree $12 N = 3 \bv_{*}$ where $\bv_{*} = \bv_{4} = 4N$ is rank at the
 trivalent node.
The standard $E_7$ quiver spectral curve hence is realized
as a specialization of the spectral curve
for $A_{55}$ quiver with ranks $(12N,12N,\dots,12N)$, or Hitchin system
with $56$ punctures on $t$-plane associated to the weights in
$\mathbf{56}$.

\subsection[$E_8$ theory]{$\boldsymbol{E_8}$ theory}
 For $E_8$ the minimal representation is adjoint
 $\mathbf{248}$.
The reduced curve in the adjoint representation for the standard conformal
$E_8$ quiver obtained from the degenerate limit of the affine theory has
 $x$-degree $60 N = 10 \bv_{*}$ where $\bv_{*} = 6N$ is rank at the
 trivalent node.
Hence the standard conformal $E_8$ quiver spectral curve is realized
as a specialization of the spectral curve
for $A_{247}$ quiver with ranks $60 (N,N,\dots, N)$, or Hitchin system
with $240$ punctures on $t$-plane
 associated to the non-zero adjoint weights in
$\mathbf{248}$.

\subsection[Class II theories of $A$ type and class II* theories]{Class II theories of $\boldsymbol{A}$ type and class II* theories}
Let us start with the simplest nontrivial examples, and then pass onto a general case.

\subsection[Class II $\hat A_{1}$ theory]{Class II $\boldsymbol{\hat A_{1}}$ theory}

For the class II theory we shift the arguments of ${\y}_{i}(x)$ by ${\mu}_i$ to get rid of the
bi-fundamental masses.

Let $g(x) \in \widehat{{\rm SL}_2}$:
\begin{equation*}
g(x) = {\qe}_{0}^{-{\hat\lam}_{0}^{\vee}}{\qe}_{1}^{-{\hat\lam}_{1}^{\vee}} {\y}_{0}(x)^{{\hat\al}_{0}^{\vee}} {\y}_{1}(x)^{\hat\al_{1}^{\vee}}.
\end{equation*}
We have: ${\qe} = {\qe}_0 {\qe}_1$,
\begin{equation*}
 g(x)^{\al_1} = \frac{{\y}_1^2}{{\qe}_1 {\y}_{0}^2 }, \qquad
 g(x)^{-\delta} = {\qe}, \qquad
 g(x)^{\hat\lam_0} = {\y}_{0}(x).
\end{equation*}
The normalized $\widehat{\mathfrak{sl}_2}$ characters \eqref{eq:affcha} of the fundamental representations
${\hat R}_0$, ${\hat R}_{1}$ are equal to
\begin{gather}
{\crf}_0 ( {\y}(x), {\bf\qe}) =
 \frac{{\y}_{0}(x)}{\phi({\qe})} \theta_3 \left( \frac{ {\y}_1(x)^2}{{\qe}_1 {\y}_{0}(x)^2} ; {\qe}^2 \right), \nonumber\\
 {\crf}_1 ( {\y}(x), {\bf\qe}) = \left( \frac{{\qe}_{1}}{\qe_{0}} \right)^{\frac 14} \frac{{\y}_{0}(x)}{\phi({\qe})} \theta_2 \left( \frac{ {\y}_1(x)^2}{{\qe}_1 {\y}_{0}(x)^2}; {\qe}^2\right)
\label{eq:ysl2}
\end{gather}
(see Appendix~\ref{appendixO} for our conventions on elliptic functions).
The characters (\ref{eq:ysl2}) are invariant under the Weyl transformations
\begin{equation*}
 {\y}_{0} \to {\qe}_0 {\y}_{0}^{-1} {\y}_1^2, \qquad
{\y}_1 \to {\qe}_1 {\y}_1^{-1} {\y}_{0}^2
\end{equation*}
and therefore we can equate them to the polynomials
\begin{alignat*}{3} 
& {\crf}_0 ( {\y}(x), {\bf\qe}) = T_{0}(x), \qquad &&
 T_{0,0} = \frac{\theta_3 \big( {\qe}_1^{-1} ; {\qe}^2 \big)}{\phi({\qe})} ,& \\
& {\crf}_1 ( {\y}(x), {\bf\qe}) = T_{1}(x) ,\qquad &&
 T_{1,0} = \bigg( \frac{{\qe}_{1}}{\qe_{0}} \bigg)^{\frac 14} \frac{\theta_2 \big( {\qe}_1^{-1} ; {\qe}^2\big)
}{\phi({\qe})} . &
\end{alignat*}
The values of characters (\ref{eq:ysl2}) and ${\qe}_0$, ${\qe}_1$ define
 ${\y}_{0}$ and ${\y}_1$ up to an affine Weyl transformation.
To recover ${\y}_{0}$ and ${\y}_{1}$ we invert the relations (\ref{eq:ysl2}):
\begin{equation*}
 {\y}_{1}(x) = {\qe}_{1}^{\frac 1 2 } {\y}_{0}(x) t ,\qquad
 {\y}_{0}(x) = \frac{\phi({\qe})}{\theta_3 \big( t^2 ; {\qe}^2 \big)} T_{0}(x)
\end{equation*}
and express
\begin{equation*}
\bigg(\frac{\qe_{0}}{{\qe}_{1}} \bigg)^{\frac 1 4}
 \frac{{\theta}_{3} \big( t^2 ; {\qe}^{2}\big)}{{\theta}_{2} \big( t^2 ; {\qe}^{2}\big)}= \frac{T_{0}(x)}{T_{1}(x)}.
\end{equation*}
Actually, the ratio
\[
{\xi} = \bigg(\frac{\qe_{0}}{{\qe}_{1}} \bigg)^{\frac 1 4} \frac{{\theta}_{3} \big( t^2 ; {\qe}^{2}\big)}{{\theta}_{2} \big( t^2 ; {\qe}^{2}\big)}
\]
is a meromorphic function on $\ec$ with two first-order poles at $t =
\pm \ii$ and two simple zeroes at $t = \pm \ii \qe$.
Therefore
\begin{equation*}
 \xi = {\xi}_{\infty} \frac { X (t,\qe) - X_0}
 { X (t,\qe) - X_1}, \qquad X_0 := X( \ii \qe, \qe), \qquad X_1
 := X( \ii,\qe)
\end{equation*}
with
\begin{equation*}
 {\xi}_{\infty} = \bigg(\frac{\qe_{0}}{{\qe}_{1}} \bigg)^{\frac 1 4} \frac{ \theta_3\big(1,\qe^2\big)}{\theta_2\big(1,\qe^2\big)}
\end{equation*}
and the explicit $\qe$-series for $X(t,\qe)$ is given in
\eqref{eq:weierx1} and \eqref{eq:weierx2}.
Hence, the algebraic Seiberg--Witten curve $C_{u}$ describing the $\hat A_1$
theory is a two-fold cover of the rational curve ${\Sigma}_{u}$
\begin{equation*}
 ( {\xi}_{\infty} T_1(x) - T_0(x)) X - ({\xi}_{\infty} T_1(x) X_0 - T_0(x) X_1) = 0
\end{equation*}
defined by the Weierstra{\ss} cubic \eqref{eq:wxy}.
There are $4N$ branch points of the $2:1$ cover $C_{u} \to {\Sigma}_{u}$:%
\begin{equation*}
\begin{aligned}
& {\xi}_{\infty} T_1(x_{\infty, {\ba}}) - T_0(x_{\infty, {\ba}}) = 0, \\
& ( {\xi}_{\infty} T_1(x_{{\al}, {\ba}}) - T_0(x_{{\al}, {\ba}})) e_{\al} - ({\xi}_{\infty} T_1(x_{{\al}, {\ba}}) X_0 - T_0(x_{{\al}, {\ba}}) X_1) = 0, \\
& {\al} =1 ,2 ,3, \qquad {\al} = 1, \dots , N, \end{aligned}
\end{equation*}
which can be split into $2$ groups of $N$ pairs, corresponding to the
cycles $A_{i{\ba}}$ with $i = 0,1$, e.g., $A_{0,{\ba}}$ is a small circle around the cut which connects $x_{1,{\ba}}$ to $x_{2,{\ba}}$, while $A_{1,{\ba}}$ is a small circle
around the cut which connects $x_{3, {\ba}}$ to $x_{\infty, {\ba}}$.
The special
coordinates are computed by the periods of
\[
{\rm d}S_{-} = x\, {\rm d} \log (t) = x \frac{{\rm d}X}{Y}.
\]
The curve $C_{u}$ is the spectral curve. The cameral curve ${\CalC}_{u}$
is a $\BZ$-cover
of spectral curve $C_{u}$, which is given by the same equations but now with $t \in {\BC}^{\times}$
as opposed to $t \in {\ec}$.
On cameral curve ${\CalC}_{u}$ we have the second differential
\[
{\rm d}S_{+} = x\, {\rm d} \log {\theta}_{3}\big(t^{2}; {\qe}\big),
\]
which would be a multi-valued differential on spectral curve $C_{u}$ whose periods are
defined up to the periods of ${\rm d}S_{-}$, similar to the polylogarithm motives~\cite{Cartier:1987}.

\subsection[Class II* $\hat A_0$ theory]{Class II* $\boldsymbol{\hat A_0}$ theory}

This is a (noncommutative) $U(1)$ ${\CalN}=2^*$ theory.
This theory was solved in~\cite{Nekrasov:2003rj} by the similar method.
There is only one amplitude ${\y} (x) = {\y}_{0}(x)$, with the single interval $I$ as its branch cut, the single function
\[
t(x) \equiv t_{0}(x) = \frac{{\y}(x)}{{\y}(x+ {\mathfrak{ m}})}
\]
with two branch cuts $I$ and $I - {\mathfrak{m}}$.
Crossing the $I$ cut maps $t(x) \mapsto {\qe} t(x - {\mathfrak{m}})$. Crossing the cut $I - {\mathfrak{m}}$
has the opposite effect: $t(x) \mapsto {\qe}^{-1} t(x + {\mathfrak{m}})$.
The extended functions
\[
t_{j}(x) = {\qe}^{j} t (x - j {\mathfrak{m}}).
\]
The analytically continued function $t(x)$ has cuts at $I +{\mathfrak{m}}\BZ$. The sheets of the Riemann surface of $t(x)$ are labeled by $j \in \BZ$, so that on the sheet $j$ the cuts are
at $I - j {\mathfrak{m}}$, and $I - (j+1) {\mathfrak{m}}$. Upon crossing $I+j {\mathfrak{m}}$ the $t_{j}(x)$ function transforms to $t_{j+1}(x)$ function. As $x \to \infty$ on this sheet the corresponding branch of $t(x)$ approaches ${\qe}^{j}$. These conditions uniquely fix the inverse function
to be the logarithmic derivative of $\theta_{1}$:
\begin{equation*}
x = a + {\mathfrak {m}} t \frac{{\rm d}}{{\rm d}t} \log {\theta}_{1} (t; {\qe}).
\end{equation*}

\subsection[Class II $A_{r}$ theories]{Class II $\boldsymbol{A_{r}}$ theories}

In order to solve the general rank $r$ theory, it is convenient to form a linear combination of fundamental characters of
${\hat A}_{r}$. Ultimately we would like to define
a regularized version of the characteristic polynomial of $g(x)$, where, as in the general
case, after the shift of the arguments of ${\y}_{i}(x) \to {\y}_{i}(x+{\mu}_{i})$:
\begin{equation*}
g(x) = \prod_{i=0}^{r} {\qe}_{i}^{-{\hat\lam}_{i}^{\vee}} {\y}_{i}(x)^{{\hat\al}_{i}^{\vee}}.
\end{equation*}
Using $t_{i}(x) = g(x)^{e_{i}}$ (see the appendix), we compute
\begin{equation}
t_{i}(x) = {\ct}_{i} \frac{{Y}_{i}(x)}{{Y}_{i-1}(x)},\qquad i = 1,\dots,
r + 1,
\label{eq:tfromyaff}
\end{equation}
where we extended the amplitude functions ${\y}_{j}(x)$ defined for
$j = 0, \dots , r$ to be defined for all $j \in \BZ$ by periodicity
${Y}_{j}(x) = {\y}_{j+(r+1)}(x)$ and where
\[
{\bf t}(x) = ( t_{1}(x), t_{2}(x), \dots , t_{r+1}(x) )
\]
represents an element of the maximal torus of ${\rm SL}(r+1, {\BC})$, i.e.,
\[
\prod_{i=1}^{r+1} t_{i}(x) = 1.
\]
The $\ct_i$ are the asymptotic values at $x \to \infty$ of $t_i(x)$ and
are given by
\begin{equation*}
 \ct_i = (\qe_{i} \dots \qe_{r})^{-1} \big(\qe_1 \qe_{2}^{2} \dots
 \qe_{r}^{r}\big)^{\frac{1}{r+1}}, \qquad i = 1, \dots, r+1,
\end{equation*}
and
\begin{equation*}
 g(x)^{-\delta} = {\qe}, \qquad g(x)^{\hat\lam_0} = {\y}_{0}(x).
\end{equation*}
Now we shall explore the relation between the conjugacy classes in Kac--Moody
group and the holomorphic bundles on elliptic curve $\ec$.
We will consider a family of bundles on $\ec$ parametrized by the $\Cx$-plane,
 e.g., as in~\cite{Friedman:1997ih}. We start with individual bundles.

Let $V$ be a rank $r+1$ polystable vector bundle of degree zero over the elliptic
curve ${\ec} = \BC^{\times}/\qe^{\BZ}$, with trivial determinant,
\[
\det V \approx {\CalO}_{\ec}.
\]
Such bundle always splits as a direct
sum of line bundles
\[
V = \bigoplus_{i=1}^{r+1} L_i.
\]
Each summand is a degree zero line bundle $L_i$
which can be represented as $L_i =
 {\CalO}(p_{0})^{-1} {\CalO}(t_i)$ where ${\CalO}(p)$ is the degree one line bundle whose divisor is a single point
 $p \in E$ and $p_{0}$ denotes the point $t=1$ corresponding to
 the identity in the abelian group law on the elliptic curve $\ec$.
 A
 meromorphic section $s_{i}$ of $L_i$ with a simple pole at $t = 1$ and zero at
 $t = t_i$ can be written explicitly using the theta-functions:
 \begin{equation*}
 s_i(t) = \frac{\theta(t/t_{i};\qe)}{\theta(t;\qe)}
 \end{equation*}
 and is unique up to a multiplicative constant.
To each degree zero vector bundle $V$ with
the divisor
\[
D_{V} = - (r+1) p_{0} + t_1 + \dots +
t_{r+1}
\]
of $\det V$ we associate a projectively unique
section $s$ of its determinant $\det V$
which has zeroes at $t_1, \dots, t_{r+1}$ and a pole of the order not
greater than $r+1$ at $t=1$:
\begin{equation}
\label{eq:s}
 s(t; {\bf t}) =\prod_{i=1}^{r+1} \frac{ \theta(t/t_i;\qe)}{\theta(t;\qe)},
\end{equation}
where we explicitly indicate the $\bf t$ dependence of the section $s$.
Now set $t_{i} = t_{i}(x)$ given by~\eqref{eq:tfromyaff}.
The meromorphic sections $s(t; {\mathbf{t}}(x); \qe)$
can be expanded in terms of the theta-functions
$\Theta_j(\y_0(x);{\mathbf{t}};\qe)$ and characters of $\hat A_{r}$
(see \eqref{eq:Archar} and \eqref{eq:Ar-theta}) as follows
\begin{align*}
 \y_0(x) \prod_{i=1}^{r+1} \frac{\theta(t/t_i(x);\qe)}{ \theta(t, {\qe})} & =
 \sum_{i=0}^{r}
 \qe^{-\frac {i}{2}} \qe^{\frac{i^2}{2(r+1)}}
 \Theta_i(\y_0(x);{\mathbf{t}(x)};\qe) \phi_i(t;\qe) \\
& = \phi(\qe)^r \sum_{i=0}^{r} \chi_i (\y_0(x);{\mathbf{t}(x)};\qe) \phi_i(t;\qe),
\end{align*}
where the functions $\phi_i(t; \qe)$ are normalized meromorphic elliptic functions defined in Ap\-pen\-dix~\ref{subsubsec:phi}.
Hence we find from \eqref{eq:tfromy} and \eqref{eq:T-matrix} that the section $s(t,x)$ \eqref{eq:s} obeys
\begin{equation*}
{\y}_{0}(x) s(t,x) = \phi(\qe)^{r} \sum_{i=0}^{r}
 \chi_{i}(\y_0(x);{\mathbf{t}(x)};\qe) M_{i{\tilde j}}(\qe) \tilde \phi_{{\tilde j}}(t;\qe),
\end{equation*}
where ${\tilde\phi}_{\tilde j}(t; {\qe})$ denotes the Weierstra{\ss}
monomials of Weierstra{\ss} elliptic functions~$X(t,\qe)$ and $Y(t,\qe)$;
and $M_{i {\tilde j}}$ is a certain modular matrix as defined in
 Appendix~\ref{subsubsec:phi}.
Recalling \eqref{eq:master} that the characters $(\chi_{i}(\y_0(x);{\mathbf{t}(x)};\qe))$ evaluated on the solutions $(\y_{i}(x))$
are polynomials in $x$, from~\eqref{eq:affcha} and \eqref{eq:hchi} we get
\begin{equation}
\frac{{\y}_{0}(x) s(t,x)}{\phi(\qe)^{r}} = \sum_{i=0}^{r} \Bigg( \prod_{j=0}^{r}
 \qe_{j}^{-\hat \lambda_i(\hat \lambda_j^{\vee})} \Bigg)T_{i}(x) \sum_{{\tilde j}} M_{i \tilde j}(\qe) \tilde \phi_{{\tilde j}}(t;\qe).
 \label{eq:Arhat-curvei}
\end{equation}
The section $s(t,x)$ vanishes at the $r+1$ points $t_1(x), \dots, t_{r+1}(x)$ for
each $x \in \Cx$, and hence defines the $(r+1)$-folded spectral cover of
$\Cx$ plane by the equation
\begin{equation}
 R(t,x) = 0,
 \label{eq:Arhat-curve}
\end{equation}
where $R(t,x)$ is the right-hand side of \eqref{eq:Arhat-curvei}. The curve (\ref{eq:Arhat-curve}) coincides with the
curve in~\cite{Witten:1997sc} constructed from by lifting to $M$-theory the IIA brane
arrangement realizing the elliptic model with $\mathfrak{m} =0$.

\subsection{Class II* theory}

Recall that in \eqref{eq:extamp} we defined an infinite set of functions $Y_i(x)$, $i \in
\BZ$ .
The analogue of the formula (\ref{eq:bgofx}) is the matrix $g(x) \in {\Gli}$, (cf.~\eqref{eq:tfromy}):
\begin{equation}
g(x) = Y_{0}(x)^{K} \times \operatorname{diag} ( t_{i}(x) )_{i \in {\BZ}}, \qquad t_{i}(x) = {\ct}_{i}
\frac{Y_{i}(x)}{Y_{i-1}(x)},
\label{eq:india}
\end{equation}
where ${\ct}_{i}$, $i \in \BZ$ solve
\[
{\ct}_{i+1} = {\qe}_{i \, {\rm mod}\, (r+1)} {\ct}_{i},
\]
and are normalized as in \eqref{eq:ctistar}
\[
\prod_{j=1}^{r+1} {\ct}_{j} =1
\]
so that for $i = 1, \dots, r+1$ the ${\ct}_{i}$ coincide with those in \eqref{eq:ctistar}, and
\begin{equation}
{\ct}_{i+b(r+1)} = {\ct}_{i} {\qe}^{b}, \qquad
{\ct}^{[i+b(r+1)]} = {\ct}^{[i]} \left( {\qe}^{r+1} \right)^{\frac{b(b-1)}{2}}.
\label{eq:tiiper}
\end{equation}
The fundamental characters of $\Gli$ evaluated on $g(x)$, ${\chi}_{i}(g(x))$
are
associated with
representations ${\CalR}_{i}$ of $\Gli$ with the highest weight taking value (cf.~\eqref{eq:hwgli}):
\begin{equation*}
g(x)^{\tilde\lam_{i}} = Y_i (x) {\ct}^{[i]} = Y_{0}(x) t(x)^{[i]}.
\end{equation*}
The characters are given by the infinite
sums over all partitions ${\lam} = ( {\lam}_{1} \geq {\lam}_{2} \geq \dots \geq {\lam}_{{\ell}({\lam})} > 0 )$ and so are the normalized invariants
\begin{align}
 \crf_{i}(\{Y_j (x) \},\qe) &{} = \frac{1}{{\ct}^{[i]} } {\chi}_{i}(g(x)) =
\sum_{{\lam}}
\prod_{j=1}^{{\ell}({\lam})} \bigg( {\qe}_{i-j+1}^{[{\lam}_{j}]}
\frac { Y_{i + {\lambda}_j - j + 1}(x)}
{ Y_{i + \lambda_j - j}(x)} \bigg)
Y_{i - {\ell}({\lam})}(x) \nonumber\\
&{} = Y_{i}(x) + {\qe}_{i} \frac{Y_{i+1}(x)Y_{i-1}(x)}{Y_{i}(x)} + \cdots,
\label{eq:glchar}
\end{align}
where we use the notation Section~\ref{se:sumpro}.

The invariant $\crf_{i}$ in (\ref{eq:glchar}) is a convergent
series for $|{\qe}_i| < 1$ like the theta-series, if $t_{i}(x)$ is uniformly bounded. In fact, for the periodic chain of arguments, i.e., for $Y_{i}(x) = Y_{i+r+1}(x)$ the~$\mathfrak{gl}_{\infty}$
character \eqref{eq:glchar} reduces to the usual affine character of $\widehat
{\mathfrak{g}\mathfrak{l}}_{r}$. The convergence of $\crf_{i}$ in the class II* case is more subtle. We shall comment on this below. For the moment let us view the invariants
as the formal power series in $\qe$ with coefficients in Laurent polynomials in $Y_{i}(x)$.

For the class II* theory the extended amplitudes
$Y_i(x)$ are quasi-periodic in $i$, cf.~\eqref{eq:shfper},
so
\begin{equation}
{\crf}_{i+r+1}(\{Y_j (x) \},\qe) = {\crf}_{i}(\{Y_j (x - (r+1){\ma}) \},\qe).
\label{eq:chiper}
\end{equation}
The cameral curve ${\CalC}_{u}$ for the class II* $A_{r}$ theory is defined by the system of
$r+1$ functional equations
\begin{equation*}
 \crf_{i}( \{ Y_j(x) \},\qe) = T_{i}(x),
 \qquad i = 0,\dots, r,
\end{equation*}
with
\begin{equation*}
 T_{i}(x) = T_{i,0} x^{N} + T_{i,1} x^{N-1} + \sum_{{\ba}=2}^{N} u_{i, {\ba}} x^{N-\ba} , \qquad
 T_{i,0} = \sum_{\lam}
\prod_{j=1}^{{\ell}({\lam})} {\qe}_{i-j+1}^{[\lam_j]}.
\end{equation*}
Let us now describe the II* analogue of the spectral curve, and find its realization in terms of some version of the Hitchin's system. Along the way we shall get an alternative derivation of~\eqref{eq:Arhat-curvei}
with the benefit of getting its Hitchin's form as well.

We form the generating function of $\crf_i$'s and study its automorphic properties.
The idea is to regularize the infinite product
\[
\prod_{i \in \BZ} ( 1- t_{i}(x) / t) / \big( 1 - {\ct}_{i}/t\big),
\]
while keeping the same set of zeroes and poles. Thus, we define
\begin{equation}
R(t,x) = \frac{Y_{0}(x)}{D_{0}(t; {\bf\qe})}
\prod_{k=1}^{\infty} \big( 1 - t_{k}(x)t^{-1} \big)\big(1- t t_{1-k}(x)^{-1}\big),
\label{eq:deltx}
\end{equation}
where
\begin{align*}
D_{0}(t; {\bf\qe}) = \prod_{k = 1}^{\infty} \big( 1 - {\ct}_{k} t^{-1}\big) \big( 1- t {\ct}_{1-k}^{-1} \big)
 = \prod_{i=1}^{r+1} \frac{{\theta}\big( t/{\ct}_{i} ; {\qe} \big)}{{\phi}({\qe})}.
\end{align*}
First of all, given that at large $x$ the eigenvalues $t_{k}(x)$ approach ${\ct}_{k}$ which,
in turn, behave as~${\qe}^{\frac{k}{r+1}}$, we expect~\eqref{eq:deltx} to define the
converging product, at least for large enough $x$.

Secondly,
let us check that \eqref{eq:deltx} is $^{i}{\CalW}$-invariant.
Let $i = 0, \dots , r$, ${\ba} = 1, \dots , N$. While crossing the $I_{i, {\ba}}$ cut the
``eigenvalue'' $t_{i}(x)$ maps to $t_{i+1}(x)$, which,
in case $i \geq 1$ or $i < 0$, leaves~\eqref{eq:deltx}
manifestly invariant. For $i = 0$ several factors in $\Delta (t,x)$
transform, altogether conspiring to make it invariant
\begin{equation*}
\begin{aligned}
& Y_{0}(x) \mapsto {\qe}_{0} Y_{-1}(x)Y_{1}(x)/Y_{0}(x) = t_{1}(x)/t_{0}(x), \\
& \big( 1 - t_{1}(x) t^{-1}\big) \big( 1 - t t_{0}(x)^{-1} \big) \\
& \qquad
\mapsto \big( 1 - t_{0}(x) t^{-1} \big) \big( 1 - t t_{1}(x)^{-1} \big) = \frac{t_{0}(x)}{t_{1}(x)} \big( 1 - t_{1}(x) t^{-1}\big) \big( 1 - t t_{0}(x)^{-1} \big). \\
\end{aligned}
\end{equation*}
Thirdly, let us introduce the analogues of
the spectral determinants for all fundamental representations
${\CalR}_i$:
\begin{equation*}
\begin{aligned}
& {\Delta}_{i}(t,x) = \frac{Y_{i}(x)}{D_{i}(t; {\bf\qe})}
\prod_{k=i+1}^{\infty} \big( 1 - t_{k}(x)t^{-1} \big)\big(1- t t_{2i+1-k}(x)^{-1}\big),
\\
& D_{i}(t; {\bf\qe}) =
\prod_{k = i+1}^{\infty} \big( 1 - {\ct}_{k} t^{-1}\big) \big( 1- t {\ct}_{2i+1-k}^{-1} \big).
\end{aligned}
\end{equation*}
Using $D_{i+1}(t ; {\qe}) = - t {\ct}_{i+1}^{-1} D_{i}( t ; {\qe})$,
$Y_{i+1}(x) = t_{i+1}(x) {\ct}_{i+1}^{-1} Y_{i}(x)$ we derive:
${\Delta}_{i}(t, x) = R(t,x)$ for all $i \in \BZ$.

{}Then, the quasi-periodicity \eqref{eq:tiiper} and \eqref{eq:chiper} implies
\begin{equation}
R( {\qe} t, x +{\ma} ) = {\Delta}_{r+1}(t,x) = R (t, x).
\label{eq:deltxii}
\end{equation}
Given the large $x$ asymptotics of $Y_{0}(x)$ and
$t_{i}(x)$, we conclude
\begin{equation*}
R(t,x) = x^{N} + \sum_{k=1}^{N} {\delta}_{k}(t) x^{N-k},
\end{equation*}
where ${\delta}_{k}(t)$ are the quasi-elliptic functions, which
have the first-order poles at $t = {\ct}_{i}$, $i=0, \dots, r$
on the elliptic curve ${\ec} = {\BC}^{\times}/ {\qe}^{\BZ}$.
Indeed, the poles come from the $D_0 (t ; {\qe})$ denominator,
while the quasi-ellipticity of $\delta_{k}(t)$ follows
from \eqref{eq:deltxii}:
\begin{equation*}
{\delta}_{i}({\qe}t) - {\delta}_{i}(t) = {\ma}^{i} + \text{polynomial in $\ma$ linear in $\delta_{k}({\qe}t)$}, \qquad k < i.
\end{equation*}
Now use
\eqref{eq:fermch} and \eqref{eq:chari} to rewrite $R(t,x)$ as
\begin{equation}
R(t,x) = \frac{\sum_{i\in \BZ} (-t)^{i} {\ct}^{[i]} {\crf}_{i}\left(\{Y_j (x) \},\qe\right)}{D_{0}(t; {\bf\qe})} = \frac{1}{D_{0}(t; {\bf\qe})} \sum_{i\in \BZ} (-t)^{i} {\ct}^{[i]} T_{i} (x),
\label{eq:deltxi}
\end{equation}
where we extended the definition of gauge polynomials $T_{i}(x)$ to $i \in \BZ$
by quasi-periodicity implied by \eqref{eq:chiper}:
\begin{equation}
T_{ i + r+1} (x) = T_{i} ( x - {\ma}).
\label{eq:tiper}
\end{equation}
Armed with \eqref{eq:tiiper} and \eqref{eq:tiper},
we reduce \eqref{eq:deltxi} to
a finite sum: let
\[
r(t,x) = \sum_{i=0}^{r} (-t)^{i} {\ct}^{[i]} T_{i} (x),
\]
then (cf.~\eqref{eq:phi_p})
\begin{align*}
 R(t,x) &{}= \frac{1}{D_{0}(t; {\qe})} \sum_{b \in {\BZ}} r ( t, x - b {\ma}) \Big( (-t)^{b} {\qe}^{\frac{b(b-1)}{2}} \Big)^{r+1}\nonumber\\
&{} = \frac{1}{D_{0}(t; {\qe})} \big( {\theta}\bigl( -(-t)^{r+1}; {\qe}^{r+1} \bigr) \ast_{{\ma}} r(t,x) \big) ,
\end{align*}
where the $\ast_{\hbar}$-product is defined by the usual Moyal formula:
\begin{equation*}
\left( f\ast_{\hbar} g\right) (t,x) = {\rm e}^{\hbar \frac{\pa^{2}}{{\pa}{\xi}_{1} {\pa}{\eta}_{2}} - \hbar\frac{\pa^{2}}{{\pa}{\xi}_{2} {\pa}{\eta}_{1}}} \vert_{{\xi}= {\eta} = 0} f( t + {\eta}_{1}, x+ {\xi}_{2} ) g (t + {\eta}_{2}, x + {\xi}_{2}).
\end{equation*}
The appearance of the $\ast$-product is the first hint that the class II* theory has something to do with the noncommutative geometry. We shall indeed soon see that a natural interpretation
of the solution to the limit shape equations of the class II* theory
involves instantons on the noncommutative four-manifold
${\BR}^{2} \times {\BT}^{2}$, where the noncommutativity
is ``between'' the $\BR^2$ and the $\BT^2$ components.

\subsection[Hitchin system on $T^2$]{Hitchin system on $\boldsymbol{T^2}$}
The above solution can be represented by the affine ${\rm GL}(N)$ Hitchin system on $\ec$:
\begin{equation*}
{\Phi}({\qe}t) = {\Phi}(t) + N{\ma}\cdot {\bf 1}_{N}
\end{equation*}
 with $r+1$ rank $1$ punctures ${\ct}_{j}$:
 \begin{equation*}
\begin{aligned}
& {\Phi}(t) \sim {\Phi}_{j} \frac{{\rm d}t}{t - {\ct}_{j}}, \qquad j = 1, \dots , r+1, \\
& {\Phi}_{j} = {\bu}_{j} \otimes {\bv}_{j}^{t}, \qquad {\bu}_{j}, {\bv}_{j} \in {\BC}^{N}, \\
\end{aligned}
\end{equation*}
whose eigenvalues are fixed in terms of masses
\begin{equation}
{\bv}_{j}^{t}{\bu}_{j} = \tr {\Phi}_{j} = N m_{j}.
\label{eq:momorb}
\end{equation}
Actually, the vectors and covectors ${\bv}_{j}$, ${\bu}_{j}$ are defined up to
the ${\BC}^{\times}$-action
\begin{equation*}
( {\bv}_{j}, {\bu}_{j} ) \mapsto \big(z_{j}{\bv}_{j}, z_{j}^{-1}{\bu}_{j}\big), \qquad z_{j} \in {\BC}^{\times}
\end{equation*}
and \eqref{eq:momorb} is the corresponding moment map equation, defining the coadjoint orbit ${\CalO}_{j}$ of ${\rm SL}(N, {\BC})$.
We can shift ${\Phi}(t)$ by the meromorphic scalar matrix
\[
{\bf\Phi}(t) = {\Phi}(t) - \sum_{j=1}^{r+1} m_{j} {\xi}\big( t/ {\ct}_{j}\big) \frac{{\rm d}t}{t} {\bf 1}_{N} ,
\]
which gives the following traceless meromorphic Higgs field (see~\cite{Nekrasov:1995nq}):
\begin{equation*}
{\bf\Phi}({t}) = \left\Vert p_{a} {\delta}_{a}^{b} + \sum_{j=0}^{r} u_{j}^{b}v^{j}_{a}\big(1 - {\delta}_{a}^{b}\big) \frac{{\theta}_{1}( {t}/{t}_{j} w_{b}/w_{a}) {\theta}_{1}^{\prime} (1)}{{\theta}_{1}( {t}/{t}_{j} ){\theta}_{1}( w_{b}/w_{a})} \right\Vert_{a,b=1}^{N},
\end{equation*}
which depends, in addition to the ${\rm SL}(N, {\BC})$-orbits ${\CalO}_{1}, \dots , {\CalO}_{r+1}$
on the choice $(w_{1}, \dots , w_{N})$ of a~holomorphic ${\rm SL}(N, {\BC})$ bundle on $\ec$, and the dual variables $(p_{1}, \dots , p_{N})$, subject to
\[
\sum_{a=1}^{N} p_{a} = 0, \qquad \prod_{a=1}^{N} w_{a} = 1.
\]
There are additional constraints
\begin{equation*}
 \sum_{j=1}^{r+1} u_{j}^{a}v_{a}^{j} = {\ma},
\end{equation*}
which generate the action of the residual gauge transformations in the maximal torus ${\bf T} = ({\BC}^{\times})^{N-1}$ of ${\rm SL}(N, {\BC})$.
The dimension of the corresponding phase space $\pv$, whose open subset~${\pv}^{\circ}$ is isomorphic to
\begin{equation*}
{\pv}^{\circ} \approx
\big( T^{*}\Bun_{{\rm SL}(N, {\BC})}({\ec}) \times \times_{j=1}^{r+1} {\CalO}_{j} \big) // {\bf T}
\end{equation*}
is equal to
\begin{equation*}
\dim {\pv} = 2(N-1) + (r+1)(2(N-1)) - 2(N-1) = 2(r+1)(N-1) = 2 {\bf r},
\end{equation*}
which is twice the dimension of the moduli space $\mv$ of vacua of the class II* $A_r$ theory with the gauge group ${\Gg} = {\rm SU}(N)^{r+1}$. The remaining $r+1$ mass parameters
are encoded in the symplectic moduli of the coadjoint orbits ${\CalO}_{j}$, as expected.

The relation to our solution is in the equality of two spectral determinants
\begin{equation}
R(t, x) = \operatorname{Det}_{N} \bigg[ \bigg( x - \sum_{j} m_{j} {\xi}(t/t_{j}) \bigg) \cdot {\bf 1}_{N} - {\bf\Phi}({t}) \bigg] = 0,
\label{eq:laxop}
\end{equation}
which is established by comparing the modular properties and the residues of the left and the right-hand sides.

Note the duality of the twisted periodicities of the gauge theory and Hitchin's system Lax operators
\begin{equation*}
\begin{aligned}
& {\Phi}({\qe}t) = w^{-1} {\Phi}(t) w + {\ma} \cdot {\bf 1}_{N} \in {\mathfrak{sl}}(N, {\BC}), \\
& {\qe} \cdot g(x - {\ma}) = S^{-1} g(x) S \in \Gli, \end{aligned}
\end{equation*}
where $S$ is the shift operator $S = \sum_{i \in \BZ} E_{i, i+r+1}$, and $w = \operatorname{diag}(w_{1}, \dots , w_{N})$. The equation~\eqref{eq:laxop} can be suggestively written as
\begin{equation*}
\operatorname{Det}_{N} ( x - {\Phi}(t)) \approx \operatorname{Det}_{H} ( t - g(x) ),
\end{equation*}
where $H$ is the single-particle Hilbert space of a free fermion $\psi$.

\subsection{Relation to many-body systems and spin chains}

The parameters of the spectral curve \eqref{eq:laxop} are holomorphic functions on ${\pv}^{\circ}$, which Poisson-commute, and define the integrable system. One way of enumerating the Hamiltonians of the integrable system is to mimic the construction of
Hamiltonians \eqref{eq:hitcham} of the higher genus Hitchin system. For example,
the quadratic Casimir
is a meromorphic $2$-differential on $\ec$ with the fixed second-order poles at $t = {\ct}_{j}$
\[
 \tr {\bf\Phi}(t)^2 = \sum_{j=1}^{r+1} N^2 m_{j}^{2} {\wp}\big(t/ {\ct}_{j}\big) {\rm d}t^2 + \sum_{j=1}^{r+1} U_{2,1,j} {\xi}\big(t/{\ct}_{j}\big) + U_{2,0}.
\]
The Hamiltonians $U_{2,0}$, $U_{2,1,j}$ are computed explicitly in~\cite{Nekrasov:1995nq}.
They describe the motion of $N$ particles on $\ec$ with the coordinates $w_{1}, \dots , w_{N}$, which have additional ${\rm GL}(r+1, {\BC})$-spin degrees of freedom.
However, in view of our gauge theory analysis, it seems more natural to view this system
as the $\Gli$-spin chain. We conjecture that the deformation quantization of the properly compactified phase space $\pv$ will contain the subalgebra ${\CalA}_{\ma}$ of the Yangian $Y(\Gli)$ algebra, which is a deformation of the Yangian of the affine ${\hat A}_{r}$.

The relation of many-body systems and spin chains based on finite-dimensional symmetry groups was discussed in the context of Hecke symplectic correspondences in~\cite{Levin:2001nm, Olshanetsky:2008uu}. One can also interpret
the results of~\cite{Felder:1995iv} as the quantum version of this correspondence.

\subsection[Class II theories of $D$ type\label{sec:affineD}]{Class II theories of $\boldsymbol{D}$ type}

In this section $\gq = \hat D_{r}$.

The fundamental weights of $\hat D_{r}$ are $\lambda_{0}$, $\hat
\lambda_{i} = a_i^\vee \lambda_{0} + \lambda_i$, $i = 1, \dots, r$ where
$\lambda_i$ are fundamental weights of $D_r$, and Dynkin labels are
$(a_0,\dots, a_r)=(1,1,2,\dots, 2, 1,1)$ (see Appendix~\ref{sec:Dr-conventions}).
Correspondingly,
\begin{gather*}
 t_{1}(x) = \ct_{1} \y_{1}(x)/\y_{0}(x), \\
 t_{2}(x) = \ct_{2} \y_{2}(x)/(\y_{1}(x)\y_{0}(x)) , \\
 t_{i}(x) = \ct_{i} \y_{i}(x)/\y_{i-1}(x), \qquad i = 3, \dots , r-2, \\
 t_{r-1}(x) = \ct_{r-1} \y_{r-1}(x)\y_{r}(x)/\y_{r-2}(x), \\
 t_{r}(x) = \ct_{r} \y_{r}(x)/\y_{r-1}(x)
\end{gather*}
with
\begin{gather}
 \ct_{i} = \left( \qe_i \qe_{i+1} \dots \qe_{r-2} \right)^{-1} \left( \qe_{r-1} \qe_{r} \right)^{-\frac 12}, \qquad i = 1, \dots , r-2, \nonumber\\
 \ct_{r-1} = (\qe_{r-1} \qe_{r})^{-\frac 12} , \qquad \ct_{r} = (\qe_{r-1}/ \qe_{r} )^{\frac 12} .
\label{eq:qDafffine}
\end{gather}

There are $4$ irreducible $\hat D_{r}$ highest weight modules
$\hat R_{0}$, $\hat R_{1}$, $\hat R_{r-1}$, $\hat R_{r}$ at level $1$,
and~${r-3}$ irreducible $\hat D_{r}$ highest weight modules $\hat R_{2},
\dots, \hat R_{r-2}$ at level $2$. In this section, to shorten formulae,
 we are using not the characters of $\hat R_{i}$ themselves but the
 closely related affine Weyl invariant functions
 $\crfDvec_j$ at level $2$ and $\crfDspin_j$ at level $1$
expressed terms of theta-functions explicitly as given in
 (\ref{eq:Dr-inv}). Such functions $\crfDspin_j$ and $\crfDvec_j$
 differ from the actual characters by a simple power of Euler function
 $\phi(\qe)$ and some $\qe$-dependent constant, also $\crfDspin_{0}$, $\crfDspin_{1}$
 appear as a linear combination of $\hat R_{0}$ and $\hat R_{1}$
 characters, while $\crfDspin_{r}$, $\crfDspin_{r-1}$ appear as linear combination of
 $\hat R_{r-1}$ and $\hat R_{r}$ characters (see (\ref{eq:Dr-inv}))
\begin{gather}
 \crfDspin_{j}(\y_0 (x), {\mathbf{t}(x)}; \qe) = T_{j}(x), \qquad
 \crfDvec_{j}(\y_0 (x), {\mathbf{t}(x)}; \qe) = T_j(x), \label{eq:AffD-polynomials}
\end{gather}
where polynomials $ T_{j}(x)$ are of degree $N$ for $j=0,1,r-1,r$
and of degree $2N$ for $j=2,\dots, r-2$. so that $\crfDspin$
is of degree $1$ in $\y_0$ for $j=0,1,r-1,r$ and $\crfDvec$ is of degree $2$ in $\y_0$
for $j=2, \dots, r-2$. Also, in this section the highest coefficient of
the polynomial $T_j(x)$ is normalized differently then in
\eqref{eq:affcha}; one can find it as theta-series evaluating
(\ref{eq:Dr-inv}) on $\ct_i$.

Using the standard embedding $\mathfrak{so}(2r) \subset
\mathfrak{sl}(2r)$, we construct the algebraic equation of the spectral curve of the
$\hat D_{r}$ theory as the specialization of the spectral curve for $\hat
A_{2r-1}$ theory. Indeed, a vector bundle V associated to the
vector representation of ${\rm SO}(2r)$ splits as the sum of~$r$ pairs of line
bundles
\[
L_{t_i} \oplus L_{t_i^{-1}}
\]
with the degree zero line bundle $L_{t}$ being
\begin{equation*}
 L_{t} = \CalO(p_0)^{-1} \CalO(t)
\end{equation*}
and $p_0 \in {\ec}$ is our friend $t = 1$. Then we proceed as in
(\ref{eq:phi_p}), (\ref{eq:s-ThetaD}), (\ref{eq:s}) by considering a~meromorphic section of the determinant bundle $\det V\approx {\CalO}_{\ec}$
\begin{equation}
\label{eq:section-st}
 s(t,x) = \prod_{i=1}^{r} \frac{ \theta(t/t_i(x);\qe)}{\theta(t;\qe)} \frac{\theta\big(t/t_{i}(x)^{-1};\qe\big)}{\theta(t;\qe)}.
\end{equation}
From Section~\ref{se:phiD}, we find
\begin{equation}\label{eq:spectralD}
 \y_0^2 s(t,x) = \sum_{ i=0}^{r} \Xi_{ i}(\y_0; \mathbf{t}(x);\qe)
M_{ij}(\qe) X^{j} (t;\qe),
\end{equation}
where $X^j(t,\qe)$ are powers of Weierstra{\ss} monomials forming a basis
in the space $H^{0}_{\text{even}}({\ec}, \CalO(2 r p_0))$
of meromorphic functions on elliptic curve symmetric under the reflection
$t \to t^{-1}$ and with a~pole of order no greater then $2r$ at the
origin, and $M_{ij}(\qe)$ is a certain modular matrix.

The linear relations (\ref{eq:D-theta-relations}) allow to express
$\Xi_i$
in terms of
\begin{gather*}
 \tilde \Xi_i = \crfDvec_i, \qquad i = 2,\dots, r-2, \\
 \tilde \Xi_i = \big(\crfDspin_i\big)^2, \qquad i = 0,1,r-1,r,
\end{gather*}
as
\begin{equation*}
 \Xi_{i} = \sum_{{\tilde i} = 0}^{r} {\tilde \Xi}_{\tilde i} {\tilde M}_{{\tilde i}i }(\qe),
\end{equation*}
where $ \tilde M_{i \tilde i}(\qe)$ is a certain modular transformation matrix.
Using the character equations (\ref{eq:AffD-polynomials}), the
spectral curve \eqref{eq:spectralD}
turns into
\begin{equation}
\label{eq:curve-AffD}
 \y_0^2 s(t,x) = \sum_{\tilde i, j} \tilde T_{\tilde i}(x) \tilde {\tilde M}_{{\tilde
 i}j}(\qe) X^{j}(t, \qe),
\end{equation}
where $\tilde {\tilde M}_{{\tilde i}j}(\qe) = \tilde M_{\tilde i i}
(\qe) M_{i j} (\qe)$
and
 \begin{gather*}
\tilde T_{\tilde i}(x) = T_{\tilde i}(x), \qquad \tilde i =2,\dots, r-1, \\
\tilde T_{\tilde i}(x) = (T_{\tilde i}(x))^2, \qquad \tilde i = 0,1, r-1,r,
 \end{gather*}



The spectral curve of the $\hat D_r$ theory
is the algebraic equation $R(t,x) = 0$ where $R(t,x)$ is the right-hand
side of~\eqref{eq:curve-AffD} combined
with the
 Weierstra{\ss} cubic equation (\ref{eq:wxy}). The $\hat D_{r}$ curve is
 the specialization of the $\hat A_{2r -1}$ curve in two ways. First,
 there are no odd in $Y$ monomials in~\eqref{eq:curve-AffD}, and,
 second, the polynomial coefficients $\tilde
 T_{\tilde i}(x)$ of degree $2N$ in $x$ satisfy factorization condition: they are full squares
 for $\tilde i = 0,1,r-1,r$.

To interpret the curve in Hitchin--Gaudin formalism, we will rewrite it in
a slightly different form.
First, notice that\footnote{Indeed, the left-hand or right-hand sides is the meromorphic elliptic function with $2r$ zeroes at
points $X_i$, $Y$ and $X_i$, $-Y$ and the pole of order $2r$ at $t = 1$, or
$X = \infty$. Such function is unique up to a normalization which is
fixed by the asymptotics at $t = 1$.}
\begin{equation}
\label{eq:section-norm}
 \prod_{i=1}^{r}\frac{ \theta_1\big(t/\ct_i;\qe\big)}{\theta_1(t;\qe)} \frac{
 \theta_1\big(t/\ct_i^{-1};\qe\big)}{\theta_1(t;\qe)} = \prod_{i=1}^{r}
 \theta_1\big(\ct_i;\qe\big) \theta_1\big(\ct_i^{-1};\qe\big) \big(X - \check{X}_i\big).
\end{equation}
We used here the notations \eqref{eq:weierx2} and \eqref{eq:eal} for the Weierstra{\ss} functions and
\begin{align*}
\check{X}_i = X\big(\ct_i ; {\qe}\big), \qquad \check{Y}_{i}^{2} = 4 \prod_{{\alpha}=1}^{3} \big( \check{X}_{i} - e_{\alpha} \big).
\end{align*}
Then, if we divide (\ref{eq:section-st}) by (\ref{eq:section-norm}), we find\footnote{And use $\theta_1(t,\qe)$ in lieu of
 $\theta(t,\qe)$ as the basic function, so that strictly speaking there is
 slightly different transformation matrix
 $\tilde {\tilde M}_{\tilde i j}$ compared to \eqref{eq:section-st} and \eqref{eq:s-ThetaD}.}
\begin{gather}
\y_{0}^2(x) \prod_{i=1}^{r} \theta_1\big(\ct_i;\qe\big) \theta_1\big(\ct_i^{-1};\qe\big) \prod_{i=1}^{r}
\frac{{\theta}_{1}\big(t/t_i(x);\qe\big)}{{\theta}_{1}\big(t/\ct_i;\qe\big)}\frac{{\theta}_{1}\big(
 t/t_i^{-1}(x);\qe\big)}{{\theta}_{1}\big(t/\ct_i^{-1};\qe\big)} = R(x,X(t, {\qe})),\nonumber\\
 R(x,X): =\frac {\sum_{\tilde i, j =0}^{r} \tilde T_{\tilde i}(x) \tilde {\tilde M}_{{\tilde
 i}j}(\qe) X^{j} }{
 \prod_{i=1}^{r} \big(X - \check X_i\big)}.
\label{eq:longD}
\end{gather}
Now, at the order two points on $\ec$,\footnote{For example, the points $\big( 1, -1, -\qe^{-1/2},
\qe^{1/2}\big)$ in the $t$-parametrization, where vanish the respective theta functions
$\theta_1(t;\qe)$, $\theta_2(t;\qe)$, $\theta_3(t;\qe)$, $\theta_4(t;\qe)$, or,
equivalently,
at the four branch points in the $X$ plane: $(\infty, e_1,e_2,e_3 )$.} the
value of the section $R(x,X)$ can be expressed in terms of
the weight $1$ invariants
$\crfDspin_{r-1}$, $\crfDspin_{r}$, $\crfDspin_{0}$, $\crfDspin_{1}$ (compare
with \eqref{eq:Dr-inv} and
\eqref{eq:D1-theta}, \eqref{eq:D1-theta-jacobi}), and it factorizes as
\begin{gather}
 R(x,X)|_{X\to\infty} = ( T_{r-1}(x))^2 ,\nonumber\\
 R(x,X)|_{X\to e_1} = c_2 ({\tqe}) ( T_{r}(x))^2, \nonumber\\
 R(x,X)|_{X\to e_2} = c_3 ({\tqe}) (T_{0}(x))^2, \nonumber\\
 R(x,X)|_{X\to e_3} = c_4 ({\tqe}) (T_{1}(x))^2,
\label{eq:consD}
\end{gather}
where
\begin{equation}
\label{eq:ci-affined}
 c_{k}({\tqe}) = \prod_{i=1}^{r} \frac{\theta_1\big(\ct_i;\qe\big)
 \theta_1\big(\ct_i^{-1};\qe\big)}
{\theta_{k}\big(\ct_i;\qe\big)
 \theta_{k}\big(\ct_i^{-1};\qe\big)}, \qquad k=2,3,4.
\end{equation}

The Seiberg--Witten differential is given by
\begin{equation*}
{\lam} = x \frac{{\rm d}X}{Y}.
\end{equation*}
It is defined on the two fold cover $C_{u}$ of the curve $R(x,X) =0$, which is a curve in the product ${\BC\BP}^{2}_{(X:Y:Z)} \times \Cx$, given by the equations
\begin{gather}
 Y^2 Z = 4 (X-e_{1} Z)(X-e_{2}Z)(X-e_{3}Z), \qquad
 F(x, Z, X) = Z^{r} R(x, X/Z) = 0 .
\label{eq:SWcurveD}
\end{gather}
The curve $C_{u}$ can be interpreted at the spectral curve of ${\rm GL}(2
N)$ Hitchin--Gaudin system on the orbifold ${\ec}/\BZ_2$, such that at the
fixed point $X = \infty, e_1, e_2, e_3$ the ${\rm GL}(2N)$ system reduces to the
${\rm Sp}(2N)$ system. For more details on the Hitchin system, Nahm transform
and the brane construction of the spectral curve for the ${\hat D}_r$
quiver see~\cite{Kapustin:1998fa,Kapustin:1998xn}. Our main result is the rigorous
derivation
of the spectral curve and its periods from the gauge theory considerations.

\subsection[Deforming the $N_f = 4$ {\rm SU}(2) theory]{Deforming the $\boldsymbol{N_f = 4}$ $\boldsymbol{{\rm SU}}$(2) theory}\label{se:formDhattoA}

The $\hat D_{4}$ theory can be interpreted as the theory obtained from
gauging the flavor group of the~$D_4$ theory with $({\bv}_{1}, {\bv}_{2}, {\bv}_{3}, {\bv}_{4}) = (N,2N,N,N)$ theory, and with $({\bw}_{1}, {\bw}_2, {\bw}_{3}, {\bw}_{4}) =(0,N,0,0)$
matter multiplets. In the limit $\qe_0 \to 0$ the elliptic curve $\ec$
degenerates to the cylinder $\BC_{\la t \ra}^{\times}$, while
Seiberg--Witten curve \eqref{eq:SWcurveD} degenerates to the
Seiberg--Witten curve of the $D_4$ theory \eqref{eq:crd}.

Let us consider the case $N=1$. Let us parametrize the polynomials $T_{0}$, $T_{1}$, $T_{3}$, $T_{4}$ as
\begin{equation}
\label{eq:Tidef}
T_{i}(x) = T_{i,0}({\tqe}) (x - m_{i}), \qquad i =0,1,3,4,
\end{equation}
and
\[
T_{2}(x) = T_{2,0} ({\tqe})\big(x^{2} - m_{2}x + u \big),
\]
where parameters $q_{i}$, $m_{i}$ and $u$ are related to the microscopic couplings ${\qe}_{i}$
and the $U(1)^{4} \times {\rm SU}(2)$ Coulomb moduli
\begin{gather}
T_{3,0} ({\tqe})= \prod_{i=1}^{4} \theta_1\big(\ct_{i}\big),\qquad
T_{4,0} ({\tqe})= \prod_{i=1}^{4} \theta_2\big(\ct_{i}\big), \nonumber\\
T_{0,0} ({\tqe})= \prod_{i=1}^{4} \theta_3\big(\ct_{i}\big), \qquad
T_{1,0} ({\tqe})= \prod_{i=1}^{4} \theta_4\big(\ct_{i}\big), \nonumber\\
T_{2,0} ({\tqe}) = \Xi_2\big(1, \mathbf{\ct},\qe\big),
\label{eq:qDseries}
\end{gather}
where $\ct_i$ are defined in \eqref{eq:qDafffine}.
Then the spectral curve of the ${\hat D}_{4}$ theory
\eqref{eq:longD} and \eqref{eq:consD} has the generic form
\begin{equation}
\label{eq:RsimpleD}
R(x, X) = T_{3}^{2}(x) + \sum_{i=1}^{4} \frac{b_{i}(x)}{X - \check{X}_{i}},
\end{equation}
where $b_{i}(x)$ are some polynomials of degree $2$ that we want to
relate to the coupling constants and Coulomb parameters. The first thing
to notice is that $R(x,X)$ in \eqref{eq:RsimpleD} obtained from~\eqref{eq:longD} does not have poles at $X = \check X_i$ at $x \to
\infty$ in the leading order $x^2$.
Therefore, the polynomials~$b_i(x)$ are actually degree 1~polynomials
containing $8$ coefficients. There are 6~linear
equations on these coefficients
coming from 3 factorization equations \eqref{eq:consD} viewed as coefficients at
 $x^1$ and~$x^0$ (and notice that the equations at $x^2$ are identically satisfied because of~\eqref{eq:ci-affined} and~\eqref{eq:qDseries})
 \begin{gather}
\sum_{i=1}^{4} \frac{b_{i}(x)}{e_{1} - \check{X}_{i}} = c_2 T_{4}^{2}(x) - T_{3}^{2}(x), \nonumber\\
\sum_{i=1}^{4} \frac{b_{i}(x)}{e_{2} - \check{X}_{i}} = c_3 T_{0}^{2}(x) - T_{3}^{2}(x), \nonumber\\
\sum_{i=1}^{4} \frac{b_{i}(x)}{e_{3} - \check{X}_{i}} = c_4 T_{1}^{2}(x) - T_{3}^{2}(x).
\label{eq:D4rel}
\end{gather}
The above three equations determine four linear functions $b_i(x)$ up to a
single linear function, which depends on two parameters ${\tilde m}_{2}$, ${\tilde u}$:
\begin{equation}
 \label{eq:D4kernel}
\tilde b_j (x) = (-1)^{j} (-\tilde m_2 x + \tilde u) \operatorname{Det} \left\Vert \begin{matrix}
 \frac{1}{e_a - \check X_{b}}
 \end{matrix} \right\Vert_{a=1,\dots, 3}^{b=1, \dots, 4, b \neq j}.
\end{equation}
From \eqref{eq:longD}, it is clear that $\tilde m_2$, $\tilde u$
are proportional to $m_2$, $u_2$. To summarize,
we can describe the spectral curve \eqref{eq:RsimpleD} of $\hat D_{4}$ theory
by the coupling
constants $\qe_i$, $i = 0,\dots, 4$, which define the elliptic curve $\ec(\qe)$
with modulus $\qe = \qe_0 \qe_1 \qe_2^2 \qe_3 \qe_4$ and positions of 4
punctures $\check X_i$ in the~$\BC_{\la X \ra}$ plane for Weierstra{\ss} cubic using
\eqref{eq:qDafffine}, the 4 parameters $m_i$, $i = 0,1,3,4$ entering into
relations~\eqref{eq:D4rel} through \eqref{eq:Tidef} and 2 parameters $\tilde m_2$, $\tilde u$ in \eqref{eq:D4kernel}.

Now consider the limit $\qe_{0} \to 0$ which turns the $\hat D_{4}$ class II quiver
theory to the $D_{4}$ class I quiver theory. In this limit
 the Weierstra{\ss} cubic degenerates: $e_1 = - 2 e_{3}$, $e_{2} = e_{3} =1/12$,
 \begin{equation*}
 Y^2 = 4 \lb X - e_{3} \rb^{2} \lb X + 2 e_{3} \rb^2
 \end{equation*}
with
\begin{equation*}
 X = \frac{t}{(1-t)^2} + \frac{1}{12},\qquad
 Y = \frac{t(1 + t)}{ (1 - t)^3}.
\end{equation*}
The Seiberg--Witten differential $x \frac{{\rm d}X}{Y}$ becomes $x\frac{{\rm d}t}{t}$.
The elliptic curve $\ec$ degenerates to the rational curve which is the double cover $t \mapsto X$ of the complex projective line
$\CP^1_{X}$. To make contact with the Seiberg--Witten curve
of the $D_{4}$ quiver theory it is convenient to work
in the coordinate which is related to the coordinate
$X$ by rational transformation
\begin{equation*}
 \eta = 2 + \frac{1}{X - e_{3}} = t + t^{-1}.
\end{equation*}
The function $\eta(X)$
is degree two meromorphic function on $\ec$ with values at the four
$\BZ_2$ invariant points given by
\begin{equation*}
 \eta(e_2) = \eta(e_3) = \infty, \qquad \eta(e_1) = -2, \qquad \eta(\infty) = 2.
\end{equation*}
Rewriting \eqref{eq:longD} in terms of $\eta(x)$, we find the equation of
spectral curve $\tilde \CalR^{D_4}(\eta, x) = 0$ for
\begin{equation*}
 \tilde \CalR^{D_4}(\eta, x) = \sum_{i=0}^{4} \eta^{i} p_i(x),
\end{equation*}
where $p_{i}(x)$ are some polynomials of degree $2$ in $x$. Moreover, the
factorization conditions~\eqref{eq:consD} translates to the statement
 that $\tilde \CalR^{D_4}(\eta, x)$ is full square at $\eta = \infty$ and at
 $\eta = \pm 2$ in the polynomial ring of $x$. Notice that this is
 precisely the factorization conditions
 \eqref{eq:D4factorization} and~\eqref{eq:D4factorization-inf} of the curve
 \eqref{eq:D4-curve} for the $D_4$ quiver. \big(The variables $t$
 and $\eta$ in the equations \eqref{eq:D4-curve} and~\eqref{eq:eta}
correspond to $t$ and $\eta$ of this section multiplied by a factor $\qe_1 \qe_2 \qe_3^{\frac 1 2}
 \qe_{4}^{\frac 1 2}$.\big)

\begin{figure}[t] \centering
 \includegraphics[width=12cm]{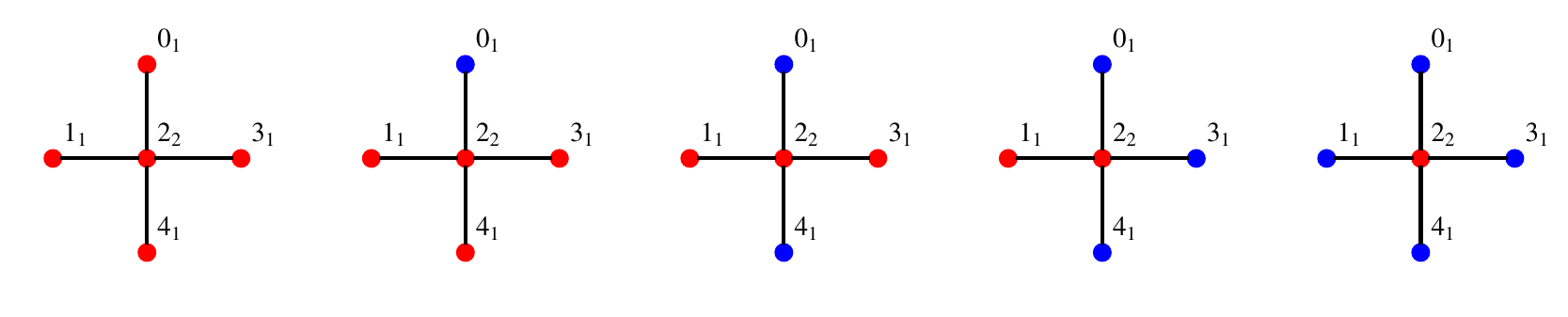}
 \caption{The freezing $\hat D_{4} \to D_{4} \to A_{3} \to
 A_{2} \to A_{1}$. The live nodes are denoted by red, the
 frozen nodes are denoted by blue. The nodes are labeled as
 $i_{\bv_{i}}$.} \label{fig:freezing}
\end{figure}

Given the above discussion and Section~\ref{se:freezing}, let us
summarize the freezing hierarchy $\hat D_{4} \to D_{4} \to A_{3} \to A_{2}
\to A_{1}$.
For $\hat D_{4}$ theory, we start with elliptic curve $\ec(\qe)$ with
$\BZ_2$ reflection symmetry $ t \to t^{-1}$ (or $Y \to -Y$)
and $8$ $\BZ_{2}$-symmetrically located punctures in 4 pairs $\big(\ct_{i},
\ct_{i}^{-1}\big)$. As we freeze $\qe_0 \to 0$, the elliptic curve $\ec(\qe)$
degenerates to a $\BZ_2$-symmetrical cylinder $\BC^{\times}_{t}$ with 4-fold pairs $(\ct_{i},
\ct_{i}^{-1})$ of punctures.
The cylinder $\BC^{\times}_{t}$ double covers
its $\BZ_{2}$-quotient $\CP^{\times}_{\eta}$. This is the situation of
$D_{4}$ quiver theory \eqref{eq:D4-curve}. As we freeze $\qe_{4} \to 0$
the second sheet of the double cover $\BC^{\times}_{t} \to
\BC^{\times}_{\eta}$ is removed to infinity and we are left with $4$
punctures of $A_3$ quiver at $ \big(\qe_1^{-1} , 1, \qe_2, \qe_2
\qe_3\big)$.\footnote{Keeping in mind the ultimate
configuration of the $A_1$ quiver dynamical at node ``2'' we
have rescaled the position of punctures by a factor of $\qe_1$.} Notice, that as discussed after \eqref{eq:A3-curve} the ${\rm SL}(2,\BC)$
residues of the Higgs field vanish at the punctures in $0$ and
$\infty$. As we freeze $\qe_3 \to 0$ the puncture at $\qe_2 \qe_3$
(with non-trivial~${\rm SL}(2,\BC)$ residue of Higgs field) merges with the
puncture $0$ and we are in the situation of the~$A_{2}$ quiver with
${\rm SL}(2,\BC)$ punctures at $\big(\qe_{1}^{-1}, 1, \qe_2, 0\big)$ and ${\rm GL}(1,\BC)$
puncture at $\infty$. Finally
as we freeze $\qe_1 \to 0$ the puncture at $\qe_{1}^{-1}$ (with
non-trivial residue of the ${\rm SL}(2,\BC)$ Higgs field) is merged with
the puncture at $\infty$ and we are left with $\CP^{1}$ with ${\rm SL}(2,\BC)$
punctures at $(\infty, 1, \qe_2, 0)$ for the $A_1$ quiver theory defined
at the dynamical node ``2''. See Figure~\ref{fig:freezing}.

\subsection[Class II theories of $E$ type]{Class II theories of $\boldsymbol{E}$ type}\label{sec:affineE}

The main technical tool is the natural
isomorphism between the moduli space
of the $E_k$-bundles on elliptic curve $\ec$
and the moduli space of del Pezzo surfaces
$\CalS_{k}$, which are obtained by blowing up
$k$ points in ${\BC}{\BP}^2$, and have
the fixed elliptic curve $\ec$ as the anticanonical
divisor.
The spectral curve is found using the ``cylinder map''~\cite{Kanev},
and see~\cite{Curio:1998bva,Donagi:2008kj,Donagi:2008ca} for applications.

Another way of encoding the geometry of the moduli
space the
$E_k$-bundles is in the unfolding of the parabolic unimodular
singularities~\cite{Arnold:1985} ${\hat T}_{a,b,c}$ with
\[
\frac 1a + \frac 1b + \frac 1c = 1,
\]
which are
\begin{alignat*}{3}
& {\tilde E}_{6} = P_{8} = {\hat T}_{3,3,3} \colon \ x^{3} + y^{3} + z^{3} + m xyz, \qquad&&
m^3 + 27 \neq 0, & \nonumber\\
& {\tilde E}_{7}= X_{9} = {\hat T}_{2,4,4} \colon \ x^{4} +y^{4} + z^{2} + m x yz , \qquad&& m^4 - 64 \neq 0,& \nonumber\\
& {\tilde E}_{8} = J_{10} = {\hat T}_{2,3,6} \colon \ x^6 + y^3 + z^2 + m x y z, \qquad&& 4 m^6 - 432 \neq 0.&
\end{alignat*}

We shall not pursue this direction in this work.

\begin{Remark}
Another important question left for future work is the connection between our description
of the special geometry via the periods of ${\rm d}{\BS}$ and the periods of non-compact
Calabi--Yau threefolds of~\cite{Katz:1997eq}.
\end{Remark}

\newcommand{\pe}{p}
\subsection[Del Pezzo and $E_6$ bundles]{Del Pezzo and $\boldsymbol{E_6}$ bundles}

The Del Pezzo surface $\CalS_{6} \subset \WP^{1,1,1,1} = {\BC\BP}^{3}$ is a zero locus of a homogeneous degree $3$ polynomial
\begin{equation*}
{\Ga}_{3}(X_{0}, X_{1}, X_{2}, X_{3}) = \sum_{i=0}^{3} X_{0}^{3-i} {\CalG}_{i} (X_{1}, X_{2}, X_{3}),
\end{equation*}
where ${\CalG}_{i}$ is the degree $i$ homogeneous polynomial in $X_{1}$, $X_{2}$, $X_{3}$. In particular,
\begin{equation*}
{\CalG}_{3}(X_{1}, X_{2}, X_{3}) = -X_{1}X_{3}^{2} + 4 X_{2}^{3} - g_{2} X_{1}^{2}X_{2} - g_{3} X_{1}^{3}
\end{equation*}
defines the elliptic curve ${\ec}$, which determines the gauge coupling $\qe = \exp (2\pi \ii \tau)$, cf.~(\ref{eq:wxy}),
\begin{equation*}
{\tau} = \frac{\oint_{B} {\rm d}X / Y}{\oint_{A} {\rm d}X/Y}, \qquad X = X_{2}/X_{1}, \qquad Y = X_{3}/X_{1} .
\end{equation*}
The rest of the coefficient functions ${\CalG}_{0,1,2}$ is parametrized as follows:
\begin{align*}
& {\CalG}_{2} (X_{1}, X_{2}, X_{3}) = \pe_{0} X_{1}^{2} + \pe_{1} X_{1}X_{2} +
\pe_{6} X_{2}X_{3} ,\\
& {\CalG}_{1} (X_{1}, X_{2}, X_{3}) = \pe_{2} X_{1} + \pe_{3} X_{2} + \pe_{5} X_{3} ,\\
& {\CalG}_{0} (X_{1}, X_{2}, X_{3}) = \pe_{4}.
\end{align*}
The isomorphism classes of $\CalS_{6}$ surfaces containing the fixed elliptic curve ${\ec}$ are in one-to-one correspondence with the points
\begin{equation*}
[p] = (\pe_{0}: \pe_{1}: \pe_{6}: \pe_{2}: \pe_{3}: \pe_{5}: \pe_{4}) \in {\CalM}
\end{equation*}
in the weighted projective space ${\CalM} = {\WP}^{1,1,1,2,2,2,3}$, which is also isomorphic, by E.~Loojienga's theorem~\cite{Loojienga:1976}, to the moduli space
$\Bun_{E_{6}({\BC})}^{ss}(\ec)$
of holomorphic semi-stable principal $E_6$-bundles on $\ec$. We label the
projective coordinates $\pe_i$ in such a way that the projective weight
of $\pe_i$ equals Dynkin mark $a_i$ in our conventions Appendix~\ref{se:mckay}.
The correspondence between the $E_{6}$-bundles on $\ec$ and the del Pezzo surfaces $\CalS_{6}$ is geometric:
there are precisely $27$ degree $1$ rational curves (``the $(-1)$-lines'') $C_{a}$ on $\CalS_{6}$,
$a = 1, \dots, 27$, which are the divisors of the line bundles ${\mathscr L}_{a}$ on $\CalS_{6}$. The direct sum
\begin{equation*}
{\CalU} = \bigoplus_{a=1}^{27} {\mathscr L}_{a}
\end{equation*}
has no infinitesimal deformations as a bundle on~$\CalS_{6}$. The mapping class group of $\CalS_{6}$ acts on the $(-1)$-lines by the $E_6$ Weyl transformations. As a result, the bundle ${\CalU}$ is a vector bundle associated to a canonical principal $E_{6}({\BC})$-bundle ${\CalP}_{\CalS_{6}}$ over $\CalS_{6}$ with the help of a ${\bf 27}$ representation:%
\begin{equation*}
{\CalU} = {\CalP}_{\CalS_{6}} \times_{E_{6}({\BC})} {\bf 27}.
\end{equation*}
The restriction of ${\CalP}_{\CalS_{6}} \vert_{E}$ is the holomorphic
principal $E_{6}({\BC})$ bundle over $E$ which corresponds to the point $[s]$ in Loojienga's theorem.
Again, the associated rank $27$ vector bundle ${\CalU} \vert_{E}$
splits
\begin{equation*}
{\CalU} \vert_{\ec} = \bigoplus_{a=1}^{27} {\mathscr L}_{a}.
\end{equation*}
The line subbundles ${\mathscr L}_{a}$ can be expressed as
\begin{equation*}
{\mathscr L}_{a} = \bigotimes_{i=1}^{6} {\BL}_{i}^{w_{a,i}},
\end{equation*}
where $w_{a,i}$, $i = 1, \dots , 6$, $a = 1, \dots , 27$
are the components of the weight vector.
The line bundles~${\BL}_{i}$, $i=1, \dots , 6$ are defined up to
the action of the $E_6$ Weyl group. Let us now compute the~${\mathscr L}_{a}$'s. The rational curve of degree one in $\CalS_{6}$ is a rational curve of degree one in ${\BC\BP}^{3}$ which is contained in $\CalS_{6}$. A parametrized rational curve of degree one in ${\BC\BP}^{3}$ is a collection
of $4$ linear functions: ${\ze} \mapsto {\bX}({\ze})$,
\begin{equation*}
{\bX}({\ze}) = ( X_{0} + {\ze} v_{0} , X_{1} + {\ze} v_{1} , X_{2} + {\ze} v_{2} , X_{3} + {\ze} v_{3} ).
\end{equation*}
The two quadruples
\[
{\bX}({\ze}) \qquad {\rm and} \qquad ( c{\ze} + d ) {\bX} \bigg( \frac{a {\ze} + b}{c{\ze} + d} \bigg)
\]
for
\begin{equation*}
\left(
\begin{matrix}
a & b \\
c & d \\
\end{matrix} \right) \in {\rm GL}_{2}({\BC})
\end{equation*}
define identical curves in ${\BC\BP}^{3}$. We can fix the ${\rm GL}_{2}({\BC})$ gauge by choosing the parameter $\ze$ so that
\begin{equation*}
{\bX}({\ze}) = ( {\ze} , 1 , X + {\ze} v_{X}, Y + {\ze} v_{Y} ).
\end{equation*}
The requirement that the curve lands in $\CalS_{6} \subset {\BC\BP}^{3}$
reads as
\begin{equation*}
{\Ga}_{3} \left( {\ze} , 1 , X + {\ze} v_{X}, Y + {\ze} v_{Y} \right) =
\sum_{i=0}^{3} {\ze}^{i} {\Xi}_{i}(X, Y ; v_{X}, v_{Y}) \equiv 0,
\end{equation*}
which is a system of $4$ equations
\[
{\Xi}_{i} (X, Y; v_{X}, v_{Y} ) = 0, \qquad i = 0, \dots , 3
\]
on $4$ unknowns $X$, $Y$, $v_{X}$, $v_{Y}$:
\begin{gather*}
\Xi_0 = -Y^2+4 X^3 -g_2 X-g_3,\\
\Xi_1= -g_2 v_X+p_6 X Y+p_1 X+p_0+12 X^2 v_X-2 Y v_Y,\\
\Xi_2= p_6 Y v_X+p_1 v_X+p_6 X v_Y+p_3 X+p_5 Y+p_2+12 X v_X^2-v_Y^2,\\
\Xi_3= p_6 v_X v_Y+p_3 v_X+p_5 v_Y+p_4+4 v_X^3.
\end{gather*}
The equation $\Xi_0 =0$ in the above system is the equation of the elliptic
curve $\ec$. To find the equation of the spectral cover associated
with the vector bundle $\CalU|_{\ec}$ in the $\mathbf{27}$
representation we can express $v_Y$ from the equation $\Xi_1 = 0$, then plug
it into the remaining equations $\Xi_2 = 0$ and $\Xi_3 = 0$, compute the resultant
of these two polynomials with respect to the variable $v_X$, reduce
modulo the equation $\Xi_0 = 0$ defining the elliptic
curve $\ec$, arriving at
\begin{equation}
\label{eq:DelPezzo-E6curve}
 C^{E_6}(X,Y; g_2,g_3, p_0,\dots, p_6) =-4 Y^4
 \mathrm{res}_{v_{X}} ( \Xi_2|_{v_Y: \Xi_1 = 0},
 \Xi_3|_{v_Y: \Xi_1 = 0}) \mod \Xi_0.
\end{equation}
The resultant $C^{E_6}(X,Y; g_2,g_3, p_0,\dots, p_6)$ is a polynomial in $X$, $Y$ with polynomial coefficients in $g_2,
g_3, p_{0}, \dots, p_{6}$ of the form
\begin{gather*}
 C^{E_6}(X,Y; g_2,g_3,p_0,\dots,p_6) = \big(p_0^6+\cdots\big) + \big(6 p_0^5 p_1 +
 \cdots\big) X + \cdots + \bigl(-256 p_3^3 + \cdots \bigr) X^{12} \\
\hphantom{C^{E_6}(X,Y; g_2,g_3,p_0,\dots,p_6) =}{} + \big(12 g_3 p_0^4 p_5 + \cdots\big) Y + \big(32 g_3 p_0^4 p_5 +\cdots\big) X Y +
 \cdots \\
\hphantom{C^{E_6}(X,Y; g_2,g_3,p_0,\dots,p_6) =}{} + \bigl(-256 p_5^3 + \cdots\bigr) X^{12} Y.
\end{gather*}
(A short \verb|Mathematica| version of this formula
is given in Appendix~\ref{sec: E6-delPezzo-code}.)
Now let us imagine having a family ${\CalU}$ of the $E_6$-bundles on
$E$.

In our solution the vacuum $u$ of the gauge theory is identified with the degree $N$ quasimap:
\begin{align*}
p\colon \ \Cpx \to \Bun_{E_6(\BC)}(\ec) \simeq \WP^{1,1,1,2,2,2,3}
\end{align*}
given by the polynomials $p_i(x)$ of degree $N a_i$
\begin{equation*}
 p_i = p_i(x), \qquad i = 0,\dots, 6.
\end{equation*}
Together with the equation of the
Weierstra{\ss} cubic $\Xi_0(X,Y,g_2,g_3) = 0$, the equation
\begin{equation*}
 C^{E_6}(X,Y; g_2,g_3, p_0(x),\dots, p_6(x)) = 0
\end{equation*}
defines the Seiberg--Witten curve of the affine $E_6$ theory as an
algebraic curve. Given that the degree of $X$ on $\ec$ is $2$ and
the degree $Y$ on $\ec$ is $3$
the polynomial $C^{E_6}$ is of degree $27$, i.e., the equation
 $C^{E_6} = 0$ defines $27$ points on the elliptic curve $\ec$.

The top degree coefficients of the polynomials $p_i(x)$ are determined
explicitly in terms of the coupling
constants $\qe_i$. Indeed, the $E_6$ characters, or more conveniently in
the present case, the $E_6$ theta-functions, $c_i(\y_0, \mathbf{t}, \qe)$
as set in \eqref{eq:c-def}, define a set of projective coordinates
 on $\WP^{1,1,1,2,2,2,3}$ (which differs slightly from the set $({\hat\chi}_{i})_{i = 0}^{6}$):
 \begin{equation*}
 (c_0:c_1:c_6:c_2:c_3:c_5:c_4).
 \end{equation*}
In these coordinates the solution of the theory has the canonical form \eqref{eq:hchi}
\begin{equation*}
 c_i(\y_0,\mathbf{t},\qe) = \Bigg( \prod_{j=0}^{r} \qe_{j}^{-\lambda_{i}(\lambda_{j}^{\vee})} \Bigg) T_i(x).
\end{equation*}
The ``del Pezzo projective coordinates'' $(p_i)_{i=0}^{6}$
 are related to the theta-function coordinates $(c_i)$ on $\Bun_{E_6(\BC)}^{ss}(\ec)$
by a polynomial map of the form
 \begin{equation*}
 p_{i} = \sum_{j_1 \leq j_2 \leq j_3 \dots} M_{i, \{j_1,j_2,j_3, \dots\}}(\qe)
 c_{j_1} c_{j_2} c_{j_3} \dots,
 \end{equation*}
where $M_{i,\{j_1,j_2,\dots\}}(\qe)$ is certain modular
transformation matrix. This matrix can be explicitly computed
by comparing the spectral curve \eqref{eq:DelPezzo-E6curve}
and the $\hat A_{26}$ spectral curve \eqref{eq:Arhat-curvei} specialized to the
 the embedding $\hat E_{6} \subset \hat A_{26}$ by fundamental
 representation. The coefficients $ M_{i, \{j_1,j_2,j_3, \dots\}}$ are
 modular forms for modular group $\Gamma(6)$ with a certain
 modular weights that can computed by
 observation that the weights of variables $(X_0, X_1, X_2, X_3)$ under
 the modular transformation $\tau \to -\tau^{-1}$ on $\ec(\qe)$ for
 $\qe = \exp(2 \pi \ii \tau)$ are
\begin{equation*}
 \begin{pmatrix}
X_0 & X_1 & X_2 & X_3 \\
6 & 1 & 2 & 3
 \end{pmatrix}.
\end{equation*}
This implies that the modular weights of $p_{i}$ are
\begin{equation*}
 \begin{pmatrix}
& p_0 & p_1 & p_6 & p_2 & p_3 & p_5 & p_4\\
 & 0 & -2 & -5 & -6 & -8 & -9 & -12
 \end{pmatrix}.
\end{equation*}
The $(c_i)$ have modular weight $3$ because they are rank $6$ lattice
theta-functions. From this assignment of weights one finds the modular
weights of all coefficients $M_{i,\{j_1, \dots, j_{k}\}}(\qe)$;
 for example~$M_{4,\{4\}}$ has modular weight $15$. The space of modular
 forms for $\Gamma(N)$ of a given weight~$k$ is a~finite-dimensional
 vector space. (For any integer $k > 0 $ the dimension is $k+1$ for
 $\Gamma(3)$ and~$6k$ for~$\Gamma(6)$). Matching a finite number of the
 coefficients in the $\qe$ expansion one finds explicitly the modular
 coefficients $M_{i,\{j_1,\dots \}}(\qe)$, see Appendix~\ref{sec:E6-modular-matrix}.

\subsection[Del Pezzo and $E_7$ bundles]{Del Pezzo and $\boldsymbol{E_7}$ bundles}
The story for the $E_7$-bundles is similar. There is a family of del Pezzo surfaces $\CalS_{7} \subset {\BW}{{\BP}}^{1,1,1,2}$, described by the
degree $4$ equation
\begin{equation*}
{\Ga}_{4}(X_{0}, X_{1}, X_{2}, X_{3} ) = \sum_{i=0}^{4} X_{0}^{i} {\CalG}_{4-i}(X_{1}, X_{2}, X_{3})
\end{equation*}
with
\begin{gather*}
{\CalG}_{4}(X_{1}, X_{2}, X_{3}) =-X_3^2 +4 X_1 X_2^3 -g_2 X_1^3 X_2 -g_3 X_1^4\big(
 X_{3}^{2} - 4 X_{1} X_{2}^{3} + g_{2} X_{2} X_{1}^{3} + g_{3} X_{1}^{4} \big),
\\
 {\CalG}_{3}(X_{1}, X_{2}, X_{3}) = \pe_0 X_1^3+\pe_7 X_1^2 X_2\big(a_{1} X_{1}^{3} + a_{2} X_{2}X_{1}^{2}\big), \\
 {\CalG}_{2}(X_{1}, X_{2}, X_{3}) = \pe_1 X_1^2+\pe_2 X_1 X_2+\pe_6 X_2^2 \big(b_{1} X_{1}^{2} + b_{2} X_{1}X_{2} + b_{3} X_{2}^{2} \big),\\
 {\CalG}_{1} (X_{1}, X_{2}, X_{3}) = \pe_3 X_1+\pe_5 X_2(c_{1}X_{1} +
c_{2} X_{2}) ,\\
 {\CalG}_{0} = \pe_4.
\end{gather*}
Again, the divisor $X_{0} =0$ is the elliptic curve ${\ec}$, which is realized as
a zero locus of the degree $4$ polynomial equation
${\CalG}_{4}(X_{1}, X_{2}, X_{3}) =0$ in ${\BW}{\BP}^{1,1,2}$.

The isomorphism classes of $\CalS_{7}$, containing the fixed elliptic curve ${\ec}$, are in one-to-one correspondence with the points
\begin{equation*}
[p] = (\pe_{0}: \pe_{7} : \pe_{1} : \pe_{2} : \pe_{5} : \pe_{3} : \pe_{5} : \pe_4 ) \in {\WP}^{1,1,2,2,2,3,3,4}.
\end{equation*}
Again, we study the ``(-1)-curves'', which in a particular gauge look like
\begin{equation*}
{\bx}({\ze}) = \bigg( {\ze} , 1 , X + {\ze} v_{X} , Y + {\ze} v_{Y} + \frac 12 {\ze}^{2} w_{Y}\bigg),
\end{equation*}
where $(X,Y; v_{X}, v_{Y}, w_{Y})$ obey a system of $5$ equations
${\Xi}_{i} (X,Y; v_{X}, v_{Y}, w_{Y}) = 0$, $i = 0, \dots, 4$:
\begin{equation*}
 {\Ga}_{4} \bigg( {\ze} , 1 , X + {\ze} v_{X} , Y + {\ze} v_{Y} + \frac 12
 {\ze}^{2} w_{Y}\bigg) =
\sum_{i=0}^{4} {\ze}^{i} {\Xi}_{i}( X, Y; v_{X}, v_{Y}, w_{Y}),
\end{equation*}
where
\begin{equation*}
\begin{aligned}
&\Xi_0 = -Y^2 + 4 X^3- g_2 Y -g_3, \\
&\Xi_1 = p_0+X p_7+12 X^2 v_X-g_2 v_X-2 Y v_Y,\\
&\Xi_2 = p_1+X p_2+X^2 p_6+p_7 v_X+12 X v_X^2-v_Y^2-Y w_Y,\\
&\Xi_3 = p_3+X p_5+p_2 v_X+2 X p_6 v_X+4 v_X^3-v_Y w_Y,\\
&\Xi_4 = p_4+p_5 v_X+p_6 v_X^2-\frac{w_Y^2}{4}.\\
\end{aligned}
\end{equation*}
We proceed similarly to the $E_6$ case: we solve for $v_Y$ and $w_Y$ from
the equations $\Xi_1$ and $\Xi_2$, plug the solution into the polynomial
$\Xi_3$ and $\Xi_4$ and compute the resultant
\begin{gather*}
 C^{E_7}(X,Y; g_2,g_3, p_0,\dots, p_7) = -2^{10} Y^{18}
 \mathrm{res}_{v_{X}}( \Xi_3|_{v_Y,w_Y: \Xi_{1,2} = 0},
 \Xi_4|_{v_Y,w_Y: \Xi_{1,2} = 0})\mod \Xi_0,
\end{gather*}
which has the structure of the degree $28$ polynomial in $X$ with the
coefficients polynomial in $(p_{0},\dots,p_{7})$ of total degree 12:
\begin{gather}
 C^{E_7}(X,Y; g_2,g_3, p_0,\dots, p_7) =\big(p_0^{12}+24 g_3 p_0^{10} p_1 +
 \cdots\big) + \big(24 g_2 p_0^{10} p_1 + \cdots\big) X + \cdots \nonumber\\
\hphantom{C^{E_7}(X,Y; g_2,g_3, p_0,\dots, p_7) =}{} + \big(2^{16} p_5^4- 2^{19} p_4 p_5^2 p_6+ 2^{20} p_4^2 p_6^2\big) X^{28}.
\label{eq:E7-delPezzo}
\end{gather}
(A short \verb|Mathematica| version of this formula
is given in Appendix~\ref{sec: E7-delPezzo-code}.)
The polynomial $C^{E_7}(X,Y; g_2,g_3, p_0(x),\dots, p_7(x))$ together with the
Weierstra{\ss} cubic $\Xi_0(X,Y; g_2,g_3)$ defines the algebraic
Seiberg--Witten curve for $E_7$ quiver theory.
Since on the elliptic curve $\ec$ the degree of $X$ is $2$,
at each $x \in \Cx$ the spectral curve \eqref{eq:E7-delPezzo} defines
$56 = 2 \times 28$ points on $\ec$ encoding the vector bundle in the
$\mathbf{56}$ representation of $E_{7}$.

The relation between the del Pezzo parametrization $(p_i)$ and the theta-function
parametrization $(c_i)$ of $\Bun_{E_7(\BC)}^{ss}(\ec)$
 can be in principle written in terms of a certain modular matrix $M(\qe)$, as in the
 the $E_{6}$ theory (see Appendix~\ref{sec:E6-modular-matrix}). We do not record this
 transformation in this work.

\subsection[Del Pezzo and $E_8$ bundles]{Del Pezzo and $\boldsymbol{E_8}$ bundles}
To get an effective description of $E_{8}$-bundles
we study the family of del Pezzo surfaces
${\CalS}_{8} \subset {\BW}{\BP}^{1,1,2,3}$:
\begin{gather*}
 {\Gamma}_{6}(X_{0}, X_{1}, X_{2}, X_{3}) =
- X_{3}^{2} + 4 X_{2}^{3} - X_{2}
G_{2}(X_{0}, X_{1}) - G_{3}(X_{0}, X_{1}) = 0 ,\\
 G_{2} (X_{0}, X_{1}) = g_{2} X_{1}^{4} +
\sum_{j=1}^{4} a_{j} X_{0}^{j}X_{1}^{4-j}, \\
 G_{3}(X_{0}, X_{1}) = g_{3} X_{1}^{6} +
\sum_{j=2}^{6}
b_{j} X_{0}^{j} X_{1}^{6-j} .
\end{gather*}
The isomorphism classes of the del Pezzo surfaces
${\CalS}_{8}$ containing the fixed elliptic curve
\[
Y^2 = 4 X^3 - g_{2} X - g_{3}
\]
are parametrized by
\begin{equation*}
[s] = (a_{1}: a_{2}: b_{2}: a_{3}: b_{3}: a_{4}: b_{4}: b_{5}: b_{6}) \in
{\WP}^{1,2,2,3,3,4,4,5,6}.
\end{equation*}
The ``$-1$''-curves in ${\CalS}_{8}$ are described
by the parametrizations ${\bx}({\ze}) = \big({\ze}, 1, X + {\ze} v_{X}
+ \frac 12 {\ze}^{2} w_{X},
Y + {\ze} v_{Y} + \frac 12 {\ze}^{2} w_{Y} + \frac 16
{\ze}^{3} u_{Y} \big)$,
 where $(X,Y, v_{X}, v_{Y}, w_{X}, w_{Y}, u_{Y})$
to be found from the equations
\begin{equation*}
{\Xi}_{i}(Y,Y, v_{X}, v_{Y}, w_{X}, w_{Y}, u_{Y}) = 0, \qquad i = 0, \dots , 6,
\end{equation*}
where
\begin{gather*}
{\Gamma}_{6}\left({\ze}, 1, X + {\ze} v_{X}
+ \frac 12 {\ze}^{2} w_{X},
Y + {\ze} v_{Y} + \frac 12 {\ze}^{2} w_{Y} + \frac 16
{\ze}^{3} u_{Y} \right) \\
\qquad = \sum_{i=0}^{6} {\ze}^{i}
{\Xi}_{i}(X,Y, v_{X}, v_{Y}, w_{X}, w_{Y}, u_{Y}).
\end{gather*}

To find explicitly the equation of affine $E_{8}$ spectral curve, one shall proceed in
spirit similarly to the $E_{6}$, $E_{7}$ cases considered above. However
the explicit computation becomes much more tedious as the minimal
representation of $E_8$ is $\bf{248}$, and the expected $x$-degree of
the curve is~$60 N$ (see below). We leave this task for future
investigation.

\section{The integrable systems of monopoles and instantons}\label{se:integrable}

As we reviewed above, the ${\CalN}=2$ gauge theory compactified on a circle
${\BS}^{1}$ becomes, at low energy, the ${\CalN}=4$ supersymmetric sigma model
with the hyperk\"abler target space $\pv$. The triplet of complex structures on $\pv$ is in correspondence with the choices of a supercharge ${\sc}$, which is nilpotent up to an infinitesimal translation along ${\BS}^{1}$. The one supercharge which is nilpotent even in the decompactified theory (it corresponds to the topological supercharge of the Donaldson theory, for pure ${\CalN}=2$ super-Yang--Mills theory) corresponds to the complex structure $\bf I$. In this complex structure ${\pv}$ has the structure of an algebraic integrable
system $({\pv}, {\Omega}, h)$:
\begin{equation*}
h\colon \ {\pv} \longrightarrow {\mv}, \qquad {\Omega} \vert_{h^{-1}(u)} = 0, \qquad u \in {\mv}.
\end{equation*}
The $\bf I$-holomorphic $(2,0)$ form ${\Omega}$ is the form which we previously denoted by ${\Omega}_{\bf I}$.

We shall now describe these systems for the class I, II and II* theories we studied so far.
For some theories several presentations of the same integrable system are possible.

In all cases we study the phase spaces $\pv$ have parameters corresponding to the masses $m$ of the matter field in the gauge theory. The cohomology class of $[{\Omega}]$ is linear in $m$. An explanation of this fact in the symplectic geometry is the
existence of a ``larger'' symplectic manifold $\pve$ with the holomorphic Hamiltonian torus
${\BT} = ({\BC}^{\times})^{M}$ action, whose holomorphic symplectic quotient of $\pve$ at some level $m$ of the moment map produces $\pv$.

The explanation in gauge theory is the three-dimensional mirror symmetry. Our phase space~$\pv$ is the Coulomb branch of the three-dimensional ${\CalN}=4$ gauge theory, obtained by the~${\BS}^{1}$ compactification of the four-dimensional ${\CalN}=2$ theory. The masses
of the matter fields are the vacuum expectation values of the scalars in the vector multiplet.
Under the three-dimensional mirror symmetry~\cite{Intriligator:1996ex} these are
exchanged with the Fayet--Iliopoulos terms, which are the levels of the three moment maps in the hyperk\"ahler quotient construction~\cite{Hitchin:1986ea} of the Higgs branch of the mirror theory.

It is amusing to identify $\pve$ and the action of the torus in the examples below. We shall treat the case of the class II theories in some detail, leaving other examples to the interested reader.

\subsection{Periodic Monopoles and the phase space of class I theories}\label{se:monopole}

We shall now demonstrate that for the class I theories the
phase space $\pv$ is the moduli space of the charge ${\bv}$
$G$-monopoles on ${\BR}^{2}\times {\BS}^{1}$
with Dirac singularities, whose location and
the embedding of the Dirac $U(1)$-monopoles into $G$ is parametrized by
${\bw}$ and the masses $m_{i, {\mathfrak {f}}}$.

Let us discuss the monopole moduli space in more detail.

The ordinary $G$-monopoles are the solutions of
Bogomolny equation on ${\BR}^{3}$
\begin{equation}
D_{A}{\phi} + \star F_{A} = 0
\label{eq:bog}
\end{equation}
with finite $L^2$-energy
\begin{equation}
{\CalE}(A, {\phi}) = \int_{{\BR}^{3}} \langle F_{A}, \star F_{A} \rangle + \langle D_{A}{\phi}, \star D_{A}{\phi} \rangle.
\label{eq:monene}
\end{equation}
One shows that as ${\vec x} \to \infty$, the conjugacy class of $\phi ({\vec x})$ approaches
a fixed value. Equivalently,
${\phi}({\vec x} ) \longrightarrow g^{-1}({\vec x} ) {\phi}_{\infty} g({\vec x} )$, for some fixed ${\phi}_{\infty} \in \h_{\BR}$, the Cartan subalgebra of the maximal compact subgroup $G$. Actually, ${\phi}_{\infty} \in \h_{\BR} / W({\g})$, but, since ${\BS}^{2}_{\infty}$
is simply connected, one can choose a uniform representative
${\phi}_{\infty} \in \h_{\BR}$. This lift from
$\h_{\BR}/W({\g})$ to $\h_{\BR}$ is going to be trickier in the case of periodic monopoles we shall study below.

Suppose $\phi_{\infty}$ is generic, i.e., the only gauge transformations which commute with it belong to the normalizer $N(T)$ of a maximal torus $T$.
The restriction of $\phi$ onto a two-sphere ${\BS}^{2}_{\infty}$ of a very large radius
defines a~map
\begin{equation*}
{\varphi}\colon \ {\BS}^{2}_{\infty} \longrightarrow G/T
\end{equation*}
and a $T$-subbundle $\BT$ of the trivial $G$-bundle $P = G \times {\BS}^{2}_{\infty}$. The latter is characterized by its Chern classes, which can be also identified with the class $[{\varphi}]$ of ${\varphi}$ in
\[
{\pi}_{2} ( G/T ) \approx {\pi}_{1}(T) \approx {\rl}^{\vee},
\]
also known as the magnetic charges of the monopole solution.
The magnetic charge can also be read off the solution of (\ref{eq:bog}) by projecting the curvature $F_{A}$
to the Cartan subalgebra $\h$ defined by $\phi\vert_{{\BS}^2_{\infty}}$, and by taking the corresponding integrals
\begin{equation*}
{\bf m}_{\varphi} = \frac 1{2\pi \ii} \int_{{\BS}^{2}_{\infty}} F_{A}^{\h}.
\end{equation*}
Now let us compactify one of the spatial directions, i.e., replace ${\BR}^{3}$ by $M^{3} = {\BS}^{1} \times {\BR}^{2}$. Let $\psi \in [ 0, 2\pi )$ be the angular coordinate on
$\BS^1$ and let $x = x_{1} + \ii x_{2}$ be the complex coordinate on~$\BR^2$. Let us
normalize the metric on $M^3$ so that the circumference of $\BS^1$ is
equal to one.

Consider the complex connection in the $\BS^1$ direction (we use the physical convention where $A$ is represented by Hermitian matrices, i.e., by a real-valued one-form for the $U(1)$ gauge group):
\begin{equation*}
{\nabla} = {\pa}_{\psi} + \ii A_{\psi} - {\phi}.
\end{equation*}
The equation~(\ref{eq:bog}) implies that
the ${\xb}$-variation of $\nabla$ is an infinitesimal gauge transformation
\begin{equation*}
{\bar\pa}_{\xb} {\nabla} + [A_{\xb}, {\nabla}] = 0.
\end{equation*}
Therefore, the conjugacy class of the holonomy $g(x,{\xb})$ of $\nabla$ around $\BS^1$
varies holomorphically with $x$, and, in the gauge where $g(x,{\xb}) \in \bT$, it
is locally holomorphic
\begin{equation}
[g(x)] = \bigg[ P \exp \bigg( \oint_{0}^{2\pi} \ii \, {\rm d}\psi ( A_{\psi} ({\psi}, x, {\xb}) + \ii {\phi} ({\psi}, x, {\xb} ) )\bigg) \bigg] \in {\Bg}.
\label{eq:mono}
\end{equation}
As we shall clarify later, when $x \to \infty$,
\[
[g(x)] \to \Bigg[ \prod_{i=1}^{r} {\qe}_{i}^{-{\lam}_{i}^{\vee}} \Bigg] = b_{\infty} \in {\Bgad}.
\]
One is left with the quasimap $u\colon {\Cpx} \to {\Bgad}$.
It is instructive to calculate (\ref{eq:mono}) for the Dirac monopole on $M^3$.

Recall that the Dirac monopole at ${\vec p} \in {\BR}^{3}$ is the connection in the $U(1)$ bundle
over ${\BR}^{3} \backslash {\vec p}$, which is a pullback of the constant curvature connection on the Hopf bundle over ${\BS}^2$ via the projection map
\[
{\pi}_{\vec p}\colon \ {\BR}^{3} \backslash {\vec p} \longrightarrow {\BS}^{2}, \qquad {\pi}_{\vec p}({\vec r}) = \frac{{\vec r}-{\vec p}}{| {\vec r}-{\vec p} |}.
\]
The corresponding curvature two-form $F$ is given by
\[
F = 2\pi \ii {\pi}_{\vec p}^{*} {\varpi}_{2} = \frac{\ii}{2} \frac{ ({\vec x} - {\vec p}) \cdot {\rm d}{\vec x} \times {\rm d}{\vec x} } { | {\vec x} - {\vec p} |^{3}},
\]
where
\[
\int_{{\BS}^{2}} {\varpi}_{2} = 1.
\]
The fact that up to the $|{\vec r} - {\vec p}|^2$ rescaling the two-form ${\varpi}_2$ coincides with the volume form on~$\BS^2$ obtained from the flat metric on ${\BR}^{3}$ implies that
$F$ solves Maxwell equations in ${\BR}^{3} \backslash {\vec p}$ and moreover there is a
magnetic potential $\phi$, such that
\[
{\rm d}{\phi} + \star_{3} F = 0.
\]
Moreover, if $\phi$ is normalized to approach zero at infinity, then
\begin{equation}
{\phi} = - \frac{1}{2 |{\vec r} - {\vec p}|}.
\label{eq:basicdirmo}
\end{equation}
The periodic Dirac monopole, i.e., the solution of Maxwell equations on $M^3 \backslash p = ({\psi}_0, x_0, {\xb}_0)$ can be obtained from the basic monopole in $\BR^3$ by taking the superposition of the fields of an infinite periodic array of monopoles, living on the universal cover $\widetilde{M^3 \backslash p} = {\BR}^{3} \backslash ( {\psi}_0 + 2\pi {\BZ}, x_0, {\xb}_0 )$.
The magnetic potential
is given by the regularized sum of potentials (\ref{eq:basicdirmo})
\begin{gather*}
 {\phi}({\psi}, x, {\xb}; p) = {\phi}_{\infty} + \frac{\log (\pi) - {\gamma}}{2\pi} +\sum_{n \in {\BZ}} \bigg( \varphi ({\psi}- {\psi}_{0} - 2\pi n, x - x_{0}) + \frac{1- {\delta}_{n,0}}{4\pi |n|} \bigg) ,\nonumber\\
 {\varphi} ({\psi}, x ) = - \frac{1}{2 \sqrt{{\psi}^{2} + x{\xb}}}.
\end{gather*}
We calculate
\begin{equation*}
\frac 1{2\pi}\int_{0}^{2\pi} {\phi} ({\psi}, x, {\xb}; p) {\rm d}{\psi} = {\phi}_{\infty} + \frac 1{2\pi}\log | x - x_0 |.
\end{equation*}
The calculation of
\[
\int_{0}^{2\pi} A_{\psi} ({\psi}, x, {\xb}; p)\, {\rm d}{\psi}
\]
is a bit tricky. Fortunately, its derivative is easy to compute
\begin{align*}
{\rm d} \int_{0}^{2\pi} A_{\psi} ({\psi}, x, {\xb}; p) \, {\rm d}{\psi} &{}= \oint_{{\BS}^{1}} F = \frac{\ii}{4} \sum_{n \in {\BZ}} \int_{0}^{2\pi}\, {\rm d}\psi \wedge \frac{ ( x - x_0 ) {\rm d}{\xb} - ( {\xb} - {\xb}_0 ) {\rm d}x }{\big( |x - x_{0} |^2 + ( {\psi} - {\psi}_{0} + 2\pi n )^2 \big)^{3/2}} \\
 &{}= \frac{\ii}{2}\, {\rm d} \log \left( \frac{x-x_0}{{\xb}-{\xb}_{0}} \right).
\end{align*}
Thus
\[
{\rm d} \oint ( A_{\psi} + \ii {\phi} ){\rm d}\psi = \ii \, {\rm d} \log (x-x_0 )
\]
and the monodromy is equal to
\begin{equation*}
g(x) = (x-x_0)^{-1}
\end{equation*}
up to some multiplicative constant.
Now, if we have a superposition of several Dirac monopoles, in the theory with the gauge group $T$, with the monopoles of the type $i$, i.e., corresponding to the
coweight ${\lam}_{i}^{\vee} \in {\Hom}(U(1), T)$ located at the points
$({\psi}_{i, {\fe}}, m_{i, \fe}, {\bar m}_{i, \fe})$, then the monodromy of the corresponding complexified connection $A + \ii {\phi}\, {\rm d}\psi$ is equal to
\begin{equation*}
g(x) \propto \prod_{i=1}^{r} \prod_{{\fe}=1}^{{\bw}_{i}} ( x - m_{i, \fe})^{-{\lam}^{\vee}_{i}}.
\end{equation*}
Now let us consider the nonabelian Bogomolny equation on $M^3$. Instead of solving the equation~(\ref{eq:bog}) modulo $G$-gauge transformations, let us solve two out of three equations in (\ref{eq:bog}), namely the equation
\begin{equation}
[ D_{\xb} , {\nabla} ] = 0
\label{eq:cmplx}
\end{equation}
modulo the action of the group ${\mathcal G}^{\BC}$ of $\bG$-valued (complex) gauge transformations. In fact, (\ref{eq:cmplx}) can be viewed as the complex moment map for ${\mathcal G}^{\BC}$, acting on the space of $(A, {\Phi})$, endowed with the ${\mathcal G}^{\BC}$-invariant holomorphic symplectic form:
\begin{equation}
{\Omega} = \frac 1{2\pi} \int_{M^{3}} \langle {\delta} {\nabla} \wedge {\delta} A_{\xb} \rangle \, {\rm d}\psi {\rm d}x{\rm d}{\xb}.
\label{eq:cmplom}
\end{equation}
Let us now try to analyze the solutions to (\ref{eq:cmplx}) in some domain $D \times {\BS}^{1} \subset M^3$ over $D \subset \Cx$.
We fix the $\bG$-gauge where $A_{\psi} + \ii \phi = {\xi} (x, {\xb})$, ${\pa}_{\psi}{\xi} = 0$. This gauge leaves some residual gauge freedom. Passing to
\[
g(x) = \exp (2\pi \ii \xi (x))
\]
partially reduces the residual gauge invariance. The equation~\eqref{eq:cmplx} implies that ${\bar\pa}_{\xb} {\xi} = 0$, and $A_{\xb} = a_{\xb} \in \h$. We then proceed with constructing the cameral curve ${\CalC}_{u}$ which is the union of the $W({\g})$-orbits of $g(x)$ over all $x \in {\Cpx} = {\Cx} \cup \{ \infty \}$. The fiber ${\mathscr A}_{u}$ of the projection $h\colon {\pv} \to {\mv}$ is the space of $W({\g})$-equivariant $\bT$-bundles over
$\CalC_{u}$ of fixed multi-degree. We shall discuss in more detail
the analogous situation for the class II theories in the next section.

The asymptotics of the solution to (\ref{eq:bog}) is characterized by a vector of magnetic charges. Namely, over ${\BT}^{2}_{\infty} = {\BS}^{1} \times {\BS}^{1}_{\infty}$ where ${\BS}^{1}_{\infty}$ is a large radius circle in ${\BR}^{2}$, the gauge group $G$ is broken down to $T$. The gauge bundle is therefore characterized by the vector of the first Chern classes:%
\begin{equation*}
{\bf m} \in H^{2} \big( {\BT}^{2}_{\infty} , {\pi}_{1}({\bT})\big) = {\rl}^{\vee}.
\end{equation*}
One can compute ${\bf m}$ by analyzing the behavior of the
conjugacy class of the holonomy $g(x)$ of the complexified connection
${\nabla}$. One can show that the finite energy (\ref{eq:monene}) solutions with non-trivial magnetic charge do not exist. Indeed, macroscopically the system looks two-dimensional, and asymptotically it looks like a charged vortex, whose energy diverges logarithmically
at large distances.

However, infinite $3$-dimensional energy solutions may correspond to the finite tension higher-dimensional objects. For example, the noncommutative $U(1)$ monopole describes a finite tension string, which is attached to the worldvolume of the gauge theory~\cite{Gross:2000wc}. Similarly, the infinite energy periodic monopole solutions have interpretation in the higher-dimensional theory, e.g., in the brane realization~\cite{Witten:1997sc} of the pure ${\CalN}=2$ ${\rm SU}(N)$
gauge theory in four dimensions~\cite{Cherkis:2000cj}.

One can actually make the energy finite in the infrared by allowing
point-like singularities in~$M^3$. The idea is to screen the asymptotic magnetic charge
of the non-abelian solution by the opposite charge of the Dirac monopole singularities.

Let us study the general situation. Suppose ${\gamma} \subset
{\Cx}$ is a closed contour. For each point $x \in \gamma$ compute the
holonomy $g(x, {\xb})$ of the complexified connection $\nabla$, e.g., starting at the
point ${\psi}=0$ on the fiber ${\BS}^{1}$. It is a functional of
the gauge field $A$ and the Higgs field $\phi$: $(A, {\phi}) \mapsto g(x, {\xb})$.
The gauge transformed $(A, \phi)$ leads to the similarity-transformed
function $g(x, {\xb})$: $(A^{h}, {\phi}^{h}) \mapsto h^{-1}(0 , x, {\xb})
g(x, {\xb}) h(0, x, {\xb})$.
We have a well-defined map
\begin{equation*}
[g_{\gamma} ] \colon \ {\gamma} \longrightarrow,
B({\g}),
\end{equation*}
which is a restriction on $\gamma$ of the holomorphic (cf.~\eqref{eq:cmplx})
map $U\colon {\Cpx} \to {\Bgad}$, $U\colon x \mapsto [g(x)]$.

Now let ${\Xi}({\g}) \subset {\Bgad}$ denote
the set of irregular orbits of $W({\g})$ in ${\bT}/Z$. Generically the image $\Sigma_u$
of $[g(x)]$ crosses ${\Xi}({\g})$ at some isolated points,
which are the branch points of the cameral curve $\CalC_{u}$:
\[
{\Xi}_{x} = U^{-1}( {\Sigma}_{u} \cap {\Xi}({\g}) ).
\]

Let us consider the subvariety
$B({\g})^{\rm reg} = {\Bgad} \backslash {\Xi} ({\g})$. The fundamental group
${\pi}_{1}(B({\g})^{\rm reg})$ is related to Artin braid group associated
with the $\g$ root system. Let us also define ${\bT}^{\rm reg}$ to be the subvariety
in $\bT$ consisting of the regular elements, i.e., the elements
of the maximal torus whose stabilizer in $\bG$ is the maximal torus $\bT$. It
is invariant under the translations by $Z$.
We have a~map: ${\pi}_{*}\colon {\pi}_{1}( {\bT}^{\rm reg}) \to {\pi}_{1}(B({\g})^{\rm reg})$, induced
by the projection ${\pi}\colon {\bT}^{\rm reg} \to {\bT}^{\rm reg}/\left( Z \times W({\g}) \right)$.

Now, let us go back to the loop $\gamma$. We would like
to define a $\bT$-bundle over ${\BS}^{1} \times {\gamma}$, by choosing a
gauge where $A_{\psi} + \ii \phi$ is $\psi$-independent
element of the Lie algebra $\h \subset \g$. Then Bogomolny equations imply that
$A_{\xb}$ also belongs to $ \h$, assuming that $ A_{\psi} + \ii \phi$ is regular,
i.e., its stabilizer in $\bG$ is the maximal
torus $\bT$.

There is an obstruction for such a gauge being possible throughout
$\gamma$. Namely the class of $\gamma$ in ${\pi}_{1}
(B({\g})^{\rm reg})$ should lie in the image of ${\pi}_{*}$.
This is related to our solution of the four-dimensional ${\CalN}=2$
theory of the class I as follows.

For the asymptotically conformal theories, with the assignments
of dimensions ${\bv}$, ${\bw}$, we have defined a ${\bG}/Z$-valued function on $\Cx$ minus
a finite number of points:
\begin{equation}
g(x) = \prod_{i=1}^{r} {\bq}_{i}(x)^{-{\lam}^{\vee}_{i}} {\y}_{i}(x)^{{\al}_{i}^{\vee}}.
\label{eq:gofxiagain}
\end{equation}
We identify $g(x)$ in \eqref{eq:gofxiagain} with the holonomy in \eqref{eq:mono}. The factor
\[
g_{\infty}(x) = \prod_{i=1}^{r} {\bq}_{i}(x)^{-{\lam}^{\vee}_{i}}
\]
clearly corresponds to the Dirac monopoles sitting at some points $({\psi}_{i, {\mathfrak{f}}},
m_{i, {\mathfrak{f}}}, {\bar m}_{i, \mathfrak{f}})$ with the charges~${\lam}^{\vee}_{i}$. The remaining
factor has to do with the nonabelian monopoles.

Recall that the map $U\colon \Cx \backslash {\Xi}_{x} \to B({\g})$ is determined by the collection of gauge polynomials
$T_{i}(x)$, $i = 1, \dots , r$. The singular locus ${\Xi}_{x}$ of $U$
are at the zeroes and poles of the discriminant
\begin{equation*}
{\Delta}(x) = g(x)^{-2{\rho}}\prod_{{\al} \in R_{+}}\big( g(x)^{\al} -1 \big)^{2},
\end{equation*}
where $R_{+}$ is the set of positive roots of $\g$, and
\[
{\rho} = \frac 12 \sum_{{\al} \in R_{+}} {\al}.
\]
The discriminant
${\Delta}(x)$ is a rational function in $T_{i}(x)$'s and ${\bq}_{i}(x)$'s.
Now, given a loop ${\gamma}$ in ${\Cx} \backslash {\Xi}_{x}$, when can
we lift $[g(x)] \vert_{\gamma}$ to the $\bT$-valued loop?

For the simple root ${\al}_{i}$ let us denote by ${\Xi}_{x,i}$ the set
of solutions to the equation $g(x)^{{\al}_{i}} = 1$ on the physical sheet of
$\CalC_{u}$.
\begin{equation*}
{\Xi}_{x, i} = \big\{ x \mid g( x)^{{\al}_{i}} = 1 \big\}
\end{equation*}
so that{\samepage
\begin{equation*}
{\Xi}_{x} = \bigcup_{i = 1}^{r} {\Xi}_{x, i}.
\end{equation*}
Actually the points of $\Xi_{x,i}$ are the endpoints of the cuts $I_{i, {\ba}}$.}

Our claim is that for the loops ${\gamma} = A_{i{\ba}}$ which encircle the
individual cuts $I_{i , \ba}$ the class $[g(x)] \in B({\g})^{\rm reg}$
lifts to ${\bT}^{\rm reg}$, and defines a ${\bT}$-bundle over ${\BS}^{1} \times
{\gamma}$. Its characteristic class is equal to ${\al}_{i}^{\vee} \in {\rl}^{\vee}$.

Thus the magnetic monopoles which correspond to the
limit shape of the ${\CalN}=2$ theory have the Dirac monopole charges
${\bf q}_{\rm Dir} = - \sum_{i=1}^{r} {\bw}_{i} {\lam}_{i}^{\vee}$
which are distributed at the points
$({\psi}_{i , {\mathfrak{f}}}, m_{i, \mathfrak{f}}, {\bar m}_{i, {\mathfrak{f}}})$, and the nonabelian
monopole charges ${\bf q}_{\rm 'tHP} = \sum_{i=1}^{r}
{\bv}_{i} {\al}_{i}^{\vee}$
which are located over the cuts $I_{i, {\ba}}$.
The net charge at infinity is equal to
\begin{equation*}
{\bf q}_{\rm Dir} + {\bf q}_{\text{'t\,HP}} = 0
\end{equation*}
for the asymptotically conformal theories.

For the asymptotically free theories the net charge at infinity is equal to (cf.~\eqref{eq:betf}):
\begin{equation*}
{\bf q}_{\rm Dir} + {\bf q}_{\text{'t\,HP}} = - \sum_{i=1}^{r} {\be}_{i} {\lam}_{i}{^\vee}.
\end{equation*}
The fact that ${\be}_{i} \leq 0$ should follow from the positivity of energy (as it does in the $A_1$ case) but we could not find a simple proof for general $\bG$.

The relation of the monopole picture of $\pv$ to the Hitchin system picture we had in Section~\ref{subsubsec:gaudin} goes via the Nahm transform, or, since we are ultimately working only in a particular complex structure of the moduli space, via a version of Fourier--Mukai transform. The $U(k)$ monopoles on $\BR^3$
are mapped via Nahm's transform to the solutions of a one-dimensional system of Nahm's equations. The $U(k)$ monopoles on ${\BR}^{2} \times {\BS}^{1}$ are mapped via Nahm's transform to the solutions of a two-dimensional system of Hitchin's equations, with singularities. Indeed, our spectral curve in the form \eqref{eq:laxop} captures the solutions to
two, ${\bar\pa}_{A}{\Phi} = 0$, ${\pa}_{A} {\bar\Phi}$ out of three Hitchin's equations.
The remaining equation $F_{A} + [ {\Phi}, {\bar\Phi} ] = 0$ away from the punctures would fix the hyperk\"ahler metric on $\pv$. Unfortunately we are not in the position to discuss the metric on~$\pv$ as long as we stay within the realm of the four-dimensional gauge theory.
See~\cite{Braam:1988qk, Corrigan:1983sv, Hitchin:1983ay,Nahm:1981nb}.

We do need to discuss the holomorphic symplectic geometry of $\pv$. The symplectic form on $\pv$ descends from the two-form \eqref{eq:cmplom} via the Hamiltonian reduction with \eqref{eq:cmplx} being the moment map.
The textbook construction of the action-angle variables of the integrable system produces the special coordinates ${\ac}^{\CalI}$, ${\ac}_{\CalI}^{D}$
of the gauge theory. We claim this construction is equivalent to the one using the periods of the
differentials $x \, {\rm d} \log {\y}_{i}$ on the cameral curve. The essential points of the
demonstration are identical for the class I and for the class II theories.

We thus return to this question in Section~\ref{sec:double}.

Now let us study the $A_1$ case in some more detail. We wish to present yet another
 perspective on the phase space $\pv$.

Consider the product of $N$ $A_1$ surfaces ${\CalO}_{a}$, $a = 1, \dots , N$,
the complex coadjoint orbits of~${\rm SL}(2, {\BC})$. Each surface ${\CalO}_{i}$ is a quadric in ${\BC}^{3}$, given by the equation
\begin{equation*}
{\xi}^{2}_{a} + {\eta}_{a} ^{2} + {\zeta}_{a} ^{2} = s_{a}^2
\end{equation*}
with some fixed constant $s_{a} \in \BC$.
The surface $\CalO_a$ has the holomorphic symplectic form
\[
{\varpi}_{a} = \frac{{\rm d}{\xi}_{a} \wedge {\rm d}{\eta}_{a} }{\zeta_{a} },
\]
which has the period $s_{a}$ along a non-contractible two-sphere in $\CalO_a$.
The moment map for the action of ${\rm SL}(2, {\BC})$ on ${\CalO}_{a}$ is a $2\times 2$
traceless matrix. Let us extend it into a general $2 \times 2$ matrix (which corresponds
to the conformal extension of the group)
\begin{equation*}
L_{a}(x) = \left( \begin{matrix}
 x + {\xi}_{a} & {\eta}_{a} + \ii {\zeta}_{a} \\
- {\eta}_{a} + \ii {\zeta}_{a} & x - {\xi}_{a} \end{matrix}\right)
\end{equation*}
and define the ``monodromy matrix''
\begin{equation*}
L(x) = \left( \begin{matrix}
{\qe} & 0 \\
0 & 1 \end{matrix}\right) \times L_{N}(x - {\mu}_{N}) L_{N-1}(x - {\mu}_{N-1}) \cdots L_{1}(x - {\mu}_{1}).
\end{equation*}
It is actually better to work with the somewhat differently normalized ``local Lax operators''
\begin{equation}
g_{a}(x) = {\bf 1}_{2} + \frac{1}{x - s_{a}} \left( \begin{matrix}
u_{a}^{+} & v_{a}^{+} \\
v_{a}^{-} & u_{a}^{-} \end{matrix} \right),
\label{eq:loclax}
\end{equation}
where
\[
u_{a}^{\pm} = s_{a} \pm {\xi}_{a}, \qquad v_{a}^{\pm} = \ii {\zeta}_{a} \pm {\eta}_{a}
\]
obey
\begin{equation*}
u_{a}^{+}u_{a}^{-} - v_{a}^{+} v_{a}^{-} = 0
\end{equation*}
and define
\begin{equation}
g(x) = \left( \begin{matrix}
{\qe} & 0 \\
0 & 1 \end{matrix}\right) \times g_{N}(x - {\mu}_{N}) g_{N-1}(x - {\mu}_{N-1}) \cdots g_{1}(x - {\mu}_{1}).
\label{eq:gmon1}
\end{equation}
Then, cf.~\cite{Gorsky:1996hs},
\begin{equation*}
\operatorname{Det}(g(x)) = {\qe} \frac{{\bq}^{+}(x)}{{\bq}^{-}(x)},
\end{equation*}
where we defined
\[
{\bq}^{\pm}(x) = \prod_{a=1}^{N} ( x - {\mu}_{a} \pm s_{a} ).
\]
Now, define the phase space to be the complex symplectic quotient
of the product of the $A_1$ surfaces by the diagonal action of the
${\BC}^{\times}$,
\begin{equation}
{\pv} = \times_{a=1}^{N} {\CalO}_{a} // {\BC}^{\times}
\label{eq:quotor}
\end{equation}
generated by
\[
H_{1} = \sum_{a=1}^{N} {\xi}_{a}.
\]
The variety \eqref{eq:quotor}, defined by fixing the level ${\xi}$
of $H_1$ and dividing the corresponding level set $H_{1}^{-1}({\xi})$ by ${\BC}^{\times}$, carries the induced holomorphic symplectic structure.
The functions $h_{k}$ defined~as
\begin{equation*}
 \tr _{2} g(x) = (1 + {\qe} ) \bigg( 1 + \sum_{k=1}^{\infty} x^{-k} h_{k} \bigg).
\end{equation*}
Poisson-commute with respect to the induced Poisson structure.
Moreover,
\[
h_{1} = H_{1} + \frac{{\qe}-1}{{\qe} +1 } \sum_{a} s_{a}
\]
and the next $N-1$ $h_{k}$'s are independent. The rest of the expansion coefficients can be expressed in terms of the first~$N$.

We argue that the normalized Lax operator $g(x)$ is the complexified monodromy field $g(x)$ in the corresponding periodic singular monopole problem, with the monopole group $U(2)$ (the compact form of C$\bG$) and ${\mu}_a \pm s_a$ are the locations of $2N$ Dirac monopoles. {\em Of course, our methods
do not allow to establish the correspondence with monopoles outside the identification of complex
symplectic manifolds, for complex structure $I$ and $(2,0)$-form $\Omega_I$}.

It is not difficult to convince oneself that the ``local Lax operator'' \eqref{eq:loclax} is indeed the complexified monodromy of a single $U(2)$ monopole screened by two Dirac monopoles of the opposite $U(1)$ charges, located at $\pm s_{a}$.

What is amusing is that the equation~\eqref{eq:gmon1} suggests that the complexified monodromy of the charge $N$ $U(2)$ monopole screened by $2N$ Dirac monopoles factorizes as the product of
$N$ elementary monodromy matrices.

Note in passing that if we do not perform the reduction with respect to ${\BC}^{\times}$ generated by~$H_1$, i.e., work with the
$2N$-dimensional phase space $\tilde\pv$ (this is a first step towards the extended phase space $\pve$), then Sklyanin's separation of variables~\cite{Sklyanin:1987ih, Sklyanin:1995bm} identifies its open subset
${\tilde\pv}^{\circ}$ with the $N$-th symmetric product of (which is most likely~\cite{Gorsky:1999rb} resolved into the Hilbert scheme of~$N$ points on)
${\Cx} \times {\Ct}^{\times}$. Incidentally~\cite{Donaldson:1985id}, this manifold is symplectomorphic to the moduli space of regular charge $N$ ${\rm SU}(2)$ monopoles on ${\BR}^{3}$.

In recent years the connection between the gauge theories and the spin chains, and the inspired duality between the Gaudin-like integrable systems and the Heisenberg spin chain was discussed in~\cite{Chen:2011sj, Dorey:2011pa, Mironov:2012ba, Muneyuki:2011qu}, building on the earlier work in~\cite{Gerasimov:2006zt,Gerasimov:2007ap, Moore:1997dj, Nekrasov:2009zz, Nekrasov:2009ui}.

Before concluding this section, note that the masses of the bi-fundamental matter hypermultiplets are encoded in the next-to-leading terms in the asymptotics of the complexified connection $A_{\psi} + {\rm i} {\phi}$ near $x \to \infty$. We shall explain this in more detail in the similar context in Section~\ref{sec:double}.

The moduli space of singular $\bG$-monopoles with fixed conjugacy class of the monodromy of $A_{\psi} + {\rm i} {\phi}$ at $x \to \infty$, with unspecified location of Dirac
singularities of specified charges defines our extended phase space $\pve$. It is acted upon
by the torus $T \times T_{M}$, where $T_{M} \subset {\GM}$ is the maximal torus of the flavor group. The action of $T$ is via the constant gauge transformations preserving the
gauge field and the Higgs field at infinity, while $T_{M}$ acts by changing the glueing data
for the grafted Dirac monopoles. Fixing the level of the corresponding moment maps and reducing with respect to the complexification of the $T \times T_M$ action produces $\pv$.

Summarizing, we conclude with the conjecture:
\emph{the moduli space of singular $\bG$-monopoles on ${\BR}^{2} \times {\BS}^{1}$ is acted upon by the Poisson--Lie group, which is a quasi-classical limit of the Yangian~$Y({\gq})$. The deformation quantization of $\pve$ $($in the complex structure $I)$ produces the
Yangian~$Y({\g})$. Fixing the asymptotics of the complexified gauge
field at infinity as well as the locations of the Dirac singularities
would specify.}
It should be interesting to explore the holomorphic symplectic geometry of $\pv$ in all of its complex structures and find the analogue of the variety of opers (cf.~\cite{Nekrasov:2011bc, Nekrasov:2010ka}), which
in the $A_r$ case is the variety of local systems on the genus zero curve coming from the $N$-th order differential operators with regular singularities at $r+3$ punctures with the monodromies whose eigenvalues coincide with our description of the residues of the Higgs field just above the equation~\eqref{eq:specdet} (cf.~\cite{Frenkel:2003qx}).

Note that in~\cite{Gerasimov:2005qz} a connection between the representation theory of Yangians and the moduli spaces of monopoles on ${\BR}^{3}$ was found. Also, in~\cite{Atiyah:1991gd, Atiyah:1996ij} the relations between the monopole solutions and the solutions to the Yang--Baxter equations were discussed. It remains to be seen, whether any of these connections carries over to the
${\BR}^{2} \times {\BS}^{1}$ case and whether it is the one we need.

What happens if one tries to study periodic $\hat \bG$-monopoles where $\hat \bG$ is
Kac--Moody affine group? One naturally finds double periodic instantons
similar to relation between $\hat \bG$-monopoles and periodic $\bG$ instantons
\cite{Garland_1988}.

\subsection{Double-periodic instantons and class II theories}\label{sec:double}

Let us now discuss the class II theories.
Recall from Section~\ref{se:cameral2} that an elliptic curve ${\ec} = {\BC}^{\times}/{\qe}^{\BZ}$, with $\qe$ given by~\eqref{eq:quell},
is associated with the class II gauge theory.
Also recall that the gauge group is the product of special unitary groups
\[
{\Gg} = \times_{i=0}^{r} {\rm SU}(Na_{i})
\]
for some number $N$.
Using \eqref{eq:hchi}, we have identified the extended moduli space
$\mve_{N}$ of vacua of the theory \eqref{eq:mve}, $\dim_{\BC} \mve_{N} = Nh^{\vee}$, with the moduli space of degree $N$ framed
quasimaps of $\big( \Cpx, {\infty} \big) $ to the moduli space $\Bnq$ of holomorphic $\bG$-bundles on $\ec$, sending ${\infty}$ to a~particular bundle $[ {\CalP}_{\tqe}]$.
This space of quasimaps
 is a natural base of the projection from the moduli space
 \begin{equation*}
 {\pve}_{N} \approx
 \Bunss_{\bG; N}(S_{\ec})^{{\CalP}_{\tqe}}_{\infty}
 \end{equation*} of framed
 semi-stable holomorphic principal $\bG$-bundles $\CalP$, $c_{2}({\CalP}) = N$, on the surface
 $S_{\ec} \equiv {\Cpx} \times {\ec}$, where the framing is the identification
 of the restriction of $\CalP$ at the fiber at infinity
 \begin{equation*}
 {\CalP} \big\vert_{\{ \infty \} \times {\ec}} \approx {\CalP}_{\tqe}.
 \end{equation*}
 In what follows we drop the subscript $N$.

 {}The projection ${\pve} \to {\mve}$ is defined
on the open dense subset of $\pve$ by restricting the bundle ${\CalP} \in \pve$ to the
fiber ${\ec}_{x}$ and taking its equivalence class ${\bf t}(x) = [ {\CalP}\vert_{{\ec}_{x}} ]$ in $\Bnq$. This gives the desired quasimap $U\colon x \mapsto {\bf t}(x)$.
This quasimap
is a map near $x = \infty$, approaching a particular holomorphic
bundle $[{\CalP}_{\tqe}] \in \Bnq$ over $\ec_{\infty} = \{ {\infty} \}
\times {\ec}$. The reason $U$ is not, in general, a map, has to do with the usual difference
between the stability condition in two complex dimensions and the fiberwise stability condition, cf.~\cite{Losev:1999tu}.

The semi-stable framed holomorphic principal bundles
$\CalP$ on $S_{\ec}$, are in one-to-one correspondence with the $G$-instantons on ${\BR}^{2}\times {\BT}^{2}$ with the flat metric (cf.~\cite{Donaldson:1985zz}), i.e., connections
on a~principal $G$-bundle $P$ over ${\BS}^{2} \times {\BT}^{2}$,
endowed with some metric, which is conformally flat on ${\BR}^{2} \times {\BT}^{2}$,
which obey
\[
F_{A} = - \star F_{A}
\]
and have finite action,
\[
\int_{{\BR}^{2}\times {\BT}^{2}} \la F_{A} \wedge \star F_{A} \ra < \infty
\]
(as we explain later, these instantons correspond to the M2 branes and the instanton action is the tension of the stack of M2 branes)
and this forces the curvature $F_{A}$ of the $G$-gauge field tend to zero as $| {\vec x} | \to \infty$, ${\vec x} \in {\BR}^{2}$.
We fix the instanton charge:
\begin{equation*}
N = -\frac{1}{8\pi^2 h} \int_{{\BR}^{2}\times {\BT}^{2}} \la F_{A} \wedge F_{A} \ra
\end{equation*}
(remember that $\la\, ,\, \ra$ is the Killing form, which is the trace in
the adjoint representation).
The real dimension of the moduli space $\pve$ of $G$-instantons of finite
action on $\BR^{2} \times \BT^2$ with fixed framing at infinity $\BT^{2}_{\infty}$ is equal to
$4 N h $.

Actually, there is a subtlety here. The moduli space
of charge $N$ $G$-instantons on ${\BR}^{2} \times {\BT}^{2}$ may have several components. This has to do with the fact that the moduli space of flat $G$-connections
on ${\BT}^{3} = {\BS}^{1}_{\infty} \times {\BT}^{2}$, {\em which parametrize the asymptotics of
instantons on ${\BR}^{2} \times {\BT}^{2}$}, may have several components~\cite{Witten:1997bs}. We shall assume we are always in the component of the trivial connection.

Note that $\pve$ is acted upon by the maximal torus $T$ of $G$, the symmetry group
of the flat connection at infinity. This action lifts to the action of the algebraic torus
${\bT}$ on the moduli space of framed holomorphic bundles ${\pve}$.
This is entirely parallel to the action of the group~$\bG$ on the moduli space
${\mv}^{\mathrm{framed}}_{G}\big({\BR}^{4}\big)$ of
framed $G$-instantons on ${\BR}^{4}$.

A holomorphic principal $\bG$-bundle ${\CalP}$ on $S_{\ec}$ can be described using the transition functions $g_{\al\be}\colon U_{\al\be} \to {\bG}$ defined on the overlaps $U_{\al\be} = U_{\al} \cap U_{\be}$ of the open sets in the appropriate open cover $( U_{\al} )_{{\al} \in A}$ of $S_{\ec}$. The transition functions are holomorphic ${\bar\pa}g_{\al\be} = 0$, must obey the cocycle condition
$g_{\al\be} g_{\be\gamma} g_{\gamma\al} = 1$ on $U_{\al\be\gamma}
= U_{\al} \cap U_{\be} \cap U_{\gamma}$, and the cocycles differing
by the holomorphic coboundaries define equivalent bundles{\samepage
\[
g_{\al\be} \sim g_{\al} g_{\al\be} g_{\be}^{-1},
\]
where $g_{\al} \colon U_{\al} \to {\bG}$ are holomorphic $\bG$-valued
functions on the open sets $U_{\al}$ themselves.}

Another way to describe the holomorphic bundle is to introduce a connection $\nabla = {\rm d} + A$ on a smooth bundle, such that its
$(0,1)$-part is $(0,2)$-flat
\begin{equation}
{\nabla}_{\bar\pa}^{2} = 0.
\label{eq:flat}
\end{equation}
The local holomorphic sections of ${\CalP}_{\al} = {\CalP} \vert_{U_{\al}}$ are the solutions to ${\nabla}_{\bar\pa} s_{\al} = 0$. Over an intersection~$U_{\al\be}$ these solutions must differ by a holomorphic $\bG$-valued gauge transformation:
\[
s_{\al}\big({\ze}, {\bar\ze}\big) = g_{\al\be} ({\ze}) s_{\be}\big({\ze}, {\bar\ze}\big),
\]
where $\ze$ stand for local holomorphic coordinates.
This is sometimes expressed by saying that locally ${\bar A}$ is a pure $\bG$-gauge
\[
{\bar A} \big\vert_{U_{\al}} = - s_{\al}^{-1} {\bar\pa} s_{\al}.
\]
In the case at hand ${\ze} = (x, z)$, where $x \in \Cx \subset \Cpx$ and $z$ is the additive coordinate on $\ec$. The
equation~\eqref{eq:flat} reads
\begin{equation}
{\bar\pa}_{\xb} A_{\zb} - {\bar\pa}_{\zb} A_{\xb} + [ A_{\xb}, A_{\zb} ] = 0.
\label{eq:xzflat}
\end{equation}
This equation can be viewed as the complex moment map equation for the action of the group~$\CalG$ of the
$\bG$-gauge transformations on the space $\CalA$ of connections on a given smooth principal $\bG$-bundle over $S_{\ec}$, endowed with the holomorphic symplectic form
\begin{equation}
{\Omega} = \int_{S_{\ec}} {\rm d}x \wedge {\rm d}z \wedge \big\langle {\delta}{\bar A} \wedge {\delta}{\bar A} \big\rangle.
\label{eq:secdada}
\end{equation}
We have a little subtlety here. The two-form ${\rm d}x \wedge {\rm d}z$ was perfectly good on $S_{\ec}^{\circ}$ but on $S_{\ec}$ is has a pole along the divisor ${\ec}_{\infty} = \{ {\infty} \} \times {\ec}$.
We impose the condition that $A_{\zb}$ approaches a specific
value as $x\to \infty$. More specifically, recall \eqref{eq:xigauge}
that
generic $A_{\zb}$ can be $\CalG$-transformed to the normal form $A_{\zb} \to \frac{2\pi \ii}{{\tau}-{\bar\tau}}\xi \in \h$, ${\pa}_{z}{\xi} = {\pa}_{\zb} {\xi} = 0$.
We impose the boundary conditions:
\begin{equation}
A_{\zb}(x,{\xb}) \to \frac{2\pi \ii}{{\tau}-{\bar\tau}}\xi_{\infty} + o\big( |x|^{-2} \big), \qquad |x| \to \infty
\label{eq:decr}
\end{equation}
for fixed $\xi_{\infty}$, which we relate to the gauge couplings ${\qe}_{i}$, $i = 0, 1, \dots , r$, via
\begin{equation*}
{\xi}_{\infty} = -\frac1{2\pi\ii} \sum_{i=1}^{r} \log {\qe}_{i} {\lam}_{i}^{\vee}.
\end{equation*}
The limiting gauge field ${\bar\pa}_{\zb} + \frac{2\pi \ii}{{\tau}-{\bar\tau}}\xi_{\infty}$ corresponds to the holomorphic
bundle ${\CalP}_{\tqe}$ on $\ec$.
The decay rate~\eqref{eq:decr}
makes the integral \eqref{eq:secdada} convergent. One can impose weaker conditions, allowing even the $O\big(x^{-1}\big)$ (but no $x^{-1}{\xb}^{-1}$ terms!) decay, which would make \eqref{eq:secdada} convergent with the principal value prescription. In fact, these subleading
terms correspond to the bi-fundamental masses of the $\Gg$-theory. Let us interpret
them using the $\bT$-action on ${\pve}$.

The constant $\bT$-gauge transformation with the parameter ${\ve} \in \h$ acts on the $(0,1)$-gauge field~${\bar A}$ as follows:
\begin{equation}
{\delta}{\bar A} = [ {\ve}, {\bar A} ] + {\nabla}_{\bar\pa} {\eta}_{\ve},
\label{eq:infcons}
\end{equation}
where ${\eta}_{\ve}(x, {\xb}, z, {\zb})$ is the compensating gauge transformation which
sufficiently fast decays at $x \to \infty$. Contracting the vector field \eqref{eq:infcons}
with $\Omega$ in \eqref{eq:secdada} gives us a closed one-form
${\delta} {\bf m}_{\ve}$ on the space of connections obeying
\eqref{eq:xzflat}, where ${\bf m}_{\ve}$ is linear in $\ve$, ${\bf m}_{\ve}
= \la {\ve}, {\bf m} \ra$ for some ${\bf m} \in \h$
\begin{align}
{\delta} {\bf m}_{\ve} &{}= \int_{S_{\ec}} {\rm d}x \wedge {\rm d}z \wedge \la ( [ {\ve}, {\bar A} ] + {\nabla}_{\bar\pa} {\eta}_{\ve} ) {\delta}{\bar A} \ra = \int_{S_{\ec}} {\rm d}x \wedge {\rm d}z \wedge {\bar\pa} \la {\ve} , {\delta}{\bar A} \ra \nonumber\\
& = {\delta} \oint_{x = \infty} {\rm d}x \la {\ve}, \int_{{\ec}_{x}} {\rm d}^{2}z A_{\zb} \ra.
\label{eq:dme}
\end{align}
Thus the moment map ${\bf m}$
for the $\bT$ action is the residue at $x = \infty$ of the
zero mode of the $\h$-projection of the $A_{\zb}$ gauge field. The analogous statement
holds for the monopoles of Section~\ref{se:monopole}.

Now let us solve \eqref{eq:xzflat} locally over some domain $D$ in $\Cx$. We choose the gauge \eqref{eq:xigauge} over each point $x \in D$:
\begin{equation}
A_{\zb}(z, {\zb}; x, {\xb}) \longrightarrow \frac{2\pi \ii}{{\tau}-{\bar\tau}}{\xi}(x, {\xb}) \in \h.
\label{eq:zgauge}
\end{equation}
If $\infty \in D$, then, using \eqref{eq:dme},
in the gauge \eqref{eq:zgauge}
\begin{equation*}
{\xi} (x, {\xb} ) = {\xi}_{\infty} + \frac{{\bf m}}{x} + \cdots, \qquad x \to \infty,
\end{equation*}
where ${\bf m} \in \h$ is a linear function of the bi-fundamental masses
$({\bf m}_{e})_{e \in \Edg}$.

{}Recall that $\bG$ is a simple
simply-connected Lie group. Therefore, the restriction
${\CalP} \vert_{{\ec}_{x}}$ is trivial
as a smooth $\bG$-bundle. Therefore ${\bar A}$ is just a $\g$-valued $(0,1)$-form on $D \times \ec$. Now let us decompose $A_{\xb} = a_{\xb} + W_{\xb}$, with $a_{\xb} \in \h$ and $W_{\xb} \in {\h}^{\perp} \subset \g$, the orthogonal decomposition being provided by the Killing form $\langle\cdot, \cdot \rangle$.
Then \eqref{eq:xzflat} implies:
${\bar\pa}_{\xb} {\xi} (x, {\xb}) = 0$, ${\bar\pa}_{\zb}a_{\xb} = 0$, $W_{\xb} = 0$, the latter equation being valid for generic ${\xi}(x)$, corresponding to the irreducible bundles on $\ec$.
Indeed, the $F^{0,2} =0$ equation \eqref{eq:xzflat} splits in
$\g = {\h} \oplus {\h}^{\perp}$ as follows:
\begin{gather}
F^{0,2} \big\vert_{{\h}} =
{\bar\pa}_{\xb} {\xi} - {\bar\pa}_{\zb}a_{\xb} = 0, \nonumber\\
F^{0,2} \big\vert_{{\h}^{\perp}} =
{\bar\pa}_{\zb} W_{\xb} + \frac{2\pi \ii}{{\tau}- {\bar\tau}} [ {\xi}, W_{\xb} ] = 0.
\label{eq:fzt}
\end{gather}
For generic $\xi \in \h$ (irreducible ${\rm ad}({\CalP})^{\perp}$) the $\h^{\perp}$-equation \eqref{eq:fzt} does not have non-zero solutions, hence $W_{\xb} = 0$. As for the $\h$-equation \eqref{eq:fzt} in our gauge its solution is
\[
a_{\xb}(x, {\xb}, z, {\zb}) = a_{\xb}^{(0)} (x , {\xb}, z) + {\zb} {\bar\pa}_{\xb} {\xi} ( x, {\xb} ).
\]
Since on $D \times {\ec}$ the connection form $A_{\zb}\,{\rm d}{\zb} + a_{\xb}\,{\rm d}{\xb}$ is simply an
$\h$-valued $(0,1)$-form
 the component $a_{\xb}(x, {\xb}, z, {\zb})$ must be $z$-periodic. This is only possible
 if ${\bar\pa}_{\xb} {\xi} ( x, {\xb} ) = 0$, which also implies $a_{\xb}(x, {\xb}, z, {\zb})$ is
 $z$-independent.

Of course, \eqref{eq:zgauge} is not the complete gauge fixing: there remain the $z$-independent $\bT$-valued gauge transformations and the $W({\g})$-transformations \eqref{eq:reswe}, which combine into the locally $z$-constant $N({\bT})$-gauge transformations. There are also the shifts \eqref{eq:ressh} by the lattice ${\rl}^{\vee} \oplus {\tau} {\rl}^{\vee}$, generated by the $(z, {\zb})$-dependent $\bT$-valued gauge transformations with discrete, as far as the $(x, \xb)$-dependence is concerned, parameters ${\be}_{1}, {\be}_{2} \in {\rl}^{\vee}$. Let us expand:
\begin{equation*}
{\xi}(x) = \sum_{i=1}^{r} {\xi}_{i}(x) {\al}^{\vee}_{i}.
\end{equation*}
The residual shifts act on the components ${\xi}_{i}(x)$ by
\[
{\xi}_{i}(x) \mapsto {\xi}_{i} (x) + {\be}_{1,i} + {\be}_{2,i} {\tau} , \qquad i = 1, \dots , r+1,
\]
where ${\be}_{1,i}, {\be}_{2,i} \in \BZ$, for $i=1, \dots ,r$ are the
expansion coefficients:
\begin{equation*}
{\be}_{A} = \sum_{i=1}^{r} {\be}_{A, i} {\al}_{i}^{\vee}, \qquad A = 1,2.
\end{equation*}
Let ${\bf t}(x) = \big( t_{i}(x) = {\rm e}^{2\pi \ii \xi_{i}(x)} \big)_{i=1}^{r} \in
({\BC}^{\times})^{r}$ and $[{\bf t}(x)] \in {\ec}^{r}$ be
the equivalence classes for the actions of the lattices
${\rl}^{\vee}$ and ${\rl}^{\vee} \oplus {\tau}{\rl}^{\vee}$,
respectively.

Thus, dividing by all
but the locally $z$-independent $N({\bT})$-gauge transformations we arrive at the collection
of $r$ points on $\ec$, or, in a more sophisticated fashion,
a point $[{\bf t}(x)]$ in ${\ec} \otimes {\rl}$,
in addition to the $\h$-valued
gauge field $a_{\xb}(x, {\xb}){\rm d}{\xb}$.
This is all done over the generic point
$x \in {\Cpx}$.
We haven't completely fixed the gauge, though.

Let us forget for a moment about the gauge field
\begin{equation}
{\bar a} = a_{\xb}(x, {\xb})\,{\rm d}{\xb}.
\label{eq:resabel}
\end{equation}
 Then we'd divide by the action
$W({\g})$, giving as back the point $[{\CalP} \vert_{{\ec}_x}]$ in the orbispace
$\Bnq = {\ec}\otimes {\rl} / W({\g})$, the holomorphic
$\bG$-bundle on $\ec$ whose holomorphic structure is given by
${\bar\pa}_{\zb} + \frac{2\pi \ii}{{\tau}-{\bar\tau}} {\xi}(x)$. The (quasi)map
$U\colon x \mapsto U(x) = [{\CalP}_{x}]$ is point $u \in {\mve}$ in the extended
moduli space of the four-dimensional class II gauge theory.

The considerations similar to those in~\cite{Friedman:1997ih}
show that the instanton charge $c_{2}({\CalP}) = N$ bundles on $S_{\ec}$ correspond to the degree $N$ quasimaps $u\colon {\Cpx} \to {\Bnq}$. This is of course in line with the original observations on the relations between the sigma model and gauge instantons~\cite{Atiyah:1984tk}.

By following the whole $W({\g})$-orbit of $[{\bf t}(x)]$ in ${\ec}^{r}$
as $x$ varies, we obtain the curve ${\obs}_{u}$, which is a
ramified $W({\g})$-cover of ${\Cpx}$. The fiber of the projection
${\obs}_{u} \to {\Cpx}$ over a point $x$ is the orbit
\[
W({\g}) \cdot [ {\bf t}(x) ] \subset {\ec}\otimes {\rl}.
\]
This leads us to the \emph{obscured curve} which we have encountered earlier in our solution of the gauge theory using the limit shape equations.

Now, as in our prior discussion of Hitchin systems, let us recall the gauge field ${\bar a}$ in \eqref{eq:resabel}.
It looks like the $\h$-valued one-form on $\Cpx$, since it is locally
$z$-independent,
 and therefore might be a $(0,1)$
part of a
$\bT$-connection, defining a holomorphic $\bT$-bundle over ${\Cpx}$.
However, we have $|W({\g})|$ worth of choices for
${\bar\pa} + {\bar a}$ at any particular point $x$
since the residual $W({\g})$-symmetry acts both on $[{\bf t}(x)]$
and ${\bar a}$. This means that $\bar a$ becomes well-defined when
lifted to ${\obs}_{u}$. The way it transforms under the $W({\g})$-action
permuting the sheets of the cover ${\obs}_{u} \to {\Cpx}$ makes it
into the $W({\g})$-equivariant gauge field on $\obs_{u}$,
defining a $W({\g})$-equivariant holomorphic $\bT$-bundle
${\mathscr T}$ over $\obs_{u}$:
\begin{equation}
{\mathscr T} \in {\mathscr A}_{u} = {\Bun}_{\bT}({\obs}_{u})^{W({\g})} \approx
{\Hom}_{W({\g})} ( {\Lambda}, {\rm Pic}({\obs}_{u}) ) ,
\label{eq:fibu}
\end{equation}
the idea of the last equality is that every weight ${\lam} \in \Lambda = \Hom({\bT}, {\BC}^{\times})$ defines a $\BC^{\times}$-bundle ${\mathscr T}^{\lam}$ on $\obs_{u}$, in the $W({\g})$-compatible fashion.

Of course this discussion is not adequate at the branch points, yet hopefully it can be extended similarly to other
constructions of spectral covers~\cite{Donagi:1995alg,Hitchin:1987mz}. In particular,
the analysis near the branch points should demonstrate that the bundle ${\mathscr T}$ has a fixed topological type, which
we shall not attempt to determine in this paper.

 We
conjecture that \eqref{eq:fibu} is the fiber of the projection
${\pi}\colon {\pve} \to {\mve}$, sending the extended moduli space
of vacua of the class II gauge theory on
${\BR}^{3} \times {\BS}^{1}$ to that of
the infinite volume four-dimensional gauge theory.
Since the projection ${\CalP} \mapsto U$ is roughly of the form
$(A_{\xb}, A_{\zb}) \mapsto A_{\zb}$, it is Lagrangian with respect to \eqref{eq:secdada}.
This is in agreement with $\dim_{\BC} \pve = 2 \dim_{\BC} \mve = 2 N
 h$.
From now on we fix the masses, divide by the global $\bT$-action and work with $\pv$ and $\mv$.

Thus the study of holomorphic bundles on $S_{\ec}$
brought us the following picture: over each point $x \in {\Cpx} $ we hang the $W({\g})$-orbit of $[{\bf t}(x)]$ in ${\ec} \otimes {\rl}$. As $x$-varies, so does the orbit, spanning a curve ${\obs}_{u}$ in
${\Cpx} \times {\ec}\otimes {\rl}$, which projects down to ${\Cpx}$.
However ${\obs}_{u}$ is not yet the cameral curve $\CalC_{u}$.
It is however relatively straightforward to lift $\obs_{u}$ to $\CalC_{u}$ using the abelianized determinant line bundle
over ${\ec}^{r}$ we discussed in Section~\ref{sec:obscura}.

Now let us discuss the period map. The moduli space $\pv$ carries
the holomorphic symplectic structure $\Omega$ which descends from \eqref{eq:secdada}.
From the complexified textbooks on classical mechanics we learn, as we reviewed in Section~\ref{subsec:appear},
that choosing some basis $A_{\CalI}$, $B^{\CalI}$, in the integral homology $H_{1}({\mathscr A}_{u}, {\BZ})$ lattice,
such that the cup product of the basis vectors obeys \eqref{eq:dualba} and then defining ${\ac}^{\CalI}$, ${\ac}_{\CalI}^{D}$
using \eqref{eq:aper}, \eqref{eq:bper}, one verifies \eqref{eq:symlag}.
It remains to compare this definition of the special coordinates
with the periods of the differentials $x\, {\rm d} \log {\y}_{i}(x)$.

Let us work in the domain $|{\qe}_{i}| \ll 1$. The cycles
${\CalA}_{\CalI}$ and ${\CalB}^{\CalI}$ which we defined (cf.~Figures~\ref{fig:acycle} and~\ref{fig:bcycle})
on the cameral curve ${\CalC}_{u}$ define the corresponding one-cycles on ${\obs}_{u} = {\CalC}_{u}/{\rl}^{\vee}$ and consequently on
${\mathscr A}_{u}$.

Now let us compute the symplectic form $\Omega$ on the reduced phase space $\pv$.
Using that in our gauge
\[
\int_{\ec} {\rm d}^{2}z A_{\zb} (x, {\xb}) = {\xi}(x),
\]
we obtain
\begin{equation*}
{\Omega} = \int {\rm d}^{2}x \la {\delta} a_{\xb} \wedge {\delta}{\xi} \ra = \frac{1}{| W({\g} |}
\int_{{\obs}_{u}} \la {\delta} {\bar a} \wedge {\delta} {\varphi} \ra,
\end{equation*}
where we denoted by $\bar a$ the $W({\g}$-equivariant $\bT$-gauge field on $\obs_{u}$
and by ${\delta}{\varphi}$ an $\h$-valued, $W({\g})$-equivariant $(1,0)$-form on $\obs_{u}$, given by
\[
{\delta}{\varphi} = x {\delta}{\xi}.
\]
It is now easy to pin down the periods of ${\rm d}{\BS}$ among the periods of ${\delta}{\varphi}$.
One uses the cycles in~${\mathscr A}_{u}$, where the monodromy of the
gauge field $\bar a$ along the cycle ${\CalA}_{\CalI}$ or ${\CalB}^{\CalI}$ changes by $2\pi$.

To conclude this section, in parallel with the class I story we conjecture, that {\it the theories of class II lead to the representation theory of the Yangians built on the Kac--Moody algebras, i.e.,
the toroidal algebras.} It would be nice to see whether the
Kac--Moody symmetry of instanton moduli spaces~\cite{Dolan:1982dc,Licata_2007} leads to the quasi-classical limit of the Yangian action, at the level of the phase space $\pve$ symmetry.

\subsection{Noncommutative instantons and class II* theories}

In this subsection, we show that the data $t_{i}(x)$
with the cross-cut transformations (\ref{eq:classis}) correspond naturally to the charge $N$ instantons
on the noncommutative space ${\BR}^{2} \times {\BT}^{2}$, with the gauge group $U(r+1)$. We shall use the constructions analogous to the constructions~\cite{Nekrasov:2000zz} of instantons on the noncommutative ${\BR}^{4}$ adapted to the periodic case. The noncommutative solitons on a~cylinder were studied in~\cite{Demidov:2003xq}.

The idea of the construction is the following. Consider first the commutative situation. Let us denote the coordinates on ${\BR}^{2}$ by $x_{1}$, $x_{2}$. Let $( u_{1}, u_{2} ) \in {\BT}^{2} \subset {\BC}^{\times} \times {\BC}^{\times}$, $| u_{1}| = |u_{2}| = 1$. Let us denote by $z$, $x$ the holomorphic local coordinates on ${\Cx} \times {\ec}$,
\begin{equation*}
x = x_{1} + {\ii} x_{2}, \qquad z = \frac{1}{2\pi \ii} ( \log (u_{1}) + {\tau} \log (u_{2}) ).
\end{equation*}
Consider the $(0,1)$-component of $U(r+1)$ gauge field ${\bar A} = A_{\zb} \,{\rm d}{\zb} + A_{\xb} \,{\rm d}{\xb}$, where $A_{\zb}$, $A_{\xb}$ are the $(r+1) \times (r+1)$ complex matrices.
They obey \eqref{eq:xzflat} and in the gauge \eqref{eq:gauge}. we have
\begin{equation}
A_{\zb} = \frac{2\pi {\rm i}}{{\bar\tau}- {\tau}} \operatorname{diag}
( {\xi}_{i}(x, {\xb}) )_{i=1}^{r+1},
\label{eq:gauge}
\end{equation}
where, by the considerations analogous to those around \eqref{eq:fzt}, we conclude that
${\bar\pa}_{\xb} {\xi}_{i} = 0$. Again, there are the residual gauge transformations
\[
A_{\zb} \mapsto G^{-1} A_{\zb} G+ G^{-1} {\partial}_{\zb} G,
\]
which permute the eigenvalues ${\xi}_{i}(x)$:
$ \{ {\xi}_{i}(x) \} \mapsto \{ {\xi}_{{\sigma}(i)}(x) \}$, for some
${\si}\in {\CalS}_{r+1}$, and shift them
\begin{equation}
 {\xi}_{i}(x) \mapsto {\xi}_{i}(x) + l_{i} - {\tau} n_{i}.
\label{eq:mnai}
\end{equation}
The latter are generated by the diagonal gauge transformations
\begin{equation*}
G = \operatorname{diag} \big( u_{1}^{n_{i}} u_{2}^{l_{i}} \big)_{i=1}^{r+1}.
\end{equation*}
We can partially reduce the ambiguity \eqref{eq:mnai} by passing to
the exponential variables
\begin{equation*}
t_{j}(x) = {\rm e}^{2\pi \ii {\xi}_{j}(x) }, \qquad j = 1, \dots , r+1.
\end{equation*}
The residual gauge transformations (\ref{eq:mnai}) become
\begin{equation*}
t_{j}(x) \mapsto {\qe}^{n_{j}} t_{j}(x), \qquad t_{j}(x) \mapsto t_{{\sigma}(j)}(x)
\end{equation*}
for some integers $n_{i} \in {\BZ}$ and permutations ${\sigma} \in {\CalS}_{r+1}$.

Let us now consider the noncommutative case. Let
us replace the algebra of functions of $u_{1}$, $u_{2}$, $x_{1}$, $x_{2}$
by the noncommutative algebra, with the generators
${\hat u}_{1}$, ${\hat u}_{2}$, ${\hat x}_{1}$, ${\hat x}_{2}$, obeying
\begin{gather}
{\hat u}_{1}^{-1} {\hat x}_{i} {\hat u}_{1} = {\hat x}_{i} + {\hbar}_{i} \cdot 1, \qquad
 {\hat u}_{2}^{-1} {\hat x}_{i} {\hat u}_{2} = {\hat x}_{i}, \qquad
{\hat u}_{1} {\hat u}_{2} = {\hat u}_{2} {\hat u}_{1}, \qquad
[ {\hat x}_{1}, {\hat x}_{2} ] = 0,
\label{eq:uxux}
\end{gather}
where
\[
\frac{\ma}{r+1} = {\hbar} = {\hbar}_{1} + {\ii}{\hbar}_{2}.
\]
Then the analogue of (\ref{eq:mnai}) for
\begin{equation*}
G = \operatorname{diag} \big( {\hat u}_{1}^{n_{i}} {\hat u}_{2}^{l_{i}} \big)_{i=1}^{r+1}
\end{equation*}
gives
\begin{equation*}
t_{i}(x) \mapsto q^{n_{i}} t_{i}\Big(x- \frac{n_i}{r+1} {\ma}\Big), \qquad t_{i}(x) \mapsto t_{{\sigma}(i)}(x).
\end{equation*}
These are precisely the $\iw$-transformations of the class II* type ${\hat A}_{r}$
gauge theory, with $t_{i}(x)$ given in \eqref{eq:rhyiis} and \eqref{eq:india}.

Instead of giving more systematic discussion along the lines of~\cite{Astashkevich:1998uc,Connes:1997cr, Douglas:2001ba,Nekrasov:2000zz, Nekrasov:1998ss, Schwarz:2001ru}
let us comment on the relation to the group $\Gli$.
We claim that the $U(r+1)$ instantons on the noncommutative
${\BR}^{2} \times {\BT}^{2}$ with both ${\BR}^{2}$ and ${\BT}^{2}$ separately commutative, can be interpreted as the commutative periodic monopoles
on ${\BR}^{2} \times {\BS}^{1}$ with the gauge group $\Gli$.

The idea is to interpret the noncommutative $U(r+1)$ gauge fields on ${\BR}^{2} \times {\BT}^{2}$ as the $\Gli$-gauge fields
on ${\BR}^{2}_{x_{1}, x_{2}} \times {\BS}^{1}_{\psi}$, using the relation of the
group $\Gli$ to the quantization of the volume-preserving diffeomorphisms of a cylinder ${\BR}^{1}_{p} \times {\BS}^{1}_{q}$, i.e., the pseudo-differential operators~${\Psi}DO$ on ${\BS}^{1}_{q}$. Here
the generators
${\hat u}_{1,2}$, ${\hat x}_{1,2}$ of the algebra \eqref{eq:uxux} are related to the commutative coordinates $(x_{1}, x_{2}, \psi)$ of the monopole theory and the generators $(p, {\rm e}^{\ii q})$ of $\Gli$ in the via
\begin{equation*}
 {\hat u}_{2} = {\rm e}^{\ii \psi}, \qquad
 {\hat u}_{1} = {\rm e}^{\ii q}, \qquad
 {\hat x}_{1} = x_{1} - {\ii}{\hbar}_{1} {\pa}_{q}, \qquad
 {\hat x}_{2} = x_{2} - {\ii}{\hbar}_{2} {\pa}_{q}.
\end{equation*}

\section{Higher-dimensional theories}\label{se:higher}

In this section, we briefly go over the higher-dimensional generalizations. We shall consider the lifts of all our theories to five dimensions, and the lifts of some of our theories to six dimensions. The latter restriction comes from the possibility of encountering gauge and mixed anomalies in six dimensions, which would prohibit the decoupling of the gauge sector from the supergravity and ultimately string theory modes.

The classical ${\CalN}=2$ theories in four dimensions admit a canonical lift to the
${\CalN}=1$ theories in six dimensions. Under this lift the vector multiplets become the vector multiplets, and the hypermultiplets become the hypermultiplets. The structure of
the hypermultiplet does not change, while the structure of the vector multiplet does change, for the complex scalar $\Phi$ in the adjoint representation becomes the remaining component $A_{5} + \ii A_{6}$ of the gauge field.

We then compactify the theory on a two-torus ${\BT}^2$. In addition to the metric on the torus we shall also fix a background $B$-field, a constant two-form. The combined metric and the two-form moduli are described by a $2 \times 2$ matrix $G$, with the positive definite symmetric part~$g$. It is convenient to parametrize $G$ by two complex numbers
$T$, $U$, with $\operatorname{Im}U, \operatorname{Im}T \geq 0$. The parameter~$T$ encodes the complex structure of $\BT^2$, while the parameter $U$ is the complexified K\"ahler class:
\begin{equation*}
 U = \int_{\BT^2} B + \ii \int_{\BT^2} \sqrt{\det (g)}, \qquad
 T = \frac{g_{12} + \ii \sqrt{\det (g)}}{g_{11}}.
\end{equation*}
Five(six-)-dimensional supersymmetric gauge theory compactified on a circle (two-torus) would look four-dimensional at low energy. The microscopic gauge coupling ${\tau}$ in four dimensions is proportional to $U$ while $T$ determines the complex geometry of the $x$-plane $\Cx$.
One can study the corresponding Seiberg--Witten geometry. Its key feature compared to
special geometry of more traditional four-dimensional models is the periodicity in the $x$-variable~\cite{Nekrasov:1996cz}. In the five-dimensional theory the $x$-plane becomes the cylinder ${\Cx}^{\times}$, while in the six-dimensional theory the $x$-plane becomes the two-torus ${\Cx}/2{\pi}{\ii}({\BZ} \oplus T {\BZ})$. It has to do with the large gauge transformations.
The result of these additional symmetries is the relativistic nature of the corresponding integrable systems. For example, the periodic Toda chain describing the pure ${\CalN}=2$ theory in four dimensions becomes the relativistic Toda chain. The Hamiltonians of the relativistic systems have periodic dependence on momenta, which are the rapidities of the particles. The resulting quantized Hamiltonians are the difference operators.

Our discussion modifies in the case of five-dimensional gauge theories in two aspects. First, the notion of the amplitude function accommodates the large gauge transformations:
\begin{equation*}
{\y}_{i}(x) = \exp \big\langle \tr _{{\bv}_{i}} \log \big( {\rm e}^{{{\ii \beta}}( x - {\Phi}_{i})} - 1 \big) \big\rangle_{u},
\end{equation*}
where the dimension length $\be$ characterizes the circumference of the compactification circle.
The limit shape integral equations (\ref{eq:limshape}) generalize straightforwardly, with
the kernel
\begin{equation}
{\CalK}_{\beta} ({\xr}) = \frac{{\ii \beta}}{12} {\xr}^{3} - \frac{\log ({{\ii \beta}}{\Lambda}_{\rm UV})}{2}{\xr}^2 - \frac 1{{\ii \beta}^2} \operatorname{Li}_{3} \big( {\rm e}^{-{{\ii \beta}}{\xr}} \big).
\label{eq:fdker}
\end{equation}
Secondly, there are additional couplings in five dimensions: the levels $k_{i}$ of the Chern--Simons interactions CS$_{5}(A^{i})$, $i \in \Ver$. Effectively the Chern--Simons term changes
the gauge coupling~${\qe}_i$ to the $x$-dependent quantity
\[
{\qe}_{i} \longrightarrow {\qe}_{i} {\rm e}^{{{\ii \beta}} k_{i}x}.
\]
In the six-dimensional case the amplitude and the kernel \eqref{eq:fdker} modify to
\begin{gather*}
{\y}_{i}(x) = \exp \big\langle \tr _{{\bv}_{i}} \log {\theta} \big( {\rm e}^{{{\ii \beta}} ( x- {\Phi}_{i})} ; Q\big) \big\rangle_{u},\\
{\CalK}_{\be} ({\xr}) = \frac{{\ii \beta}}{12} {\xr}^{3} - \frac{\log ({{\ii \beta}}{\Lambda}_{\rm UV})}{2}{\xr}^2 - \frac 1{{\ii \beta}^2} \operatorname{Li}_{3} \big( {\rm e}^{-{{\ii \beta}}{\xr}} \big) - \sum_{n=1}^{\infty} \big( \operatorname{Li}_{3} \big( {\rm e}^{{{\ii \beta}}{\xr}} Q^{n} \big) +
\operatorname{Li}_{3} \big( {\rm e}^{-{{\ii \beta}}{\xr}} Q^{n} \big) \big) ,
\end{gather*}
where $Q = \exp ( 2\pi \ii T)$, and now ${\be}^{2}$ is the scale of the area of the compactification torus ${\BT}^{2}$.

\appendix

\section{McKay correspondence, D-branes, and M-theory}\label{se:mckay}
\subsection{From finite groups to Lie groups}

McKay correspondence states~\cite{McKay:1980} that the affine ADE graphs $\gamma$ can be constructed from the representation theory
of finite subgroups $\Gamma$ of ${\rm SU}(2)$.

The vertices $i \in \Ver$ in this case are the irreducible representations of $\Gamma$,
$i \mapsto {\CalR}_{i}$. One usually enumerates them, $\Ver = \{ 0, 1,
\dots,r \}$, so that $0$ corresponds to the trivial representation
${\CalR}_{0} = {\BC}$. By $r$ in this section
we denote the number of nodes in the finite
graph $\gamma_{\text{fin}}$ obtained by discarding the node ``0'' from
affine $\gamma$. In all cases $r$ is the rank of the Cartan matrix
associated to Dynkin graph.
 Moreover, the Dynkin marks $a_i$ are the dimensions of $V_i$,
and the numbers of colors in affine quivers is
\begin{equation*}
{\bv}_{i} = N a_{i},
\end{equation*}
where $a_{i} = \dim {\CalR}_{i}$, and $N$ is some non-negative
integer. As we said above, $i=0$ corresponds to the trivial
representation, so $a_{0} =1$.
In the table of McKay correspondence we write ``$i(a_i)$'' by each node to
denote its label $i$ and Dynkin mark $a_i$. We label the nodes in
Bourbaki conventions.

The number of edges $I_{ij}$ in the McKay graph between the node $i$
and the node $j$ is the multiplicity of representation $\CalR_{j}$
in the tensor product of $\CalR_i$ with the defining representation~$\BC^2$
\begin{equation*}
{\BC}^{2} \otimes {\CalR}_{i} = \bigoplus_{j} {\BC}^{I_{ij}} \otimes {\CalR}_{j}.
\end{equation*}
The equation ${\be}_{i}=0$ is verified by computing the dimensions of
the left and the right-hand sides.

The order of $\Gamma$
agrees with the dimensions of the irreducible representations / Dynkin marks of
the McKay/Dynkin affine graph computed using the standard relation from the
orthogonality of characters
\begin{equation*}
 |\Gamma| = \sum_{i} \dim \CalR_i^2 = \sum_{i} a_i^2.
\end{equation*}

McKay's observation is that $\gamma$ is affine ADE Dynkin diagram, with
the trivial representation~$\CalR_{0}$ associated with the affine node
``0''. The finite Dynkin graph $\gamma_{\text{fin}}$
 is always
a tri-star graph $T_{a,b,c}$ with one
trivalent vertex and three legs containing
$a$, $b$, $c$ vertices (where in the counting we included
the center trivalent vertex). Hence the rank of the finite quiver
$\gamma_{\mathrm{fin}}$ is $a+b+c -2$. In fact $a$, $b$, $c$ have simple
interpretation: the group $\Gamma$ is
always a Coxeter group $\Cox(a,b,c)$ defined on three generators $(x,y,z)$ subject
to relation
 \begin{equation*}
 x^a = y^b = z^c = xyz.
 \end{equation*}
The affine graph of $\hat E_{r}$ series is also a trivalent graph denoted by
$\big(\hat T_{a,b,c}\big)$ in the table of McKay correspondence. Note that for
each of the three cases
$\hat E_{6}$, $\hat E_{7}$, $\hat E_{8}$ the identity
\begin{equation*}
 \frac 1a + \frac 1b + \frac 1c = 1
\end{equation*}
holds,
which makes contact with the unimodular parabolic singularities.

Now we shall list explicitly the discrete subgroups of ${\rm SU}(2)$, the
classification going back to Plato and Klein~\cite{Klein:1884}.

\begin{table}[t]\centering\renewcommand{\arraystretch}{1.5}
\begin{tabular}{|@{}c@{}m{28pt} |@{\,\,}m{32pt}@{\,\,}|@{}c@{}@{\,\,}m{25pt}@{\,\,}|}
\hline
Polyhedron & $\Gamma$ & $T_{a,b,c}$ & Affine Dynkin graph $\gamma$ with
labels $i(a_i)$ &
Lie $\hat \g$\\
\hline
\parbox[c]{80pt}{\centering \includegraphics{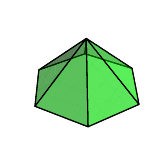}} & $\BZ_{r+1}$ $_{(r=5)}$ &
$T_{r,1,1}$ &
\parbox[c]{180pt}{\centering \includegraphics{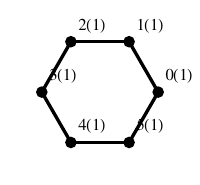}} & $\hat A_{r}$ $_{(r=5)}$
\\ 
\parbox[c]{80pt}{\centering \includegraphics{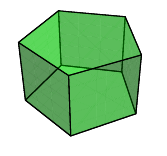}} & $\BB \BD_{r-2}$ $_{(r=7)}$ &
$T_{r-2,2,2}$ &
\parbox[c]{180pt}{\centering \includegraphics{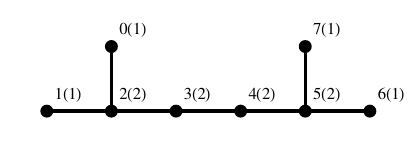}} & $\hat D_{r}$ $_{(r=7)}$
\\ 
\parbox[c]{80pt}{\centering \includegraphics{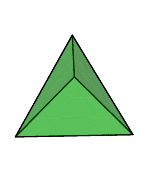}} & $\BB \BT$ &
$T_{3,3,2}$ $(\hat T_{3,3,3})$ &
\parbox[c]{180pt}{\centering \includegraphics{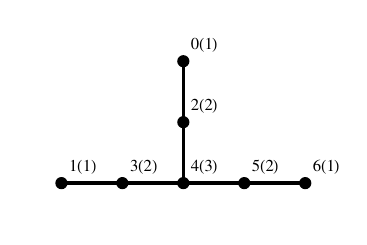}} & $\hat E_6$
 \\ 
\parbox[c]{80pt}{\centering \includegraphics{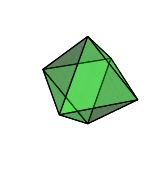}} & $\BB \BO$ &
$T_{4,3,2}$ $(\hat T_{4,4,2})$ &
\parbox[c]{180pt}{\centering \includegraphics{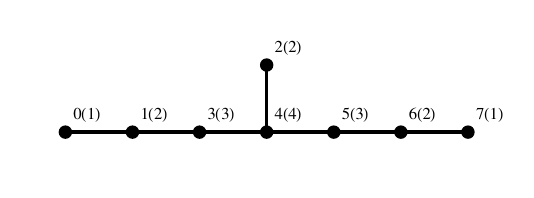}} & $\hat E_7$
\\ 
\parbox[c]{80pt}{\centering \includegraphics{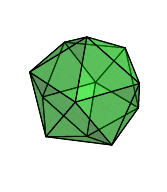}} & $\BB \BI$ &
$T_{5,3,2}$ $(\hat T_{6,3,2})$ &
\parbox[c]{265pt}{\centering \includegraphics{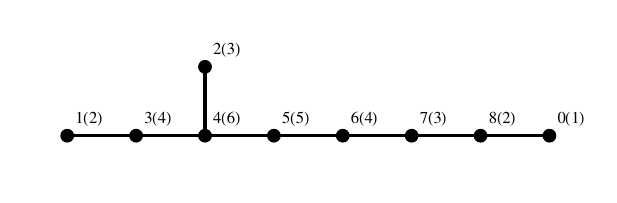}} & $\hat E_8$
\\ \hline
\end{tabular}
\end{table}

\subsection{Platonic symmetries}
\subsubsection{Cyclic group}
Let us present cyclic group $\BZ_{r+1}$ in the $(a,b,c)$
notation: consider group generated by $x$, $y$, $z$ subject to
\begin{equation*}
 x^{r} = y = z = xyz.
\end{equation*}
Then $z = y$ and $ x y = 1$ and $y = x^{r}$, hence
\begin{equation*}
 x^{r+1} = 1
\end{equation*} hence $\BZ_{r+1} = \Cox(r,1,1) = \McKay(A_{r})$.
The order of $\Gamma$ is
\begin{equation*}
 |{\BZ}_{r+1}| = r+1 =\sum_{i \in \gamma} a_i^2 = (r+1) \cdot 1.
\end{equation*}

\subsubsection{Dihedral group}
The dihedral group $\BD_{r-2} \simeq \BZ_{r-2} \rtimes \BZ_2$ of order $2(r-2)$ is the symmetry
group of regular $(r-2)$-gon considered as a subgroup of ${\rm SO}(3)$.
The lift of dihedral group $ \BZ_2 \to \BB \BD_{r-2} \to \BD_{r-2}$ to a~subgroup
of
 ${\rm SU}(2)$ is called bidihedral group $\BB \BD$, which has order $4(r-2)$.
The $\BB \BD_{r}$ is generated by $(x,y,z)$ subject to relations
\begin{equation*}
 x^{r-2} = y^2 = z^2 = xyz
\end{equation*}
hence $\BB \BD_{r-2} = \Cox(r-2,2,2) = \McKay(D_{r})$.
The order of $\Gamma$ is
\begin{equation*}
 |\BB \BD_{r-2}|= 4(r-2) = \sum_{i \in \gamma} a_i^2 = (r-3)\cdot 4 + 4 \cdot 1.
\end{equation*}

\subsubsection{Tetrahedral group}

 The tetrahedral group $\BT \simeq A_4$, of order 12,
is the subgroup of ${\rm SO}(3)$ realizing the symmetries
of tetrahedron. The lift of tetrahedral group
$ \BZ_2 \to \BB \BT \to \BT$ to a subgroup of ${\rm SU}(2)$ is called
bitetrahedral group $\BB\BT$, which has order $24$
and is generated by $(x,y,z)$ subject to the relations%
 \begin{equation*}
 x^3 = y^3 = z^2 = xyz.
 \end{equation*}
Hence $\BB \BT = \Cox(3,3,2) = \McKay(E_{6})$.
The order of $\Gamma$ is
\begin{equation*}
|\BB \BT| = 24 = \sum_{i \in \gamma} a_i^2 = 3 \cdot \big(1^2+2^2\big)+3^2 = 24.
\end{equation*}

\subsubsection{Octahedral group}
 The octahedral group $\BO \simeq S_4$, of order $24$, is the subgroup of ${\rm SO}(3)$ realizing
the symmetries of the cube/octahedron. The lift to ${\rm SU}(2)$ is called
bioctahedral group $\BB \BO$, of order $48$; the $\BB \BO$ can be generated by $(x,y,z)$
subject to relations
 \begin{equation*}
 x^4 = y^3 = z^2 = xyz
 \end{equation*}
associated to the symmetry axes of the cube/octahedron of
 the order $4,3$ and $2$.
Hence $\BB \BO = \Cox(4,3,2) = \McKay(E_{7})$.
The order of $\Gamma$ is
\begin{equation*}
|\BB \BO|= 48 = \sum_{i \in \gamma} a_i^2 = 2 \cdot \big(1^2+2^2+3^2\big)+4^2+2^2.
\end{equation*}

\subsubsection{Icosahedral group}
 The icosahedral group $\BI \simeq A_5$, of order $60$, is the subgroup of ${\rm SO}(3)$ realizing
the symmetries of the icosahedron/dodecahedron The lift to ${\rm SU}(2)$ is called
biicosahedral group $\BB \BO$, of order $120$; the biicosahedral group $\BB \BI$ can be generated by $(x,y,z)$
subject to relations
 \begin{equation*}
 x^5 = y^3 = z^2 = xyz
 \end{equation*}
associated to the symmetry axes of the icosahedron/dodecahedron of
 the order $5$, $3$ and $2$. Hence $\BB \BO = \Cox(5,3,2)= \McKay(E_8)$.
The order of $\Gamma$ is
\begin{equation*}
| \BB \BI| = 120 = \sum_{i \in \gamma} a_i^2= \big(1^2+2^2+3^2+4^2+5^2+6^2\big) + 2^2+4^2+3^2.
\end{equation*}

\subsection{D-branes at singularities}\label{sec:dbranes}

The physical explanation of the relation between the
${\BC}^{2}/{\Gamma}$-singularities~\cite{Kronheimer:1989} and the superconformal theories with
${\CalN}=2$ supersymmetry is the following. Consider the IIB string in
the background ${\BC}^{2}/{\Gamma} \times {\BC}^{1} \times {\BR}^{1,3}$,
where ${\Gamma} \subset {\rm SU}(2) \subset {\rm SO}(4)$ acts on ${\BC}^{2} \approx
{\BR}^{4}$ by the hyperk\"ahler rotations. This background preserves
half of the ten-dimensional IIB supersymmetry. Now add a stack of $N$
regular D3 branes. At the orbifold singularity the regular D3 brane
splits as a~collection of $r+1$ types of fractional branes, which
correspond to the irreducible representations of $\Gamma$
\cite{Douglas:1996sw, Johnson:1996py}. Moreover, the regular brane contains $a_i$ fractional branes of the ${\CalR}_i$ type. The stack of $N$ regular brane splits, therefore, as a collection of $Na_i$ fractional branes of $\CalR_{i}$ type, for all $i=0, \dots, r$. Each cluster of fractional branes can be classically moved anywhere in the two-plane, transverse to the singularity and the D3 brane worldvolume. To summarize, the D3 branes are located at:
\[
0 \times \big\{ {\mu}_{0}^{Na_0}, {\mu}_{1}^{Na_{1}}, \dots , {\mu}_{r}^{Na_{r}} \big\} \times {\BR}^{1,3}.
\]
Here $\big\{ {\mu}_{0}^{v_0}, \dots , {\mu}_{r}^{v_r} \big\}$ represents the positions of the $v_i$ copies of the $\CalR_i$ type of the fractional branes in the two-dimensional plane ${\BC}^{1} \approx {\BR}^{2}$ transverse to the singular ALE space ${\BC}^{2}/{\Gamma}$ and the worldvolume ${\BR}^{1,3}$ of the branes. The positions ${\mu}_{i}$ are the ones which enter the equation~(\ref{eq:bmpm}). At the low energy the worldvolume theory on the D3 branes is the quiver gauge theory we are studying. If all ${\mu}_{i}$ parameters are scaled to zero (or at least all coincide), the theory has no scale except for the string scale which is absent in the low energy description. Therefore the theory has the scaling invariance, which is promoted to the full ${\CalN}=2$ superconformal invariance.
The couplings ${\tau}_{i}$, except for
\begin{equation}
\label{eq:Btau}
\tau = \sum_{i=0}^{r}a_{i}{\tau}_{i},
\end{equation}
which corresponds to the IIB dilaton-axion field, are not of geometric origin. The gauge couplings~$\tau_i$, $i=1, \dots, r$, come from the twisted sector fields
\begin{equation*}
\tau_i = \int_{{\Sigma}_i} B_{\rm RR} + {\tau} B_{\rm NS},
\end{equation*}
where $B_{\rm NS}$, $B_{\rm RR}$ are the Neveu--Schwarz (NSNS) and the Ramond--Ramond two-form fields of the IIB supergravity, and $\Sigma_i$ stand for the non-trivial two-cycles in the resolved $\widetilde{{\BC}^2 / {\Gamma}}$ geometry. These exceptional cycles are well-known to correspond to the simple roots of~$\gq$.

The 4d gauge theory on the stack of D3 branes in the above geometry is
$\CalN=2$ supersymmetric theory of the affine ADE quiver type. To find
the algebraic integrable system associated to the IR solution of such
theory we compactify the worldvolume of the D3 branes
$(x_0,x_1,x_2,x_3)$ on a circle $S^{1}$, say, along $x_3$. Hence, our
setup is $N$ D3 branes along $\BR^{1,2}\times \BS^{1}$ inside the IIB
string theory on $\BR^{1,2}\times \BS^{1}_{\la x_3 \ra} \times \Cx \times
\BC^{2}/\Gamma$. Here $\Cx$ is the real-two-dimensional space associated
with the scalars of $\CalN=2$ vector multiplet in 4 dimensions. For
4d theories this is affine complex line $\Cx = \BC$, for 5d theories on
$S^1$ this is a cylinder $\Cx =\BC^{\times}$ and for 6d theories on
$T^2$ this is a torus $\Cx= \ec_{\la x \ra}$.

Given this type IIB string theory realization of the affine ADE quiver
theory, first we perform T-duality along $\BS^1_{\la x_3\ra}$. The
stack of $N$ D3-branes converts to the stack of $N$ D2-branes on $\BR^{2,1}$ in type IIA string theory on
$\BR^{1,2}\times \BS^{1}_{\la x_3 \ra} \times \Cx \times
\BC^{2}/\Gamma$. Next we lift the type IIA string to M-theory on
$\BR^{1,2}\times \BS^{1}_{\la x_3 \ra} \times \Cx \times
\BC^{2}/\Gamma\times \BS^{1}_{\la x_{10} \ra}$, where $\BS^{1}_{\la x_{10}
 \ra}$ is the M-theory circle. The radius of M-theory circle is
determined by IIB coupling constant \eqref{eq:Btau}. We treat the
product of two circles along $x_3$ and along $x_{10}$ as the elliptic curve $\ec =\BS^{1}_{\la x_{3}\ra} \times
\BS^{1}_{\la x_{10} \ra}$ which has elliptic modulus $\qe = {\rm e}^{2 \pi \ii
 \tau}$. The stack of $N$ D2
branes converts to the stack of $N$ M2 branes along $\BR^{1,2}$. So
finally we arrive to the M-theory picture.

That is, we consider the following configuration in M-theory: the space-time background is
\begin{equation}
{\BR}^{1,2} \times X_{4} \times {\BC}^{2}/{\Gamma}
\label{eq:sst}
\end{equation}
to which we add a stack of $N$ M2 branes whose worldvolume is ${\BR}^{1,2}$ in
(\ref{eq:sst}) localized at the orbifold singularity in ${\BC}^{2}/{\Gamma}$, and anywhere
in $X_{4} = \BC_{\la x \ra} \times \ec_{\qe}$.
The M-theory on $\BC^{2}/\Gamma$
is believed to contain the seven-dimensional $G$-gauge theory with sixteen
supersymmetries, localized at the singular locus. The maximal torus $T$ part of the gauge bosons and their superpartners descend from the supergravity modes associated
with the hyperk\"ahler deformations of the metric on~${\BR}^{4}/{\Gamma}$. In particular, the gauge fields are the modes of the three-form field $C_3$ integrated along the collapsed two-cycles $\Sigma_{i}$, $i=1, \dots , r$ in the deformed geometry ${\widetilde{{\BR}^{4}/{\Gamma}}}$. The remaining
$W$-bosons come from the M2-branes wrapped on various two-cycles in
$H_2 ({\widetilde{{\BR}^{4}/{\Gamma}}}, {\BZ}) \approx {\rl}$.

In the IR, the stack of $N$ M2 branes stretched along $\BR^{1,2}$
 in this 7d $G$-gauge theory on $\BR^{1,2} \times X_{4}$ dissolve
 into the charge $N$ $G$-instantons on $X_{4} = \Cx \times \ec_{\qe}$
~\cite{ Blum:1997mm,deBoer:1996ck,Intriligator:1996ex,Kapustin:1998fa,Kapustin:1998xn,Katz:1997eq,Witten:1997kz}. The moduli
 space of framed charge $N$ $G$-instantons on $X_{4}$ is the phase
 space of the algebraic integrable system (see Section~\ref{se:low-energy}) we have found from the microscopic four-dimensional instanton
 counting in the quiver gauge theory.

\section{Partitions and free fermions}\label{se:partitions}

Recall that a partition $\lam$ is a non-increasing sequence of integers, stabilizing at zero,
\begin{equation*}
{\lam} = ( {\lam}_{1} \geq {\lam}_{2} \geq \dots \geq {\lam}_{{\ell}({\lam})} > 0 = 0 = 0= \cdots ).
\end{equation*}
The number ${\ell}({\lam}) \geq 0$ is called the \emph{length} of the partition $\lam$, while
\begin{equation*}
| {\lam} | = \sum_{i=1}^{\ell ({\lam})} {\lam}_i
\end{equation*}
is called the \emph{size} of the partition $\lam$.

Consider the theory of a single chiral complex fermion, which is a $(1/2,0)$-differential, in two dimensions
\begin{equation}
{\CalL} = \int {\tilde\psi} {\bar\pa} {\psi}.
\label{eq:clff}
\end{equation}
The theory (\ref{eq:clff}) has a $U(1)$-symmetry
\begin{equation}
\big( {\psi}, {\tilde\psi} \big) \mapsto \big( {\rm e}^{{\rm i}{\al}} {\psi}, {\rm e}^{-{\rm i}{\al}} {\tilde\psi} \big).
\label{eq:uoneps}
\end{equation}
One can couple the fermion to the background abelian gauge field corresponding to this symmetry:
\begin{equation*}
{\bar\pa} \mapsto {\bar\pa} + {\bar A},
\end{equation*}
so that the Lagrangian \eqref{eq:clff} deforms to
\[
{\CalL} \to {\CalL} + \int {\bar A} J,
\]
where $J = {\tilde\psi}{\psi}$ is the $U(1)$ current, which we define as an operator below.

In studying the space of states ${\CalH}$ corresponding to the quantization of the theory (\ref{eq:uoneps}) one can distinguish various sectors, corresponding to twisted boundary conditions.
Expand
\begin{gather*}
 {\psi}(z) = \bigg(\frac{{\rm d}z}{z}\bigg)^{\half} \sum_{r \in {\BZ} + {\al}} {\psi}_{r} z^{-r}, \qquad
 {\tilde\psi}(z) = \bigg(\frac {{\rm d}z}{z}\bigg)^{\half} \sum_{r \in {\BZ} - {\al}} {\tilde\psi}_{r} z^{-r},
\end{gather*}
where ${\al} = \int_{{\BS}^{1}} {\bar A}$.
The fermion modes ${\psi}_{r}$, ${\tilde\psi}_{s}$ form the Clifford algebra
\begin{equation*}
\big\{ {\psi}_{r}, {\tilde\psi}_{s} \big\} = {\delta}_{r+s,0}, \qquad \big\{ {\psi}_{r}, {\psi}_{s} \big \} = \big\{ {\tilde\psi}_{r}, {\tilde\psi}_{s} \big\} = 0.
\end{equation*}
The vacuum state $| {\varnothing}; {\al} \rangle$ is annihilated
by all the ${\al}$-positive harmonics
\begin{gather*}
{\psi}_{r} | {\varnothing}; {\al} \rangle = 0, \qquad r > {\al}, \\
{\tilde\psi}_{r} | {\varnothing}; {\al} \rangle = 0, \qquad r > -{\al}.
\end{gather*}
The space of states is built by acting on $| {\varnothing}; {\al} \rangle $
with creation operators ${\psi}_{r}$, $r \leq {\al}$, and ${\tilde\psi}_{s}$, $s \leq - {\al}$.
The resulting Hilbert space ${\CalH}_{\al}$ has a basis labeled by partitions:
\begin{gather*}
 {\CalH}_{\al} = \bigoplus_{\lam} {\BC} \vert {\lam}; {\al} \rangle, \\
\vert {\lam}; {\al} \rangle = {\psi}_{-{\lam}_{1} + {\al}}
{\psi}_{-{\lam}_{2} + 1 + {\al}} \cdots {\psi}_{-{\lam}_{{\ell}({\lam})}+{\ell}({\lam}) - 1 + {\al}}, {\tilde\psi}_{-{\al}} {\tilde\psi}_{-{\al}-1} \cdots {\tilde\psi}_{-{\al}+ 1 - {\ell}({\lam})} \vert {\varnothing}; {\al} \rangle.
\end{gather*}
Define the normal ordering with respect to the ${\al} = 0$ vacuum:
\begin{equation*}
{:} {\psi}_{i} {\tilde\psi}_{j}{:} = \begin{cases}
{\psi}_{i} {\tilde\psi}_{j}, & j > 0, \\
- {\tilde\psi}_{j}{\psi}_{i}, & j \leq 0. \end{cases}
\end{equation*}
The $U(1)$ symmetry of the Lagrangian \eqref{eq:clff} is promoted to the ${\hat u(1)}$ current algebra symmetry.
It is generated by the operator
\begin{gather*}
 J(z) = {:} {\tilde\psi}(z) {\psi}(z) {:} = \sum_{n\in \BZ} J_{n} z^{-n-1} {\rm d}z ,\\
 J_{n} = \sum_{r \in {\BZ} + {\al}} {:} {\tilde\psi}_{r}{\psi}_{n-r}{:}.
\end{gather*}
The generating function
\begin{equation*}
\sum_{{\lam}} {\qe}^{|{\lam}|} t^{{\ell}({\lam})} = \prod_{n=1}^{\infty} \frac{1}{1 - t {\qe}^{n}}
\end{equation*}
is a character of the fundamental $\hat{u(1)}$ module, ${\CalH}_{0}$.

\section{Lie groups and Lie algebras}\label{se:Lie}

In this section we fix our notations for the notions from the Lie group and Lie algebra theory we are using in the work. In this section we work over the field $\BC$ of complex numbers.

\subsection{Finite-dimensional Lie algebras}

Let $\g$ be a finite-dimensional simply-laced simple Lie algebra, $\h$ its Cartan subalgebra. Let $\bG$ be the corresponding \emph{simply-connected} simple Lie group, and $\bT \subset \bG$ the corresponding to $\h$ maximal torus.
We have the exponential map $ \exp _{\h}\colon {\h} \to \bT$, which is a restriction on $\h$ of the exponential map $ \exp _{\g}\colon {\g} \to \bG$. We shall only use the $ \exp _{\h}$ map in this work and will omit the~$\h$ subscript in what follows. We also use the notation
\begin{equation}
{\ex}({\bf x}) = \exp (2\pi \ii {\bf x}) = {\rm e}^{2\pi \ii {\bf x}} \in \bT, \qquad {\rm for} \ {\bf x} \in \h. \label{eq:expm}
\end{equation}

\subsubsection{The coroots}

The kernel of the map (\ref{eq:expm}), i.e., the set ${\rl}^{\vee} \subset \h$ which is mapped to the identity element in~$\bT$, is called the coroot lattice (it is obviously an abelian group).
The coroot lattice ${\rl}^{\vee}$ has the rank $r = {\rm rk}\bG = \dim {\h}$. Let us denote by ${\al}_{i}^{\vee}$, $i=1, \dots , r$ its integral basis
\[
{\rl}^{\vee} = {\BZ} {\al}_{1}^{\vee} \oplus {\BZ} {\al}_{2}^{\vee} \oplus \dots \oplus {\BZ} {\al}_{r}^{\vee}.
\]
The generators ${\al}_{i}^{\vee}$ are called the simple coroots.

Now, let $z \in {\BC}^{\times}$, and ${\al}^{\vee} \in {\rl}^{\vee}$, then
\begin{equation}
z^{{\al}^{\vee}} \equiv {\ex} \left( \frac{ \log (z)}{2\pi \ii} {\al}^{\vee} \right) \in \bT
\label{eq:ztoal}
\end{equation}
is independent of the choice of the branch of the $ \log (z)$. This identifies the coroot lattice with the lattice of homomorphisms:
\begin{equation}
{\rl}^{\vee} = {\Hom} \big({\BC}^{\times}, \bT\big).
\label{eq:trrl}
\end{equation}
Using equation~(\ref{eq:ztoal}), we can parametrize $\bT$ by
\begin{equation*}
{\bz} = ( z_{1}, \dots , z_{r} ) \mapsto t_{\bz} = \prod_{i=1}^{r} z_{i}^{{\al}_{i}^{\vee}},
\end{equation*}
where $z_{i} \in {\BC}^{\times}$.
\subsection{The weights}

The dual lattice is called the lattice of weights
\begin{equation*}
{\Lambda} = {\Hom} \big(\bT, {\BC}^{\times}\big).
\end{equation*}
Let us represent element $t \in \bT$
as
\begin{equation}
t = {\ex}( {\bx})
\label{eq:tex}
\end{equation}
for some $\bf x \in \h$, which is defined, as we recall, up to an a shift by an element of the coroot lattice~$\rl^{\vee}$. Then for any $\lam \in \Lambda$, the value of the homomorphism ${\lam}$ on $t$, which we denote by $t^{\lam} \in {\BC}^{\times}$, can be computed as
\begin{equation}
t^{\lam} = {\rm e}^{2\pi \ii \lam ({\bx})},
\label{eq:tlam}
\end{equation}
where ${\lam}({\bf x})$ is a linear function of $\bf x$. In this way we view ${\lam}$ as an element of $\h^*$, so that
 $\Lambda \subset \h^*$. In order for (\ref{eq:tlam}) be independent of the choice of $\bf x$ in (\ref{eq:tex}), the value of $\lam$ on any element of the coroot lattice must be an integer
\begin{equation*}
 {\lam} \big({\rl}^{\vee}\big) \subset {\BZ},
\end{equation*}
which is another definition of the dual lattices. Let us fix the basis $({\lam}_{i})_{i=1}^{r}$ of $\Lambda$:
\[
{\Lambda} = {\BZ} {\lam}_{1} \oplus {\BZ} {\lam}_{2} \oplus \dots \oplus {\BZ} {\lam}_{r}
\]
dual to the basis $({\al}_{i}^{\vee})_{i=1}^{r}$ of simple coroots, so that
\begin{equation*}
{\lam}_{i} \big({\al}_{j}^{\vee}\big) = {\delta}_{ij}.
\end{equation*}

\subsubsection{The roots}

The torus $\bT$ acts on $\bG$ via the adjoint action
\[
\operatorname{Ad}\colon \ \bT \times \bG \to \bG, \qquad (t, g) \mapsto t^{-1} g t.
\]
Infinitesimally, it acts linearly on the Lie algebra
\begin{equation}
\operatorname{Ad}\colon \ \bT \longrightarrow \operatorname{Aut}({\g}), \qquad \operatorname{Ad}_{t}({\xi}) = \frac{{\rm d}}{{\rm d}s}\biggr\vert_{s=0} t^{-1} \exp (s {\xi} ) t \in \g
\label{eq:rootsi}
\end{equation}
for $\xi \in \g$, and finally, this defines an action of $\h$ on $\g$:
\begin{equation*}
\operatorname{ad} \colon \ {\h} \to \operatorname{End}({\g}), \qquad \operatorname{ad}_{\bf x} ({\xi} ) = [{\xi}, {\bf x}] \in \g
\end{equation*}
for ${\bf x}\in \h$, ${\xi} \in \g$.

This action gives us several structures: the root decomposition of $\g$:
\begin{equation*}
{\g} = {\h} \oplus \bigoplus_{{\al} \in R} {\BC} e_{\al},
\end{equation*}
where $R \subset \Lambda$ is a set of non-vanishing weights of the adjoint representation
\[
\operatorname{ad}_{t} (e_{\al}) = t^{\al} e_{\al}.
\]
These weights are called roots, and the lattice $\rl \subset \Lambda$ they generate is a sublattice of $\Lambda$, called the \emph{root lattice}. It has a basis $({\al}_{i})_{i=1}^{r}$ of \emph{positive simple roots}
\begin{equation*}
{\rl} = {\BZ}{\al}_{1} \oplus {\BZ}{\al}_{2} \oplus \dots \oplus {\BZ} {\al}_{r},
\end{equation*}
which allows us to define the \emph{Cartan matrix} $C^{\g}$ of $\g$:
\begin{equation}
C_{ij}^{\g} = {\al}_{i} \big({\al}_{j}^{\vee}\big),
\label{eq:cartm}
\end{equation}
which is non-degenerate. The additional requirement we impose on $\g$ is that it is \emph{simply-laced}, i.e., by an appropriate choice of integral bases one can make $C^{\g}$ symmetric:
\[
C_{ij}^{\g} = C_{ji}^{\g}.
\]
\subsubsection{The center}
The quotient
\begin{equation*}
Z = {\Lambda}/{\rl}
\end{equation*}
is an abelian group, which is a subgroup both of $\bT$ and $\bG$. In fact, it is the center of $\bG$. Clearly, the center does not act in the adjoint representation, so that in the equation~(\ref{eq:rootsi}) it is the quotient $\bT/Z$ which acts faithfully. Hence the root lattice
\[
{\rl} \subset {\Lambda} \subset {\h}^{*}
\]
can be also identified with the lattice of $\bT/Z$ characters
\begin{equation*}
{\rl} = {\Hom} \big(\bT/Z , {\BC}^{\times}\big).
\end{equation*}
Finally, the coweight lattice ${\Lambda}^{\vee} \subset \h$ is both
the integral dual to $\rl$, and the lattice of $\bT/Z$ cocharacters:
\begin{equation}
{\Lambda}^{\vee} = {\Hom} \big( {\BC}^{\times}, \bT/Z \big)
\label{eq:lamvee}
\end{equation}
with the basis $\big({\lam}_{i}^{\vee}\big)_{i=1}^{r}$, dual to
that of $\rl$:
\begin{equation*}
{\al}_{i}\big({\lam}_{j}^{\vee}\big) = {\delta}_{ij}.
\end{equation*}
The expression
\[
w^{{\lam}^{\vee}} = {\ex} \bigg( \frac{ \log (w)}{2\pi \ii} {\lam}^{\vee} \bigg) \in \bT
\]
\emph{does} depend on the choice of the branch of the logarithm $ \log (w)$, however, the ambiguity is in the multiplicative $Z$-valued factor, since for any ${\lambda}^{\vee} \in {\rl}^{\vee}$:
\[
{\ex} \big({\lam}^{\vee}\big) \in Z \subset \bT.
\]
Thus, for any $w \in {\BC}^{\times}$,
$w^{{\lam}^{\vee}}$ is well-defined as an element of $\bT/Z$, as is claimed by (\ref{eq:lamvee}).

{}The center $Z$, being a finite abelian group, is isomorphic to a product of cyclic groups
\begin{equation*}
Z \approx \bigotimes_{{\xi}=1}^{z_{\g}} {\BZ}/{\ell}_{\xi}{\BZ}
\end{equation*}
for some $z_{\g}$, which is equal to $0$, $1$, or $2$.

\begin{table}\centering\renewcommand{\arraystretch}{1.2}
\begin{tabular}{ l | c || c }
 \hline
 $\bf G$ & $z_{\g}$ & ${\ell}_{\xi}$, ${\xi} = 1, \dots, a_{\g}$ \\
 \hline
 $A_{r}$ & $1$ & $r+1$ \\
 $D_{2s}$ & $2$ & $2$, $2$ \\
 $D_{2s+1}$ & $1$ & $4$ \\
 $E_{6}$ & $1$ & $3$ \\
 $E_{7}$ & $1$ & $2$ \\
 $E_{8}$ & $0$ & \\
 \hline
\end{tabular}
\caption{The number of cyclic factors in $Z$ with their orders.}\label{table:ade}
\end{table}
Table \ref{table:ade} shows the values of $z_{\g}$'s and ${\ell}_{\xi}$'s for all simple simply-laced Lie groups.

Let ${\varpi}_{\xi} \in \bT$ be the generator of the ${\BZ}/{\ell}_{\xi}{\BZ}$ factor in $Z \subset \bT$. In other words,
\begin{equation}
 {\varpi}_{\xi}^{k} \neq 1\quad {\rm for}\ k = 1, 2, \dots , {\ell}_{\xi}-1 \qquad \text{and} \qquad
{\varpi}_{\xi}^{{\ell}_{\xi}} = 1 \in T.
\label{eq:wygen}
\end{equation}
Of course, the equation~(\ref{eq:wygen}) does not characterize ${\varpi}_{\xi}$ uniquely. Indeed, for any integer
$s_{\xi}$, which is mutually prime with ${\ell}_{\xi}$,
i.e., $\big(s_{\xi}, {\ell}_{\xi}\big) = 1$, the $Z$ element
$\widetilde{{\varpi}_{\xi}} = {\varpi}_{\xi}^{s_{\xi}}$ also generates
${\BZ}/{\ell}_{\xi}{\BZ}$.
We write
\begin{gather}
{\ex} \big( {\lam}_{i}^{\vee} \big) = \prod_{{\xi}=1}^{z_{\g}} {\varpi}_{\xi}^{l_{i{\xi}}}, \qquad i = 1, \dots , r, \nonumber\\
 {\varpi}_{\xi} = \prod_{j=1}^{r} {\ex} \bigg( \frac{w_{{\xi}j}}{\ell_{\xi}} {\al}_{j}^{\vee} \bigg)
\label{eq:elami}
\end{gather}
for some integers $l_{i{\xi}}, w_{{\xi}j} \in \BZ$, which are normalized
\begin{equation}
0 \leq l_{i{\xi}} < {\ell}_{{\xi}}, \qquad 0 \leq w_{{\xi}j} < {\ell}_{\xi}
\label{eq:wyalj}
\end{equation}
and
\[
\big( w_{{\xi}j}, {\ell}_{{\xi}} \big) = 1.
\]
Note that
\begin{equation}
{\lam}_{i}\big({\lam}_{j}^{\vee}\big) = (C^{\g})^{-1}_{ij} \in {\BQ}.
\label{eq:invca}
\end{equation}
By combining equations~\eqref{eq:invca} with \eqref{eq:elami}, we derive
\begin{equation}
(C^{\g})^{-1}_{ij} = \sum_{{\xi}=1}^{z_{\g}} \frac{l_{i{\xi}}w_{{\xi}j}}{{\ell}_{{\xi}}} + {\CalL}^{\g}_{ij},
\label{eq:cinv}
\end{equation}
where ${\CalL}^{\g}$ is some integral matrix. Note in passing that if we were to study the group $G$ over the field ${\BF}_{p^{n}}$ where $p$ divides ${\ell}_{y}$ then $C^{\g}_{ij}$
would correspond to an affine (or even double affine) root system.

\subsubsection{Killing metric}

Another structure we gain from the adjoint action of $T$ on $\g$ is the Killing metric on $\h$,
\begin{equation}
\langle x, x \rangle = \frac 1{h_{\g}} \tr _{\g} \operatorname{ad}_{x}^2,
\label{eq:killh}
\end{equation}
which identifies ${\h}$ with ${\h}^{*}$, ${\rl}^{\vee}$ with ${\rl}$, and ${\Lambda}^{\vee}$
with $\Lambda$, and the constant $h_{\g}$ is chosen so that
\begin{equation*}
\langle {\lam}_{i}, {\al}_{j} \rangle = {\delta}_{ij}, \qquad
\langle {\al}_{i}, {\al}_{j} \rangle = C_{ij}^{\g}, \qquad
\langle {\lam}_{i}, {\lam}_{j} \rangle = (C^{\g}_{ij})^{-1}.
\end{equation*}
\subsubsection{The Weyl group}

The torus $\bT$ has some symmetries within $\bG$. Namely,
for any $t \in \bT$ there are transformations of the form: \begin{equation}
t \mapsto g^{-1} t g
\label{eq:adjg}
\end{equation}
for some $g \in \bG$, for which
$g^{-1} t g \in \bT$ also. Such transformations form a group, which is called the \emph{normalizer $N(\bT)$ of $\bT$}. This group obviously contains $\bT$, since the transformation (\ref{eq:adjg}) with $g \in \bT$ doesn't even move $t$. It turns out that there are additional nontrivial transformations. These additional transformations form \emph{the Weyl group} $W({\g}) = N(\bT)/\bT$.

For $t \in \bT$ and $w \in W({\g})$ let us denote by $^{w}t$ the result of the application of $g_{w}$ representing~$w$ in $N(\bT) \subset \bG$:
\begin{equation*}
^{w}t = g_{w}^{-1} t g_{w}.
\end{equation*}
By taking the limit $t \to 1$ we get the action of $W({\g})$ on $\h$: for $\xi \in \h$
\begin{equation*}
{\xi} \mapsto {}^{w}{\xi} = \frac{{\rm d}}{{\rm d}s}\bigg\vert_{s=0} {} ^{w} \big( \exp s {\xi} \big).
\end{equation*}
It is clear that $W({\g})$
acts on $\h$ by the orthogonal transformations preserving the metric (\ref{eq:killh}).
The action of $W({\g})$ on $\h^*$ ($\h$)
preserves both the (co)root lattice and the (co)weight lattice.

{}The less trivial result is that the group $W({\g})$ is generated by reflections $r_{i}$ at the simple roots. The corresponding transformations on $\h$ and $\bT$ are
\begin{equation*}
^{r_{i}}x = x - {\al}_{i}(x) {\al}_{i}^{\vee}, \qquad ^{r_{i}}t = t \big( t^{-{\al}_{i}} \big)^{{\al}_{i}^{\vee}}.
\end{equation*}
The $W({\g})$-action on $\bT$ can be also described in the $z$-coordinates
\[
 ^{r_{i}}t_{\bf z} = t_{\bf z} \big(t_{\bf z}^{-{\al}_{i}}\big)^{{\al}_{i}^{\vee}} = t_{\tilde{\bf z}} ,
\]
 where
\begin{gather}
 {\tilde z}_{j} = z_{j}, \qquad j \neq i, \qquad
 {\tilde z}_{i} = z_{i} \prod_{k=1}^{r} z_{k}^{-C^{\g}_{ik}}.
\label{eq:riz}
\end{gather}
\subsection{Weyl group and the center}
Since $W({\g})$ acts on $\bT$ by similarity transformations in $\bG$, the center $Z \subset \bT \subset \bG$ is fixed by any $w \in W({\g})$.
In particular, $r_{i}$ preserves ${\varpi}_{\xi}$ for any $i$ and $\xi$, which is equivalent to
\begin{equation}
{\varpi}_{\xi}^{{\al}_{i}} = 1.
\label{eq:wyri}
\end{equation}
Substituting into equation~(\ref{eq:wyri}) the representation (\ref{eq:wyalj}) and recalling the definition
(\ref{eq:cartm}) of~$C^{\g}$, we get
\begin{equation*}
\frac 1{{\ell}_{\xi}} \sum_{j=1}^{r} w_{{\xi}j} C_{ji}^{\g} = {\fC}_{{\xi}i}^{\g} \in {\BZ}, \qquad {\xi} = 1, \dots , z_{\g}, \quad i = 1, \dots , r.
\end{equation*}
Combining this relation with equation~(\ref{eq:cinv}), we obtain
\begin{equation*}
{\delta}_{ij} = \sum_{{\xi}=1}^{z_{\g}} l_{i{\xi}} {\fC}_{{\xi}j}^{\g} + \sum_{k=1}^{r} {\CalL}^{\g}_{ik} C^{\g}_{kj}.
\end{equation*}
On the other hand, using the equation~(\ref{eq:elami}), and the relations $\sum_{j=1}^{r} C^{\g}_{ij}{\lam}_{j}^{\vee} = {\al}_{i}^{\vee}$, and
${{\ex}\big({\al}_{i}^{\vee}\big) = 1}$ for any $i$, we derive
\[
\prod_{j=1}^{r} \prod_{{\xi}=1}^{z_{\g}} {\varpi}_{\xi}^{C_{ij}^{\g} l_{j\xi}} = 1,
\]
which implies
\begin{equation*}
\frac 1{{\ell}_{\xi}} \sum_{j=1}^{z_{\g}} C_{ij}^{\g} l_{j\xi} = {\fc}_{i\xi}^{\g} \in {\BZ}.
\end{equation*}
\subsubsection{Langlands dual, adjoint, and conformal groups}\label{se:conformalgroup}
The simple Lie group $\bG/Z$ has a trivial center, but it is not simply-connected. This group
is also denoted by $\bG^{\text{ad}}$, since it is represented faithfully in the adjoint representation $\g$. The maximal torus $\bT^{\text{ad}}$ of $\bG^{\text{ad}}$, is the quotient $\bT/Z = {\h}/ {\rl}$.
Also, since the lattices of weights and roots of $\bG$ and $\bG^{\text{ad}}$ are dual to each other, these groups are Langlands duals, $^{\rm L}\bG = \bG^{\text{ad}}$.

The group $\bG^{\text{ad}}$ is not very convenient to work with. For one thing, the center $Z$ looks differently for different groups.
One defines the \emph{conformal extension} ${\rm C}\bG $ of $\bG$ as the
group (see~\cite{Morgan:2000math, Levin:2010mz, Levin:2010ve} for recent applications in a
related context),
\begin{equation*}
{\mathrm C}\bG = ( {\cc} \times \bG ) / Z,
\end{equation*}
where
\begin{equation*}
{\cc} = ({\BC}^{\times})^{z_{\g}},
\end{equation*}
where the center $Z$ acts on $\bG$ in the usual way, and on $\cc$ via some character
\begin{equation*}
{\chi} \in {\Hom}(Z, {\cc}).
\end{equation*}
For example, we can choose
\[
{\chi}({\varpi}_{x}) = \big({\ze}_{{\ell}_{y}}^{{\de}_{x,y}}\big)_{y=1}^{z_{\g}},
\]
where
${\ze}_{l}$ is the primitive $l$-th root of unity
\begin{equation*}
{\ze}_{l} = {\rm e}^{2\pi \frac{\ii}{ l} }.
\end{equation*}
The elements of the group ${\rm C}\bG $ are the classes of pairs $[({\fb}; g)]$, where ${\fb} \in \cc$ is the $z_{\g}$-tuple of
non-zero complex numbers,
${\fb} = ({\fb}_{1}, \dots , {\fb}_{z_{\g}})$, and $g \in \bG$, under the equivalence
$({\fb}; g) \sim ({}^{x}{\fb}; {\varpi}_{x}g )$, where
\[
^{x}{\fb} = \big( {\fb}_{y} {\ze}_{{\ell}_{x}}^{-{\de}_{x,y}} \big)_{y=1}^{z_{\g}}
\]
with the multiplication law
\[
[({\fb}_{1}; g_{1})] \cdot
[({\fb}_{2}; g_{2})] =
[({\fb}_{1}{\fb}_{2}; g_{1}g_{2})].
\]
The maximal torus ${\rm C}\bT$ of ${\rm C}\bG $ has rank $r+z_{\g}$.
The center ${\rm C}Z$ of ${\rm C}\bG $ is the subgroup of ${\rm C}\bG $ which consists of
the equivalence classes containing $({\fb}; 1)$, with
$\fb \in {\cc}$. Clearly, ${\rm C}Z \approx {\cc}$. In this way, the centers of the conformal extensions look the same for all groups of equal $z_{\g}$.

The lattice
\begin{equation}
{\rm C}{\rl}^{\vee} = {\Hom}({\BC}^{\times}, {\rm C}\bT)
\label{eq:trlvee}
\end{equation}
is an extension of ${\rl}^{\vee}$ by the rank
$z_{\g}$ lattice with the generators ${\be}_{\xi}^{\vee}$, ${\xi} = 1, \dots, z_{\g}$:
\begin{equation*}
{\rm C}{\rl}^{\vee} = \bigoplus_{i=1}^{r} {\BZ}{\al}_{i}^{\vee} \oplus \bigoplus_{{\xi}=1}^{z_{\g}} {\BZ} {\be}_{\xi}^{\vee}.
\end{equation*}
The root subspace decomposition of ${\fc\g} = {\rm Lie}({\rm C}\bG ) = {\g} \oplus {\BC}^{z_{\g}}$ is easy to compute.
There are $r$ simple roots which we denote by ${\al}_{i}$. They act on ${\fc}{\h} = {\rm Lie}({\rm C}\bT)$ as follows:
\begin{gather}
{\al}_{i} \big({\al}_{j}^{\vee}\big) = C_{ji}^{\g}, \qquad
{\al}_{i} \big({\be}_{\xi}^{\vee}\big) = {\fC}_{{\xi}i}^{\g},
\label{eq:albe}
\end{gather}
so that if we define
\begin{equation}
K_{\xi} = -{\ell}_{\xi} {\be}_{\xi}^{\vee} + \sum_{j=1}^{r} w_{{\xi}j}{\al}_{j}^{\vee},
\label{eq:kx}
\end{equation}
then
\begin{equation}
{\al}_{i}(K_{\xi}) = 0
\label{eq:cengen}
\end{equation}
for any $i$, $\xi$.
The meaning of (\ref{eq:kx}) is the following. The lattice of cocharacters
of the maximal torus of $\mathrm{C}\times \bG$
is the direct sum of the lattice ${\rl}^{\vee}$ of coroots of $G$
and the lattice
\begin{equation*}
 {\Hom}({\BC}^{\times}, {\rm C} \times \bT) = L \oplus {\rl}^{\vee}, \qquad
 L = {\pi}_{1}( {\rm C} ) = \bigoplus_{{\eta}=1}^{z_{\g}} {\BZ} K_{\eta},
\end{equation*}
so that a generic element of $\mathrm{C} \times \bT$ can be represented by
\begin{equation}
{\hat t}_{\fb, {\bf g}} = \prod_{{\xi}=1}^{z_{\g}} {\fb}_{\xi}^{K_{\xi}} \times \prod_{i=1}^{r}
g_{i}^{{\al}_{i}^{\vee}}
\label{eq:elct}
\end{equation}
(again, in writing (\ref{eq:elct}) we assume normalization ${\ex}(K_{\xi})=1 \in $ C.)
The action of $Z$ on $\mathrm{C} \times \bG$ translates to the action on ${\fb}, {\bf g}$:
\begin{equation*}
{\varpi}_{\xi}\colon \ {\hat t}_{{\fb}, {\bf g}} \mapsto {\hat t}_{{\fb}, {\bf g}} {\zeta}_{{\ell}_{\xi}}^{- K_{\xi} + \sum_{i=1}^{r} w_{{\xi}i}{\al}_{i}^{\vee}}.
\end{equation*}
The quotient ${\rm C}\bT = ({\rm C} \times \bT)/Z$ is coordinatized by
\begin{gather}
 [ ({\fb}; t_{\bz} ) ] = {\ct}_{{\fu}, {\bz}} = \prod_{{\xi}=1}^{z_{\g}} {\fu}_{\xi}^{{\beta}_{\xi}^{\vee}} \prod_{i=1}^{r} z_{i}^{{\alc}_{i}^{\vee}} \in {\rm C} \bT , \qquad
 {\fu}_{\xi} = {\fb}_{\xi}^{-{\ell}_{\xi}} , \qquad z_{i} = g_{i} \prod_{{\xi}=1}^{z_{\g}} {\fb}_{\xi}^{w_{{\xi}i}}
\label{eq:paramq}
\end{gather}
in agreement with (\ref{eq:trlvee}). Simply put, ${\be}_{\xi}^{\vee} \in {\Hom}({\BC}^{\times}, {\rm C}\bT)$, ${\ell}_{\xi}{\be}_{\xi}^{\vee} \in {\Hom}({\BC}^{\times}, {\rm C} \times \bT)$, and ${\be}^{\vee}_{\xi} \notin {\Hom}({\BC}^{\times}, {\rm C} \times \bT)$.

Clearly, there are other choices of the $Z$-invariant coordinates, which differ by multiplication of $z_{i}$ by any function of ${\fu}_{\xi}$'s.

Comparing the equations~(\ref{eq:paramq}) and (\ref{eq:elct}), we arrive at (\ref{eq:kx}).
The equation~(\ref{eq:cengen}) simply reflects the fact that ${\rm C}$ acts trivially
in the adjoint representation. The lattice ${\rm C}{\rl}$ generated by~${\al}_{i}$ is isomorphic to the root lattice ${\rl}$.

The weight lattice $\widetilde{{\rm C}\Lambda}$ of $\mathrm{C} \times \bG$ is the direct sum of the weight lattice of ${\rm C}$ and that of $\bG$:
\begin{equation*}
\widetilde{{\rm C}\Lambda} = {\Hom} ({\rm C} \times \bT , {\BC}^{\times}) = L^{\vee} \oplus {\Lambda} .
\end{equation*}
The weight lattice ${\rm C}\Lambda$ of ${\rm C}\bG $ is a sublattice of $\widetilde{{\rm C}\Lambda}$
which consists of the weights which are trivial on the elements of
$\mathrm{C} \times \bT$ of the form
\[
{\zeta}_{{\ell}_{\xi}}^{- K_{\xi} + \sum_{i=1}^{r} w_{{\xi}i}{\al}_{i}^{\vee}}
\]
for any $\xi$. In the $\widetilde{{\rm C}\Lambda}$-basis $({\mu}_{\xi}; {\lam}_{i})$, ${\xi}=1, \dots, z_{\g}$, $i = 1, \dots , r$ such that
\begin{alignat*}{3}
&{\mu}_{\xi}(K_{\eta}) = {\de}_{{\xi},{\eta}}, \qquad &&{\mu}_{\xi} \big( {\al}_{i}^{\vee}\big) = 0, &\nonumber\\
&{\lam}_{i} (K_{\eta}) = 0, \qquad &&{\lam}_{i}\big({\al}_{j}^{\vee}\big) = {\de}_{ij}&
\end{alignat*}
the lattice ${\rm C}\Lambda$ is spanned by
\begin{equation*}
{\cmu}_{\xi} = - {\ell}_{\xi}{\mu}_{\xi}, \qquad
{\cla}_{i} = {\lam}_{i} + \sum_{\xi} w_{{\xi}i} {\mu}_{\xi}.
\end{equation*}
An easy computation gives
\begin{alignat*}{3}
& {\cmu}_{\xi}\big({\beta}_{\eta}^{\vee}\big) = {\de}_{\xi\eta}, \qquad &&{\cla}_{i}\big({\al}_{j}^{\vee}\big) = {\de}_{ij}, &\\
& {\cmu}_{\xi}\big({\al}_{i}^{\vee}\big) = 0, \qquad && {\cla}_{i}\big({\be}_{\eta}^{\vee} \big) = 0. &
\end{alignat*}
Finally, the lattice ${\rm C}\Lambda^{\vee}$ of coweights of ${\rm C}\bG $, which is dual to the root lattice ${\rm C}{\rl}$, is generated by the fundamental coweights ${\cla}_{i}^{\vee}$, which obey
\begin{equation}
{\al}_{i}\big({\cla}_{j}^{\vee}\big) = {\de}_{ij}.
\label{eq:dualcow}
\end{equation}
The equations (\ref{eq:dualcow}) define ${\cla}_{i}^{\vee}$ up to the shifts by
the integer multiples of $K_{\xi}$. We choose the representative
\begin{equation*}
{\cla}_{i}^{\vee} = \sum_{j=1}^{r} {\CalL}_{ij}^{\g} {\al}_{j}^{\vee} + \sum_{{\xi}=1}^{z_{\g}}
l_{i{\xi}} {\be}_{\xi}^{\vee}.
\end{equation*}
Shifting ${\cla}_{i}^{\vee}$ by ${\rm d}_{\xi} K_{\xi}$ would change the coefficient
$l_{i{\xi}} \mapsto l_{i{\xi}}+{\ell}_{\xi} {\rm d}_{\xi}$.

As we see, the coweights ${\cla}_{i}^{\vee}$ are the integral linear combinations of
the coroots ${\al}^{\vee}$ and ${\be}^{\vee}$, as if ${\rm C}\bG$ were the simple simply-connected
group with the trivial center.

\subsubsection[The $D$-homomorphism]{The $\boldsymbol{D}$-homomorphism}

By construction of the conformal group, there is a homomorphism
\begin{gather*}
 D \colon \ {\rm C}\bG \longrightarrow {\rm C}/Z \approx {\rm C}, \\
 D [ ( {\fb} ; g ) ] = \big( {\fb}_{\xi}^{{\ell}_{\xi}} \big)_{{\xi}=1}^{z_{\g}}.
\end{gather*}
In terms of the weights, the restriction of the homomorphism $D$ onto
$C\bT$ is given by
\begin{equation*}
D\big( {\ct}_{{\fu}, {\bz}} \big) =
\big( {\ct}_{{\fu}, {\bz}}^{{\cmu}_{\xi}} \big)_{{\xi}=1}^{z_{\g}} = ({\fu}_{\xi})_{{\xi}=1}^{z_{\g}}.
\end{equation*}

\subsubsection{Twisted Weyl group action}

Now we can define the twisted action of Weyl group $W({\g})$. It is nothing but the
natural action of $W({\g})$ on ${\rm C}\bT$. However, we shall encounter a somewhat redundant parametrization of ${\rm C}\bT$, by $\bT \times \bT$:
\begin{equation}
{\bg}_{{\bq}, {\y}} = \prod_{i=1}^{r} {\bq}_{i}^{-{\cla}_{i}^{\vee}}
{\y}_{i}^{{\alc}_{i}^{\vee}},
\label{eq:param}
\end{equation}
which is equal to ${\ct}_{{\fu}, {\bz}}$ with
\begin{equation*}
{\fu}_{\xi} = \prod_{j} {\bq}_{j}^{-l_{j\xi}} , \qquad z_{i} = {\y}_{i} \prod_{j} {\bq}_{j}^{-{\CalL}_{ji}}.
\end{equation*}
Using (\ref{eq:albe}), we compute the Weyl group $W({\g})$ action on $C\bT$: under the reflection $r_i$ the group element ${\bg}_{{\bq},{\y}}$ maps to
$^{r_{i}}{\bg}_{{\bq}, {\y}} = {\bg}_{{\bq}, {\tilde{\y}}}$, where
\begin{gather}
 {\tilde \y}_{j} = {\y}_{j}, \qquad j \neq i, \qquad
 {\tilde \y}_{i} = {\bq}_{i}{\y}_{i} \prod_{j} {\y}_{j}^{-C_{ij}^{\g}}.
\label{eq:twwac}
\end{gather}
The homomorphism $D$ is $W({\g})$-invariant
\[
D ( {\bg}_{{\bq}, {\y}} ) = D ( {\bg}_{{\bq}, {\tilde{\y}}} ) = \Bigg( \prod_{j=1}^{r} {\bq}_{j}^{-l_{j{\xi}}} \Bigg)_{{\xi}=1}^{z_{\g}}.
\]

\subsection{Affine Lie algebras}\label{se:affineLie}

The affine Lie algebras $\hat\g$ show up in the solution of the theories of the class II.
In preparing this section we consulted with~\cite{Bardakci:1970nb, Kac:1984}.
Given a simple Lie group $\bG$ with its Lie algebra $\g$ one defines the loop group ${\rm L}\bG$ and the loop algebra ${\rm L}\g$ of (suitably defined, analytic, formal, polynomial) maps of the neighborhood of a circle in $\BC^{\times}$ into $\bG$ and $\g$, respectively,
\begin{equation}
{\rm L}\bG = {\rm Maps}({\BC}^{\times}, \bG), \qquad {\rm L}{\g} = {\rm Maps} ({\BC}^{\times}, {\g}) ,
\label{eq:loopg}
\end{equation}
with the point-wise product and Lie brackets, respectively. Then one defines the central extension $\widetilde{{\rm L}{\g}} = {\rm L}{\g} \oplus {\BC}K$
by
\[
[ f_1(t) \oplus a_1 K , f_2(t) \oplus a_2 K ] = [ f_1(t), f_2(t)] \oplus \oint_{{\BS}^{1}}
\langle {\rm d}f_1 (t), f_2(t) \rangle K
\]
 and the additional (non-central) extension $\hat\g$ of $\widetilde{{\rm L}{\g}} $ by $\BC$, which acts on ${\rm L}\g$ by the infinitesimal rotation of ${\BC}^{\times}$ in (\ref{eq:loopg}). The elements of $\hat\g$ can be represented as the $\g$-valued first-order differential operators on $\BS^1$ plus a constant:
\[
 {\tau} {\rm d} + f(t) \oplus a K \in \hat\g , \qquad {\tau}, a \in \BC, f(t) \in {\g}
\]
 with the commutation relations
\begin{align}
& [ {\tau}_1 {\rm d} + f_1(t) \oplus a_1 K , {\tau}_2 {\rm d} + f_2(t) \oplus a_2 K ] \nonumber\\
&\qquad= 0 \cdot {\rm d} + t \left( {\tau}_{1} f_{2}^{\prime}(t) - {\tau}_{2} f_{1}^{\prime}(t) \right) + [ f_{1}(t), f_{2}(t) ] \oplus \oint_{{\BS}^{1}} \langle f_{1} (t), {\rm d} f_{2}(t) \rangle K.\label{eq:affkm}
\end{align}
 Analogously, one defines the central extension $\widetilde{{\rm L}\bG}$ which is a non-trivial ${\BC}^{\times}$-bundle over ${\rm L}\bG$
(this is analogous to the construction of the conformal group ${\rm C}\bG $ in the previous section), and then the extension $\hat{\bG}$ of $\widetilde{{\rm L}\bG}$ by ${\BC}^{\times}$ which acts by rotation of ${\BC}^{\times}$ in (\ref{eq:loopg}).

The algebra $\hat\g$ can also be defined by the general construction of the generalized Kac--Moody algebras associated with the Cartan matrix $C_{ij}^{\hat\g}$. The affine Cartan matrix has exactly one eigenvector with the eigenvalue zero:
\begin{equation}
\sum_{j=0}^{r} C_{ij}^{\hat\g} a_{j} = 0.
\label{eq:nule}
\end{equation}
The Cartan subalgebra $\hat\h$ of the corresponding affine Lie algebra $\hat\g$ with the underlying finite-dimensional simply-laced Lie algebra $\g$ of rank $r$ is the complex vector space of
dimension $r+2$. The dual space $\hat\h^{*}$ contains the root lattice ${\hat\rl} \subset \hat\h^{*}$, which is generated by the simple roots
${\hat\al}_{i}$, $i=0, \dots , r$, of which the simple roots ${\hat\al}_{i}$ with $i > 0$ generate the root lattice of $\g$.
 Likewise, the Cartan subalgebra $\hat\h$ contains the coroot lattice ${\hat\rl}^{\vee} \subset \hat\h$, generated by the simple coroots ${\al}_{i}^{\vee}$, which obey:
\begin{equation*}
 {\hat\al}_{i}\big( {\hat\al}_{j}^{\vee} \big) = C_{ij}^{\hat\g}.
\end{equation*}
The following linear combination of the simple roots:
\begin{equation}
{\delta} = \sum_{i=0}^{r-1} a_{i} {\hat\al}_{i} \in \hat\h^{*}
\label{eq:imroot}
\end{equation}
is called the imaginary root. It annihilates the simple coroots, cf.~(\ref{eq:nule}),
\[
 {\delta} \big( {\hat\al}_{i}^{\vee} \big) = 0 , \qquad i = 0, \dots , r-1.
\]
The analogous linear combination of coroots
\begin{equation}
K = \sum_{i=0}^{r-1} a_{i} {\hat\al}_{i}^{\vee} \in \hat\h
\label{eq:cenele}
\end{equation}
obeys
\begin{equation}
 {\hat\al}_{i} ( K ) = 0 , \qquad i = 0 , \dots , r
\label{eq:cent}
\end{equation}
and generates the center of the affine Kac--Moody algebra
since (\ref{eq:cent}) implies it commutes with everything in $\hat\g$.
Then
\begin{equation*}
{\hat\rl}^{\vee}_{\BC} = {\h} \oplus {\BC}K , \qquad {\hat \rl}_{\BC} = {\h}^{*} \oplus {\BC}{\delta},
\end{equation*}
where $\h$ and $\h^*$ are the Cartan subalgebra and its dual space of the corresponding finite-dimensional simply-laced Lie algebra
$\g$, respectively.
In order to generate $\hat\h$ and $\hat\h^{*}$ as the vector spaces over $\BC$ we need to add one more generator in addition to the simple roots and the simple coroots, respectively:
\begin{equation*}
{\hat\h}^{*} = {\hat\rl}_{\BC} \oplus {\BC}{\lambda}_{0} , \qquad {\hat\h} = {\hat\rl}^{\vee}_{\BC} \oplus {\BC} {\delta}^{\vee},
\end{equation*}
which obey:
\begin{equation*}
 {\hat\lam}_{0} \big( {\hat\al}_{i}^{\vee} \big) = {\hat\al}_{i} \big( {\delta}^{\vee} \big) = {\delta}_{i0}.
\end{equation*}
The generator ${\de}^{\vee}$ is equal to the generator of the infinitesimal loop rotation $d$ we used in (\ref{eq:affkm}).

The weight lattice ${\hat\Lambda}$ of $\hat\g$ is generated by the fundamental weights ${\lam}_{i} \in \hat\h^*$, $i = 0, 1, \dots, r$:
\begin{equation*}
{\hat\Lambda} = {\Lambda} \oplus {\BZ}{\lam}_{0} \subset \hat\h^{*},
\end{equation*}
which obey the following basic relations:
\begin{equation*}
 {\hat\lam}_{i} \big( {\hat\al}_{j}^{\vee} \big) = {\delta}_{ij}, \qquad
{\hat\lam}_{i} \big( {\delta}^{\vee} \big) = 0, \qquad {\delta}\big({\delta}^{\vee} \big) = a_{0} = 1.
\end{equation*}
The fundamental coweights ${\hat\lam}_{i}^{\vee} \in \hat\h$ obey
\[
 {\hat\al}_{j} \big( {\hat\lam}_{i}^{\vee} \big) = {\delta}_{ij}
\]
and form the coweight lattice
\[
{\hat\Lambda}^{\vee} = \bigoplus_{i=0}^{r} {\BZ} {\lam}_{i}^{\vee} \subset {\h}.
\]
The level of a weight ${\hat\lam} \in \hat\Lambda$ is defined as $k = {\hat\lam}( K )$ so that the level of the $i$-th fundamental weight is equal to
\begin{equation*}
{\hat\lam}_{i}( K ) = a_i = {\delta} \big({\hat\lam}^{\vee}_{i}\big).
\end{equation*}
Note that ${\hat\rl}^{\vee}_{\BC}$ is the Cartan subalgebra of the central extension $\widehat{{\rm L}{\g}}$ of the loop algebra $L{\g}$. Adding ${\delta}^{\vee}$ makes up the Cartan of the affine Kac--Moody algebra of $\g$, which is an extension of~$\widehat{{\rm L}{\g}}$ by the operator of the infinitesimal loop rotation, the zero mode of the Virasoro generator. In physics literature the more common notation for ${\delta}^{\vee}$ is $L_0$.

 Also we notice that
\begin{equation*}
 {\lam}_{i} \big({\lam}_{j}^{\vee}\big) = (C^{\g})^{-1}_{ij}, \qquad i,j = 1, \dots, r,
\end{equation*}
hence
\begin{equation*}
{\hat\lam}_{i} \big({\hat\lam}_{j}^{\vee}\big) =
\begin{pmatrix}
 0 & 0 \\
 0 & (C^{\g})^{-1}
\end{pmatrix}_{ij}
\end{equation*}
A useful relation obtained from identity operator $|\Lambda_i \rangle
\big\langle \alpha_i^\vee\big| + |\delta \rangle \big\langle \delta^{\vee}\big|$ is
\begin{equation*}
\sum_{i=0}^{r} {\hat\lam}_{i}\big({\hat\lam}_{j}^{\vee}\big) {\hat\al}_{k} \big({\hat\al}_{i}^{\vee}\big) =
{\hat\al}_{k}\big({\hat\lam}_{j}^{\vee}\big) - {\delta}\big({\hat\lam}_{j}^{\vee}\big) {\al}_{k}\big({\delta}^{\vee}\big)
 = \delta_{jk} - a_{j} \delta_{k0}.
\end{equation*}

Let $\lambda \in \h^{*}$ for $\hat\lam \in \hat\h^{*}$ denote the image
 of the projection $\hat\h^{*} \to \h^{*}$, i.e., forgetting components
 spanned by $\delta$ and $\hat\lam_0$. Then
 \begin{equation*}
 \hat\lam_i^\vee= a_{i} \delta^{\vee} + \lam_i^{\vee},\qquad
 \hat\lam_j = a_{j} \hat\lam_0 + \lam_j.
 \end{equation*}

\section{ADE Cartan matrices, roots and weights}
Our ADE conventions are summarized in the table in Appendix~\ref{se:mckay}.

\subsection[$A_r$ series]{$\boldsymbol{A_r}$ series}

 The Cartan matrix $C_{ij}^{\hat A_{r}}$ with $i, j = 0, \dots, r$
 \begin{equation*}
 C^{\hat A_{r}} =
 \begin{pmatrix}
 \hphantom{-}2 & -1 & \hphantom{-}0 & \dots &\dots & -1 \\
 -1 & \hphantom{-}2 & - 1 & \dots &\dots & \hphantom{-}0\\
 \hphantom{-}0 & -1 & \hphantom{-}2 & -1 &\dots & \hphantom{-}0 \\
 \dots& \dots&\dots&\dots &\dots & \dots \\
 \hphantom{-}0 & \hphantom{-}0 &\dots & \dots& \hphantom{-}2 & - 1\\
 -1 & \hphantom{-}0 & \dots &\dots & -1 & \hphantom{-}2
 \end{pmatrix}
 \end{equation*}
The affine simple roots are
\[
{\hat\al}_0 = \delta - \theta, \quad\hat\al_1 = e_1 - e_2,\quad
\hat\al_2 = e_2 -e_3, \quad \dots, \quad \hat\al_{r} = e_{r}
- e_{r+1},
\]
where $\theta = e_1 - e_{r+1}$ is the highest root.
The Dynkin marks are $a_i =
1$, $i = 0,\dots, r$.
The fundamental weights are
\begin{equation*}
\begin{aligned}
&\hat\lam_0, \\
&\hat\lam_1 = \hat\lam_0 + e_1 - \frac 1r {\bf e},\\
&\hat\lam_2 = \hat\lam_0 + e_1 + e_2 - \frac 2r {\bf e},\\
& \cdots\cdots\cdots\cdots\cdots\cdots\cdots\cdots\cdots\cdots \\
&\hat\lam_{r} = \hat\lam_0 + e_1 + e_2 + \dots + e_{r} - \frac{r}{r+1} {\bf e},
\end{aligned}
\end{equation*}
where
\[
{\bf e} = e_1 + \dots + e_{r}.
\]
The inverse Cartan matrix of the block $i, j = 1\dots r$, i.e., the inverse Cartan matrix $\big(C^{A_{r}}\big)^{-1}$ of ${\g} = A_{r}$ is
\begin{equation*}
{\lam}_{i}\big({\lam}_{j}^{\vee}\big) = \big(C^{A_{r}}\big)^{-1}_{ij} =
 {\rm max}(i,j) - \frac{i j}{r+1}, \qquad i,j = 1, \dots, r.
\end{equation*}

\subsection[$D_r$ series]{$\boldsymbol{D_r}$ series}\label{sec:Dr-conventions}

In the standard basis $\{e_{i}\mid i = 1,\dots , r \}$ the simple roots of $D_{r}$ are
\begin{equation*}
\al_i = e_i - e_{i+1}, \qquad i = 1, \dots , r-1 ,\qquad
\al_r = e_{r-1} + e_r
\end{equation*}
and the fundamental weights are
\begin{gather*}
 \lam_i = e_1 + e_2 + \dots +e_{i}, \qquad i = 1 \dots r-2, \\
 \lam_{r-1} = \frac 1 2 (e_1 + e_2 + \dots + e_{r-1} - e_r),\\
 \lam_{r} = \frac 1 2 ( e_1 + e_2 + \dots + e_{r-1} + e_r).
\end{gather*}
The highest root is $\theta =\sum_{i=1}^{r} \alpha_i a_i = e_1
+ e_2$.

In the basis $(e_{i})$ the root (coroot) lattice of $D_{r}$ is given by
\begin{equation*}
Q = Q^{\vee} = \{\mathbf{n} \in \BZ^{r}\mid | {\bf n} | \in
2 \BZ\}.
\end{equation*}
 The weight (coweight) lattice of $D_{r}$ is
 \begin{equation*}
 \oplus_{i=1}^{r} \BZ \lambda_i = \bigcup_{{\ve} = 0, \frac 12}
( {\BZ} + {\ve} )^{r}.
 \end{equation*}
\newcommand{\fra}[2]{#1/#2}
The inverse Cartan matrix is
\begin{equation*}
 \big(C^{D_{r}}\big)^{-1} = \Vert \lam_{i} \big( \lam_{j}^{\vee}\big) \Vert_{i,j}^{r}=
 \begin{pmatrix}
1 & 1 & 1 & 1 &\cdots & 1 & \frac 12 & \frac 12 \vspace{1mm}\\
1 & 2 & 2 & 2 &\cdots & 2 & \frac 22 & \frac 22\vspace{1mm}\\
1 & 2 & 3 & 3 &\cdots & 3 & \frac 32 & \frac 32 \vspace{1mm}\\
1 & 2 & 3 & 4 &\cdots & 4 & \frac 42 & \frac 42 \vspace{1mm}\\
\cdots&\cdots&\cdots&\cdots&\cdots&\cdots & \cdots & \cdots \vspace{1mm}\\
1 & 2 & 3 & 4 & \cdots& r-2 & \frac{r-2}{2} & \frac{r-2}{2}
\vspace{1mm}\\
\frac 12 & \frac 22 & \frac 32 & \frac 42 & \cdots &
\frac{r-2}{2} & \frac r4 & \frac{r-2}{4} \vspace{1mm}\\
 \frac 12 & \frac 22 & \frac 32 & \frac 42 & \cdots &
\frac{r-2}{2} & \frac{r-2}{4} & \frac {r}4
 \end{pmatrix}.
\end{equation*}
The basis of the affine Cartan dual is $\{\lambda_0, \delta , e_1, \dots,
e_{r}\}$. The simple roots are $\hat\al_0 = \delta - \theta$ and the roots
$\al_i$ of $D_{r}$. The fundamental weights
are
\begin{equation*}
\hat\lam_0, \qquad
\hat\lam_{i} = a_i \hat\lam_0 +{\lam}_{i}, \quad i = 1,
 \dots , r,
\end{equation*}
where $\lam_{i}$ denote the fundamental weights of
$D_{r}$, and $a_{i}$ are the Dynkin labels:
\begin{equation*}
 (a_{0}, \dots, a_{r}) = (1,1,2,\dots,2,1,1).
\end{equation*}

 \subsection[$E_6$]{$\boldsymbol{E_6}$}
 The Cartan matrix of $E_6$ is
\begin{equation*}
C^{E_6} = \begin{pmatrix}
 \hphantom{-}2 & \hphantom{-}0 & -1 & \hphantom{-}0 & \hphantom{-}0 & \hphantom{-}0 \\
 \hphantom{-}0 & \hphantom{-}2 & \hphantom{-}0 & -1 & \hphantom{-}0 & \hphantom{-}0 \\
 -1 & \hphantom{-}0 & \hphantom{-}2 & -1 & \hphantom{-}0 & \hphantom{-}0 \\
 \hphantom{-}0 & -1 & -1 & \hphantom{-}2 & -1 & \hphantom{-}0 \\
 \hphantom{-}0 & \hphantom{-}0 & \hphantom{-}0 & -1 & \hphantom{-}2 & -1 \\
 \hphantom{-}0 & \hphantom{-}0 & \hphantom{-}0 & \hphantom{-}0 & -1 & \hphantom{-}2
\end{pmatrix}
\end{equation*}
and the inverse is
\begin{equation*}
\big(C^{E_6}\big)^{-1} =\begin{pmatrix}
 \frac{4}{3} & 1 & \frac{5}{3} & 2 & \frac{4}{3} & \frac{2}{3} \\
 1 & 2 & 2 & 3 & 2 & 1 \\
 \frac{5}{3} & 2 & \frac{10}{3} & 4 & \frac{8}{3} & \frac{4}{3} \\
 2 & 3 & 4 & 6 & 4 & 2 \\
 \frac{4}{3} & 2 & \frac{8}{3} & 4 & \frac{10}{3} & \frac{5}{3} \\
 \frac{2}{3} & 1 & \frac{4}{3} & 2 & \frac{5}{3} & \frac{4}{3}
 \end{pmatrix}
\end{equation*}
In the affine Cartan matrix the affine node ``0'' of $\hat E_{6}$ connects to the node ``2'' of
$E_{6}$. The Dynkin marks
\begin{equation*}
 (a_{0}, \dots, a_{6}) = (1,1,2,2,3,2,1).
\end{equation*}

 \subsection[$E_7$]{$\boldsymbol{E_7}$}
The Cartan matrix of $E_7$ is
\begin{equation*}
C^{E_7} = \begin{pmatrix}
 \hphantom{-}2 & \hphantom{-}0 & -1 & \hphantom{-}0 & \hphantom{-}0 & \hphantom{-}0 & \hphantom{-}0 \\
 \hphantom{-}0 & \hphantom{-}2 & \hphantom{-}0 & -1 & \hphantom{-}0 & \hphantom{-}0 & \hphantom{-}0 \\
 -1 & \hphantom{-}0 & \hphantom{-}2 & -1 & \hphantom{-}0 & \hphantom{-}0 & \hphantom{-}0 \\
 \hphantom{-}0 & -1 & -1 & \hphantom{-}2 & -1 & \hphantom{-}0 & \hphantom{-}0 \\
 \hphantom{-}0 & \hphantom{-}0 & \hphantom{-}0 & -1 & \hphantom{-}2 & -1 & \hphantom{-}0 \\
 \hphantom{-}0 & \hphantom{-}0 & \hphantom{-}0 & \hphantom{-}0 & -1 & \hphantom{-}2 & -1 \\
 \hphantom{-}0 & \hphantom{-}0 & \hphantom{-}0 & \hphantom{-}0 & \hphantom{-}0 & -1 & \hphantom{-}2
\end{pmatrix}
\end{equation*}
and the inverse is
\begin{equation*}
\big(C^{E_7}\big)^{-1} = \begin{pmatrix}
 2 & 2 & 3 & 4 & 3 & 2 & 1 \\
 2 & \frac{7}{2} & 4 & 6 & \frac{9}{2} & 3 & \frac{3}{2} \\
 3 & 4 & 6 & 8 & 6 & 4 & 2 \\
 4 & 6 & 8 & 12 & 9 & 6 & 3 \\
 3 & \frac{9}{2} & 6 & 9 & \frac{15}{2} & 5 & \frac{5}{2} \\
 2 & 3 & 4 & 6 & 5 & 4 & 2 \\
 1 & \frac{3}{2} & 2 & 3 & \frac{5}{2} & 2 & \frac{3}{2}
 \end{pmatrix}.
\end{equation*}
In the affine Cartan matrix the affine node ``0'' of $\hat E_{7}$ connects to the node ``1'' of
$E_{7}$.
The Dynkin marks
\begin{equation*}
 (a_{0}, \dots, a_{7}) = (1,2,2,3,4,3,2,1).
\end{equation*}

 \subsection[$E_8$]{$\boldsymbol{E_8}$}
The Cartan matrix of $E_8$ is
\begin{equation*}
C^{E_8} =
\begin{pmatrix}
 \hphantom{-}2 & \hphantom{-}0 & -1 & \hphantom{-}0 & \hphantom{-}0 & \hphantom{-}0 & \hphantom{-}0 & \hphantom{-}0 \\
 \hphantom{-}0 & \hphantom{-}2 & \hphantom{-}0 & -1 & \hphantom{-}0 & \hphantom{-}0 & \hphantom{-}0 & \hphantom{-}0 \\
 -1 & \hphantom{-}0 & \hphantom{-}2 & -1 & \hphantom{-}0 & \hphantom{-}0 & \hphantom{-}0 & \hphantom{-}0 \\
 \hphantom{-}0 & -1 & -1 & \hphantom{-}2 & -1 & \hphantom{-}0 & \hphantom{-}0 & \hphantom{-}0 \\
 \hphantom{-}0 & \hphantom{-}0 & \hphantom{-}0 & -1 & \hphantom{-}2 & -1 & \hphantom{-}0 & \hphantom{-}0 \\
 \hphantom{-}0 & \hphantom{-}0 & \hphantom{-}0 & \hphantom{-}0 & -1 & \hphantom{-}2 & -1 & \hphantom{-}0 \\
 \hphantom{-}0 & \hphantom{-}0 & \hphantom{-}0 & \hphantom{-}0 & \hphantom{-}0 & -1 & \hphantom{-}2 & -1 \\
 \hphantom{-}0 & \hphantom{-}0 & \hphantom{-}0 & \hphantom{-}0 & \hphantom{-}0 & \hphantom{-}0 & -1 & \hphantom{-}2
\end{pmatrix}
\end{equation*}
and the inverse is
\begin{equation*}
\big(C^{E_8}\big)^{-1} =
 \begin{pmatrix}
4 & 5 & 7 & 10 & 8 & 6 & 4 & 2 \\
 5 & 8 & 10 & 15 & 12 & 9 & 6 & 3 \\
 7 & 10 & 14 & 20 & 16 & 12 & 8 & 4 \\
 10 & 15 & 20 & 30 & 24 & 18 & 12 & 6 \\
 8 & 12 & 16 & 24 & 20 & 15 & 10 & 5 \\
 6 & 9 & 12 & 18 & 15 & 12 & 8 & 4 \\
 4 & 6 & 8 & 12 & 10 & 8 & 6 & 3 \\
 2 & 3 & 4 & 6 & 5 & 4 & 3 & 2
 \end{pmatrix}.
\end{equation*}
In the affine Cartan matrix the node ``0''
of $\hat E_{8}$ connects to the node ``8'' of $E_{8}$.
The Dynkin marks
\begin{equation*}
 (a_{0}, \dots, a_{8}) = (1,3,2,4,6,5,4,3,2).
\end{equation*}

\section{Affine Weyl group}

For the class II theories the relevant reflection group turns out to be the affine Weyl group~$W({\hat\g})$. As a group, it is a semi-direct product of the finite Weyl group~$W({\g})$ and the root lattice~$\rl$ of~$\g$ (recall that for $\g$ the root and the coroot lattices are identified):
\begin{equation*}
W({\hat\g}) = W({\g}) \ltimes \rl.
\end{equation*}
We can view $W({\hat\g})$ as the group acting in $\hat\h$, preserving the non-degenerate scalar product
$( \cdot , \cdot )$ which extends the Killing form on $\h$
by the pairing between $K$ and ${\delta}^{\vee}$ as follows:
\begin{gather}
 x = {\tau} {\delta}^{\vee} + {\si} K + {\xi}, \qquad {\tau}, {\si} \in {\BC}, \quad {\xi} \in \h,
 \label{eq:tsxi}
\\
(x,x) = \langle {\xi}, {\xi} \rangle + 2 {\tau}{\si}.\nonumber
\end{gather}
The group $W({\hat\g})$ is generated by simple reflections $r_{i}$, $i = 0, 1, \dots, r$.
The action of $r_i$ on $\hat\h$ is given by
\begin{equation}
r_{i}\colon \ x \mapsto x - {\al}_{i} ( x ) {\al}_{i}^{\vee}, \qquad i = 0, \dots , r
\label{eq:affre}
\end{equation}
for $x \in {\hat\h}$.
Similarly, the action of $r_i$
on $\hat\h^*$ is given by
\begin{equation}
r_{i}\colon \ p \mapsto p - p\big({\al}_{i}^{\vee}\big) {\al}_{i}, \qquad i = 0, \dots , r.
\label{eq:affred}
\end{equation}
Note that $K$ is invariant under the reflections (\ref{eq:affre}), while ${\delta}$ is invariant under the reflections~(\ref{eq:affred}), cf.~(\ref{eq:imroot}).

On the hyperplane
\begin{equation}
H_{\tau} = \{ x\mid {\delta}(x) = {\tau} \} \subset \hat\h
\label{eq:htau}
\end{equation}
the group
 $W({\hat \g})$ acts by orthogonal transformations, generated by the ordinary reflections, and by translations. In the
 decomposition (\ref{eq:tsxi}), we have
 \begin{equation*}
 r_{0}\big({\tau} {\delta}^{\vee} + {\si} K + {\xi}\big) = {\tau} {\delta}^{\vee} + {\si} K + {\xi} - ( {\tau} - {\theta}({\xi})) \big(K - {\theta}^{\vee}\big),
 \end{equation*}
where we introduced the highest root $\theta \in \h^{*}$, and the highest coroot $\theta^{\vee} \in \h$:
\begin{equation*}
\theta = {\delta} - {\alpha}_{0} = \sum_{i=1}^{r} a_{i}{\alpha}_{i}, \qquad \theta^{\vee} = K - {\alpha}_{0}^{\vee} = \sum_{i=1}^{r} a_{i}{\alpha}_{i}^{\vee},
\end{equation*}
which obey ${\theta}({\theta}^{\vee}) = \big\langle {\theta}^{\vee}, {\theta}^{\vee} \big\rangle = 2$, and also ${\theta}({\xi}) = \big\langle \xi, {\theta}^{\vee} \big\rangle$ for any $\xi \in \h$.

Now, to make the translational part of the $W({\hat\g})$ action explicit, let us perform an additional reflection at $\theta$:
 \begin{gather*}
r_{\theta}r_{0}(x) = r_{0}(x) - {\theta} \big({\xi} - ( {\tau} - {\theta}({\xi})) \big( - {\theta}^{\vee}\big)\big) {\theta}^{\vee} = {\tau} {\delta}^{\vee} + ( {\si} - {\tau} + {\theta}({\xi})) K + {\xi} - {\tau} {\theta}^{\vee}.
 \end{gather*}
Finally, the general element of $W({\hat\g})$ can be represented as a pair
$(w, {\beta})$, where $w \in W({\g})$ and $\beta \in \rl$, with the composition rule
\[
(w_{1}, {\be}_{1}) \cdot (w_{2}, {\be}_{2}) = (w_{1}\cdot w_{2}, {\be}_{1} + {\be}_{2}^{w_{1}}).
\]
The element ${\hat w} = (w, {\beta})$ acts on $x \in H_{\tau}$ as follows:
\begin{equation}
 \big({\tau} {\delta}^{\vee} + {\si} K + {\xi}\big)^{\hat w} =
 {\tau} {\delta}^{\vee} + \Big( {\si} - \langle \xi^{w}, \beta \rangle - \frac{\tau}{2} \langle \beta, \beta \rangle \Big) K + ( {\xi}^{w} + {\tau} {\beta} ) .
\label{eq:wshft}
\end{equation}
The dual picture is the action of $W({\hat\g})$ on the fixed level hyperplane $H^{*}_{k} = \{ p \mid p (K) = k \} \subset \hat\h^{*}$. Write
 \begin{equation*}
p = k {\lam}_{0} + {\mu} {\delta} + {\eta}, \qquad {\eta} \in \h^{*}, k, {\mu} \in {\BC},
\end{equation*}
then
\[
p(x) = k {\sigma} + {\tau}{\mu} + {\eta}({\xi}).
\]
Then the affine Weyl transformation $w_{\beta}$ generated by $\beta \in \rl$
acts on $p$ as follows:
\begin{equation}
w_{\beta} (p) = p+ \left( - \langle {\eta}, {\beta} \rangle + \frac k2 \langle {\beta}, {\beta} \rangle \right) {\delta} + k {\beta}.
\label{eq:pbet}
\end{equation}
We need the affine Weyl lattice action \eqref{eq:pbet} to construct
lattice theta-functions in Appendix~\ref{se:lattice-theta}.

\section{Conjugacy classes and moduli of bundles}\label{se:conjugacy}

In this section we recall the relation of the moduli space of holomorphic $G$-bundles on elliptic curve and the space of conjugacy classes in
$\hat\g$, see, e.g.,~\cite{Baranovsky-Ginzburg_1996}.

Let $H_{\tau}$ be the hyperplane (\ref{eq:htau}) and let $B_{\tau}$ be the quotient
$H_{\tau}/{\BC}K$, i.e., we forget about the~${\si}K$ part in the expansion
\begin{equation*}
x = {\tau} {\delta}^{\vee} + {\si} K + \sum_{i=1}^{r} {\xi}_{i} {\al}_{i}^{\vee}.
\end{equation*}
We can identify the quotient with the Cartan subalgebra $\h$ of the finite-dimensional Lie algebra~$\g$, $B_{\tau} \approx {\h}$. In this way we arrive at the ${\tau}$-dependent action of $W({\hat\g})$ on $\h$:
\begin{equation*}
{\xi} \mapsto {\xi}^{\hat w} = {\xi}^{w} + {\tau}{\beta}.
\end{equation*}
Note that this action does not preserve the coroot lattice $\rl^{\vee} \subset \h$. However, it preserves the ``doubled'' coroot lattice ${\rl}^{\vee} + {\tau} {\rl}^{\vee} \subset \h$.

Recall that \emph{Bernstein--Schwarzman group} $W_{\tau}({\hat\g})$
\cite{Bernstein_1978} is the semi-direct product of $W({\g})$ and the lattice ${\rl}^{\vee} + {\tau} {\rl}^{\vee}\subset \h$. The quotient
\begin{equation}
{\h}/W_{\tau}({\hat\g})
\label{eq:quobs}
\end{equation}
is identified with the coarse moduli space ${\Bun}_{\bG}({\ec})$ of
holomorphic principal semi-stable $\bG$-bundles on the elliptic curve
${\ec} = {\BC}/({\BZ} + {\tau}{\BZ})$~\cite{Narasimhan_1965,Ramanathan_1975}. Let us denote by $z$ the additive coordinate on $\ec$:
\[
z \sim z + m + n{\tau}, \qquad m,n \in {\BZ}.
\]
We can perform the quotient (\ref{eq:quobs}) in two steps: first, divide by ${\rl}^\vee$, and
then, divide by the action of $W({\hat\g})$. The first step is accomplished by the exponential map from $\h$ to ${\h}/{\rl}^{\vee} \approx \bT$. In the second step we divide by the lattice $\tau \rl^{\vee}$, giving us ${\ec} \otimes {\rl}^{\vee}$ (as an abelian group), and finally we divide by the Weyl group $W({\g})$:
\begin{equation*}
\bT/W({\hat\g}) = \big(\bT/{\tau}{\rl}^{\vee}\big)/W({\g}) = \big({\ec} \otimes {\rl}^{\vee}\big)/W({\g}).
\end{equation*}
It is instructive to recall ``Dolbeault'' realization of the moduli space of bundles, since it will be useful in our further analysis. A holomorphic structure in the $G$-bundle is a $(0,1)$-connection ${\bar\nabla}_{\zb} = {\pa}_{\zb} + A_{\zb}$ on a principal smooth $G$-bundle $P$, which is topologically trivial since we assume~$G$ simply connected, and therefore $A_{\zb}$ is a $\g$-valued
$(0,1)$-form on $\ec$. The gauge equivalent connections ${\nabla}_{\zb}$ correspond to the isomorphic bundles:
\begin{equation*}
A_{\zb} \mapsto g^{-1}A_{\zb} g + g^{-1} {\pa}_{\zb} g, \qquad g \in C^{\infty} ({\ec}, G).
\end{equation*}
Recall that the holomorphic structure on $P$ is determined by the condition that the local holomorphic sections of $P$ are annihilated by $\nabla_{\zb}$, e.g., in an open neighborhood $U_{\al} \subset {\ec}$,
the holomorphic sections $s_{\al}$ obey
\begin{equation*}
{\nabla}_{\zb} s_{\al} = 0 \ \Leftrightarrow \ A_{\zb} \vert_{U_{\al}} = - ({\pa}_{\zb}s_{\al})s_{\al}^{-1}.
\end{equation*}
One can find a gauge for $A_{\zb}$, where it is a given by a constant $\h$-valued $(0,1)$-form
\begin{equation}
{\nabla}_{\zb} = {\pa}_{\zb} + \frac{2\pi \ii}{{\tau} - {\bar\tau}} {\xi},\label{eq:xigauge}
\end{equation}
which still leaves a room for the residual gauge transformations,
which preserve the fact that ${\xi \in \h}$ and ${\pa}_{z} {\xi} = {\pa}_{\zb} {\xi} = 0$. There are two kinds of transformations (the general such transformation is a composition of these two):
\begin{equation}
{\xi} \mapsto {\xi}^{w} = g_{w}^{-1} {\xi} g_{w}, \qquad
g_{w} \in N(\bT)/\bT = W({\g})
\label{eq:reswe}
\end{equation}
and
\begin{equation}
{\xi} \mapsto {\xi} + \frac{{\tau}-{\bar\tau}}{2\pi \ii} {\pa}_{\zb} \log g = {\xi} - {\be}_{1} - {\tau} {\be}_{2},
\label{eq:ressh}
\end{equation}
where
\begin{equation*}
g(z, {\zb}) = \exp 2\pi \ii \left( \frac{z - {\zb}}{{\tau} - {\bar\tau}} {\be}_{1} + \frac{z{\bar\tau} - {\zb} {\tau}}{{\tau} - {\bar\tau}} {\be}_{2} \right)
\end{equation*}
with ${\be}_{1}, {\be}_{2} \in \rl^{\vee}$, cf.~(\ref{eq:trrl}). Thus the space of all the $(0,1)$-connections ${\nabla}_{\zb}$ modulo the smooth gauge transformations is isomorphic to the quotient $({\ec} \otimes {\rl})/W({\g})$, as we claimed.

E.~Loojienga's theorem~\cite{Loojienga:1976} identifies the moduli space of $\bG$-bundles on $\ec$ with the weighted projective space:
\begin{equation}
{\Bun}_{\bG}({\ec}) \approx {\WP}^{a_{0}, a_{1}, \dots, a_{r}}.
\label{eq:loojienga}
\end{equation}
There are several mathematical interpretations of this theorem~\cite{Donagi:1997dp,Friedman:1997yq, Friedman:2000ze, Friedman:1998si, Friedman:1997ih}. We shall give yet another, more physical explanation of this result, using $\hat\g$ representation theory. This realization of (\ref{eq:loojienga}) is closer related to our problem.

\section{Infinite matrices and their Weyl group}

We shall also encounter the group $\Gli$. It is the group whose Lie algebra
is the central extension of the Lie algebra ${\g\fl}({\infty})$ of infinite matrices which have only a finite number
of non-vanishing elements away from the main diagonal:
\[
A = \sum_{i,j \in \BZ} a_{ij} E_{ij}, \qquad a_{ij} = 0 \quad {\rm for} \ | i - j | \gg 0,
\]
where $E_{ij}$ is the matrix with all the entries vanishing except for the
$(i,j)$ entry.
The central extension is given by the cocycle
\begin{equation*}
[ A\oplus {\al}K , B \oplus {\be}K ] = [A,B] \oplus {\gamma}(A,B) K , \qquad
{\gamma}(A, B) = \tr A [ J, B ],
\end{equation*}
where
\[
J = \frac 12 \sum_{i \in \BZ} \operatorname{sign} \bigg( i - \frac{1}{2} \bigg) E_{ii}.
\]
The Cartan subalgebra $\hat\h_{\infty} \subset \widehat{{\g\fl}({\infty})}$
consists of the diagonal matrices plus the span of the central generator:
\begin{equation*}
 \hat\h_{\infty} = \bigoplus_{i \in \BZ} {\BC} E_{ii} \oplus {\BC}K.
\end{equation*}
The Weyl group $W(\widehat{{\g\fl}({\infty})})$ is the group of finite permutations of the eigenvalues
of the infinite diagonal matrix.

\section{The representation theory}

Let us start over with the finite-dimensional simple Lie group $\bG$.
Recall that the weight ${\mu} \in {\Lambda}$ is \emph{dominant}, ${\mu} \in {\Lambda}_{+}$ iff ${\mu}\big({\al}_{i}^{\vee}\big) \geq 0$ for all $i = 1, \dots , r$.

Recall that to every dominant weight ${\mu} \in {\Lambda}_{+}$ an irreducible highest weight finite-dimensional representation ${\CalV}_{\mu}$ of the group $\bG$ is associated. In particular, the fundamental representations $R_i = {\CalV}_{{\lam}_{i}}$ of the group $\bG$ correspond to the fundamental weights ${\lam}_{i}$. Moreover, the Grothendieck ring ${\rm Rep}(\bG)$
of finite-dimensional representations of $\bG$ is generated by $R_i$.

\begin{Example}
For the $A_{r}$ series, the group $\bG = {\rm SL}(r+1, {\BC})$, the fundamental representations $R_{i} = {\bigwedge}^{i}{\BC}^{r+1}$, $i = 1,\dots , r$, are the exterior powers of the defining $(r+1)$-dimensional representation. The center $Z = {\BZ}_{r+1}$, the adjoint group $G^{\text{ad}} = {\rm PGL}(r+1,{\BC}) = {\rm SL}(r+1, {\BC})/{\BZ}_{r+1}$.
\end{Example}

\begin{Example}
For the $D_r$ series, the group $\bG = {\rm Spin}(2r, {\BC})$, the fundamental representations $R_{i} = {\bigwedge}^{i}{\BC}^{2r}$, $i = 1, \dots , r-2$, i.e., antisymmetric tensors of the corresponding rank, while $R_{r-1} = {\BS}_{-}$, $R_{r} = {\BS}_{+}$ are the chiral spinors. The center $Z = {\BZ}_{4}$ for~$r$ odd, and $Z = {\BZ}_{2} \times {\BZ}_{2}$ for~$r$ even.
\end{Example}

\subsection{Weights of the representation}

The representations of $\bG$ produce the $W({\g})$-invariant functions on $\bT$. To every finite-di\-men\-sion\-al representation $V$ of $\bG$, ${\pi}_{V}\colon \bG \to {\rm End}(V)$, one assigns a function on $\bT$, the character
\begin{equation*}
{\chi}_{V}(t) = \tr _{V} {\pi}_{V}(t) , \qquad t \in \bT \subset \bG,
\end{equation*}
which is the trace of the element of $\bT$ viewed as the element of $\bG$ in the representation $V$. By definition of the Weyl group $W({\g}) = N(\bT)/\bT$ it acts on $\bT$ by conjugation in $\bG$, therefore the character $\chi_{V}$ is $W({\g})$-invariant.
The representation $V$, viewed as a representation of $\bT \subset \bG$, splits as a direct
sum of one-dimensional representations $L_{\lam}$, i.e., the sum of weight subspaces~$L_{\lam}$ for some ${\lam} \in {\Lambda}_{V} \subset {\Hom}(\bT, {\BC}^{\times}) = {\Lambda}$. Here
${\Lambda}_{V}$ is the (finite, for finite-dimensional $V$) set of weights of the representation $V$. The character ${\chi}_{V}(t)$ can be written as
\begin{equation}
{\chi}_{V}(t) = \sum_{{\lam} \in {\Lambda}_{V}} {\chi}_{V,\lam} t^{\lam},
\label{eq:wmult}
\end{equation}
where ${\chi}_{V,\lam} \in {\BZ}_{+}$ is the multiplicity of the given weight $\lam$ in the decomposition of
$V$:
\[
V = \bigoplus_{{\lam} \in {\Lambda}_{V}} {\BC}^{{\chi}_{V,\lam}} \otimes L_{\lam}
\]
and $t^{\lam} \in {\BC}^{\times}$ is the value of ${\lam} \in {\Hom}(\bT, {\BC}^{\times})$ on $t \in T$.
For the highest weight representations~${\CalV}_{\mu}$ with the highest weight $\mu$ the set $\Lambda_{V}$ contains $\mu$, with multiplicity ${\chi}_{V,\mu} = 1$, while all the other elements ${\lam}$ of $\Lambda_{V}$ obey
\begin{equation}
{\mu} - {\lam} = \sum_{i=1}^{r} n_{i} {\al}_{i}, \qquad n_{i} \geq 0, \qquad \sum_{i} n_{i} > 0.
\label{eq:poswe}
\end{equation}
The Weyl group $W({\g})$ action on $t \in \bT$ can be traded for the $W({\g})$-action on the set of weights~$\Lambda_{V}$:
\[
(t^{w})^{\lam} = t^{{\lam}^{w}}.
\]
In other words, the set ${\Lambda}_{V} \subset \Lambda$ is $W({\g})$ invariant, including the multiplicities
\begin{equation*}
{\chi}_{V,\lam} = {\chi}_{V,{\lam}^{w}} \qquad \text{for any}\ w \in W({\g}).
\end{equation*}
The ring of $W({\g})$-invariant rational functions on $\bT$ is generated, as a vector space,
by the characters of all the irreducible representations of $G$, and polynomially by the characters ${{\chi}_{i} = {\chi}_{R_{i}}}$ of~$r$ fundamental representations of $\bG$ (Chevalley):
\begin{equation*}
{\BC}[\bG]^{{\rm Ad}(G)} = {\BC}[\bT]^{W({\g})} \equiv {\BC}\big[z_{1}, \dots , z_{r}, z_{1}^{-1}, \dots, z_{r}^{-1} \big]^{W({\g})} = {\BC} [{\chi}_{1}, \dots , {\chi}_{r} ].
\end{equation*}

\subsection{The action of the center}

Any element $z$ of the center $Z$ of $\bG$ acts in the irreducible representation $(V , {\pi}_{V})$ of $\bG$ by multiplication by a scalar character:
\begin{equation*}
{\pi}_{V}(z) = {\mu}_{V}(z) \cdot {\bf 1}_{V}
, \qquad
 {\mu}_{V} \in {\Hom}(Z, {\BC}^{\times}) \approx {\Lambda}/{\rl}.
 \end{equation*}
For the highest weight module $V = {\CalV}_{\lam}$ this character is easy to compute
\begin{equation*}
{\mu}_{{\CalV}_{\lam}} ({\varpi}_{y}) = {\varpi}_{y}^{\lam}.
\end{equation*}
In particular, the character ${\bm}_{i} \equiv {\mu}_{R_{i}}$ corresponds to the image of the fundamental weight ${\lam}_{i}$ in the coset ${\Lambda}/{\rl}$:
\begin{equation*}
{\bm}_{i} ({\varpi}_{y}) = {\ze}_{y}^{w_{yi}}.
\end{equation*}

\subsection{Representation theory of affine Kac--Moody algebras}

The affine Kac--Moody algebra $\hat\g$ has the positive energy
 highest weight irreducible representations ${\hat R}_i$, $i=0, 1, \dots , r$, which
 are integrable, i.e., lift to the representations of the corresponding Kac--Moody group
 $\widetilde{{\rm LG}}$. The highest weight of ${\hat R}_i$ is ${\hat\lam}_i$. The representations ${\hat R}_i$ are infinite-dimensional of finite type: the dimensions of the eigenspaces of ${\delta}^{\vee}$ are finite, so that the character
\begin{equation*}
 {\hat\CalX}_{i}({\hat t}) = \tr _{{\hat R}_{i}} {\pi}_{{\hat R}_{i}} ( {\hat t} ) = \sum_{{\hat\nu}} {\hat\chi}_{{\hat R}_{i}, {\hat\lam}_{i} - {\hat\nu}} {\hat t}^{{\hat\lam}_{i} - {\hat\nu}}
\end{equation*}
 makes sense as a function in the domain
 $|{\qe} | < 1$, where
\begin{equation*}
{\qe} = {\hat t}^{-\delta} \in {\BC}^{\times}.
\end{equation*}

\subsection{On E.~Loojienga's theorem}

In view of what we explained above, this theorem has the following interpretation.
The fundamental characters ${\hat\chi}_{i}$ of
$\widetilde{{\rm L}\bG}$, with $i=0,1, \dots, r$ are the $W({\hat\g})$-invariant functions on
the maximal torus $\tilde \bT$ of $\widetilde{{\rm L}\bG}$. The $\tilde \bT$ is a
trivial fibre bundle over ${\BC}^{\times}_{{\delta}^{\vee}}$
\begin{equation}
\label{eq:ttor}
 \tilde \bT = {\BC}^{\times}_{{\delta}^{\vee}} \times \hat \bT
\end{equation}
with the
fiber over the point ${\qe} = {\rm e}^{2\pi \ii \tau} \in {\BC}^{\times}_{{\delta}^{\vee}}$
being the $(r+1)$-dimensional torus
\begin{equation}
{\hat \bT} = {\BC}^{\times}_{K} \times \bT = H_{\tau}/{\hat\rl}^{\vee},
\label{eq:toruslg}
\end{equation}
where
${\BC}^{\times}_{K}$ is the center of $\widetilde{{\rm L}\bG}$. Note that the central extension of ${\rm L}\bG$ is topologically nontrivial. However, its restriction on any torus of ${\rm L}\bG$ (which we can identify with the space of constant loops valued in $\bT \subset \bG$) is topologically trivial, hence (\ref{eq:toruslg}). Now, since $K$ generates the center of $\widetilde{{\rm L}\bG}$, the
$c^{K}$-dependence of ${\hat\CalX}_{i}$ is a simple factor
\[
{\hat\CalX}_{i}\big(c^{K}{\hat t}\big) = c^{a_{i}} {\hat\CalX}_{i}({\hat t}).
\]
On the other hand, as we explained above, forgetting the ${\BC}^{\times}_{K}$-factor
in the quotient
\begin{equation*}
{\widetilde B}_{\tau} = \big(H_{\tau}/{\hat\rl}^{\vee} \big)/W({\hat\g}) = {\hat \bT}/W({\hat \g})
\end{equation*}
leads to the quotient $B_{\tau}/W({\hat\g}) = {\Bun}_{\bG}({\ec})$.
In other words, the space ${\widetilde B}_{\tau}$
is the total space of a~${\BC}^{\times}_{K}$-bundle over ${\Bun}_{G}({\ec})$. The ring of
holomorphic functions on ${\widetilde B}_{\tau}$ is identified with the ring of polynomials in the fundamental characters ${\hat\chi}_{i}$:
\begin{equation*}
{\BC} \big[ {\widetilde B}_{\tau} \big] = {\BC} [ {\hat\chi}_{0}, {\hat\chi}_{1}, \dots , {\hat\chi}_{r}].
\end{equation*}
Dividing (or, better to say, grading) by ${\BC}^{\times}_{K}$
makes ${\Bun}_{\bG}({\ec})$ out of ${\widetilde B}_{\tau}$ on the left-hand side, and the space of holomorphic sections of the bundles ${\CalO}(k)$ over the weighted projective space ${\WP}^{a_{0}, a_{1}, a_{2}, \dots , a_{r}}$ on the right-hand side.

\section{Affine Weyl group, characters and theta-functions}\label{se:lattice-theta}
The affine Weyl group of affine Lie algebra $\hat \g$ is the semi-direct product $W ({\hat\g}) = W ({\g})
\ltimes {\rl}^{\vee}$.
For each weight $\hat \lambda \in {\hat \h}^{\vee}$ the affine
transformation $w_\alpha$ by $\alpha \in {\rl}^{\vee} $ is a parabolic transformation
which linearly shifts the $\h$ component of $\hat \lambda$ by $\alpha$ and
quadratically adjusts the
$\delta$-component of $\hat \lambda$ keeping the norm $\big(\hat \lambda, \hat \lambda\big)$
invariant. Concretely, if $\hat \lambda$ is of level $k$ we represent it
as $\hat \lambda = k \lambda_0 + \lambda$ with $\lambda \in \h^\vee$,
and then we have (cf.~equation~\eqref{eq:pbet})
\begin{equation*}
 w_\alpha \big(\hat \lambda\big) = \hat \lambda + k \alpha - \tfrac{\delta}{2k}
\big( \big(k\alpha + \hat \lambda\big)^2 - \hat \lambda^2\big).
\end{equation*}

For each weight $\hat \lambda$ we introduce theta-function
$\Theta_{\hat \lambda}$
\begin{equation}\label{eq:thetaL}
 \Theta_{\hat \lambda} = {\rm e}^{-\frac{\lambda^2}{2k} \delta} \sum_{\alpha \in
 {\rl}^{\vee}} {\rm e}^{w_{\alpha}(\lambda)}
\end{equation}
equivalently
\begin{equation}
\label{eq:theta-lat}
 \Theta_{\hat \lambda} = {\rm e}^{k\Lambda_0} \sum_{\gamma \in \lambda
 + k {\rl}^{\vee}} {\rm e}^{ \gamma - \frac{ \delta \gamma^2}{2k}},
\end{equation}
where formally omitted the arguments in the exponentiated Cartan of
$\hat \g$, i.e., ${\tilde T}$.
The theta function $\Theta_{\hat \lambda}$ can be evaluated on an element ${\bf g} = (c,
\mathbf{t}, \qe)$:
\begin{equation*}
 \Theta_{\lambda}(c,\mathbf{t},\qe) = c^{k} \sum_{\gamma \in \hat\lambda
 + k {\rl}^{\vee}} t^{ \gamma} \qe^{ \frac{ \gamma^2}{2k }}.
\end{equation*}
Sometimes it is convenient not to include the prefactor in the
definition of the theta-function~\eqref{eq:thetaL} and for all
fundamental weights we set
\begin{equation}
\label{eq:c-def}
 c_{i}(c, \mathbf{t}, \qe) := \qe^{- \frac{ \lambda^2}{2k}}
 \Theta_{\hat\lambda_i}(c, \mathbf{t}, \qe).
\end{equation}
From (\ref{eq:theta-lat}), it is clear that $\Theta_{\hat\lam}$ depends only
on the affine Weyl conjugacy class of $\hat\lambda$.

The ring of affine Weyl group invariants is generated by
the theta-functions $\Theta_{\hat \lambda_0}, \dots, \Theta_{\hat
 \lambda_r}$ for the fundamental weights $\hat \lambda_{0}, \dots, \hat \lambda_{r}$.

Another important function is the Weyl skewsymmetric function
\begin{equation*}
 R(t) = \prod_{\alpha \in{\Delta^+}} (1 - t^{-\alpha}) = \sum_{w \in W} (-1)^w t^{w (\rho)
 - \rho},
\end{equation*}
where $\rho$ is the Weyl covector defined by $\rho ( {\al}_i ) = 1$
for all $i = 0, \dots, r$.
The equality of the two representations above for
Weyl denominator for affine algebras leads to the famous Macdonald
identity.
By definition, $R(h)$ is Weyl skewsymmetric.

\section{Characters}

For the weight $\hat\lambda \in {\hat \Lambda}$ let us denote by ${\hat \CalR}_{\hat\lam}$ the irreducible
$\hat \g$-module with the highest weight $\hat \lambda$. Let $\chi_{\hat \CalR}(t)$ be
the character evaluated on an element $t$ of ${\tilde T}$ \eqref{eq:ttor}:
\begin{equation*}
 \chi_{\hat \CalR}(t) = \sum_{\lambda \in \h^{*}} t^{\lambda}
 \mathrm{mult}_{\hat \CalR} \lambda.
\end{equation*}
The Weyl character formula literally holds for the affine Kac--Moody algebras
\begin{equation*}
 \chi_{\hat \CalR}(t) = \frac {\sum_{w \in W} \ep(w) t^{ w( \Lambda + \rho) - \rho} } { \prod_{\alpha \in \Delta^{+}} (1 - t^{-\alpha})},
\end{equation*}
where $\Delta_{+}$ is the set of positive roots in $\g$, $\ep(w) = \pm 1
$ is the parity of Weyl transformation $w \in W$.

If $\hat \g$ is simply laced affine ADE algebra and $\hat \CalR$ is an
irreducible
$\g$-module of level $1$, then the character can be also computed in
terms of theta-functions~\cite{Kac:1984}
 \begin{equation}
 \chi_{\hat \CalR} (t) = \phi(\qe)^{-r} \sum_{\alpha \in {\rl}^\vee} t^{
 w_{\alpha} (\hat \lambda )}.
 \label{eq:charl}
\end{equation}
More generally,
\begin{equation}\label{eq:KP}
 \ch_{\hat \CalR}(t) = \sum_{ \lambda \in P_{k} \, \mathrm{mod} \, k {\rl}^{\vee}}
 c^{\Lambda}_{\lambda} \Theta_{\lambda}(t),
\end{equation}
where the string function $c^{\Lambda}_{\lambda}(\qe)$ depends only on
$\qe=t^{-\delta}$ and counts multiplicities of weights on the vertical
rays $ \hat \lambda + \BZ \delta$ in the $\g$-module $L(\Lambda)$
\begin{equation*}
 c^{\Lambda}_{\lambda}(\qe) =\qe^{-\frac{ \lambda^2}{2k}} \sum_{n \in
 \BZ} \qe^{n} \mathrm{mult}_{\hat \CalR}(\lambda -
n \delta).
\end{equation*}
The formula (\ref{eq:KP}) follows from the invariance of the
multiplicities under affine Weyl group, with non-trivial information
now contained in the finite set of the functions
$c^{\Lambda}_{\lambda}$. If $\lambda$ is of level $k=1$ and $\hat \g$ is
simply laced affine ADE then there is only one non-zero inequivalent string
function
\begin{equation*}
c^{\Lambda}_{\lambda}(\qe) = \qe^{-\frac{\lambda^2}{2}} \phi(\qe)^{-r}
\end{equation*}
reproducing formula (\ref{eq:charl}).

In what follows, we use the notations
\begin{gather*}
\mathbf{n} = (n_1, \dots, n_{r}) \in \BZ^{r},\qquad \mathbf{\tilde n} = ( {\tilde n}_1, \dots, {\tilde n}_{r}) \in \left(\BZ + \frac 12 \right)^{r},
\\
| {\bf n} | = \sum_{i=1}^{r} n_{i}, \qquad | {\bf\tilde n} | = \sum_{i=1}^{r} {\tilde n}_{i}, \qquad
{\bf n}^{2} = \sum_{i=1}^{r} n_{i}^{2}, \qquad {\bf\tilde n}^2 = \sum_{i=1}^{r} {\tilde n}_{i}^{2}.
\end{gather*}
\subsection[$\hat A_{r}$ theta-functions]{$\boldsymbol{\hat A_{r}}$ theta-functions}

If we specialize (\ref{eq:thetaL})
to fundamental weight $\hat\lambda_j$, we get
 \begin{equation*}
 \qe^{-\frac{j}{2}} \qe^{\frac{j^2}{2 (r+1)}} \Theta_{\hat\lambda_j} =
 \sum_{\alpha \in Q^{\vee}} {\rm e}^{w_\alpha (\hat\lambda_j)}.
 \end{equation*}
For each fundamental weight ${\hat\lam}_j$, $j = 0,\dots, r$ we define the
 theta-function $\Theta_j \equiv \Theta_{\Lambda_j}$. We use the basis of multiplicative coordinates on
 $\exp(\hat \h) \ni {\bf g}$ as follows: $c = {\bf g}^{\hat\lam_{0}}$, $\qe =
 {\bf g}^{-\delta}$, $\mathbf{t} = (t_1, \dots, t_{r+1}) $ subject to
$\prod_{i=1}^{r+1}{t_i} =1 $ in our conventions for the
basis in the root and weight spaces of $\hat A_r$. We find
 \begin{equation}
\label{eq:Ar-theta}
 \Theta_{j} ({\bf g}) = c {\qe}^{-\frac{ j^2}{2 (r+1)}} \sum_{\mathbf {n} \in
 \Lambda_j} \mathbf{t}^{\mathbf{n}} {\qe}^{\frac 12 \mathbf{n}^2},
 \end{equation}
where
\begin{equation*}
 \Lambda_j = \{ \mathbf{n} \in \BZ^{r+1} \mid |{\bf n}| = j \}.
\end{equation*}
Consequently, from \eqref{eq:charl} and \eqref{eq:thetaL} the character of the $j$-th fundamental representation for~$\hat A_{r}$~is
\begin{equation}
\label{eq:Archar}
 \chi_j( c; {\mathbf{t}}; {\qe}) = \phi(\qe)^{-r} {\qe}^{\frac{-j}{2}+\frac{j^2}{2 (r+1)}} \Theta_j (c; {\mathbf{t}}; {\qe} ).
\end{equation}

\subsection[$\hat D_{r}$ theta-functions]{$\boldsymbol{\hat D_{r}}$ theta-functions}
 Let $t \in {\tilde \bT}$, $\mathbf{1} =
 (1,1,\dots, 1,1) = 2 \lam_{r}$, $t_i = t^{e_i}$, $\qe = t^{-\delta}$,
 $c =t^{\lam_0}$.
 The level $1$ theta-functions are associated to the fundamental weights
 $\lam_0$, $\lam_1$, $\lam_{r-1}$, $\lam_{r}$.
 According to our
 definition (\ref{eq:theta-lat}) we get
 \begin{gather}
 \Theta_{0} (c;\mathbf{t};\mathbf{q})=c \sum_{\mathbf{n}, |
 \mathbf{n}| \in 2 \BZ}
 \mathbf{t}^{\mathbf{n}} \qe^{\frac 1 2 \mathbf{n}^2} ,\nonumber\\
 \Theta_{1} (c;\mathbf{t};\mathbf{q})= c \sum_{\mathbf{n}, | \mathbf{n}| - 1 \in 2 \BZ }
 \mathbf{t}^{\mathbf{n}} \qe^{\frac 1 2 \mathbf{n}^2} ,\nonumber\\
 \Theta_{r-1}(c;\mathbf{t};\mathbf{q}) = c \sum_{{\bf\tilde n} , |{\bf\tilde n}| - \frac {r}{2} + 1 \in 2 \BZ}
 \mathbf{t}^{{\bf\tilde n}} \qe^{\frac 12 {\tilde{\mathbf{n}}^2}} ,\nonumber\\
 \Theta_{r}(c;\mathbf{t};\mathbf{q}) = c \sum_{{\bf\tilde n} , |{\bf\tilde n}| - \frac {r}{2} \in 2 \BZ}
 \mathbf{t}^{{\bf\tilde n}} \qe^{\frac 12 {\bf\tilde n}^2}.
\label{eq:D1-theta}
 \end{gather}
We can express $\hat D_{r}$ theta-functions of level $1$
in terms of Jacobi theta-functions as follows
\begin{gather}
 \Theta_{0}(c;\mathbf{t};\qe) = \frac {c} 2
\Bigg( \prod_{i=1}^{r} \theta_3(t_i;\qe) + \prod_{i=1}^{r}
\theta_4(t_i;\qe) \Bigg),\nonumber\\
 \Theta_{1}(c;\mathbf{t};\qe) = \frac {c} 2
 \Bigg( \prod_{i=1}^{r} \theta_3(t_i;\qe) - \prod_{i=1}^{r}
\theta_4(t_i;\qe)\Bigg),\nonumber\\
\Theta_{r}(c;\mathbf{t};\qe) = \frac {c} 2
 \Bigg( \prod_{i=1}^{r} \theta_2(t_i;\qe) + \ii^{-r}
\prod_{i=1}^{r} \theta_1(t_i;\qe)\Bigg),\nonumber\\
\Theta_{r-1}(c;\mathbf{t};\qe) = \frac {c} 2
\Bigg( \prod_{i=1}^{r} \theta_2(t_i;\qe) - \ii^{-r}
\prod_{i=1}^{r} \theta_1(t_i;\qe)\Bigg).
\label{eq:D1-theta-jacobi}
\end{gather}

To construct $D_{r}$ affine Weyl invariant functions at level $2$ we use
the embedding $\mathfrak{so}(2r) \subset \mathfrak{sl}(2r)$, i.e., $D_{r} \subset A_{2r -1}$
in which the maximal tori are mapped as
\begin{gather}
 T_{{\rm SO}(2r)} \to T_{{\rm SL}(2r)}, \nonumber\\
 (t_1, \dots, t_{r}) \mapsto \big(t_{1}, \dots, t_{r}, t_{r}^{-1}, \dots, t_{1}^{-1}\big).
\label{eq:D-A-embedding}
\end{gather}
It is easy to check that $\hat D_{r}$ affine Weyl group action on
$\hat {\h}_{D_{r}}$
 at level $2$ is a subgroup of $\hat A_{2r-1}$
 affine Weyl group action on $\hat {\h}_{A_{2r-1}}$ at level
 1.
Therefore, using (\ref{eq:D-A-embedding}), we can construct $\hat D_{r}$ affine Weyl
invariant functions at level $2$ using $\hat A_{2r-1}$ theta-functions
at level~1. From (\ref{eq:Ar-theta}) and (\ref{eq:D-A-embedding}), we get
\begin{equation}
\label{eq:DXi}
 \Xi_{i}(c;\mathbf{t};\qe) = c^2 \qe^{-\frac{i^2}{4r}} \sum_{\mathbf{n,m} \in
 \BZ^{r},\, |\mathbf{m}| = i} \mathbf{t}^{\mathbf{2n-m}} \qe^{ \mathbf{n}\cdot \left( \mathbf{n} - \mathbf{m}\right) + \frac 1 2 \mathbf{m}^2}
\end{equation}
with
\begin{equation*}
 \Xi_{i} = \Xi_{-i} = \Xi_{2r +i}.
\end{equation*}
The theta-functions $\Xi_{i}$ for $i = 2 ,\dots, r-2$
compute the characters of the level $2$ fundamental~$\hat D_{r}$
modules
$\hat R_{i}$ with the
highest weight $\hat \lambda_i = 2 \lambda_0 + \bar \lambda_i$ and the
character is
\begin{equation*}
 \chi_{\hat R_{i}}(c; \mathbf{t}; \qe) = c^2 \frac{1}{ \phi(\qe)^{2r-1}} \qe^{-\frac i 2}
 \qe^{\frac{i^2}{4r}} \Xi_{i}(\y_0; \mathbf{t}; \qe).
\end{equation*}

There are relations obtained immediately from the definitions
(\ref{eq:D1-theta}),~(\ref{eq:D1-theta-jacobi}),~(\ref{eq:Jacobi-theta})
\begin{gather}
(\Theta_0 + \Theta_1)^2 = c^2 \prod_{i=1}^{r}{\theta_3}(t_i;\qe)^2 = c^2\sum_{j \in \BZ} \qe^{
 \frac{j^2}{4r}} \Xi_j(\mathbf{t};\qe), \nonumber\\
(\Theta_0 - \Theta_1)^2 = c^2 \prod_{i=1}^{r}{\theta_4}(t_i;\qe)^2 = c^2 \sum_{j \in \BZ} \qe^{
 \frac{j^2}{4r}}(-1)^j \Xi_j(\mathbf{t};\qe),\nonumber\\
(-1)^r (\Theta_r - \Theta_{r-1})^2 = c^2 \prod_{i=1}^{r}{\theta_1}(t_i;\qe)^2 = c^2 \sum_{j \in \BZ} \qe^{
 \frac{j^2}{4r}}(-1)^{j-r} \Xi_{j-r}(\mathbf{t};\qe), \nonumber\\
(\Theta_r + \Theta_{r-1})^2 = {c}^2 \prod_{i=1}^{r}{\theta_2}(t_i;\qe)^2 = c^2\sum_{j \in \BZ} \qe^{
 \frac{j^2}{4r}} \Xi_{j-r}(\mathbf{t};\qe).
\label{eq:D-theta-relations}
\end{gather}
Notice
\begin{align*}
c^{-2} \phi(\qe)^{2r} (\chi_{r} + \chi_{r-1})^2 &=
\qe^{-\frac{r}{4}} (\Theta_r + \Theta_{r-1})^2 \\
&=\sum_{j \in \BZ} q^{
 \frac{j^2 }{4r} + \frac{j}{2} } \Xi_{j}(\mathbf{t};q)
=c^{-2} \phi(\qe)^{2r -1} \sum_{s \in \BZ} \sum_{j'=0 }^{2r -1} q^{
 rs(s+1) + (s+1)j' }
 \chi_{j'}
\end{align*}
is similar the relation
\begin{equation*}
 S \otimes S = \bigoplus_{i \in {0 \dots 2r}} {\bigwedge}^{i} V ,
\end{equation*}
where $S = S^{+} \oplus S^{-}$ is the spin representation of $D_{r}$.
As the fundamental invariants of $\hat D_{r}$ we will take the set of
theta-functions associated to the irreducible representations with the
fundamental highest weight. For $i = 2, \dots, r-2$, we define
the invariants of level $2$, and for $s=1, \dots, 4$, we define invariants
of level 1 as follows:
\begin{gather}
 \big(\crfDvec_i \big) := \Xi_{i}, \qquad i = 2,\dots, r-2, \nonumber\\
 \big(\crfDspin_s \big) :=
 \begin{cases}
 \Theta_0 + \Theta_1 , & s = 0,\\
 \Theta_0 - \Theta_1 , & s = 1,\\
 \ii^r ( \Theta_{r} - \Theta_{r-1} ), & s = r-1,\\
 \Theta_{r} + \Theta_{r-1} , & s = r.
 \end{cases}
\label{eq:Dr-inv}
\end{gather}
The linear relations (\ref{eq:D-theta-relations}) allow to express
$\Xi_{0}$, $\Xi_{1}$, $\Xi_{r-1}$, $\Xi_{r}$ as linear combinations of
$\crfDvec_2, \dots, \allowbreak \crfDvec_{r-2}$ and $\big(\crfDspin_1\big)^2, \dots, \big(\crfDspin_4\big)^2$.

\subsection{Infinite matrices}
The fundamental representations ${\CalR}_{i}$ = of $\Gli$ are labeled by a
single integer $i \in \BZ$. They are highest weight ${\tilde\lam}_{i}$
representations,
 also infinite-dimensional of the finite type. Let us compute their characters (we shall remind the fermion construction of these representations in the next section).
Let ${\bf t} = (t_{i})_{i \in {\BZ}}$ be the generic infinite diagonal matrix, with entries behaving as ${\qe}^{i}$ for large $i$, for some $| {\qe} | < 1$. In other words ${\bf t}$ is close (in the topology of Hilbert space operators) to the evolution operator ${\qe}^{L_{0}}$ (with $L_0$ being the energy operator), and let $c^{K}$, for $c \in {\BC}^{\times}$ represent the center of $\Gli$.
Let
\[
{\bf g} = c^{K} \times {\bf t}.
\]
The characters
${\chi}_{i}( {\bf g} )$ can be arranged into the generating function
\begin{equation}
\sum_{i \in {\BZ}} {\zeta}^{i} {\chi}_{i} ( {\bf g}) = c \prod_{n=1}^{\infty}
( 1 + {\zeta} t_{n} ) \big( 1 + {\zeta}^{-1} t_{1-n}^{-1} \big),
\label{eq:fermch}
\end{equation}
which is explicitly $W(\hat{{\mathfrak gl}({\infty})})$-invariant, with the Weyl group acting on the
eigenvalues $t_{i}$ of ${\bf t}$ by the finite permutations,
\begin{equation*}
(t_{i})_{i \in {\BZ}} \mapsto (t_{{\si}(i)})_{i \in {\BZ}}, \qquad {\si}(i) = i, \quad | i |
\gg 0,
\end{equation*}
while the central element $c^{K}$ transforming via
\begin{equation*}
c \mapsto c \prod_{i = - \infty}^{0} \frac{t_{{\si}(i)}}{t_{i}}.
\end{equation*}

\begin{Example} The transformation $t_{0} \leftrightarrow t_{1}$ acts on $c$ by
\[
c \mapsto c t_{1}/t_{0}.
\]
The character ${\chi}_{i}$ of ${\CalR}_i$ is given by the sum over the partitions. The formula is most simply obtained by expanding
 (\ref{eq:fermch}) as a sum over partitions
\begin{equation}
{\chi}_{j}({\bf g}) = c
 \sum_{\lam} \prod_{i=1}^{\infty} \frac{t_{{\lam}_{i}-i+1+j}}{t_{-i+1}}.
 \label{eq:chari}
 \end{equation}
 In \eqref{eq:chari} the infinite product is to be understood in the following way
\[
 \prod_{i=1}^{\infty} \frac{t_{{\lam}_{i}-i+1+j}}{t_{-i+1}} \equiv t^{[j]} \prod_{i=1}^{{\ell}({\lam})}
 \frac{t_{{\lam}_{i}-i+1+j}}{t_{-i+1+j}}.
\]
 The value of the highest weight ${\tilde\lam}_{i}$ on ${\bf g}$
is equal to
\begin{equation}
{\bf g}^{\tilde\lam_{i}} = c t^{[j]}.
\label{eq:hwgli}
\end{equation}
\end{Example}

\section{Free fermion representation}\label{sectionK}

In describing the representations of ${\widetilde{{\rm LG}}}$ and $\Gli$
it is convenient to use the free fermions.

One can realize the level 1 representations of $\g$
using $r$ flavors of chiral complex fermions in two dimensions:
\begin{equation*}
L_{1} = \sum_{a=1}^{r} {\tilde \psi}_{a} {\bar\partial} {\psi}^{a}.
\end{equation*}
The currents of $\widehat{{\mathfrak{so}}(2r)}$ are
\begin{equation}
J^{ab} = {\psi}^{a}{\psi}^{b} , \qquad
 J^{a}_{b} = {:} {\tilde\psi}_{b}{\psi}^{a}{:}, \qquad
J_{ab} = {\tilde\psi}_{a}{\tilde\psi}_{b}.
\label{eq:spintn}
\end{equation}
The currents (\ref{eq:spintn}) do not preserve the $U(1)$ charge $J_0 = \oint {:} {\tilde\psi}_{a}{\psi}^{a}{:}$, however they preserve its parity. Therefore,
one can construct the following four representations: take the fermions
in the NS or R sector and project onto the even or odd $J_0$ subsectors.
Recall that the NS sector fermions are anti-periodic on the circle $|z| = 1$,
\begin{equation*}
\psi^{a} = \sum_{n \in {\BZ} + \frac 12} {\psi}_{n}^{a} z^{-n} \bigg( \frac{{\rm d}z}{z} \bigg)^{\frac 12} , \qquad
\tilde\psi_{a} = \sum_{n \in {\BZ} + \frac 12} {\tilde\psi}_{a,n} z^{-n} \bigg( \frac{{\rm d}z}{z} \bigg)^{\frac 12}
\end{equation*}
and their modes obey the anti-commutation relations
\begin{equation*}
\big\{ {\tilde\psi}_{a,n}, {\psi}^{b}_{m} \big\} = {\delta}_{a}^{b} {\delta}_{n+m,0}.
\end{equation*}

The full space of states in the NS sector for the theory with $r$ fermions is
\begin{equation*}
{\BH}^{\rm NS} = {\BH}^{\rm NS}_{\rm even} \oplus {\BH}^{\rm NS}_{\rm odd},
\end{equation*}
where
\begin{equation*}
{\BH}^{\rm NS}_{\rm even/odd} = \bigoplus_{\stackrel{
 {\vec M} \in {\BZ}^{r} }
{ \sum_{i} M_{i} \,\, {\rm even/odd} }
} \otimes_{i=1}^{r} \mathcal{H}^{\rm NS}_{M_{i}}.
\end{equation*}
The character of the ${\BH}^{\rm NS}_{\rm even/odd}$ representation
is given by
\begin{equation*}
{\chi}^{\rm NS}_{\rm even / odd} = \frac{1}{ {\phi}^{r}(q)} \Theta_{r / r-1}.
\end{equation*}
In the Ramond sector the fermions are periodic on the circle $|z|=1$,
\begin{equation*}
\psi^{a} = \sum_{n \in {\BZ}} {\psi}_{n}^{a} z^{-n} \bigg( \frac{{\rm d}z}{z} \bigg)^{\frac 12} , \qquad
\tilde\psi_{a} = \sum_{n \in {\BZ}} {\tilde\psi}_{a,n} z^{-n} \bigg( \frac{{\rm d}z}{z} \bigg)^{\frac 12}
\end{equation*}
and their modes obey the anti-commutation relations
\begin{equation*}
\big\{ {\tilde\psi}_{a,n}, {\psi}^{b}_{m} \big\} = {\delta}_{a}^{b} {\delta}_{n+m,0}.
\end{equation*}
The characters of the Ramond representations are
\begin{equation*}
{\chi}^{R}_{\rm even(odd)} = \frac{1}{\phi(q)}\Theta_{0 / 1}.
\end{equation*}
The remaining fundamental representations are constructed
using $2r$ complex fermions (hence level two)
\begin{equation*}
L_{2} = \sum_{a=1}^{r}\big( {\tilde\psi}_{a} {\bar\partial} {\psi}^{a} +
{\tilde\psi}^{a} {\bar\partial} {\psi}_{a} \big),
\end{equation*}
whose $\hat \gq$ symmetry is generated by
\begin{equation*}
J^{ab} = {\tilde\psi}^{a}{\psi}^{b} - {\tilde\psi}^{b}{\psi}^{a} , \qquad
 J^{a}_{b} ={:} {\tilde\psi}_{b}{\psi}^{a} - {\psi}_{b}{\tilde\psi}^{a}{:} , \qquad
J_{ab} = {\tilde\psi}_{a}{\psi}_{b} - {\tilde\psi}_{b}{\psi}_{a},
\end{equation*}
which is actually a subalgebra of the level one $\widehat{\mathfrak{su}(2r)}$
symmetry. Accordingly, $\hat \gq$ commutes with the ${\widehat{\mathfrak{u}(1)}}$ Kac--Moody generated by
\begin{equation*}
\mathcal{J} = {:} {\tilde \psi}_{a}{\psi}^{a} {:} + {:} {\tilde \psi}^{a}{\psi}_{a} {:}
\end{equation*}
we can construct the representations
$\mathcal{V}_{p}$, $p = 0, \dots, 2r-1$ by taking the charge $p$ subspace
in the space of states of $2r$ fermions. Unlike the case of $r$ complex
or $2r$ real fermions, for the~$2r$ complex fermions there is no difference between the NS or R boundary conditions (in fact, one can
continuously interpolate between them using the spectral flow).
The corresponding characters are given by~$\Xi_p$.

The exceptional affine Kac--Moody algebras are realized using spin operators which
map the NS sectors to R and vice versa. For example, to realize ${\hat E}_{8}$, we take $8$ complex fermions ${\tilde\psi}_{i}$, ${\psi}_{i}$.

The $\Gli$ representations also have the free fermion realization.

The generators of $\Gli$ are given by the expressions
\begin{equation*}
{\CalO}_{a} = \sum_{i,j \in \BZ} a_{i,j}\, {:} {\psi}_{i} {\tilde\psi}_{j} {:}.
\end{equation*}

\section{Examples of conformal extensions}

We list here the examples of the groups $\bG$, their conformal extensions ${\rm C}\bG $, and the homomorphisms
\begin{equation*}
D\colon \ {\rm C}\bG \longrightarrow {\rm C} = ({\BC}^{\times})^{z_{\g}}.
\end{equation*}
\begin{enumerate}\itemsep=0pt
\item The $A_{r}$ series.
The conformal group C$\bG= {\rm GL}(r+1, {\BC})$, i.e., the group of
 non-degenerate $(r+1) \times (r+1)$ matrices $h$. The homomorphism $D$ is given by the determinant
\[
 D(h) = \det (h).
\]

\item The $D_{r}$ series.
First we describe the conformal extension of the group ${\rm SO}(2r)$. This is a~group of the conformal linear transformations
of ${\BC}^{2r}$, i.e., the group of $2r \times 2r$ matrices~$h$, such that
\begin{equation*}
h h^{t} = D(h) \cdot {\bf 1}_{2r}, \qquad D(h) \in {\BC}^{\times}
\end{equation*}
with the homomorphism $D(h_{1}h_{2}) = D(h_{1}) D(h_{2})$, such that
$D(h)^{r} = \det (h)$.
In other words, these are the linear transformations which preserve the non-degenerate symmetric bilinear form up to a scalar multiple. The group C${\rm Spin}(2r)$
for $r$ even has two independent central elements $c_{+}$, $c_{-}$,
so that $c_{+}$ acts trivially in $S_{-}$ and $c_{-}$ acts trivially in $S_{+}$
which act
non-trivially in the spin representations $S_{+}$, while in the tensor
representations $\wedge^i V$ both act as $c_{\pm}^{i}$, and their product
is equal to $D(h)$. For $r$ odd there is one central generator~$c$ which acts
as $c^{i}$ on $\wedge^{i}V$ and as $c^{\pm 1}$ on $S_{\pm}$.
\item The $E_6$, $E_7$ case.
The $E_{6}$, $E_{7}$ groups can be characterized as the symmetries of certain poly-linear forms in the $27$ and the $56$-dimensional vector spaces, respectively. Their conformal versions preserve these forms up to a scalar multiple. For example,
$E_6$ is a~stabilizer of the cubic form
\[
Q(z,x,y) = {\rm Pf}(z) + z( x \wedge y),
\]
where $x, y \in V \approx {\BC}^{6}$, $z \in {\Lambda}^{2}V^{*} \approx {\BC}^{15}$.
\end{enumerate}

\subsection[The maximal torus $\bT$ and the action of $Z$]{The maximal torus $\bT$ and the action of $\boldsymbol{Z}$}

The maximal torus $\Tq$ can be coordinatized by
$({\BC}^{\times})^{r}$ via
\begin{equation*}
(g_{1}, \dots , g_{r} )\in ({\BC}^{\times})^{r} \mapsto \prod_{i=1}^{r} g_{i}^{{\al}_{i}} \in {\bT}
\end{equation*}
since in our conventions $ \exp 2\pi \ii {\al}_{j} = 1 \in \bT$.
\begin{enumerate}\itemsep=0pt
\item
The maximal torus of the $A_r$ series is the group of diagonal matrices of size $r+1$
\begin{equation}
g = \operatorname{diag}\big(g_{1}, g_{2}g_{1}^{-1}, g_{3}g_{2}^{-1}, \dots , g_{r}g_{r-1}^{-1}, g_{r}^{-1} \big)
\label{eq:arc}
\end{equation}
with unit determinant. The formula (\ref{eq:arc}) corresponds to the standard choice of the simple roots
\[
{\al}_{j} = e_{j} - e_{j+1}, \qquad j = 1, \dots , r,
\]
where $e_{j}$ represent the vectors of some orthonormal basis in ${\BC}^{r+1}$.
The center $Z = {\BZ}/{(r+1)}{\BZ}$ acts on $\bT$, in the $g_{j}$ coordinates, via
\begin{equation*}
g_{j} \mapsto {\omega}^{j} g_{j},
\qquad {\omega} = {\rm e}^{\frac{2\pi \ii}{r+1}}
\end{equation*}
so that the matrix $g$ in (\ref{eq:arc}) is multiplied by $\omega$.
The generator ${\zeta} \equiv {\zeta}_{1}$ of the center $Z$ can be chosen as follows:
\begin{equation*}
{\zeta} = \frac{1}{r+1} \sum_{j=1}^{r} j {\al}_{j}.
\end{equation*}
\item
The maximal torus of the $D_r$ series can be identified with the group of block-diagonal matrices of size $2r \times 2r$
\begin{equation}
g = \operatorname{diag}\big( R(g_{1}), R\big(g_{2}g_{1}^{-1}\big), \dots, R\big(g_{r-1}g_{r}g_{r-2}^{-1}\big), R\big(g_{r}g_{r-1}^{-1}\big) \big),
\label{eq:drc}
\end{equation}
where for $t \in \BC$:
\[
R(t) = \frac{t}{2} \left( \begin{matrix} 1 & -\ii \\ \ii & \hphantom{-}1 \\ \end{matrix} \right) + \frac{t^{-1}}{2}
\left( \begin{matrix} \hphantom{-}1 & \ii \\ -\ii & 1 \\ \end{matrix} \right).
\]
Of course, equation~(\ref{eq:drc}) defines only an element of the
group ${\rm SO}(2r,{\BC})$, in order to define the element
of the spin cover ${\rm Spin}(2r,{\BC})$ we have to specify
the way $g$ acts in the spin representations $S_{+}$, $S_{-}$.
This is equivalent to the choice of the square root of $g_{r-1}^2$ and $g_r^2$, respectively, and the parametrization
\begin{equation}
g = \prod_{i=1}^{r} g_{i}^{{\al}_{i}}
\label{eq:drots}
\end{equation}
corresponds to these square roots being $g_{r-1}$ and
$g_{r}$, respectively. In equation~(\ref{eq:drots}) the simple roots of the $D_r$ algebra are
\begin{equation*}
 {\al}_{i} = e_{i} - e_{i+1}, \qquad i = 1, \dots , r-1, \qquad
{\al}_{r} = e_{r-1} + e_{r}.
\end{equation*}
The center acts on $\bT$ as follows. If $r = 2s+1$, then
the transformation generated by the $4$-th root of unity $\omega$, ${\omega}^4 = 1$, acts by
\begin{equation*}
\begin{aligned}
& g_{2m+1} \mapsto {\omega}^{2} g_{2m+1}, \\
& g_{2m} \mapsto g_{2m}, \qquad m = 1, \dots, s-1, \\
& g_{2s} \mapsto {\omega}^{-1} g_{2s} , \qquad
g_{2s+1} \mapsto {\omega} g_{2s+1} \end{aligned}
\end{equation*}
and multiplies $g$ by ${\omega}^{2}$. The center is $Z = {\BZ}/4{\BZ}$ in this case.
If $r = 2s$, then the transformations generated by
${\omega}_{1}$, ${\omega}_{2}$, such that ${\omega}_{1}^{2} ={\omega}_{2}^{2} = 1$, act on $g_{j}$ via
\begin{gather*}
 g_{2m+1} \mapsto {\omega}_{1}{\omega}_{2} g_{2m+1}, \\
 g_{2m} \mapsto g_{2m}, \qquad m = 1, \dots, s-1, \\
 g_{2s-1} \mapsto {\omega}_{1} g_{2s-1} , \qquad g_{2s} \mapsto {\omega}_{2} g_{2s}
\end{gather*}
and multiply $g$ by ${\omega}_{1}{\omega}_{2}$. The center is $Z = ({\BZ}/2{\BZ}) \times ({\BZ}/2{\BZ})$ in this case.
\item
In order to describe the action of the center $Z = {\BZ}/3{\BZ}$ on the maximal torus of the group~$E_6$ let us
choose the canonical basis of simple roots of $E_6$:
\begin{gather*}
 {\al}_{1} = e_{1} - e_{2}
, \qquad {\al}_{3} = e_{2} - e_{3} , \\
 {\al}_{4} = e_{3} - e_{4} ,
\qquad {\al}_{5} = e_{4} - e_{5} , \\
 {\al}_{6} = e_{5} - e_{6} , \qquad
 {\al}_{2} = - e_{1} - e_{2} - e_{3} + {\eta} e, \qquad
 e = \sum_{i=1}^{6} e_{i},
\end{gather*}
where ${\eta} = \frac {\sqrt{3}-1}{2\sqrt{3}} $.
Then the element $\om \in Z$, $\om^3 = 1$ acts on the element of the maximal torus of $E_6$
\[
g = \prod_{i=1}^{6} g_{i}^{{\al}_{i}} \in \Tq
\]
by
\[
g \mapsto
\prod_{i=1}^{6} \big( {\om}^{i}g_{i} \big)^{{\al}_{i}}
= {\om}^{{\al}_{1} - {\al}_{2} + {\al}_{4} - {\al}_{5}} g.
\]
\item
For the group $E_7$, the simple roots are
\begin{gather*}
 {\al}_{1} = e_{1} - e_{2}
, \qquad {\al}_{3} = e_{2} - e_{3} , \qquad {\al}_{4} = e_{3} - e_{4}, \\
 {\al}_{5} = e_{4} - e_{5} , \qquad {\al}_{6} = e_{5} - e_{6} , \qquad {\al}_{7} = e_{6} - e_{7} ,\\
 {\al}_{2} = - e_{1} - e_{2} - e_{3} + {\eta} e, \qquad e = \sum_{i=1}^{7} e_{i},
\end{gather*}
where $\eta = \frac {3-\sqrt{2}}{7}$.
The generator ${\om}$, ${\om}^{2}=1$, of the center $Z = {\BZ}/2{\BZ}$ acts
on the element of the maximal torus of $E_7$
\[
g = \prod_{i=1}^{7} g_{i}^{{\al}_{i}} \in \bT
\]
by
\[
g \mapsto {\om}^{{\al}_{5} + {\al}_{7} + {\al}_{2}} g.
\]
\end{enumerate}

\section{The elliptic curve and elliptic functions}\label{se:elliptic}

The combination
\begin{equation}
{\qe} = \prod_{i \in \Ver} {\qe}_{i}^{a_{i}} = {\rm e}^{2\pi \ii \tau}
\label{eq:qivi}
\end{equation}
of the multiplicative couplings ${\qe}_{i}$ plays a special role.
It defines an elliptic curve
\begin{equation}
{\ec} = {\BC}^{\times}/{\qe}^{\BZ},
\label{eq:ellcur}
\end{equation}
which underlies many constructions related to the class II and class II* theories.

The infinite product
\begin{equation*}
 \phi(\qe) = \prod_{n=1}^{\infty} (1 - \qe^n) \equiv (\qe,\qe)_{\infty}
\end{equation*}
is the Euler product. Dedekind eta-function
\begin{equation*}
{\eta}({\qe}) = \frac{1}{{\qe}^{\frac 1{24}} {\phi}({\qe})}.
\end{equation*}

\section[Coordinates on $\ec$]{Coordinates on $\boldsymbol{\ec}$}

The elliptic curve $\ec$ can be described in several coordinate systems.
The multiplicative coordinate system
\[
t \sim {\qe}^{n} t, \qquad t \in {\BC}^{\times},\qquad n \in \BZ
\]
is used in (\ref{eq:ellcur}). The additive coordinate system
\[
z \sim z + m + n {\tau}, \qquad z \in {\BC},
\]
where
\[
\tau \in {\BC}, \qquad \operatorname{Im}{\tau} > 0
\]
is defined in (\ref{eq:qivi}),
and $m,n \in {\BZ}$, is related to the multiplicative one via:
\begin{equation*}
t = {\rm e}^{2\pi \ii z}.
\end{equation*}

\section{Elliptic functions}\label{appendixO}

\subsection{Jacobi theta-functions}
Our conventions for the theta-functions, associated to
the elliptic curve ${\ec} = {\BC}^{\times}/{\qe}^{\BZ}$, are
\begin{gather}
 \theta_1(t;{\qe}) = {\rm i}\sum_{n \in \BZ + \frac 1 2}
 (-1)^{n-\frac 12} t^{n} {\qe}^{\frac 1 2n^2}, \nonumber\\
 \theta_2(t;{\qe}) = \sum_{n \in \BZ + \frac 1 2}
 t^{n} {\qe}^{\frac 1 2n^2}, \nonumber\\
 \theta_3(t;{\qe}) = \sum_{n \in \BZ }
 t^{n} \qe^{\frac 1 2n^2}, \nonumber\\
 \theta_4(t;{\qe}) = \sum_{n \in \BZ}
 (-1)^{n} t^{n} {\qe}^{\frac 1 2n^2}.
\label{eq:Jacobi-theta}
\end{gather}
Note that in older literature on elliptic theta-functions $\qe$ is
 used to denote the nome'' $q_{\mathrm{nome}} = {\rm e}^{\pi{\rm i} \tau} = {\qe}^{\frac
 1 2}$. In order not to be confused we shall always specify
the $\qe$ modulus in the formula for~$\theta_{\al}(t; {\qe})$.

It is often convenient to use basic pseudo-elliptic $\theta$-function
\begin{equation*}
 \theta(t;{\qe}) =
 \sum_{b \in \BZ} (-t)^{b} {\qe}^{ \frac 1 2 b(b-1)}
 = (1-t) + {\qe} \big(t^2 - t^{-1}\big) + \cdots,
\end{equation*}
which is related in a simple manner to $\theta_1(t;\qe)$:
\begin{equation*}
 \theta(t; \qe) = -\ii \qe^{-\frac 1 8}
 t^{\frac 12} \theta_1(t; \qe)
\end{equation*}
and also to the ${\hat A}_{1}$ Weyl denominator
in the affine Kac--Weyl character formula
\begin{equation*}
 \theta(t;{\qe}) = R(h) = \prod_{n \geq 0} \big(1 -t {\qe}^n\big)\big(1-{\qe}^{n+1}\big)\big(1-t^{-1} {\qe}^{n+1}\big).
\end{equation*}

The transformation rules for $\theta(t;\qe)$ are
\begin{equation*}
 \theta( {\qe}t ; {\qe}) = \theta\big( t^{-1} ; {\qe}\big) = -t^{-1} \theta( t ; {\qe})
\end{equation*}
and transformation properties of $\theta_1(t,\qe)$ are
\begin{equation*}
 \theta_1(\qe t; \qe) = -\qe^{-\frac{1}{2}} t^{-1} \theta_1( t; \qe).
\end{equation*}

The relation of Jacobi theta functions to the Weierstra{\ss} functions
is described below.

\subsection[The $\xi$- and $\wp$-functions]{The $\boldsymbol{\xi}$- and $\boldsymbol{\wp}$-functions}

Define the function
\begin{equation*}
 \xi (t; \qe) = t \pa_{t} \log \theta_1 (t ; \qe),
\end{equation*}
which satisfies
\begin{equation*}
 \xi (\qe t ; \qe) = \xi (t; \qe) - 1, \qquad {\xi}\big(t^{-1}; {\qe}\big) = - {\xi}(t ; {\qe})
\end{equation*}
and has simple poles at $ t = \qe^\BZ$. Therefore,
up to a shift by a linear function of $ \log (t)$, $\xi(t; \qe)$ is the Weierstra{\ss} $\zeta$-function.

From the product representation
of $\theta_{1}(t; q)$ we immediately get
\begin{equation*}
 \xi(t; \qe) =
 \frac 12 \frac{t+1}{t-1} - \sum_{n \geq 1} \qe^n \sum_{ d|n } \big(t^d - t^{-d}\big).
\end{equation*}
The expansion of $\xi(t; \qe)$ near $t=1$ reads as
\begin{equation*}
 \xi(t,\qe) = \frac{1}{t-1} + \sum_{k \geq 0} \xi_{k} (t-1)^{k}
\end{equation*}
with $\xi_0 = 1/2$ and
\begin{equation*}
 \xi_1(\qe) = - 2 \sum_{n \geq 1, k \geq 1} k \qe^{nk} = 2 \qe \frac{{\rm d}}{{\rm d}\qe} \log \phi(\qe).
\end{equation*}

The function
\begin{equation*}
 \wp(t; \qe) = -t\pa_{t} \xi(t; \qe) + {\wp}_{0} , \qquad {\wp}_{0} = 2 {\qe} \frac{{\rm d}}{{\rm d}{\qe}} \log {\eta}({\qe})
 \end{equation*}
is elliptic function, i.e., a meromorphic function on $\ec$:
\begin{equation*}
 \wp(\qe t;\qe) = \wp(t; \qe),
\end{equation*}
which has a single second-order pole in $\ec$, and on the covering space
${\BC}^{\times}_{\langle t \rangle}$ it has a double pole at $t = \qe^\BZ$.

The function $\wp(t; \qe)$ is actually the Weierstra{\ss} ${\wp}$-function:
the double-periodic version of the~$\frac{1}{z^2}$ function in the plane $z \in \BC /(\BZ \oplus \tau
\BZ)$, with the constant ${\wp}_{0}$ chosen so that the constant
term in the expansion of
$\wp\big( t={\rm e}^{2\pi \ii z}; \qe\big)$ in Laurent series in $z$ at $z=0$
vanishes.
This gives
\begin{equation}
\label{eq:weierx1}
 \wp(t,\qe) = \frac{t}{(1-t)^2} +\frac{1}{12} + \sum_{k\geq 1} k
 \frac{\qe^k}{1 - \qe^k} \big(t^{k} + t^{-k} - 2\big).
\end{equation}

One more differentiation with respect to $\log t$ defines the Weierstra{\ss} elliptic function with
the pole of the third order
\begin{equation*}
 \wp'(t,\qe) := t \pa_{t} \wp(t,\qe)
\end{equation*}
and the expansion near $t=1$:
\begin{equation*}
 \wp'(t,\qe) = t \frac{ 1 +t}{(1-t)^3} + \sum_{k \geq 0} k^2
\frac{\qe^k}{1 - \qe^k} \big( t^k - t^{-k}\big).
\end{equation*}
Define the functions
$X(t,\qe)$ and $Y(t,\qe)$ by
\begin{gather*}
 X(t,\qe) = \wp(t,\qe), \qquad
 Y(t,\qe) = \wp'(t, \qe)
\label{eq:weierx2}
\end{gather*}
and define the weight $4$ modular form $g_2(\qe)$ and weight $6$ modular
form $g_3(\qe)$
\begin{gather*}
 g_{2}(\qe) := \frac{1}{12} + 20 \sum_{k=1}^{\infty} k^3 \frac{{\qe}^{k}}{1-{\qe}^{k}}, \\
 g_{3}(\qe) :=- \frac{1}{216} + \frac{7}{3} \sum_{k=1}^{\infty} k^{5} \frac{{\qe}^{k}}{1-{\qe}^{k}}.
\end{gather*}
The Weierstra{\ss} functions $X(t,\qe)$ and $Y(t,\qe)$ satisfy the cubic equation
\begin{equation}\label{eq:wxy}
 Y(t,\qe)^2 = 4 X(t,\qe)^3 - g_2(\qe) X(t,\qe) - g_3(\qe) = 4 \prod_{{\al}=1}^{3}( X (t, \qe) - e_{\al} ),
\end{equation}
where
\begin{equation}
e_{1} = {\wp}(-1; {\qe}), \qquad e_{2} = {\wp}\bigl(-{\qe}^{\frac 12}; {\qe}\bigr), \qquad e_{3} = {\wp} \big( {\qe}^{\frac 12}; {\qe} \big).
\label{eq:eal}\end{equation}
The periods of the holomorphic differential
${\rm d}X/Y$ on $\ec$ are equal to
\begin{equation*}
\frac{1}{2\pi \ii} \oint_{A} \frac{{\rm d} X}{Y} = 1, \qquad
\frac{1}{2\pi \ii} \oint_{B} \frac{{\rm d} X}{Y} = {\tau}.
\end{equation*}

\subsection[${\phi}_{i}$ and ${\tilde\phi}_{\tilde j}$ -functions]{${\boldsymbol{\phi}_{i}}$ and $\boldsymbol{{\tilde\phi}_{\tilde j}}$ -functions}\label{subsubsec:phi}

We shall also need to deal with meromorphic functions on $\ec$ with the higher-order poles
at $t=1$:\footnote{It is easy to verify that $\phi_i( t \qe ,\qe) = \phi_i(t,\qe)$.} on $\ec$ with
the only pole of order $\leq (r+1)$ at $t = 1$. Explicitly,
\begin{equation}
\label{eq:phi_p}
 \phi_i(t;\qe) = \frac{ (-{\qe}/t)^i \theta\big(- (- t)^{r+1} \qe^{-i}; \qe^{r+1}\big)}{
\theta(t;\qe)^{r+1} }.
\end{equation}
The $r+1$ meromorphic functions $\phi_i(t;\qe)$, $i = 0 \dots r$ with the pole of
order $r+1$ at $t=1$ form a basis of $H^{0}({\ec}, {\CalO}((r+1)p_{0})) \approx {\BC}^{r+1}$. On the
other hand, if we represent elliptic curve $\ec$ in
the Weierstra{\ss} form \eqref{eq:wxy},
$Y^2 = 4 X^3 - g_2 X - g_3$,
then the basis in the space of meromorphic elliptic functions on $\ec$ with the
pole at $t=1$ of less then $r+2$ order can also be chosen as the monomials
\begin{equation*}
{\tilde \phi}_{{\tilde j}}(t,\qe)= \begin{cases}
 X^{{\tilde j}/2} & \text{if } 2| {\tilde j}, \\
 X^{({\tilde j}-3)/2}Y & \text{if } 2 | \big({\tilde j}+1\big),
 \end{cases}\qquad {\tilde j} \in \{0,2,\dots, r+1\},
\end{equation*}
since $X(t;\qe) = \wp(t;\qe)$ is the Weierstra{\ss} elliptic function
 with the pole of second order at $t =1$
and $Y(t;\qe) = t\pa_{t} \wp(t;\qe)$ has the pole of third order at $t =1$.
There is a linear relation between the Weierstra{\ss} monomials $\tilde
\phi_{\tilde j}(t;\qe)$ and
the ratio \eqref{eq:phi_p} of theta-functions $\phi_i(t;\qe)$ which we
shall denote as
\begin{equation}\label{eq:T-matrix}
 \phi_i (t;\qe) = \sum_{{\tilde j}} M_{i{\tilde j}}(\qe) \tilde \phi_{{\tilde j}}(t;\qe),
\end{equation}
where $M_{j\tilde j}(\qe)$ is a certain $\qe$-modular $(r+1)\times (r+1)$
transformation matrix, see example in Appendix~\ref{se:modular-example}.

\subsection{Expansion of Weierstra{\ss} functions in terms of
 theta-functions}\label{se:modular-example}

The matrix $M(\qe)$ for the change of basis from meromorphic elliptic functions
\eqref{eq:phi_p} to Weierstra{\ss} functions for $r=1$ is
\begin{equation*}
 \begin{pmatrix}
 1 \\
 X(t,\qe)
 \end{pmatrix} =
 \begin{pmatrix}
 M_{11} & M_{12}\\
 M_{21} & M_{22}
 \end{pmatrix}
 \begin{pmatrix}
 \phi_0(t,\qe) \\
 \phi_1(t,\qe)
 \end{pmatrix}
\end{equation*}
with
\begin{equation*}
 \begin{aligned}
& M_{11} = \theta_3(1,\qe), \\
& M_{12} = \theta_2(1,\qe), \\
& M_{21} = M_{11} \bigg(-\frac 1 6 m_2(2; \qe) + \frac 1 2 m_2(4;\qe)^{(1)}
 - 4 m_2(4;\qe)^{(2)} \bigg), \\
& M_{22} = M_{12} \bigg(-\frac 1 6 m_2(2; \qe) - \frac 1 2 m_2(4;\qe)^{(1)}
 + 4 m_2(4;\qe)^{(2)} \bigg),
 \end{aligned}
\end{equation*}
 where $m_k(N;\qe)^{(i)}$ denotes $i$-th modular form for modular group $\Gamma_0(N)$ of
 weight $k$ in a certain basis. Concretely
 \begin{equation*}
 \begin{aligned}
& m_2(2,\qe) = \Theta_{D_4}(\qe) = \frac 1 2 \big(\theta_3(\qe)^4 +
 \theta_4(\qe)^4\big), \\
& m_2(4,\qe)^{(1)} = m_2\big(2,\qe^2\big), \\
& m_2(4,\qe)^{(2)} =\frac{1}{16} \theta_2(\qe)^4 = \frac{ \eta\big(\qe^4\big)^8}{\eta\big(\qe^2\big)^4}.
\end{aligned}
 \end{equation*}

\subsection[Even $\phi$ functions for D series]{Even $\boldsymbol{\phi}$ functions for D series}\label{se:phiD}
From \eqref{eq:DXi}, we find
\begin{equation}
\label{eq:s-ThetaD}
\y_0(x)^2
\prod_{i=1}^{r} \frac{ \theta(t/t_i(x);\qe)}{\theta(t;\qe)} \frac{\theta\big(t/t_{i}(x)^{-1};\qe\big)}{\theta(t;\qe)}
 = \sum_{i=0}^{2r-1}
 (-1)^i \qe^{\frac {i}{2}+ \frac{i^2}{4r}} \Xi_i(\y_0;\mathbf{t}(x);\qe)
\phi^{(2r)}_i(t;\qe),
\end{equation}
where $\phi_{i}^{(2r)}(t;\qe)$ are the meromorphic functions, with the pole at
$t=1$ of
order not greater than~$2r$, associated to the
holomorphic sections
$H^{0}({\ec}, \CalO( 2 r p_0))$
\begin{equation*}
 \phi_{i}^{(2r)}(t;\qe) = \frac{ t^{-i} \theta \big(-t^{2r} \qe^{-i}; \qe^{2r}\big)}{
 \theta (t;\qe)^{2r}}.
\end{equation*}
Using $\Xi_{2r -j} = \Xi_{j}$,
we get
\begin{equation}\label{eq:spectralDapp}
 \y_0^2 s(t,x) = \sum_{ i=0}^{r} \Xi_{ i}(\y_0; \mathbf{t}(x);\qe)
M_{ij}(\qe) X^{j} (t;\qe)
\end{equation}
with
\begin{equation*}
 \begin{aligned}
 &\tilde \phi_{ 0}(t;\qe) = \phi_0^{(2r)}(t;\qe), \\
 &\tilde \phi_{i}(t;\qe) = (-1)^{i} \qe^{\frac{i}{2}+ \frac{i^2}{4r}}\big( \phi_i^{(2r)}(t;\qe) + \qe^{2r - 2i} \phi^{(2r)}_{2r
 -i}(t;\qe)\big), \qquad i = 1,\dots,r-1,\\
 &\tilde \phi_{r}(t;\qe) =(-1)^{r} \qe^{\frac{3}{4} r} \phi_r^{(2r)}(t;\qe).
 \end{aligned}
\end{equation*}

The $r+1$ functions $\tilde \phi_{i}(t;\qe)$ for $j=0,\dots,r$ form the
basis in the space $H^{0}_{\text{even}}({\ec}, \CalO(2 r p_0))$
of meromorphic functions on elliptic curve symmetric under the reflection
$t \to t^{-1}$ and with a pole of order no greater then $2r$ at the
origin. Another basis in $H^{0}_{\text{even}}({\ec},\CalO(2r p_0))$ is given
by the powers of Weierstra{\ss} $\wp$-function
\begin{equation*}
 X(t;\qe)^{j}, \qquad j = 0, \dots, r.
 \end{equation*}
 Let $M_{ij}(\qe)$ be the transformation matrix between these bases in
$H^{0}_{\text{even}}({\ec},\CalO(2r p_0))$
\begin{equation*}
 \tilde \phi_{i}(t;\qe) = \sum_{j=0}^{r} M_{ij}(\qe) X^{j} (t;\qe),
\end{equation*}
then
\eqref{eq:spectralDapp} can be written as \eqref{eq:spectralD}.

\section[Affine $E$ spectral curves]{Affine $\boldsymbol{E}$ spectral curves}\label{se:appendixE}

\subsection[The $E_{r}$ spectral curves from del Pezzo]{The $\boldsymbol{E_{r}}$ spectral curves from del Pezzo}
\subsubsection[The $E_6$ spectral curve from del Pezzo]{The $\boldsymbol{E_6}$ spectral curve from del Pezzo}\label{sec: E6-delPezzo-code}
\begin{verbatim}
E6Pezzo[X0_, X1_, X2_, X3_] := -X1 X3^2 + 4 X2^3 - g2 X2 X1^2 -
 g3 X1^3 +
 X0 (p[0] X1^2 + p[1] X2 X1 + p[6] X2 X3) +
 X0^2 (p[2] X1 + p[3] X2 + p[5] X3) + X0^3 p[4];
E6Xi = CoefficientList[E6Pezzo[z, 1, X + vx z, Y + vy z ], z];
vys = First[Solve[E6Xi[[2]] == 0, vy]];
E6PezzoCurveU = -4 Y^4 Resultant[ E6Xi[[3]] /. vys, E6Xi[[4]] /. vys,
 vx];
E6PezzoCurve =
 PolynomialRemainder[E6PezzoCurveU, E6Pezzo[0, 1, X, Y], Y] //
 Expand
\end{verbatim}

\subsubsection[The $E_7$ spectral curve from del Pezzo]{The $\boldsymbol{E_7}$ spectral curve from del Pezzo}\label{sec: E7-delPezzo-code}
\begin{verbatim}
E7Pezzo[X0_, X1_, X2_, X3_] := -X3^2 + 4 X1 X2^3 - g2 X2 X1^3 -
 g3 X1^4 +
 X0 (p[0] X1^3 + p[7] X1^2 X2 ) +
 X0^2 ( p[1] X1^2 + p[2] X1 X2 + p[6] X2^2 ) +
 X0^3 ( p[3] X1 + p[5] X2) + p[4] X0^4 ;
E7Xi = CoefficientList[
 E7Pezzo[z, 1, X + vx z, Y + vy z + 1/2 z^2 wy], z];
vys = First[Solve[E7Xi[[2]] == 0, vy]];
wys = First[Solve[E7Xi[[3]] == 0, wy]] /. vys;
E7PezzoCurveU = -1024 Y^18 Resultant[(E7Xi[[4]] /. wys /. vys),
 E7Xi[[5]] /. wys /. vys, vx];
E7PezzoCurve =
 PolynomialRemainder[E7PezzoCurveU, E7Pezzo[0, 1, X, Y], Y] //
 Expand
\end{verbatim}

\subsection{The map from
del Pezzo to theta-function coordinates}\label{sec:E6-modular-matrix}

In this section, we record the modular matrix $M_{i,\{j_1,\dots,j_r\}}(\qe)$
discussed in Section~\ref{sec:affineE} for~$E_6$ bundles on
$\ec(\qe)$. The result is presented in terms of a certain basis $(m_{h})$ in the
space of modular forms of $\Gamma(6)$ of weight $1$ with the expansion
at the cusp $\qe = 0$ starting as
\begin{equation*}
 m_{h} = \qe^{h} \Bigg(1 + \sum_{n=1}^{\infty} \mu_{h,n} \qe^n\Bigg).
\end{equation*}
(If the argument of $m_{h}$ is not explicitly
spelled out it is always assumed to be $\qe$, i.e., in this section we
set $m_{h} \equiv m_{h}(\qe)$).
Concretely, the basis of weight $1$ modular forms of $\Gamma(6)$ is
\begin{gather*}
 m_0 = \big[1^{-3},2^{6},3^{1},6^{-2}\big] + 3 \big[ 1^1, 2^{-2}, 3^{-3}, 6^{6} \big]
 = 1 + 6 \qe + 6 \qe^3 + \cdots,\\
 m_{\tfrac 1 6} = \big[1^{-2},2^{3}, 3^{2}, 6^{-1}\big] = \qe^{\tfrac 1 6} \big(1 +
 2 \qe + 2 \qe^2 + 2 \qe^3 + \cdots \big), \\
 m_{\tfrac 1 3} = \big[1^{-1}, 3^{3}\big] = \qe^{\tfrac 1 3} \big( 1 + \qe + 2
 \qe^2 + 2 \qe^4 + \cdots \big),\\
 m_{\tfrac 1 2} = \big[1^{-1}, 2^{2}, 3^{-1}, 6^{2}\big] = \qe^{\tfrac 1 2} \big( 1 + \qe + 2 \qe^3 + \qe^4 + \cdots\big), \\
 m_{\tfrac 2 3} = \big[2^{-1}, 6^{3}\big] = \qe^{\tfrac 2 3} \big( 1 + \qe^2 + 2 \qe^4
+ \cdots \big),\\
 m_{1} = \big[1^{1}, 2^{-2}, 3^{-3}, 6^{6}\big] = \qe\big(1 - \qe + \qe^2 +
\qe^3 - \qe^5 + 2 \qe^6 + \cdots \big),
\end{gather*}
where we use the standard notation for the eta-products
\begin{equation*}
 \big[1^{p_1}, 2^{p_2}, 3^{p_3}, \dots, L^{p_L} \big] = \prod_{j=1}^{L} \eta\big(\qe^j\big)^{p_j},
\end{equation*}
where $\eta(\qe)$ is the Dedekind modular function
\begin{equation*}
 \eta(\qe) = \qe^{\tfrac 1 {24}} \prod_{j=1}^\infty \big( 1- \qe^{j}\big).
\end{equation*}

In particular, notice that the Weierstra{\ss} parameters $g_2(\qe)$,
$g_3(\qe)$
are expressed as
\begin{gather*}
 g_2(\qe) =m_0 \big(\tfrac 1 {12} m_0^3 + 18 m_{\tfrac 1 3}^3\big) =\tfrac{1}{12} + 20 \qe
 + 180 \qe^2 + 560 \qe^3 + 1460 \qe^4 + \cdots, \\
 g_3(\qe) = -\frac 1 {216} m_0^6 + \tfrac {15} 6 m_0^3 m_{\tfrac 1 3}^3 +
 27 m_{\tfrac 1 3}^6 = -\tfrac 1 {216} + \tfrac {7} {3} \qe + 77 \qe^2 +
 \tfrac{1708}{3} \qe^3 + \tfrac {7399} 3 \qe^4 + \cdots.
\end{gather*}

We remark that $m_{0}(\qe)$ and $m_{\tfrac 1 3}(\qe)$ are modular forms for
$\Gamma(3)$ and they are equal to the theta-constants of the $\hat A_{2}$
lattice associated respectively to the fundamental weight $\Lambda_0$ and $\Lambda_1$
\begin{gather*}
 m_{0} = \sum_{ \alpha \in Q(A_2)} \qe^{ \frac{1}{2} (\alpha, \alpha)},\qquad
 m_{\tfrac 1 3} = \sum_{ \alpha \in \Lambda_1 + Q(A_2)} \qe^{ \frac 1 2 (\alpha, \alpha)}.
\end{gather*}

With all notations in place we finally present the explicitly
computed\footnote{The authors used computer algebra
 \textsc{Mathematica} to find these expressions for modular forms by matching the first
coefficients in the $\qe$-expansion.} components of the modular matrix
$M_{i,\{j_1,\dots \}}(\qe)$ (the fixed argument $\qe$ is assumed for all
$M_{i,\{j_1,\dots \}}$ and $m_{h}$) in the following table of relations
\begin{gather*}
M_{0,\{0\}}=m_0^3+54 m_{\frac{1}{3}}^3,\\
M_{0,\{1\}}=\frac{1}{12} m_0^2 \big(m_0^3-270 m_{\frac{1}{3}}^3\big),\\
M_{0,\{6\}}=0,\\
M_{1,\{0\}}=27 m_0 m_{\frac{1}{3}}^2,\\
M_{1,\{1\}}=-\frac{3}{4} m_{\frac{1}{3}}^2 \big(5 m_0^3+108 m_{\frac{1}{3}}^3\big),\\
M_{1,\{6\}}=-m_{\frac{1}{3}}^2 \big(3 m_{\frac{1}{3}}-m_0\big){}^2 \big(m_0^2+3 m_{\frac{1}{3}} m_0+9 m_{\frac{1}{3}}^2\big){}^2,\\
M_{2,\{2\}}=-4 m_0 m_{\frac{1}{2}} (m_0-12 m_1){}^3 (m_0-4 m_1) (m_0-3 m_1) \big(m_0^2+12 m_1 m_0-72 m_1^2\big),\\
M_{2,\{3\}}=-\frac{1}{3} m_{\frac{1}{2}} (m_0-12 m_1){}^3 (m_0-4 m_1) (m_0-3 m_1) \big(m_0^2+12 m_1 m_0-72 m_1^2\big) \\
\hphantom{M_{2,\{3\}}=}{} \times \big(m_0^3-108 m_1 m_0^2+864 m_1^2 m_0-1728 m_1^3\big), \\
M_{2,\{5\}}=0, \\
M_{2,\{0,0\}}=72 m_0 m_{\frac{1}{2}} (m_0-3 m_1), \\
M_{2,\{0,1\}}=-12 m_{\frac{1}{2}} (m_0-3 m_1) \big(m_0^3+52 m_1 m_0^2-384 m_1^2 m_0+576 m_1^3\big), \\
M_{2,\{1,1\}}=-\frac{1}{2} m_{\frac{1}{2}} (m_0-3 m_1) \big(m_0^5-200 m_1 m_0^4-96 m_1^2 m_0^3+20736 m_1^3 m_0^2-110592 m_1^4 m_0 \\
\hphantom{M_{2,\{1,1\}}=}{} +165888 m_1^5\big), \\
M_{2,\{6,6\}}=-\frac{1}{144} m_{\frac{1}{2}} (m_0-12 m_1){}^3 (m_0-4 m_1) (m_0-3 m_1) \big(m_0^7-780 m_1 m_0^6 \\
\hphantom{M_{2,\{6,6\}}=}{}+40680 m_1^2 m_0^5-648000 m_1^3 m_0^4+4914432 m_1^4 m_0^3-19906560 m_1^5 m_0^2 \\
\hphantom{M_{2,\{6,6\}}=}{}+41803776 m_1^6 m_0-35831808 m_1^7\big), \\
M_{3,\{3\}}=3 m_0^2 m_{\frac{1}{6}} m_{\frac{2}{3}} (m_0-12 m_1){}^3 (m_0-4 m_1) (m_0-3 m_1) \big(m_0^2+12 m_1 m_0-72 m_1^2\big), \\
M_{3,\{5\}}=-4 m_{\frac{1}{6}} m_{\frac{2}{3}}(m_0-12 m_1){}^5 (m_0-4 m_1) (m_0-3 m_1){}^2 \big(m_0^2+12 m_1 m_0-72 m_1^2\big), \\
M_{3,\{0,0\}}=216 m_{\frac{1}{6}} m_{\frac{2}{3}} (m_0-3 m_1),\\
M_{3,\{0,1\}}=-12 m_{\frac{1}{6}} m_{\frac{2}{3}} (m_0-3 m_1) \big(7 m_0^2+48 m_1 m_0-288 m_1^2\big), \\
M_{3,\{0,6\}}=-16 m_{\frac{1}{6}} m_{\frac{2}{3}} (m_0-12 m_1){}^2 (m_0-3 m_1){}^2 \big(m_0^2+12 m_1 m_0-72 m_1^2\big), \\
M_{3,\{1,1\}}=\frac{1}{2} m_0 m_{\frac{1}{6}} m_{\frac{2}{3}} (m_0-3 m_1) \big(13 m_0^3+360 m_1 m_0^2-2592 m_1^2 m_0+3456 m_1^3\big), \\
M_{3,\{1,6\}}=\frac{2}{3} m_0 m_{\frac{1}{6}} m_{\frac{2}{3}} (m_0-12 m_1){}^2 (m_0-3 m_1){}^2 \big(m_0^3+216 m_1 m_0^2-1728 m_1^2 m_0 \\
\hphantom{M_{3,\{1,6\}}=}{}+3456 m_1^3\big), \\
M_{4,\{4\}}=16 m_{\frac{1}{3}}^3 \big(3 m_{\frac{1}{3}}-m_0\big){}^3 \big(m_0^2+3 m_{\frac{1}{3}} m_0+9 m_{\frac{1}{3}}^2\big){}^3 \big(36 m_{\frac{1}{3}}^3-m_0^3\big), \\
M_{4,\{0,2\}}=24 m_{\frac{1}{3}}^3 \big(3 m_{\frac{1}{3}}-m_0\big) \big(m_0^2+3 m_{\frac{1}{3}} m_0+9 m_{\frac{1}{3}}^2\big) \big(11 m_0^3+108 m_{\frac{1}{3}}^3\big), \\
M_{4,\{0,3\}}=2 m_0^2 \big(m_0-3 m_{\frac{1}{3}}\big) m_{\frac{1}{3}}^3 \big(m_0^2+3 m_{\frac{1}{3}} m_0+9 m_{\frac{1}{3}}^2\big) \big(13 m_0^3+864 m_{\frac{1}{3}}^3\big), \\
M_{4,\{1,2\}}=2 m_0^2 \big(m_0-3 m_{\frac{1}{3}}\big) m_{\frac{1}{3}}^3 \big(m_0^2+3 m_{\frac{1}{3}} m_0+9 m_{\frac{1}{3}}^2\big) \big(13 m_0^3+864 m_{\frac{1}{3}}^3\big), \\
M_{4,\{1,3\}}=-\frac{1}{6} m_0 \big(m_0-3 m_{\frac{1}{3}}\big) m_{\frac{1}{3}}^3 \big(m_0+6 m_{\frac{1}{3}}\big) \big(m_0^2+3 m_{\frac{1}{3}} m_0+9 m_{\frac{1}{3}}^2\big) \\ \hphantom{M_{4,\{1,3\}}=}{} \times \big(m_0^2-6 m_{\frac{1}{3}} m_0+36 m_{\frac{1}{3}}^2\big) \big(11 m_0^3+108 m_{\frac{1}{3}}^3\big), \\
M_{4,\{5,6\}}=\frac{2}{3} m_0^5 \big(m_0-3 m_{\frac{1}{3}}\big){}^3 m_{\frac{1}{3}}^3 \big(m_0^2+3 m_{\frac{1}{3}} m_0+9 m_{\frac{1}{3}}^2\big){}^3 , \\
M_{4,\{0,0,0\}}=720 m_{\frac{1}{3}}^3, \\
M_{4,\{0,0,1\}}=-540 m_0^2 m_{\frac{1}{3}}^3, \\
M_{4,\{0,1,1\}}=15 m_0 m_{\frac{1}{3}}^3 \big(7 m_0^3+54 m_{\frac{1}{3}}^3\big), \\
M_{4,\{0,6,6\}}=\frac{1}{24} m_0 \big(m_0-3 m_{\frac{1}{3}}\big) m_{\frac{1}{3}}^3 \big(m_0^2+3 m_{\frac{1}{3}} m_0+9 m_{\frac{1}{3}}^2\big) \big(133 m_0^6+5292 m_{\frac{1}{3}}^3 m_0^3 \\
\hphantom{M_{4,\{0,6,6\}}=}{}-338256 m_{\frac{1}{3}}^6\big), \\
M_{4,\{1,1,1\}}=-\frac{5}{12} m_{\frac{1}{3}}^3 \big(17 m_0^6+54 m_{\frac{1}{3}}^3 m_0^3+5832 m_{\frac{1}{3}}^6\big), \\
M_{4,\{1,6,6\}}=-\frac{1}{288} m_{\frac{1}{3}}^3 \big(3 m_{\frac{1}{3}}-m_0\big) \big(m_0^2+3 m_{\frac{1}{3}} m_0+9 m_{\frac{1}{3}}^2\big) \big(61 m_0^9-36936 m_{\frac{1}{3}}^3 m_0^6 \\
\hphantom{M_{4,\{1,6,6\}}=}{}+1014768 m_{\frac{1}{3}}^6 m_0^3+6298560 m_{\frac{1}{3}}^9\big), \\
M_{5,\{2\}}=-12 m_{\frac{1}{6}} m_{\frac{2}{3}} (m_0-12 m_1){}^3 (m_0-4 m_1) (m_0-3 m_1) \big(m_0^2+12 m_1 m_0-72 m_1^2\big), \\
M_{5,\{6,6\}}=\frac{1}{48} m_0 m_{\frac{1}{6}} m_{\frac{2}{3}} (m_0-12 m_1){}^3 (m_0-4 m_1) (m_0-3 m_1) \big(7 m_0^5-780 m_1 m_0^4 \\
\hphantom{M_{5,\{6,6\}}=}{}+11160 m_1^2 m_0^3-74304 m_1^3 m_0^2+228096 m_1^4 m_0-248832 m_1^5\big), \\
M_{6,\{0\}}=27 m_0 m_{\frac{1}{3}}^2 , \\
M_{6,\{1\}}=-\frac{3}{4} m_{\frac{1}{3}}^2 \big(5 m_0^3+108 m_{\frac{1}{3}}^3\big), \\
M_{6,\{6\}}=m_{\frac{1}{3}}^2 \big(3 m_{\frac{1}{3}}-m_0\big){}^2 \big(m_0^2+3 m_{\frac{1}{3}} m_0+9 m_{\frac{1}{3}}^2\big){}^2.
\end{gather*}

\LastPageEnding

\end{document}